# The CMS High Level Trigger

**The CMS Trigger and Data Acquisition Group**[1]

**Abstract.** At the Large Hadron Collider at CERN the proton bunches cross at a rate of 40MHz. At the Compact Muon Solenoid experiment the original collision rate is reduced by a factor of $\mathcal{O}$ (1000) using a Level-1 hardware trigger. A subsequent factor of $\mathcal{O}(1000)$ data reduction is obtained by a software-implemented High Level Trigger (HLT) selection that is executed on a multi-processor farm. In this review we present in detail prototype CMS HLT physics selection algorithms, expected trigger rates and trigger performance in terms of both physics efficiency and timing.

(page 2) The participants in the CMS Trigger and Data Acquisition Group and the contributors to the High Level Trigger by Country and Institute. Correspondance to M. Spiropulu, CERN-PH, 1211 Geneva 23, Switzerland; smaria@cern.ch

# The CMS Trigger and Data Acquisition Group [2]

**Institut für Hochenergiephysik der OeAW, Wien, AUSTRIA**
W. Adam, T. Bergauer, C. Deldicque, J. Erö, R. Fruehwirth, M. Jeitler, K. Kastner, S. Kostner, N. Neumeister[**1a,2], M. Padrta P. Porth, H. Rohringer, H. Sakulin[**1b], J. Strauss, A. Taurok, G. Walzel, C.-E. Wulz

**Vrije Universiteit Brussel, Brussel, BELGIUM**
S. Lowette, B. Van De Vyver[**1a]

**Université Libre de Bruxelles, Bruxelles, BELGIUM**
G. De Lentdecker, P. Vanlaer

**Université Catholique de Louvain, Louvain-la-Neuve, BELGIUM**
C. Delaere, V. Lemaitre, A. Ninane, O. Van der Aa

**Institute for Nuclear Research and Nuclear Energy, Sofia, BULGARIA**
J. Damgov

**Helsinki Institute of Physics, Helsinki, FINLAND**
V. Karimäki, R. Kinnunen, T. Lampén, K. Lassila-Perini, S. Lehti, J. Nysten, J. Tuominiemi

**Laboratoire Leprince-Ringuet, Ecole Polytechnique, IN2P3-CNRS, Palaiseau, FRANCE**
P. Busson

**Institut de Recherches Subatomiques, IN2P3-CNRS - ULP, LEPSI Strasbourg, UHA Mulhouse, Strasbourg, FRANCE**
T. Todorov[**1b]

**RWTH, I. Physikalisches Institut, Aachen, GERMANY**
G. Schwering

**Institut für Experimentelle Kernphysik, Karlsruhe, GERMANY**
P. Gras [**3]

**University of Athens, Athens, GREECE**
G. Daskalakis[**4], A. Sfyrla

**Institute of Nuclear Physics “Demokritos”, Attiki, GREECE**
M. Barone, T. Geralis, C. Markou, K. Zachariadou

**KFKI Research Institute for Particle and Nuclear Physics, Budapest, HUNGARY**
P. Hidas

**Tata Institute of Fundamental Research - EHEP, Mumbai, INDIA**
S. Banerjee[**1c], K. Mazumdar[**1a]

**Tata Institute of Fundamental Research - HECR, Mumbai, INDIA**
S. Banerjee

**Università di Bari, Politecnico di Bari e Sezione dell’ INFN, Bari, ITALY**
M. Abbrescia, A. Colaleo[**1a], N. D’Amato, N. De Filippis, D. Giordano, F. Loddo, M. Maggi, L. Silvestris, G. Zito

---
[2] CERN/LHCC 02-26 p.465

**Università di Bologna e Sezione dell' INFN, Bologna, ITALY**
S. Arcelli, D. Bonacorsi, P. Capiluppi, G.M. Dallavalle, A. Fanfani, C. Grandi, S. Marcellini, A. Montanari, F. Odorici, R. Travaglini

**Università di Catania e Sezione dell' INFN, Catania, ITALY**
S. Costa, A. Tricomi

**Università di Firenze e Sezione dell' INFN, Firenze, ITALY**
V. Ciulli, N. Magini, R. Ranieri

**Laboratori Nazionali di Legnaro dell' INFN, Legnaro, ITALY (associated institute)**
L. Berti, M. Biasotto, M. Gulmini[**1a], G. Maron, N. Toniolo, L. Zangrando

**Università di Padova e Sezione dell' INFN, Padova, ITALY**
M. Bellato, U. Gasparini, S. Lacaprara, A. Parenti, M Passaseo P. Ronchese, S. Vanini, S. Ventura P.L. Zotto

**Università di Perugia e Sezione dell' INFN, Perugia, ITALY**
D. Benedetti, M Biasini, L. Fanò, L. Servoli

**Università di Pisa, Scuola Normale Superiore e Sezione dell' INFN, Pisa, ITALY**
G. Bagliesi, T. Boccali, S. Dutta, S. Gennai, A. Giassi, F. Palla, G. Segneri, A. Starodumov[**5,6], R. Tenchini

**Università di Roma I e Sezione dell' INFN, Roma, ITALY**
P. Meridiani, G. Organtini

**Università di Torino e Sezione dell' INFN, Torino, ITALY**
N. Amapane, F. Bertolino, R. Cirio

**Chonnam National University, Kwangju, KOREA**
J.Y. Kim I.T. Lim

**Dongshin University, Naju, KOREA**
M.Y. Pac

**Seoul National University, Seoul, KOREA**
K.K. Joo, S.B. Kim

**Sungkyunkwan University, Suwon, KOREA**
Y.I. Choi, I.T. Yu

**Kyungpook National University, Taegu, KOREA**
K. Cho, J. Chung,S.W. Ham, D.H. Kim, G.N. Kim, W. Kim, J.C Kim, S.K. Oh, H. Park, S.R. Ro, D.C. Son, J.S. Suh

**National Centre for Physics, Quaid-I-Azam University, Islamabad, PAKISTAN**
Z. Aftab, H. Hoorani, A. Osman[**1a]

**Institute of Experimental Physics, Warsaw, POLAND**
K. Bunkowski, M. Cwiok, W. Dominik, K. Doroba, M. Kazana, J. Krolikowski, I. Kudla, M. Pietrusinski, K. Pozniak[**7], W. Zabolotny[**7], J. Zalipska, P. Zych

**Soltan Institute for Nuclear Studies, Warsaw, POLAND**
L. Goscilo, M. Górski, G. Wrochna, P. Zalewski

**Laboratório de Instrumentação e Física Experimental de Partículas, Lisboa, PORTUGAL**

R. Alemany-Fernandez, C. Almeida, N. Almeida, J. C. Da Silva, M. Santos, I. Teixeira, J.P. Teixeira, J. Varela[**1a], N. Vaz Cardoso

**Joint Institute for Nuclear Research, Dubna, RUSSIA**

V. Konoplyanikov, A. Urkinbaev

**Institute for Nuclear Research, Moscow, RUSSIA**

A. Toropin [**8]

**Institute for Theoretical and Experimental Physics, Moscow, RUSSIA**

V. Gavrilov, V. Kolosov, A. Krokhotin, A. Oulianov, N. Stepanov

**Moscow State University, Moscow, RUSSIA**

O.L. Kodolova[**1a], I. Vardanyan

**Vinca Institute of Nuclear Sciences, Belgrade, SERBIA**

J. Ilic, G. Skoro

**Universidad Autónoma de Madrid, Madrid, SPAIN**

C. Albajar, J.F. de Trocóniz

**Instituto de Física de Cantabria (IFCA), CSIC-Universidad de Cantabria, Santander, SPAIN**

A. Calderon, M.A. Lopez Virto, R. Marco, C. Martinez Rivero, F. Matorras, I. Vila

**Universitt Basel, Basel, SWITZERLAND**

S. Cucciarelli[**1b], M. Konecki

**CERN, European Organization for Nuclear Research, Geneva, SWITZERLAND**

S. Ashby, D. Barney, P. Bartalini[**9], R. Benetta, V. Brigljevic[**10], G. Bruno[**11], E. Cano, S. Cittolin, M. Della Negra, A. De Roeck, P. Favre, A. Frey, W. Funk, D. Futyan, D. Gigi, F. Glege, J. Gutleber, M. Hansen, V. Innocente, C. Jacobs, W. Jank, M. Kozlovszky, H. Larsen, M. Lenzi, I. Magrans, M. Mannelli, F. Meijers, E. Meschi, L. Mirabito, S.J. Murray, A. Oh, L. Orsini, C. Palomares Espiga, L. Pollet, A. Racz, S. Reynaud, D. Samyn, P. Scharff-Hansen, C. Schwick, G. Sguazzoni, N. Sinanis, P. Sphicas[**12], M. Spiropulu, A. Strandlie, B.G. Taylor, I. Van Vulpen, J.P. Wellisch, M. Winkler

**Paul Scherrer Institut, Villigen, SWITZERLAND**

D. Kotlinski

**Universität Zürich, Zürich, SWITZERLAND**

K. Prokofiev, T. Speer

**Cukurova University, Adana, TURKEY**

I. Dumanoglu

**University of Bristol, Bristol, UNITED KINGDOM**

D.S. Bailey, J.J. Brooke, D. Cussans, G.P. Heath, D. Machin, S.J. Nash, D.M. Newbold, M.G. Probert

**Rutherford Appleton Laboratory, Didcot, UNITED KINGDOM**

J.A. Coughlan, R. Halsall, W.J. Haynes, I.R. Tomalin

**Imperial College, University of London, London, UNITED KINGDOM**
N. Marinelli[**13], A. Nikitenko[**6], S. Rutherford, C. Seez [**1a]
**Brunel University, Uxbridge, UNITED KINGDOM**
O. Sharif
**Boston University, Boston, Massachusetts, USA**
G. Antchev[**14], E. Hazen, J. Rohlf, S. Wu
**University of California, Davis, Davis, California, USA**
R. Breedon, P.T. Cox, P. Murray, M. Tripathi
**University of California, Los Angeles, Los Angeles, California, USA**
R. Cousins, S. Erhan, J. Hauser, P. Kreuzer[**13], M. Lindgren, J. Mumford, P. Schlein, Y. Shi, B. Tannenbaum, V. Valuev, M. Von Der Mey[**15]
**University of California, Riverside, Riverside, California, USA**
I. Andreeva[**1a], R. Clare, S. Villa
**University of California, San Diego, La Jolla, California, USA**
S. Bhattacharya, J.G. Branson, I. Fisk, J. Letts, M. Mojaver, H.P. Paar, E. Trepagnier
**California Institute of Technology, Pasadena, California, USA**
V. Litvine, S. Shevchenko, S. Singh, R. Wilkinson
**Fermi National Accelerator Laboratory, Batavia, Illinois, USA**
S. Aziz, M. Bowden, J.E. Elias, G. Graham, D. Green, M. Litmaath, S. Los, V. O'Dell, N. Ratnikova, I. Suzuki, H. Wenzel
**University of Florida, Gainesville, Florida, USA**
D. Acosta, D. Bourilkov[**24], A. Korytov, A. Madorsky, G. Mitselmakher, J.L. Rodriguez, B. Scurlock
**University of Maryland, College Park, Maryland, USA**
S. Abdullin[**6,15], D. Baden, S.C. Eno, T. Grassi, S. Kunori
**Massachusetts Institute of Technology, Cambridge, Massachusetts, USA**
S. Pavlon, K. Sumorok, S. Tether
**University of Mississippi, University, Mississippi, USA**
L.M. Cremaldi, D. Sanders, D. Summers
**Northeastern University, Boston, Massachusetts, USA**
I. Osborne, L. Taylor, L. Tuura
**Princeton University, Princeton, New Jersey, USA**
W.C. Fisher[**15], J. Mans[**16], D. Stickland, C. Tully, T. Wildish, S. Wynhoff
**Rice University, Houston, Texas, USA**
B.P. Padley
**University of Wisconsin, Madison, Wisconsin, USA**
P. Chumney, S. Dasu, W.H. Smith

**1a:Also at CERN, European Organization for Nuclear Research, Geneva, SWITZERLAND

**1b:Now at CERN, European Organization for Nuclear Research, Geneva, SWITZERLAND

**1c:Now also at CERN, European Organization for Nuclear Research, Geneva, SWITZERLAND

**2:Now at Purdue University, West Lafayette, USA

**3:Now at DAPNIA, Centre d'Etudes de Saclay (CEA-Saclay), FRANCE

**4:Now at Imperial College, University of London, London, UNITED KINGDOM

**5:Now at Institutfür Teilchenphysik, Eidgenössische Technische Hochschule (ETH), Zürich, SWITZERLAND

**6:Also at Institute for Theoretical and Experimental Physics, Moscow, RUSSIA

**7:Also at Institute of Electronic Systems, Technical University of Warsaw, POLAND

**8:Also at Università di Pisa, Scuola Normale Superiore e Sezione dell' INFN, Pisa, ITALY

**9:Now at University of Florida, Gainesville, Florida, USA

**10:Now also at Institute Rudjer Boskovic, Zagreb, CROATIA

**11:Now at Université Catholique de Louvain, Louvain-la-Neuve, BELGIUM

**12:Also at MIT, Cambridge, USA and University of Athens, Athens, GREECE

**13:Now at University of Athens, Athens, GREECE

**14:Also at Institute for Nuclear Research and Nuclear Energy, Sofia, BULGARIA

**15:Now at Fermi National Accelerator Laboratory, Batavia, Illinois, USA

**16:Now at University of Minnesota, Minneapolis, Minnesota, USA

## 1. Introduction

The Large Hadron Collider (LHC) [1], is a hadron-hadron collider to be installed in the Large Electron Positron (LEP) tunnel at the CERN Laboratory (the European Laboratory for Particle Physics outside Geneva, Switzerland). It will be a unique tool for fundamental physics research and the highest energy accelerator in the world for many years following its completion. The LHC will provide two proton beams, circulating in opposite directions, at an energy of 7 TeV each (center-of-mass $\sqrt{s} = 14$ TeV). These beams upon collision will produce an event rate about 1,000 times higher than that presently achieved at the Tevatron $p\bar{p}$ collider [2]. In order to support the 7 TeV proton beams, in total 1104 8.4 Tesla superconducting dipoles and 736 quadrupoles will be installed in the underground tunnel of 26.6 km circumference formerly used by LEP.
The physics potential of the LHC is unprecedented: it will allow to study directly and in detail the TeV scale region. The LHC is expected to elucidate the electroweak symmetry breaking mechanism (EWSB) and provide evidence of physics beyond the standard model [3]. The LHC will be also a standard model precision measurements instrument [4] mainly due to the very high event rates as shown in table 1.

**Table 1.** Approximate event rates of some physics processes at the LHC for a luminosity of $\mathcal{L} = 2 \times 10^{33}\,\mathrm{cm}^{-2}\mathrm{s}^{-1}$. For this table, one year is equivalent to 20 fb$^{-1}$.

| Process | Events/s | Events/year |
|---|---|---|
| $W \rightarrow e\nu$ | 40 | $4 \cdot 10^8$ |
| $Z \rightarrow ee$ | 4 | $4 \cdot 10^7$ |
| $t\bar{t}$ | 1.6 | $1.6 \cdot 10^7$ |
| $b\bar{b}$ | $10^6$ | $10^{13}$ |
| $\tilde{g}\tilde{g}$ (m = 1 TeV) | 0.002 | $2 \cdot 10^4$ |
| Higgs (m= 120 GeV) | 0.08 | $8 \cdot 10^5$ |
| Higgs (m= 800 GeV) | 0.001 | $10^4$ |
| QCD jets $p_T >$ 200 GeV | $10^2$ | $10^9$ |

The proton beams cross at interaction points along the ring where detectors that measure the particles produced in the collisions are installed. Interaction "Point 5" hosts the multiple purpose $4\pi$ coverage CMS detector, shown in figure 1.
The CMS detector measures roughly 22 meters in length, 15 meters in diameter, and 12,500 metric tons in weight. Its central feature is a huge, high field (4 Tesla) solenoid, 13 meters in length, and 6 meters in diameter. Its "compact" design is large enough to contain the electromagnetic and hadron calorimetry surrounding a tracking system, and allows a superb muon detection system. All subsystems of CMS are bound by means of the data acquisition and trigger system.
In the CMS coordinate system the origin coincides with the nominal collision point at the geometrical center of the detector. The $z$ direction is given by the beam axis. The rest frame of the hard collision is generally boosted relative to the lab frame along the beam direction, $\theta$ is the polar angle with respect to the $z$ axis and $\phi$ the azimuthal angle with respect to the LHC plane. The detector solid angle segmentation is designed to be invariant under boosts along the $z$ direction. The *pseudorapidity* $\eta$, is related to the polar angle $\theta$ and defined as $\eta \equiv -\ln(\tan(\theta/2))$. The transverse momentum component $z$-axis is given by $p_T = p\sin\theta$ and similarly $E_T = E\sin\theta$ is the transverse energy of a physics object.

The experiment comprises a tracker, a central calorimeter barrel part for $|\eta| \leq 1.5$, and endcaps on both sides, and muon detectors. The tracking system is made of several layers of silicon pixel and silicon strip detectors and covers the region $|\eta| < 2.5$. The electromagnetic calorimeter consists of lead tungstate ($PbWO_4$) crystals covering $|\eta| < 3$ (with trigger coverage $|\eta| < 2.6$). Its resolution at the initial luminosity ($\mathcal{L} = 2 \times 10^{33}\,\mathrm{cm}^{-2}\mathrm{s}^{-1}$) is $\Delta E/E = 3\%/\sqrt{E} \oplus 0.5\%$. The surrounding hadronic calorimeter uses brass/scintillator tiles in the barrel and endcaps. Its resolution for jets, when combined with the electromagnetic calorimeter, is $\Delta E/E = 110\%/\sqrt{E} \oplus 5\%$. The region $3 < |\eta| < 5$ is covered by forward calorimeters with a resolution of $\Delta E/E = 180\%/\sqrt{E} \oplus 10\%$. Muons are measured in gas chambers in the iron return yoke. The muon momentum measurement using the muon chambers and the central tracker covers the range $|\eta| < 2.4$ with a resolution of $\Delta p_T/p_T = 5\%$ at $p_T = 1$ TeV and $\Delta p_T/p_T = 1\%$ at $p_T = 100$ GeV. The muon trigger extends over the pseudorapidity range $|\eta| < 2.1$.

The total non-diffractive inelastic cross section at the LHC is expected to be $\sigma_{\mathrm{T}} = 55\,\mathrm{mb}$ [5, 6]. The LHC will operate at a bunch crossing rate of 40 MHz. Only 80 % of the bunches will be filled [1], resulting in an effective bunch crossing rate of 32 MHz. The instantaneous luminosity in the first two years after start–up is expected to be $\mathcal{L} = 2 \times 10^{33}\,\mathrm{cm}^{-2}\mathrm{s}^{-1}$ and subsequently upgraded to $\mathcal{L} = 10^{34}\,\mathrm{cm}^{-2}\mathrm{s}^{-1}$ in a second phase. In the following these two values are referred to as “low luminosity” and “high luminosity”, respectively. The average number of inelastic non-diffractive interactions per bunch crossing $\mu$ is $\mu = 17.3$ at high and $\mu = 3.5$ at low luminosity.

In total CMS has $\sim 10^8$ data channels that are checked each bunch crossing. The design data-size per event is about 1 MB. At start-up it is essential to allow for a larger event size, up to 1.5 MB per event, in order to be able to thoroughly study and understand the detector performance.

One of the most important and difficult aspects of the experiment is the design of the trigger, the real-time selection and recording of the (judged) useful events. Minimum deadtime and maximum physics retention are required of the data acquisition (DAQ), trigger and data recording architecture. To achieve this, sophisticated and fast selection algorithms are implemented that use selective information from all the detector subsystems during real-time data taking. The rate of interesting physics processes that need be recorded compared to the inclusive $pp$ cross section is many orders of magnitude smaller.

The CMS trigger is designed to perform a data reduction from 32 MHz down to $\mathcal{O}(100)$

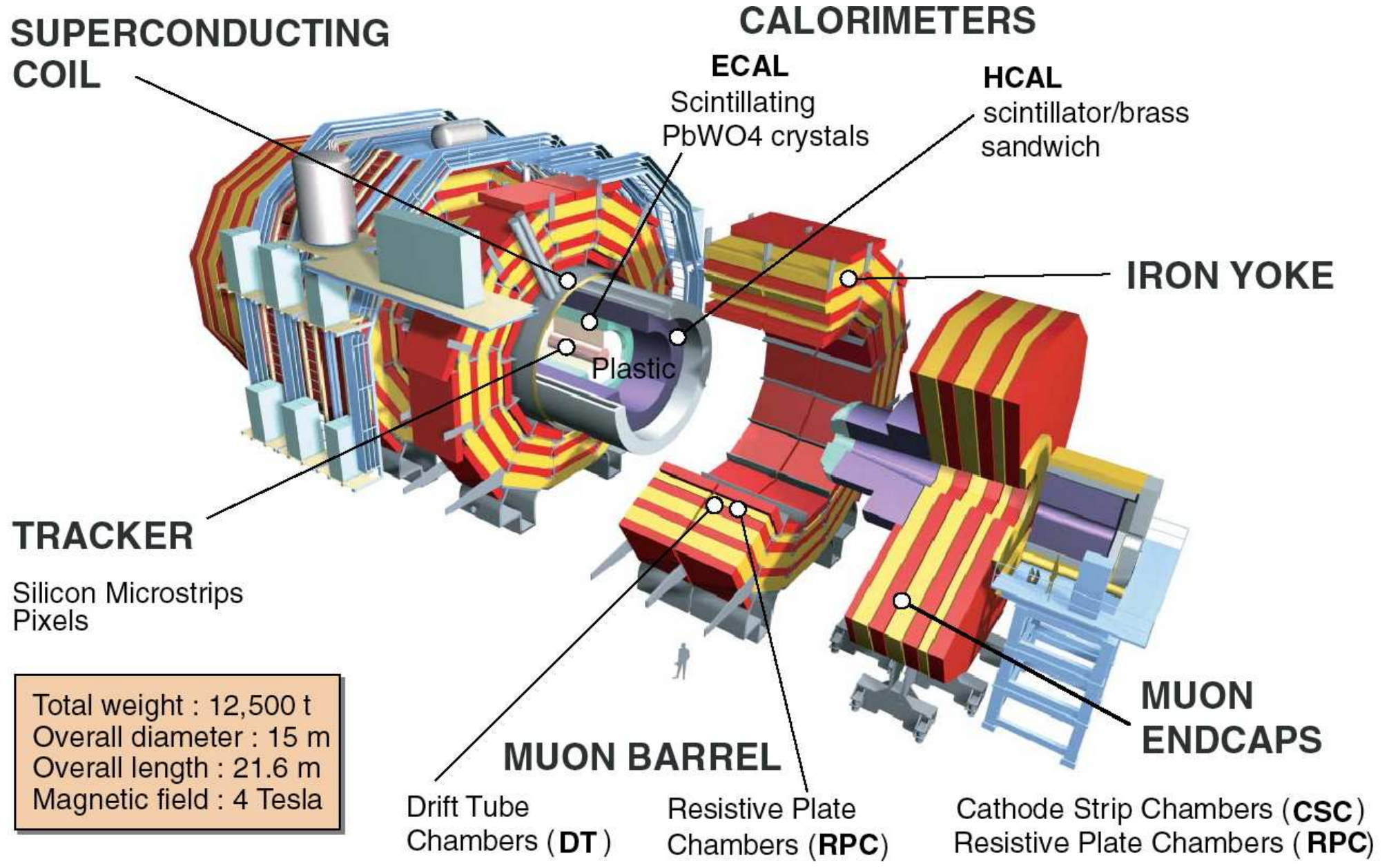

**Figure 1.** Three dimensional view of the CMS detector, and it's detector components.

Hz via different sequential trigger levels. The first trigger level of CMS,Level-1, is hardware implemented and reduces the data rate, by using specific low level analysis in custom trigger processors. All further levels are software filters which are executed on (partial) event data in a processor farm. This is the upper level of real-time data selection and is referred to as High-Level Trigger or HLT. Only data accepted by the HLT are recorded for offline physics analysis. Additionally, small samples of the rejected data are retained for monitoring the performance of the HLT while the bulk of it is discarded and dropped from any further processing without compromising physics efficiency.

This paper details how CMS proposes to achieve the necessary reduction factor, while keeping an uncompromising physics efficiency. The HLT algorithms are described in detail, typical threshold choices are given and the physics performance for several benchmark processes is demonstrated. The results reported here demonstrate that CMS has developed an efficient and agile trigger structure which will select the physics signals of interest in the harsh LHC conditions.

The report is organized as follows. Section 2 describes the CMS trigger, section 3 the electron/photon triggers and section 4 the muon triggers. The jet and missing $E_T$ related triggers are discussed in sections 5 and 6. Sections 7 and 8 discuss the $\tau$ and other heavy flavour triggers respectively. In section 9 we present the initial ideas on calibration triggers, while section 10 studies benchmark physics channels and discusses a prototype trigger table for the HLT. The Appendix describes the Level-1 trigger algorithms.

## 2. The CMS Trigger

### *2.1. Physics Requirements*

The main physics requirements on the L1 trigger and the HLT are:

- foremost, the requirement that the selection fulfill the needs of the CMS physics program; the efficiency for the physics objects must be as high as possible;
- the selection must be as inclusive as possible. The LHC represents a new energy frontier, and unexpected new phenomena may appear. The selection requirements must retain events of potential use in such exotic searches;
- all thresholds and other requirements applied by the selection should be very robust, so that detailed knowledge of calibration constants and other run conditions do not pose a stringent constraint in real time;
- the final selection of events should include data samples for the calculation of all trigger and reconstruction efficiencies offline;
- the rate of events accepted by the HLT should be within limits allowed by the data recording technology ($\mathcal{O}(100)$ Hz);
- all algorithms and their implementation should be monitored continuously for their correct functioning;
- the events selected by the HLT should be tagged to indicate the reasons for their selection, to aid the offline reconstruction;
- the HLT should include all major improvements in the offline reconstruction, and thus it should be as close as possible to the standard offline reconstruction code.

### *2.2. Selection Strategy and Reconstruction on Demand*

The CMS DAQ/HLT processes all events accepted by the Level-1 trigger in a single processor farm. There is therefore no separate Level-2 or Level-3, but a single entity, the High-Level Trigger. Nevertheless, as in a traditional multi-level trigger system, the selection of events can be optimized by rejecting events as quickly as possible. The basic event building strategy is to reconstruct those parts of each physics object that can be used for selection while minimizing the overall CPU usage.

As an example, reconstruction of an electron includes the reconstruction of a cluster in the electromagnetic calorimeter, the matching of hits in the pixel detector and the subsequent reconstruction of a full charged particle track in the tracker. At the end of each step a set of selection criteria results in the rejection of a significant fraction of the events accepted by the previous step. The rate of events that need to be processed through the remaining algorithms is decreased reducing the required CPU. Reconstruction and selection are therefore closely intertwined in the online environment of the filter farm. For an optimal system the HLT should reconstruct the minimum amount of detector information needed to apply a set of selection criteria that reject background events while keeping the desired physics events for further processing.

It is important to note that a fraction of the bandwidth, between 10-30%, (closer to 30% at start-up) needs to be reserved for calibration triggers to ensure a complete and accurate understanding of the detector performance.

### *2.3. Trigger Levels – Definitions*

The reconstruction and selection in the HLT takes place in steps which correspond roughly to what would have been distinct trigger systems, the Level-2 and Level-3 trigger systems. It is thus convenient to use the terminology, and to refer to a "Level-2 trigger" or a "Level-3 step" to describe the selection algorithms and criteria of the HLT. As mentioned previously the CMS HLT architecture does not include a sharp division between these trigger steps, other than the order in which they are applied.

In what follows, the convention used is that "Level-2" triggers, algorithms and requirements refer to the first selection step in the HLT process. Typically, a Level-2 trigger, which has the maximum rate of events input to it, uses only information from the calorimeter and muon detectors. In contrast, "Level-3" refers to selection that includes the reconstruction of full tracks in the tracker. Traditionally, because of the high number of channels, the complex pattern recognition and higher combinatorics, track reconstruction is a process that demands large amounts of CPU time. Extending the terminology, in what follows there are references to "Level-2.5" triggers, which refer to algorithms that use partial tracker information, e.g. pixel hits, for a fast confirmation of the electron candidate. The numbering, "2.5", attempts to indicate the intermediate nature of the selection, as one that occurs between the selection that is based solely on the calorimeter information, and the selection that is based on the full CMS detector information.

### *2.4. Partial Event Reconstruction*

To minimize the CPU required by the HLT, a key feature of the algorithms is to reconstruct the information in the CMS detector only partially. In many cases the decision on whether an event should be accepted by the HLT involves the reconstruction of quantities in only a limited region of the detector. As an example, for an event accepted by the Level-1 trigger in the inclusive muon stream, only the parts of the muon chambers indicated by the Level-1 trigger results, and the corresponding road in the tracker, need be considered for the validation of the muon.

The idea of partial event reconstruction has been embedded in the CMS reconstruction code from the very beginning. In the case of the HLT, the reconstruction of physics objects is driven by the corresponding candidates identified by the Level-1 trigger. All reconstruction starts from the Level-1 trigger information. This approach leads to significant CPU savings, however it also leads to rejecting events that contain "volunteer" objects, i.e. objects that did not pass the Level-1 trigger. This disadvantage is mitigated by the fact that it is, in general, very difficult to understand the properties of such objects in later offline analyses. It is clearly always feasible, albeit with an increased CPU cost, to look for additional objects in each event if a particular physics analysis introduces such a requirement.

### *2.5. Level-1 Trigger Settings and Rates*

The performance of the Level-1 trigger system is summarized in the Appendix A and further details of the system can be found in the Level-1 Trigger Technical Design Report (TDR) [7]. Some small changes in expected rates and efficiencies with respect to that document have been incorporated in this report. The changes reflect the Monte Carlo input, including an updated version of the PYTHIA Monte Carlo simulation and increased statistics in the Monte Carlo samples generated, as well as a more detailed and tuned detector and trigger simulation. The main thrust of the conclusions of reference [7] remains unaltered: the overall performance of the system satisfies the requirements of the CMS physics program. An additional change with respect to the trigger TDR is the definition of the low luminosity run conditions: "low" luminosity in this document refers to an instantaneous luminosity of $\mathcal{L} = 2 \times 10^{33}\,\mathrm{cm}^{-2}\mathrm{s}^{-1}$(twice the definition of low luminosity at the TDR) while "high" luminosity refers to the nominal LHC design value of $\mathcal{L} = 10^{34}\,\mathrm{cm}^{-2}\mathrm{s}^{-1}$. The current plans of CMS call for a phased installation of the Data Acquisition system so that at start-up the DAQ can handle a Level-1 accept event rate of up to 50 kHz and at the full design luminosity it can handle a rate of 100 kHz.

The allocation of the Level-1 trigger bandwidth must be optimized in order to ensure the widest possible physics reach for the experiment, while including all technical triggers intended for calibration and monitoring of the detector. The procedure used for this optimization starts with a determination of the maximum Level-1 trigger rate that can be allocated, which is 16 kHz at low luminosity and 33 kHz at high luminosity. These maximum rates arise from dividing the total DAQ bandwidth, i.e. 50 kHz (100kHz) at low (high) luminosity, by a safety factor of three. This safety factor is used to account for all uncertainties in the simulation of the basic physics processes, the CMS detector, and the beam background conditions.

The second step is a first allocation of this Level-1 bandwidth across the different objects. An equal allocation across the four categories of "objects" taken as (a) electrons/photons, (b) muons, (c) tau-jets and (d) jets-combined channels, is made. For example, at low luminosity about4 kHz is pre-allocated to each of these triggers above.

The third step is to determine thresholds for the single and double-object triggers within this first Level-1 rate allocation. As an example, the 8 kHz of electron and photon triggers at high luminosity has to be divided among single and double electrons/photons. For this purpose the iso-rate contours on the plane of the $E_T$ cut on the single objects versus the $E_T$ cut on the double objects are examined, as shown in fig. 2 (left). An operating point in this single-double threshold plane must be selected for a given a Level-1 trigger allocation. To determine the optimal operating point, the efficiency for triggering on selected physics channels, via the single and the double-object trigger, is considered. In the case of the electron/photon trigger the efficiency for $W \to e\nu$ decays versus the efficiency for $Z \to ee$ decays can be used, as in figure 2 (right). The optimal point can be selected as a trade-off between the two efficiencies, e.g. adopting the point at which both efficiencies drop at the same rate, as shown in the same figure. The same method is used for muons and $\tau$s.

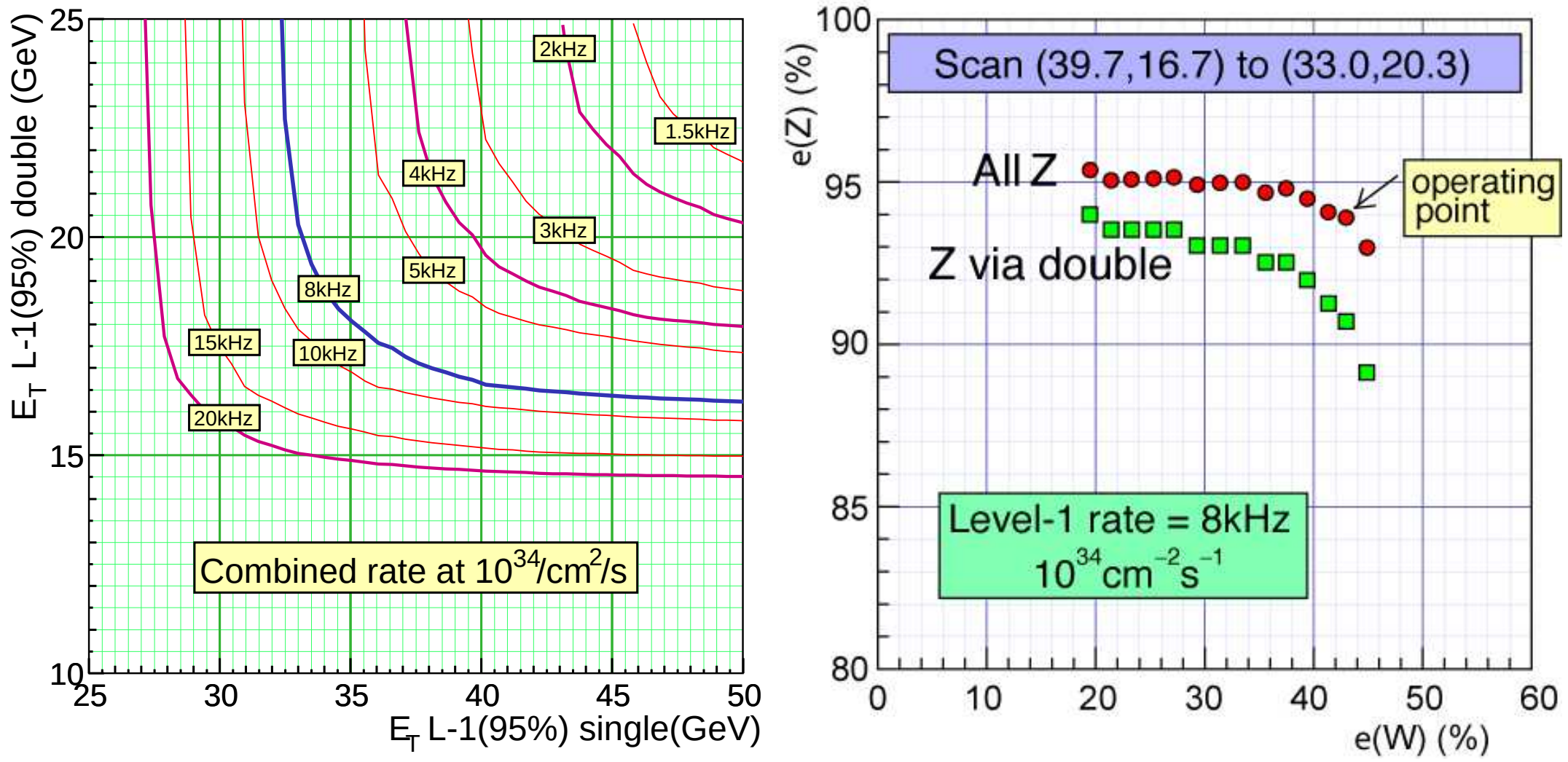

**Figure 2.** Contours of equal rate on the plane of $E_T$ thresholds for single and double-electron/photon triggers (left), and efficiency for $W$ and $Z$ electronic decays as a function of the same thresholds (right).

There is no similar optimization of the allocation for jet triggers since there is no reason to include a 2-jet trigger requirement. All QCD events have a second jet which, though by definition at a lower $E_T$ than the leading jet, is always present in the event.

The last step in this procedure is to evaluate the optimal efficiencies for each of the four object categories and to then re-iterate the entire procedure with slightly different Level-1 trigger allocations.

Note that the Level-1 triggers only consider cross-object triggers for a limited number of categories.

### *2.6. Level-1 Trigger Table*

With the constraints described previously, the prototype Level-1 trigger tables that are used in the present paper for further analysis of the HLT algorithms and performance are listed in tables 2 and 3 for low and high luminosity respectively. The thresholds quoted correspond to the $E_T$ or $p_T$ value at which the efficiency of the trigger is 95% of its maximum value. There is no entry for a $\mu$+jet trigger at low luminosity because there is little to be gained given the low inclusive muon threshold. In both cases, a 1 kHz bandwidth is allocated to minimum-bias events which will be used for calibration

**Table 2.** Level-1 trigger table at low luminosity. Thresholds correspond to values with 95% of the maximum available efficiency (except for the muon triggers which are at 90% efficiency)

| Trigger | Threshold (GeV or GeV/c) | Rate (kHz) | Cumulative Rate (kHz) |
|---|---|---|---|
| inclusive isolated electron/photon | 29 | 3.3 | 3.3 |
| di-electron/di-photon | 17 | 1.3 | 4.3 |
| inclusive isolated muon | 14 | 2.7 | 7.0 |
| di-muon | 3 | 0.9 | 7.9 |
| single $\tau$-jet | 86 | 2.2 | 10.1 |
| di- $\tau$ -jet | 59 | 1.0 | 10.9 |
| 1-jet, 3-jet, 4-jet | 177, 86, 70 | 3.0 | 12.5 |
| jet * $E_T^{miss}$ | 88 * 46 | 2.3 | 14.3 |
| electron * $\tau$-jet | 19 * 45 | 0.8 | 15.1 |
| minimum bias (calibration) | | 0.9 | 16.0 |
| TOTAL | | | 16.0 |

**Table 3.** Level-1 trigger table at high luminosity. Thresholds correspond to values with 95% of the maximum available efficiency (except for the muon triggers which are at 90% efficiency).

| Trigger | Threshold (GeV or GeV/c) | Rate (kHz) | Cumulative Rate (kHz) |
|---|---|---|---|
| inclusive isolated electron/photon | 34 | 6.5 | 6.5 |
| di-electron/di-photon | 19 | 3.3 | 9.4 |
| inclusive isolated muon | 20 | 6.2 | 15.6 |
| di-muons | 5 | 1.7 | 17.3 |
| single $\tau$-jet trigger | 101 | 5.3 | 22.6 |
| di- $\tau$-jets | 67 | 3.6 | 25.0 |
| 1-jet, 3-jets, 4-jets | 250, 110, 95 | 3.0 | 26.7 |
| jet * $E_T^{miss}$ | 113 * 70 | 4.5 | 30.4 |
| electron * $\tau$-jet | 25 * 52 | 1.3 | 31.7 |
| muon * $\tau$-jet | 15 * 40 | 0.8 | 32.5 |
| minimum bias (calibration) | | 1.0 | 33.5 |
| TOTAL | | | 33.5 |

and monitoring purposes.

### *2.7. Event Generation and Detector Simulation*

The studies of the HLT have been performed with simulated events, by tracking particles through a realistic simulation of the detector, adding pileup events (the additional 3.5(17.5) interactions at low(high) luminosity conditions), and applying the HLT algorithms to the digitized data. Standard Monte Carlo generators, such as PYTHIA [5] and ISAJET [8], are used to simulate the collisions between two protons at $\sqrt{s} = 14$ TeV.

Event samples have been generated for the muon, electron, $b$ and $\tau$ and jet signal samples as well as for the pileup of minimum bias collisions. In several cases special care has to be taken to ensure an efficient but correct sample generation. As an example the generation of the minimum bias background for HLT studies involving muons is treated somewhat differently than for studies without muons since every charged pion or kaon with a momentum above a few GeV/$c$ can potentially decay into a muon, penetrate into the muon system, and be mismeasured as a high-$p_T$ muon because of multiple scattering through the iron yoke. A complete simulation of this background would be CPU-wise prohibitive, hence a weighting procedure is used. Details are given in ref. [9].

A procedure has been developed to take into account the effects of the pileup in the muon system, as detailed in [10]. For the jet trigger sample a cross section weighting technique is applied as well [7]. Attention has also been paid for the pileup samples, to ensure that the limited statistics and re-use of events in these samples does not lead to overestimated trigger rates due to a few recurring events.

For the HLT studies presented in this paper, the CMS detector simulation package, CMSIM [11] is used. It is an application of the GEANT3 [12] detector description and simulation tool. CMSIM is used to describe the detector geometry and materials. It also includes and uses information about the magnetic field. CMSIM reads the individual generated events and simulates the effects of energy loss, multiple scattering and showering in the detector materials with GEANT3. The average storage size of a single event for these studies is about 2 MB. The CPU time required to simulate one event on a 1 GHz CPU ranges from 60 s for a minimum-bias event, to 500 s for a 1 TeV di-jet event.

## 3. Electron and Photon Identification

The HLT selection of electrons and photons proceeds in three steps. The first step (Level-2.0), uses the calorimeter information alone. The next step (Level-2.5) demands hits in the pixel detectors consistent with an electron candidate. Matching of energy in the electromagnetic calorimeter (ECAL) with hits in the pixel detector or failure to match, splits the electromagnetic triggers into two categories: electron candidates (single and double), and above significantly higher thresholds, photon candidates. In the final step (Level-3) the selection of electrons uses full track reconstruction, seeded from the pixel hits obtained at the matching step.

For the calorimeter reconstruction – clustering followed by energy and position measurement – the emphasis is predominantly on the reconstruction of electrons, because the transverse momentum thresholds for triggering on electrons are much lower than those for photons. Since the amount of material traversed in the tracker volume is almost one radiation length for certain polar angles, the first challenge for ECAL clustering is to include all energy radiated by electrons. Photons that have converted in the tracker material are adequately reconstructed by the electron algorithms. For the final higher precision reconstruction of photons it is foreseen that unconverted photons will be reconstructed using energy sums of fixed arrays of crystals and that tracker information will be used to assist the clustering for converted photons. Such algorithms that improve the photon energy resolution, are not necessary for the HLT selection.

### *3.1. Calorimeter Reconstruction: Clustering*

The first step in the reconstruction of an electron in the High-Level Trigger is the clustering of the energy deposits in the ECAL and the estimate of the electron's energy and position from this information. In the barrel section this involves the energy deposited in the lead tungstate crystals alone, in the endcap energy is also deposited in the $3X_0$ thick preshower detector.

Electrons radiate in the material between the interaction point and the ECAL. The bending of the electron in the 4T magnetic field results in an azimuthal spray of energy reaching the ECAL. The electron energy can be collected by forming a group of clusters along a $\phi$-road. This cluster of clusters is called a “super-cluster”.

*3.1.1. The Island Algorithm* The island algorithm starts by a search for seeds which are defined as crystals with an energy above a certain threshold. Starting from the seed position, adjacent crystals are examined, scanning first in $\phi$ and then in $\eta$. Along each scan line crystals are added until either a rise in energy or a crystal that has not been read out is encountered. The clustering starts from the highest $E_T$seed in the list of seeds, and the crystals included in the cluster are marked after being used so as to be only included in a single cluster

In much the same way as energy is clustered at the level of calorimeter cells, non-overlapping clusters can in turn be clustered into calorimetric “super-clusters”. The procedure is seeded by searching for the most energetic cluster and then collecting all the nearby clusters in a very narrow $\eta$-window and much wider $\phi$-window.

*3.1.2. The Hybrid Algorithm* For single showers, such as those produced by unconverted photons, or those produced by electrons in test beam conditions, energy sums of fixed arrays of crystals give a better energy resolution performance than energy sums of crystals collected dynamically according to a cluster or “bump” finding algorithm. The Hybrid algorithm uses the $\eta-\phi$ geometry of the barrel crystals to exploit the knowledge of the lateral shower shape in the $\eta$ direction (taking a fixed bar of three or five crystals in $\eta$), while searching dynamically for separated (bremsstrahlung) energy in the $\phi$ direction.

The Hybrid algorithm is used to reconstruct electrons with $p_T$>10 GeV/$c$ in the barrel. By contrast, when looking for small deposits of energy in individual clusters, for example when making a calorimetric isolation cut, the basic clusters of the Island algorithm are more appropriate objects to work with. Further details on the clustering algorithms can be found in reference [13].

### *3.2. Endcap Reconstruction with the Preshower*

Much of the endcap is covered by a preshower detector with two planes of silicon strip readout. The energy deposited in the preshower detector (which is about $3X_0$ thick) needs to be added to the crystal clusters [14]. The energy in the crystals is clustered using the Island algorithm and the clusters are associated to form super-clusters. A preshower cluster is constructed in each plane, in front of each crystal cluster of the super-cluster. The search area in the preshower is centered on the point determined by extrapolating the crystal cluster position to the preshower plane in the direction of the nominal vertex position.

### *3.3. Energy and Position Measurement*

*3.3.1. Position Measurement Using Log-weighting Technique* A simple measurement of the shower position can be obtained by calculating the energy-weighted mean position of the crystals in the cluster. Two features need to be addressed in more detail in order to obtain a precise position measurement.

The first is the precise definition of the "crystal position". The lateral position of the crystal depends upon depth because the crystals are "off-pointing" and the incident particle and shower direction is not exactly parallel to the crystal axis. The lateral position of the crystal is thus defined as the $(\eta, \phi)$ position of its axis at a particular depth. The depth at which the shower maximum occurs is taken as the longitudinal baricentre of the shower which has a logarithmic dependence on the shower energy. This depth is roughly the longitudinal center of gravity of the shower, and its optimal mean value varies logarithmically with the shower energy. There is also a dependence on particle type: electron showers have a maximum about one radiation length less deep than photon showers. In the position measurement used for both Island and Hybrid super-clusters the depth is measured from the front face of the crystals along the direction from the nominal vertex position to the approximate shower position calculated using the arithmetic energy weighted mean of the shower front face centers. The energy dependence is accounted for with a logarithmic parametrization [13].

The second feature that requires more detailed treatment is related to the lateral shower shape. Since the energy density does not fall away linearly with distance from the shower axis, but rather exponentially, a simple energy weighted mean of crystal energies is distorted and the measured position is biased towards the center of the crystal containing the largest energy deposit.

A simple algorithm, which yields adequate precision consists of using the weighted mean, calculated using the logarithm of the crystal energy:

$$x = \frac{\sum x_i \cdot W_i}{\sum W_i}$$

with the sum over all crystals of the cluster, and where $x_i$ is the position of crystal $i$, and $W_i$ is the log weight of the crystal – the logarithm of the fraction of the cluster energy contained in the crystal, calculated with the formula:

$$W_i = W_0 + \ln \frac{E_i}{\sum E_j}$$

where the weight is constrained to be positive, or is otherwise set to zero. $W_0$ then controls the smallest fractional energy that a crystal can have and still contribute to the position measurement[13].

So far what has been described refers to the measurement of the position of a single cluster. The position of a super-cluster is calculated by making the energy-weighted mean of the positions of its component clusters. For an electron that has radiated into the material, this method allows to reconstruct its position at production.

*3.3.2. Energy Measurement and Corrections* The measurement of energy in the crystals is obtained by simple addition of the deposits measured in the crystals – although more complex estimators have been proposed [15].

Even in the areas not covered by the preshower detector the energy containment of the clustered crystals is not complete. The reconstructed over the generator level energy distribution, $E_{meas}/E_{true}$, shows a peak at a few percent less than unity, and a long tail on the low side due to non-recovered bremsstrahlung energy. The Gaussian part of the distribution corresponds, roughly, to the energy that would be reconstructed from an electron in the absence of bremsstrahlung. The amount of tracker material varies strongly with $\eta$, as shown in figure 3, and thus so does the amount of bremsstrahlung radiation, so a variation in the fraction of events in the tail as a function of $\eta$ is expected. This inevitably leads to a small variation in the peak position as a function of $\eta$.

The energy scale is "calibrated" using corrections designed to place the peak in $E_{meas}/E_{true}$ at 1.0, see figure 4. The corrections are parametrized in terms of the number of crystals in the cluster ($f(N_{cry})$ corrections). This helps to minimize the residual dependence on both $E$ and $\eta$ of the energy scale. Figure 5 shows, as an example, $E_{meas}/E_{true}$ as a function of the number of crystals in a reconstructed Hybrid super-cluster, for electrons with $10 < p_T < 50$ GeV/$c$, together with a fitted polynomial function.

*3.3.3. Energy and Position Measurement Performance* For these performance figures an isolated Level-1 trigger (Appendix A.2) is requested, which results in about 8% inefficiency mainly due to bremsstrahlung radiation in the isolation region (about 6%), but no trigger threshold is applied.

Figure 4 shows the distribution of $E_{meas}/E_{true}$ for $p_T = 35$ GeV/$c$ electrons reconstructed using the Hybrid algorithm in the barrel, and using the Island algorithm and the preshower in the endcap. The energy resolution is parametrized in two ways: in terms of the fitted width of the Gaussian part of the distribution (fitted between -1.5$\sigma$ and +2.0$\sigma$), and in terms of $\sigma_{eff}$, defined as the half-width containing 68.3% of the

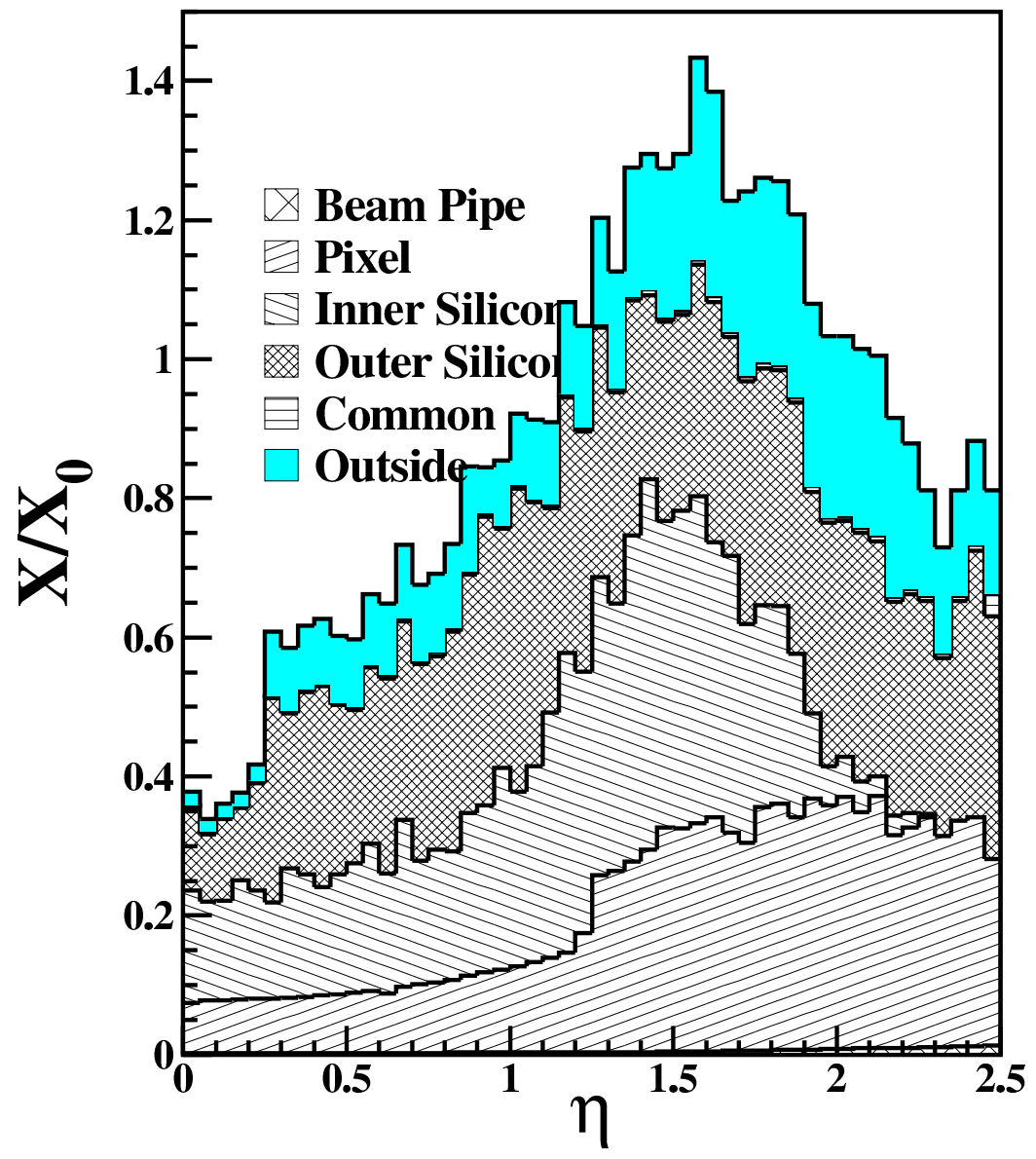

**Figure 3.** Material budget as a function of $\eta$ for the different tracker sub-units.

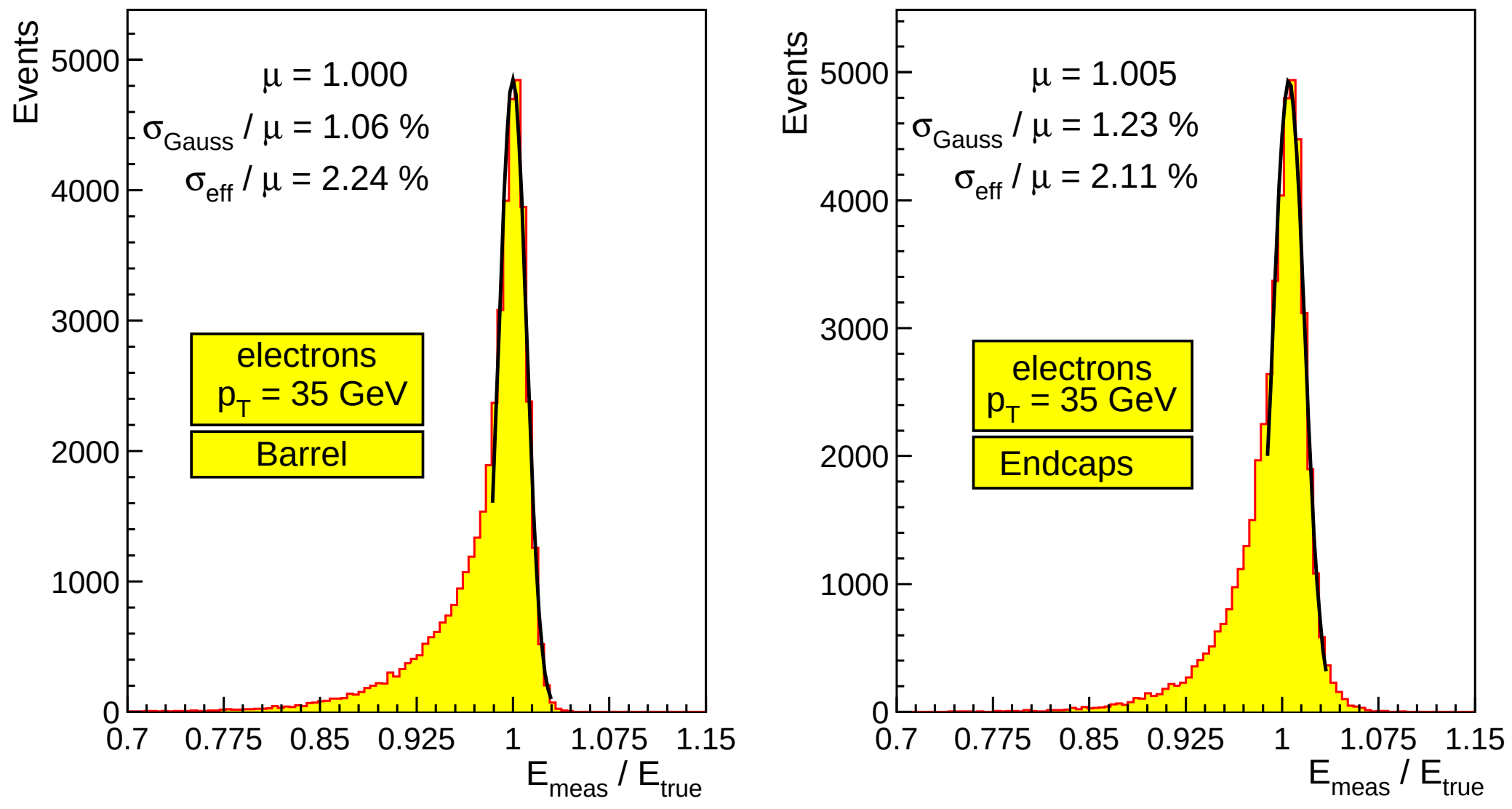

**Figure 4.** Distribution of $E_{meas}/E_{true}$ for $p_T$= 35 GeV/$c$ electrons, a) in the barrel ECAL fully digitized without pileup, and reconstructed with the Hybrid super-clustering algorithm, b) the same distribution for electrons in the endcap, reconstructed with the Island super-clustering algorithm, and with preshower energy included. $E_{meas}/E_{true}$. The variable $\mu$ is the mean value of the Gaussian fit, shown by the curve; for $\sigma_{eff}$, see text.

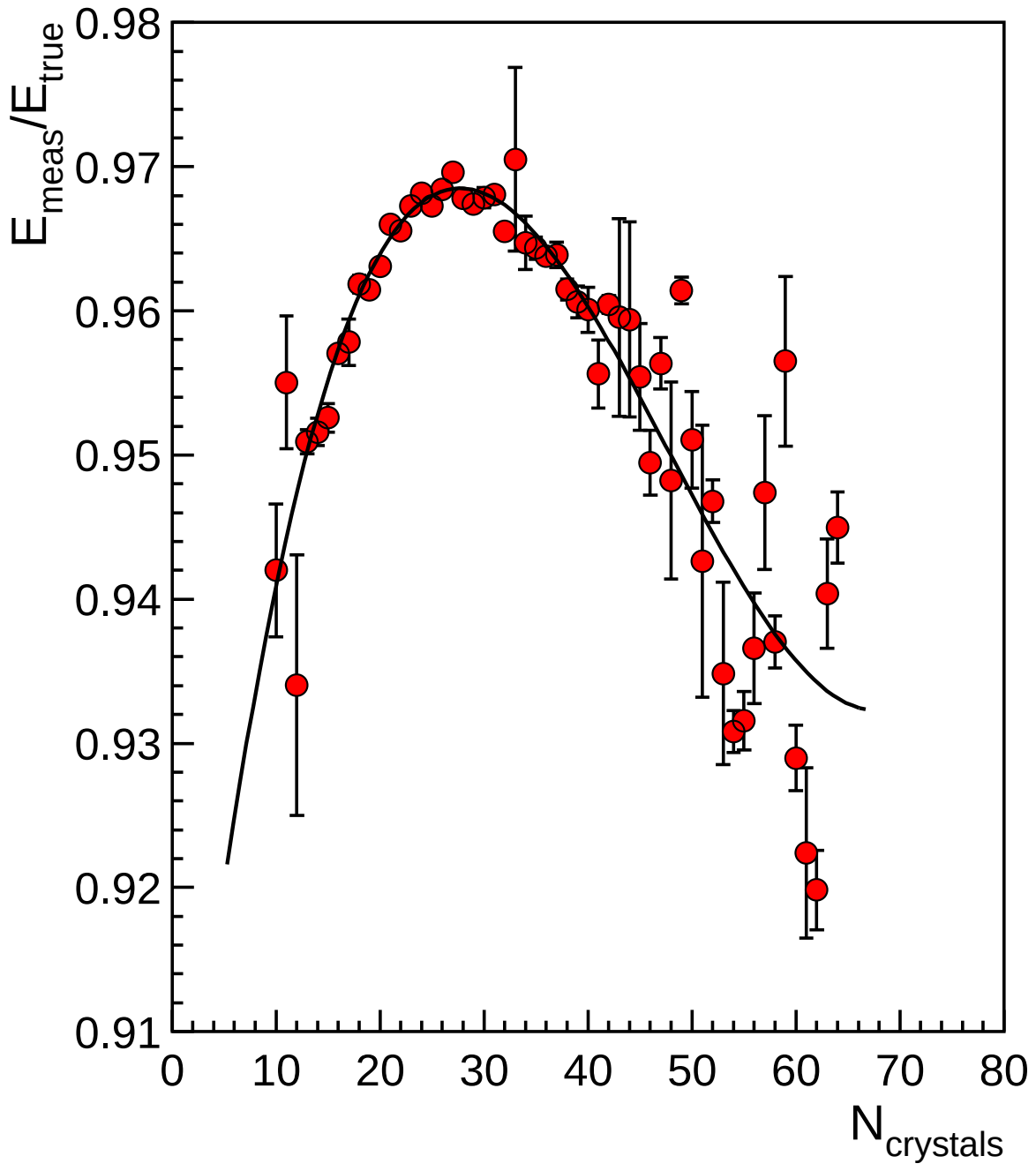

**Figure 5.** $E_{meas}/E_{true}$ as a function of the number of crystals in a Hybrid super-cluster together with a fitted polynomial function.

distribution – if the distribution is Gaussian then $\sigma_{eff}$ is just the Gaussian sigma, when the distribution has more significant tails then $\sigma_{eff}$ provides some measure of this. The parameter $\sigma_{eff}$ provides a convenient measure of performance which adequately reflects final physics performance (e.g. in extracting signal significance).

It is worth noting, as a comparison, that unconverted photons with a flat $p_T$ spectrum in the range 10< $p_T$<50 GeV/$c$ can be reconstructed, in the barrel using a fixed window in $\eta-\phi$ of $5\times5$ crystals, and in the endcap using a $3\times3$ fixed window with impact position correction, achieving in both cases a resolution of $\sigma_{eff}/E = 0.9\%$ at low luminosity.

Figure 6 shows the position resolution in $\eta$ and $\phi$ for the same sample. Table 4 gives a more complete list of performance results for electron reconstruction in both the barrel and endcap. Values for $p_T$= 35 GeV/$c$ electrons are given, together with values for electrons having a flat $p_T$ spectrum in the range $10 < p_T < 50$ GeV/$c$. The $\sigma_{eff}$ value is not given for the $\eta$ resolution because it is, in all cases, the same as the Gaussian fit value.

### *3.4. Level-2.0 Selection of Electrons and Photons*

The first step of the HLT, using only calorimeter information, is to reconstruct an ECAL super-cluster in a region specified by the Level-1 trigger. The super-cluster is first required to fall within the precision physics fiducial region of the ECAL, which is obtained by removing the barrel/endcap transition region from the overall coverage of $|\eta| <2.5$ [16]. This transition region is strongly shadowed by tracker cables exiting through the gap between the barrel and endcap. This region comprises the first trigger tower in the endcap. The last two crystals in the barrel are also removed from the trigger. The excluded area thus covers $1.4442 < |\eta| < 1.5660$.

**Table 4.** Energy and position resolution performance for barrel and endcap ECAL using electron samples simulated with different pileup conditions, using single particles.

| electron sample | Energy resolution | | Position resolution | | |
|---|---|---|---|---|---|
| | $\sigma/E$ | $\sigma_{eff}/E$ | $\sigma(\eta)$ | $\sigma(\phi)$ | $\sigma_{eff}(\phi)$ |
| Barrel reconstruction; $\|\eta\| <1.4442$ (Hybrid algorithm) | | | | | |
| $p_T$= 35 GeV/$c$ (no pileup) | 1.1% | 2.2% | $1.1 \times 10^{-3}$ | 1.7 mrad | 2.5 mrad |
| $p_T$= 35 GeV/$c$ ($2 \times 10^{33}\,\mathrm{cm}^{-2}\mathrm{s}^{-1}$) | 1.2% | 2.3% | $1.1 \times 10^{-3}$ | 1.7 mrad | 2.5 mrad |
| $p_T$= 35 GeV/$c$ ($10^{34}\,\mathrm{cm}^{-2}\mathrm{s}^{-1}$) | 1.5% | 2.7% | $1.1 \times 10^{-3}$ | 1.9 mrad | 2.7 mrad |
| $10 < p_T < 50$ GeV/$c$ ($10^{34}\,\mathrm{cm}^{-2}\mathrm{s}^{-1}$) | 1.5% | 3.4% | $1.2 \times 10^{-3}$ | 2.1 mrad | 3.4 mrad |
| Endcap reconstruction; $1.566 < \|\eta\| < 2.5$ (Island algorithm, and preshower) | | | | | |
| $p_T$= 35 GeV/$c$ (no pileup) | 1.2% | 2.1% | $1.8 \times 10^{-3}$ | 2.2 mrad | 3.4 mrad |
| $p_T$= 35 GeV/$c$ ($2 \times 10^{33}\,\mathrm{cm}^{-2}\mathrm{s}^{-1}$) | 1.6% | 2.4% | $1.8 \times 10^{-3}$ | 2.3 mrad | 3.5 mrad |
| $p_T$= 35 GeV/$c$ ($10^{34}\,\mathrm{cm}^{-2}\mathrm{s}^{-1}$) | 2.7% | 3.4% | $2.0 \times 10^{-3}$ | 2.9 mrad | 4.2 mrad |
| $10 < p_T < 50$ GeV/$c$ ($10^{34}\,\mathrm{cm}^{-2}\mathrm{s}^{-1}$) | 2.9% | 4.2% | $2.2 \times 10^{-3}$ | 2.7 mrad | 5.1 mrad |

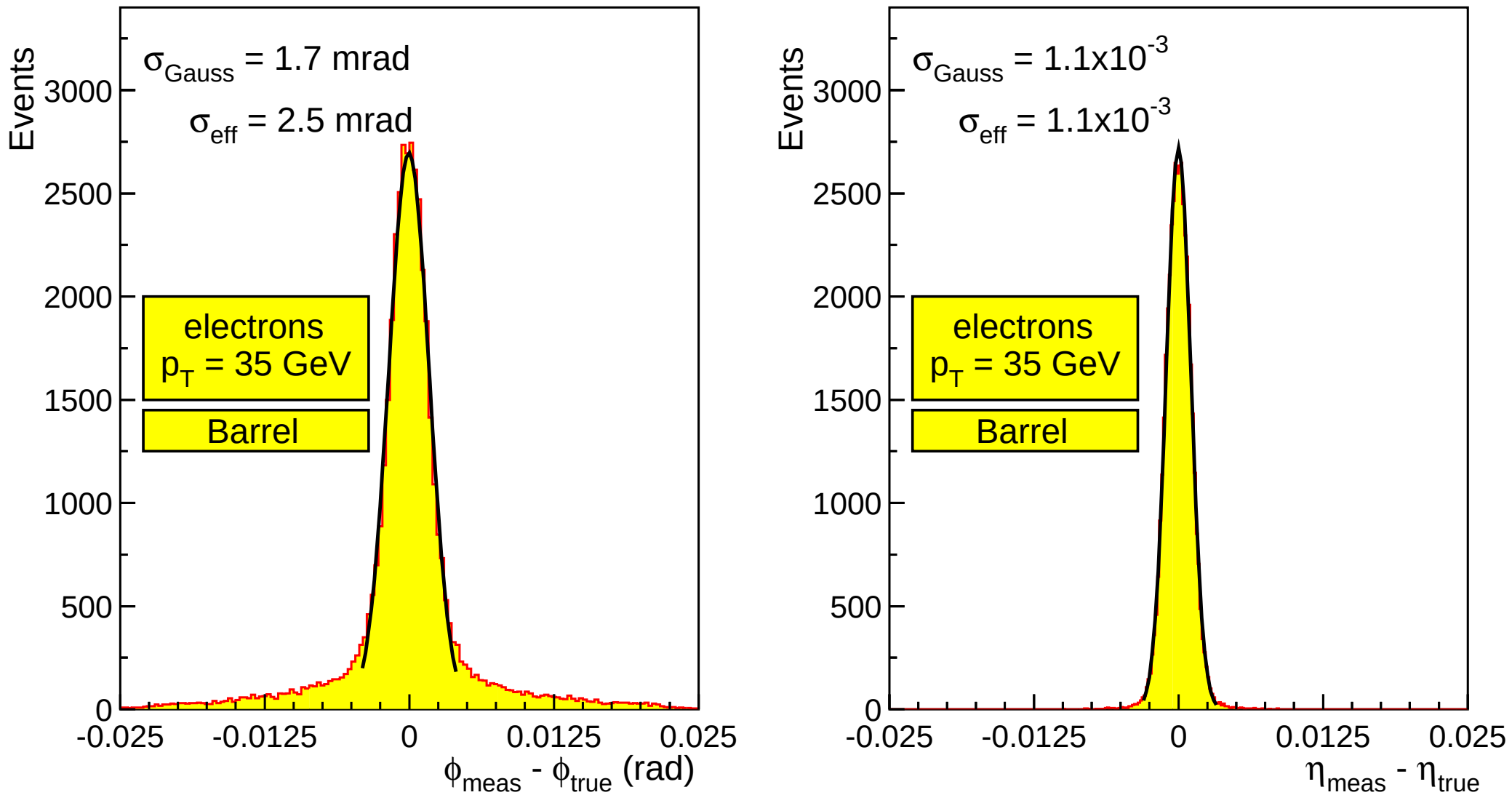

**Figure 6.** Position resolution for $p_T$= 35 GeV/$c$ electrons in the barrel ECAL, fully digitized without pileup, and reconstructed with the Hybrid super-clustering algorithm.

The super-cluster is then required to have $E_T$ above a threshold which is chosen to give 95% efficiency for electrons at the same point on the $E_T$ scale at which the Level-1 trigger has 95% efficiency . The same threshold is required for both objects in the di-electron trigger. At $\mathcal{L} = 2 \times 10^{33}\,\mathrm{cm}^{-2}\mathrm{s}^{-1}$the thresholds are 26 GeV for the single, and 14.5 GeV for the double electron trigger. The corresponding thresholds at $\mathcal{L} = 10^{34}\,\mathrm{cm}^{-2}\mathrm{s}^{-1}$ are 31 GeV and 16.9 GeV. This cut on the transverse energy reconstructed in the ECAL increases the rejection of fake electrons by about a factor of two.

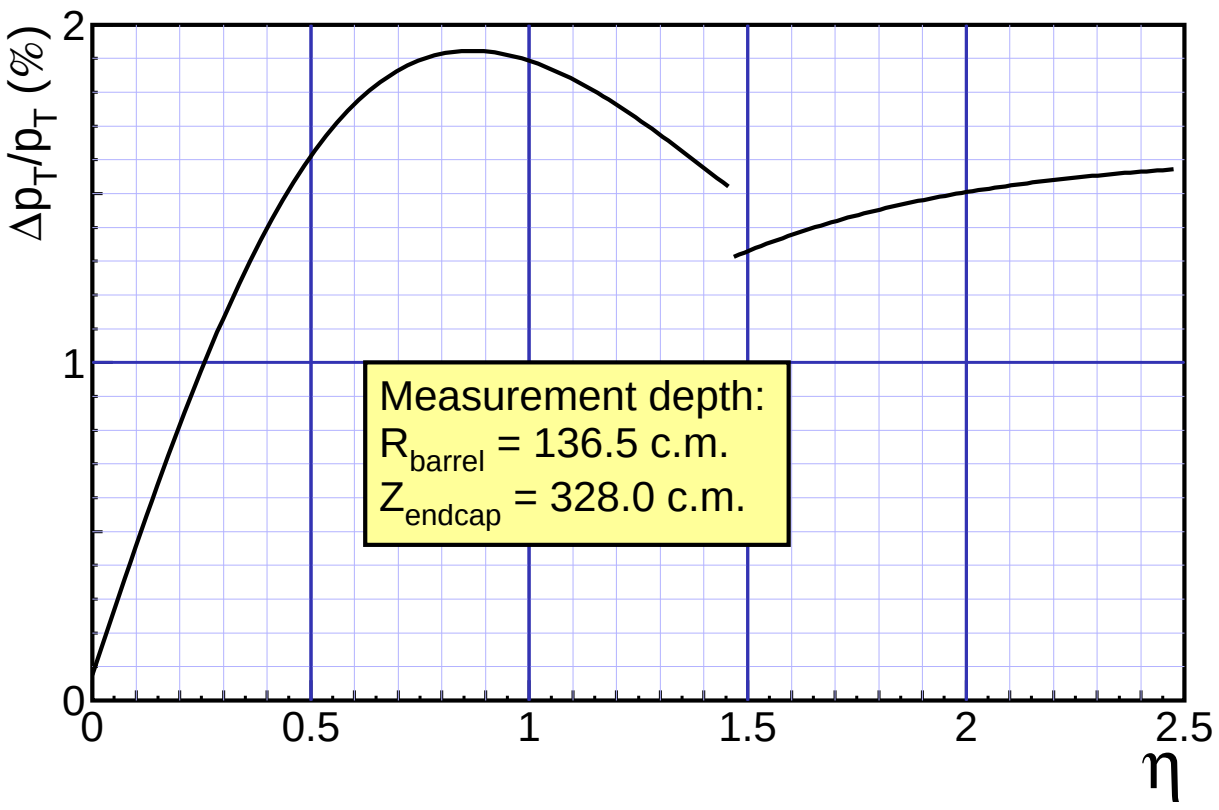

**Figure 7.** Fractional $E_T$ resolution, as a function of $\eta$, induced by a $1\sigma$ shift of the longitudinal vertex position.

Some refinements of this Level-2.0 selection can be envisaged, but have not been used in the basic selection procedure for which results are here presented. Firstly, the Level-3 selection uses cuts on purely calorimetric quantities – ECAL isolation and an H/E (hadronic/electromagnetic) cut – which could be applied at an earlier stage. Secondly, a significant contribution to the transverse energy resolution comes from the lack of knowledge, at this stage, of the longitudinal vertex position. The contribution to the fractional $E_T$ resolution, as a function of $\eta$, due to a $1\sigma$ shift of the vertex longitudinal position (5.3 cm), is shown in figure 7. A small improvement in efficiency would be obtained by making the electron threshold cuts in two stages: a looser cut at Level-2.0 followed by the full set of cuts at a later stage, after the pixel matching of Level-2.5 which gives a precise longitudinal vertex position.

### *3.5. Level-2.5: Matching of Super-clusters to Hits in the Pixel Detector*

The matching of super-clusters reconstructed in the calorimeter to hits in the pixel detector takes advantage of the fact that the energy-weighted average impact point of an electron and the bremsstrahlung photons it has radiated, is precisely where a non-radiating electron would have impacted. It is this space-point that the position measurement of the super-cluster attempts to determine. This point can be propagated back through the field to obtain an estimate of the direction of the electron at the vertex, and the hit positions expected in the pixel detector. Since most of the tracker material lies after the pixel detector, most electrons do not radiate significantly before it, and most photon conversions take place after it. So matching hits are given by most electrons and by few photons.

The layout of the pixel system is shown in figure 8. The continuous lines show the maximum values of the ECAL barrel and endcaps and the dashed lines correspond to electrons generated $z = \pm 15$ cm away from the nominal vertex. For the clusters at the ends of the ECAL barrel, pixel disks are needed as well as barrel pixel layers. For endcap clusters the mix of pixel barrel and pixel disk hits depends on both the pseudorapidity and the $z$-vertex of the electron. With the geometry used for this simulation study there is a loss of efficiency in the high $|\eta|$ region of the ECAL endcap where, in some cases, the electron goes through only one pixel layer.

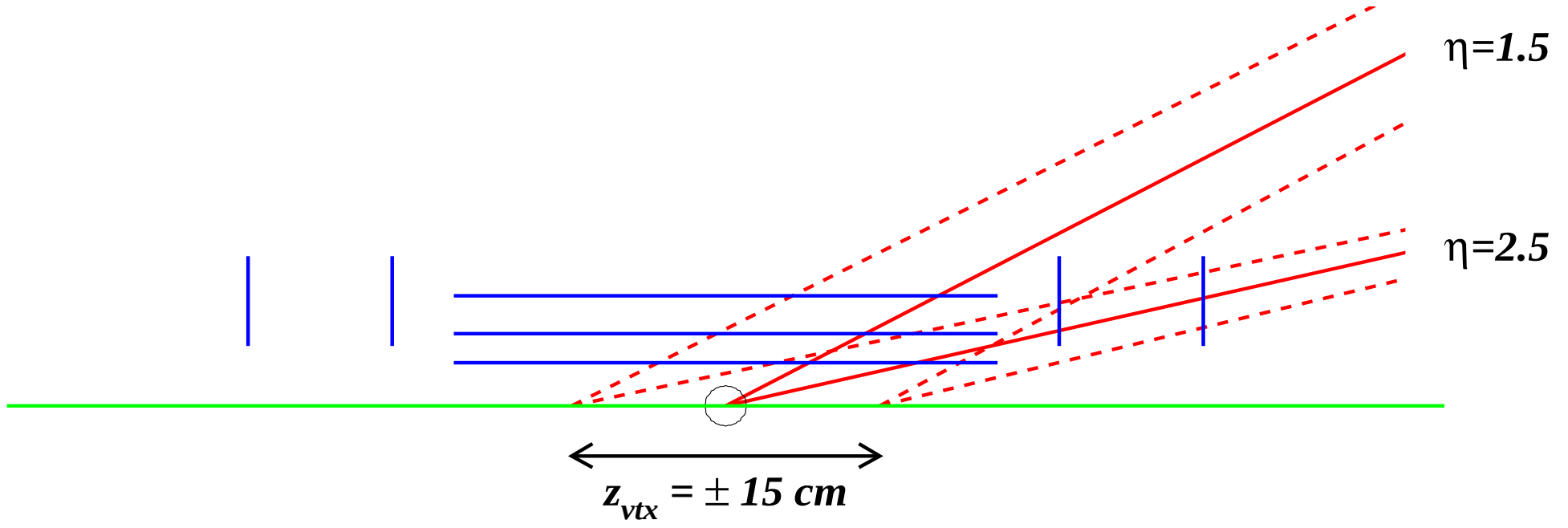

**Figure 8.** Full pixel detector (high luminosity) with continuous lines pointing from the nominal vertex to the edges of the ECAL barrel and endcap, and dashed lines pointing from $z = \pm 15$ cm. cf. Appendix A.1

The outer barrel layer and the outermost endcap disks of the pixel detector is staged and may not be present at the start-up of the experiment. The simulation of the HLT selection at low luminosity is made by discarding pixel information in these layers.The method proceeds as follows:

(i) The electromagnetic cluster gives the energy and the position of the electron candidate. The transverse momentum is computed from the measured energy and position of the cluster. The expected hit position on the pixel layers is estimated by propagating the electron inwards to the nominal vertex using the magnetic field map. A search area is defined in the innermost pixel layer. It is unrestricted in $z$, and its width in $\phi$ (typically 40 mrad) is the main parameter used to control the tightness of the pixel matching. The errors on the calculated hit positions are due to the spread in $z$ of the vertex position and the precision of the $\phi$ measurement in the calorimeter. At this stage the error on the cluster $E_T$ measurement contributes very little to the uncertainty on the expected hit position in the pixel detector.

(ii) If a compatible hit is found within the search area on the innermost pixel layer, a better estimate of the longitudinal ($z$) vertex position is obtained by interpolating the line from the cluster through the corresponding hit to the beam line. Nominal values (0,0) for $x$ and $y$ coordinates of the vertex are assumed. If no hit is found in the search area of the innermost pixel layer, the search is repeated in the next layer.

(iii) The track is propagated from the newly estimated vertex to the next pixel

layer through the compatible hit in the first (or second) layer. The dominating uncertainty, in the $r - \phi$ plane, results from the estimate of the $E_T$ of the electromagnetic cluster, and thus the radius of curvature of the electron track. But this is a very small uncertainty since the distance from one pixel layer to the next is short. Indeed, assuming a hit on the $r = 4$ cm layer, the distance between the hit predictions on the $r = 11$ cm layer for $p_T$= 15 GeV/c and $p_T$= 20 GeV/c electron would be 78 $\mu m$ – less than the size of a pixel. In the $r - z$ plane, the main uncertainty comes from the vertex $z$.

(iv) If another compatible hit is found, the cluster is identified as an electron, if not, it is rejected as a jet. If there are no compatible hits in the current layer, there may be one more pixel layer left, and the search is repeated there.

The search is made twice, once for each charge. In the first step of the search the electron and positron search areas can overlap, but in the second step, when a compatible hit is propagated to another pixel layer, the $p_T$ needed for the search areas of different charges to overlap is almost 1 TeV/$c$.

For low luminosity both the staged and full pixel detector configurations are investigated. Using the inner pixel layers alone results in a large inefficiency ($\sim$ 7%) in the staged pixel configuration, so the low luminosity staged pixel detector algorithm supplements the pixel hits with hits found in the first silicon strip layer of the tracker, and regains some of the lost efficiency. More details of the Level-2.5 pixel matching algorithm are given in reference [17].

The "rejection" versus efficiency obtained from the Level-2.5 pixel matching is shown in figure 9, at (left) low luminosity ($\mathcal{L} = 2 \times 10^{33}\,\mathrm{cm}^{-2}\mathrm{s}^{-1}$), and (right) high luminosity ($\mathcal{L} = 10^{34}\,\mathrm{cm}^{-2}\mathrm{s}^{-1}$). The efficiency is calculated using an electron sample passing the Level-2.0 threshold. The "rejection" is calculated from the single electron triggers in the jet background passing the Level-2.0 threshold and is the efficiency of the electron algorithm in a jet sample. In practice, single and double photon streams are created by applying an $E_T$ threshold to the events rejected by the pixel matching. For a 95% efficiency the rejection factors are 16.5 (12.5) at low (high) luminosity.

For the low luminosity case two different curves of the efficiency versus rejection, corresponding to the full and staged pixel detector configurations, are shown in figure 9 (left).

The performance at high luminosity, shown in figure 9 (right), is very similar to the low luminosity performance without staging with respect to electron efficiency, but there is a noticeable loss of rejection power which can be attributed to pileup hits. The high luminosity figure also shows the performance in a more restricted central region ($|\eta|$ <2.1) where the efficiency is noticeably better. A similar conclusion is reached at low luminosity (not shown).

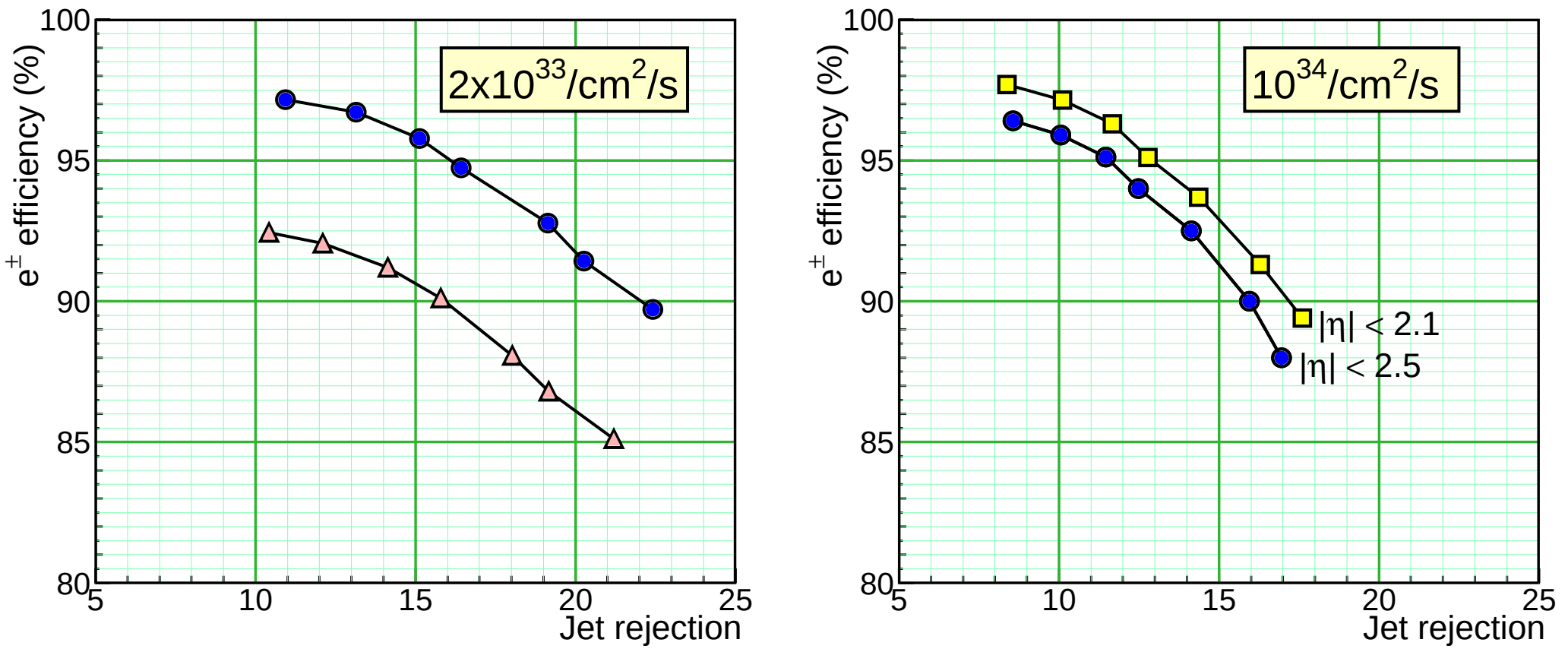

**Figure 9.** Rejection versus efficiency obtained from the Level-2.5 pixel matching. Left: at low luminosity ($\mathcal{L} = 2 \times 10^{33}\,\mathrm{cm}^{-2}\mathrm{s}^{-1}$); the top curve shows the performance when the full pixel detector is used while the lower curve shows the performance for the staged pixel scenario (see text). Right: at high luminosity ($\mathcal{L} = 10^{34}\,\mathrm{cm}^{-2}\mathrm{s}^{-1}$); the lower curve is the nominal detector configuration; the top curve corresponds to $|\eta| <2.1$. The definition of the rejection is discussed in the text.

### *3.6. Level-3 Selection of Electrons and Photons: Inclusion of Full Tracking Information*

The Level-3 selection includes all further requirements needed to reach an acceptable rate to final offline storage. The full event information, including tracks, is available, but some of the cuts used – hadronic/electromagnetic energy fraction and calorimetric isolation – use only calorimetric information.

*3.6.1. Electrons* The Level-3 selection for electrons starts with electron track-finding, seeded by the Level-2.5 pixel match. To maintain high efficiency track-finding is made with very loose cut parameters. Cuts are then made on both $E/P$ and on the distance between the super-cluster position and the extrapolated track position in the ECAL in the longitudinal coordinate, $\Delta\eta$(track - cluster), which is only slightly distorted by bremsstrahlung. In the endcap a cut on the energy found behind the super-cluster, in the HCAL, expressed as a fraction of the super-cluster energy, H/E, is found to give useful additional rejection

Figure 10 shows, as an example, the $E/P$ distribution for barrel electrons, and for jet background electron candidates in the barrel after selection at Level-2.5 followed by loose track finding seeded with the Level-2.5 pixel matches. When the distributions are broken down according to the number of hits associated in the tracks, the width and proportion of events in the tail of the distribution for electrons is found to vary (electrons which radiate little have tracks with more hits, and a better measured momentum). The ratio of signal to background also varies with the number of hits. Hence increased performance can be obtained by optimizing the $E/P$ cut as a function of the number of hits in the track.

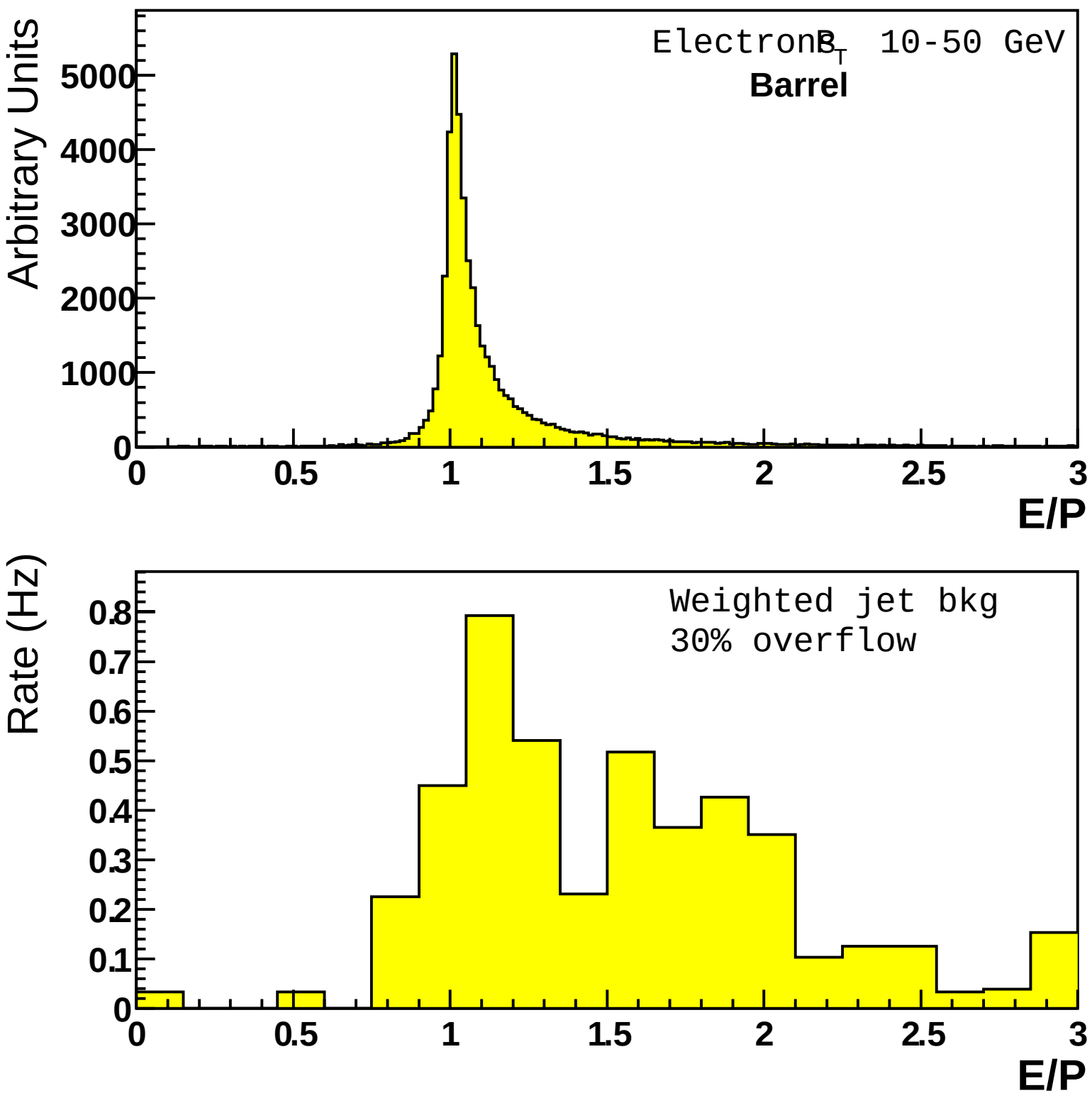

**Figure 10.** $E/P$ for (upper plot) electrons and (lower plot) jet background candidates in the barrel, after Level-2.5 selection followed by track finding seeded by the Level-2.5 pixel hits ($\mathcal{L} = 2 \times 10^{33}\,\mathrm{cm}^{-2}\mathrm{s}^{-1}$). The background distribution has 30% overflows.

Specifically, for the studies discussed below the following selection cuts were applied: $E/P < 1.5$ (2.45) for the barrel (end cap), H/E $< 0.028$ in the endcap, and in the barrel a track/ECAL position match in $\eta$ ($\Delta\eta$(track - cluster) $< 0.028$). At high luminosity the additional rejection power of isolation cuts is used to reduce the background to single electrons from jets. Three isolation techniques have been studied: ECAL isolation, pixel-track isolation, and full-track isolation. Track isolation has the advantage that it is less sensitive to pileup, which is the dominant source of signal inefficiency at high luminosity, because only tracks associated with the primary vertex are selected for the isolation cuts. The results presented in the tables below use no isolation for electrons at low luminosity, and a simple pixel-track isolation cut at high luminosity.

Figure 11 shows the rejection against jet background, plotted as the efficiency of the electron algorithm on the QCD jet sample. This is compared to the efficiency for signal electrons from $W \to e\nu$ events when a pixel-track isolation cut is applied after the Level-2.5 selection at high luminosity. In addition, only events where a track can be made from the pixel hits of the Level-2.5 candidate are used. The different points on the plot correspond to different cuts on the number of 3-hit pixel tracks with $p_T > 1$

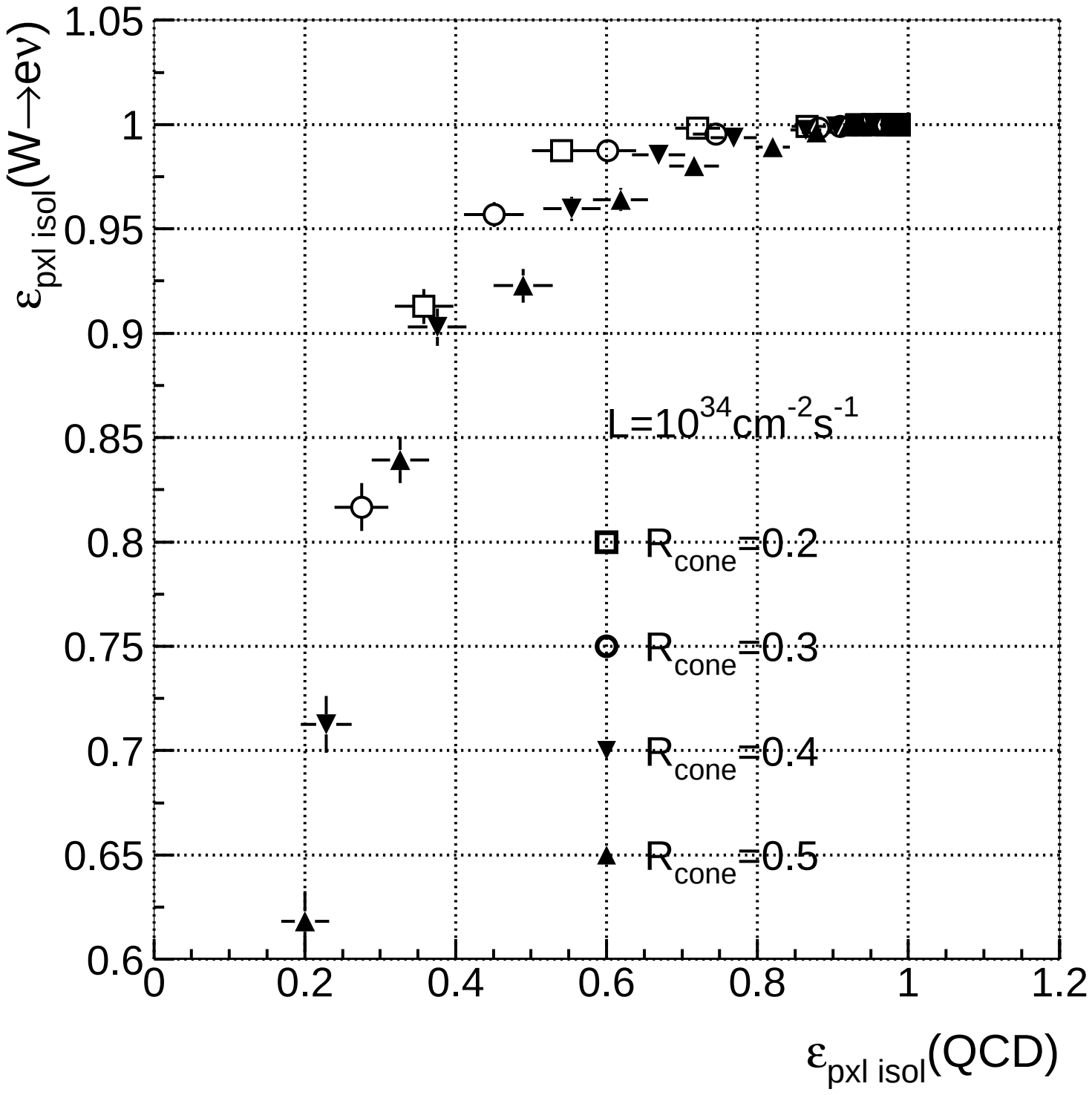

**Figure 11.** Rejection against jet background versus the efficiency for electrons from $W$s when a pixel-track isolation cut is applied after the Level-2.5 selection at $\mathcal{L} = 10^{34}\,\mathrm{cm}^{-2}\mathrm{s}^{-1}$. $R_{\mathrm{cone}}$ is the cone radius. The points represent different values of the isolation (see text).

**Table 5.** Photon stream thresholds and rates before additional Level-3 cuts.

| | Threshold | Rate ($2 \times 10^{33}\,\mathrm{cm}^{-2}\mathrm{s}^{-1}$) | Rate ($10^{34}\,\mathrm{cm}^{-2}\mathrm{s}^{-1}$) |
|---|---|---|---|
| Single photon | $E_T >$ 80 GeV ($2 \times 10^{33}\,\mathrm{cm}^{-2}\mathrm{s}^{-1}$) | 7 Hz | |
| | $E_T >$ 100 GeV ($10^{34}\,\mathrm{cm}^{-2}\mathrm{s}^{-1}$) | | 10 Hz |
| Double photons | ${E_T}^1 > 35$, ${E_T}^2 > 20$ GeV | 85 Hz | 382 Hz |

GeV/$c$ found within cones of different sizes.

*3.6.2. Photons* Further $E_T$ thresholds, higher than those applied at Level-2.0, are applied to super-clusters of single and double triggers that fail the Level-2.5 pixel matching. The events passing these cuts form the photon stream. The di-photon thresholds are asymmetric, chosen to be 5 GeV lower than the offline analysis cuts envisaged for the Standard Model $H \to \gamma\gamma$ search [16]. The single photon thresholds are chosen to give an acceptable rate. Table 5 lists the thresholds and the rates before further Level-3 selection.

Backgrounds can be rejected using track isolation cuts and by rejecting $\pi^0$'s based on the lateral shower shape. Defining the longitudinal coordinate of the vertex is a significant issue for the analysis of the $H \to \gamma\gamma$ signal channel. For events where one or more of the photons has converted in the tracker, the track segment and the ECAL cluster can be used to locate the vertex. The vertices in the remaining events can be found using algorithms that choose the track vertex associated with the largest track activity. The efficiency for $H \to \gamma\gamma$ is 80-90% for a jet rejection factor of 30-60 in the high luminosity environment.

**Table 6.** Electron and photon rates output by the HLT at low and high luminosity. (1) $\pi^{\pm}/\pi^0$ overlap; (2) $\pi^0$ conversions

| | $2 \times 10^{33}\,\mathrm{cm}^{-2}\mathrm{s}^{-1}$ | | | $10^{34}\,\mathrm{cm}^{-2}\mathrm{s}^{-1}$ | | |
|---|---|---|---|---|---|---|
| | Signal | Background | Total | Signal | Background | Total |
| Single electron | $W \to e\nu$: 10Hz | (1): 5Hz | 33 Hz | $W \to e\nu$: 35Hz | (1): 15Hz | 75 Hz |
| | | (2): 10Hz | | | (2): 19Hz | |
| | | $b/c \to e$: 8Hz | | $b/c \to e$: 6Hz | | |
| Double electron | $Z \to ee$: 1Hz | $\sim 0$ | 1Hz | $Z \to ee$: 4Hz | $\sim 0$ | 4Hz |
| Single photon | 2Hz | 2Hz | 4Hz | 4Hz | 3Hz | 7Hz |
| Double photon | $\sim 0$ | 5Hz | 5Hz | $\sim 0$ | 8Hz | 8Hz |
| TOTAL: | | | 43Hz | | | 94 Hz |

### *3.7. Summary of Electron and Photon HLT Selection*

*3.7.1. Final Rates to Permanent Storage* The electron and photon rates output by the HLT at both low and high luminosity, broken down by contribution, are listed in table 6. For the low-luminosity selection a loose calorimetric isolation has been applied to the photon streams (ECAL $E_T$ in a cone of radius 0.45, excluding the supercluster, less than 3.5 GeV), but no isolation beyond that of the Level-1 trigger has been applied to the electron streams. To control the two-photon rate the thresholds have been raised to ${E_T}^1 > 40$ GeV, ${E_T}^2 > 25$ GeV (equal to the final offline cuts envisaged for $H \to \gamma\gamma$). This reduces the rate from 11 Hz to 5 Hz, and has a negligible effect on the efficiency, as is shown in the second column in table 7. A fully optimized selection will also involve track isolation on the photon streams (wholly or partly replacing the calorimetric isolation and the raised threshold) and track isolation in the single electron stream. This can reduce the total rate to about 26 Hz, of which only half is background, with the introduction of only a small further inefficiency. For the high-luminosity selection, pixel-track isolation has been applied to the electron stream, and full track isolation has been applied to the photon streams (no track with $p_T$> 2 GeV/c in a cone of $\Delta R$=0.2).

*3.7.2. Signal Efficiencies for Electron and Photon HLT* The streams where most work is required to control the background rates are the single-electron and double-photon streams, so, the efficiencies for the decays $W \to e\nu$ and $H \to \gamma\gamma$ are used as benchmarks. Table 8 lists the efficiency for single electrons from $W$ decay through the complete selection chain, at $\mathcal{L} = 2 \times 10^{33}\,\mathrm{cm}^{-2}\mathrm{s}^{-1}$ and at $\mathcal{L} = 10^{34}\,\mathrm{cm}^{-2}\mathrm{s}^{-1}$. Events are preselected requiring the generated electrons to be within the ECAL fiducial region of $|\eta| < 2.5$, with the region $1.4442 < |\eta| < 1.5660$ excluded. The geometric acceptance is approximately 60% and is not included in the efficiency. The second and fourth columns list the efficiencies for electrons that have $p_T$ greater than the Level-1 and Level-2 95% efficiency point.

**Table 7.** Efficiency for $H \to \gamma\gamma$ ($M_H$=115 GeV/$c^2$) through the complete selection chain, at $\mathcal{L} = 2 \times 10^{33}\,\mathrm{cm}^{-2}\mathrm{s}^{-1}$.

| | Both photons in fiducial region | Photons passing offline $p_T$ cuts |
|---|---|---|
| Level-1 | 90.8% | 92.3% |
| Level-2 | 98.7% | 99.4% |
| Level-2.5 | 93.4% | 99.1% |
| Level-3 | 92% | 92% |
| Overall (Level-1 × HLT) | 77% | 83.7% |

**Table 8.** Efficiency for electrons from $W$ decay through the complete selection chain

| | $2 \times 10^{33}\,\mathrm{cm}^{-2}\mathrm{s}^{-1}$ | | $10^{34}\,\mathrm{cm}^{-2}\mathrm{s}^{-1}$ | |
|---|---|---|---|---|
| | All fiducial electrons | Fiducial electrons with $p_T$> 29 GeV/$c$ | All fiducial electrons | Fiducial electrons with $p_T$> 34 GeV/$c$ |
| Level-1 | 63.2% | 87.2% | 51.1% | 83.2% |
| Level-2 | 88.8% | 99.4% | 82.9% | 99.3% |
| Level-2.5 | 93.1% | 94.6% | 92.8% | 94.1% |
| Level-3 | 81% | 82% | 77% | 78% |
| HLT (Level-2 to Level-3) | 67% | 77% | 59% | 73% |

The efficiencies at $\mathcal{L} = 10^{34}\,\mathrm{cm}^{-2}\mathrm{s}^{-1}$ are only slightly lower than those at low luminosity. The main difference comes from the loss due to the additional isolation cuts – typically a 5% loss per object.

Table 7 lists the efficiency for $H \to \gamma\gamma$ for a Higgs with mass $M_H = 115$ GeV/$c^2$ through the complete selection chain, at $\mathcal{L} = 2 \times 10^{33}\,\mathrm{cm}^{-2}\mathrm{s}^{-1}$. As in the previous table, events are preselected, requiring that the generated photons fall within the ECAL fiducial region. The geometric acceptance is 65%. The second column shows the efficiency for events where the two generated photons satisfy, in addition, the cuts currently assumed for offline analysis in this channel – ${E_T}^1 > 40$ GeV, ${E_T}^2 > 25$ GeV.

*3.7.3. CPU Usage for Electron and Photon HLT* Table 9 shows the CPU usage of the HLT selection, benchmarked on 1 GHz processor, for jet background events at low luminosity. At high luminosity the time taken for the unoptimized global search for ECAL clusters at Level-2.0 is greatly increased and the overall total CPU time per Level-1 event, at a luminosity of $\mathcal{L} = 10^{34}\,\mathrm{cm}^{-2}\mathrm{s}^{-1}$ is about three times as large, as at $\mathcal{L} = 2 \times 10^{33}\,\mathrm{cm}^{-2}\mathrm{s}^{-1}$. Preliminary results indicate that this clustering time can be reduced by a factor $\sim 10$ using regional reconstruction.

**Table 9.** CPU usage of the HLT electron selection for jet background events at $\mathcal{L} = 2 \times 10^{33}\,\mathrm{cm}^{-2}\mathrm{s}^{-1}$, benchmarked on 1 GHz processors.

| HLT level | Mean CPU time (ms) |
|---|---|
| Level-2.0 | 154 /Level-1 event |
| Level-2.5 | 32 /Level-2 event |
| Level-3 | 100 /Level-2.5 event |
| Total | 162 ms/Level-1 event |

## 4. Muon Identification

The muon selection for the HLT proceeds in two steps: firstly, muons are reconstructed in the muon chambers, which confirms the Level-1 decision and refines the $p_T$ measurement using more precise information; secondly, the muon trajectories are extended into the tracker, which further refines the $p_T$ measurement. After each step, isolation is applied to the muon candidates – the calorimeter being used after the first step and the tracker after the second.

The muon track reconstruction algorithm used by the HLT is seeded by the – up to four– muon candidates found by the Level-1 Global Muon Trigger (see Appendix), including those candidates that did not necessarily lead to a Level-1 trigger accept by the Global Trigger. The algorithm uses the reconstructed hits built from the digitized signals in the muon system, and constructs tracks according to the Kalman filter technique [18]. The resulting trajectories are used to validate the Level-1 decision as well as to refine the muon measurement in this Level-2 muon selection. The basis of the Level-3 muon selection is to add silicon tracker hits to the muon trajectory, thus greatly improving the muon momentum measurement and sharpening the trigger threshold. Isolation criteria can be applied to the muon candidates to provide additional rejection: at Level-2 using the calorimetric energy sum in a cone around the muon, and at Level-3 using the number of pixel tracks in a region around the projected muon trajectory. This suppresses muons from $b, c, \pi$, and K decays.

### *4.1. Muon Reconstruction*

*4.1.1. Muon Standalone Reconstruction and Level-2 Selection* Reconstructed track segments from the muon chambers are used for muon identification and selection at Level-2. The state vectors (track position, momentum and direction) associated with the segments found in the innermost chambers are propagated outwards through the iron yoke using the GEANE package [19], which takes into account the muon energy loss in the material, the effect of the multiple scattering, and the non-constant magnetic field in the muon system. The estimate of the momentum from the Level-1 Global Muon Trigger is used initially for the track propagation in the magnetic field. The predicted state vector at the next measurement surface is compared with existing measured points and updated accordingly using a Kalman filter technique. In the barrel chambers, reconstructed track segments are used as measurements in the Kalman filter procedure; in the endcap chambers, where the magnetic field is inhomogeneous, it is the individual reconstructed hits belonging to the track segments that are used. Reconstructed hits from RPC chambers are also included. The procedure is iterated until the outermost measurement surface of the muon system is reached, at which point a constrained fit to the track parameters, working from the outside in, is performed under the assumption that the muon candidate originated from the interaction region (defined by the beam spot size: $\sigma_{xy} = 15\,\mu$ m and $\sigma_z = 5.3$ cm; the beam spot can be determined from minimum bias events at the very start of the run). In both the forward and backward propagation just described, a hit is not added to the muon trajectory if its contribution to the total $\chi^2$ exceeds 25. The resulting track parameters, propagated inward to the collision vertex, are used to reject or accept the event for further Level-3 processing. More details on this stand-alone muon reconstruction can be found in reference [20].
Figure 12 shows the resolution of the transverse momentum determined by the Level-2 constrained fit for muons from $W$ decays at high luminosity. The distributions are broken up into three pseudorapidity intervals: barrel ($|\eta|$ <0.8), overlap (0.8< $|\eta|$ <1.2) and endcap (1.2< $|\eta|$ <2.1). In these three regions, the fitted $p_T$ resolutions are 10%, 15%, and 16%, respectively.

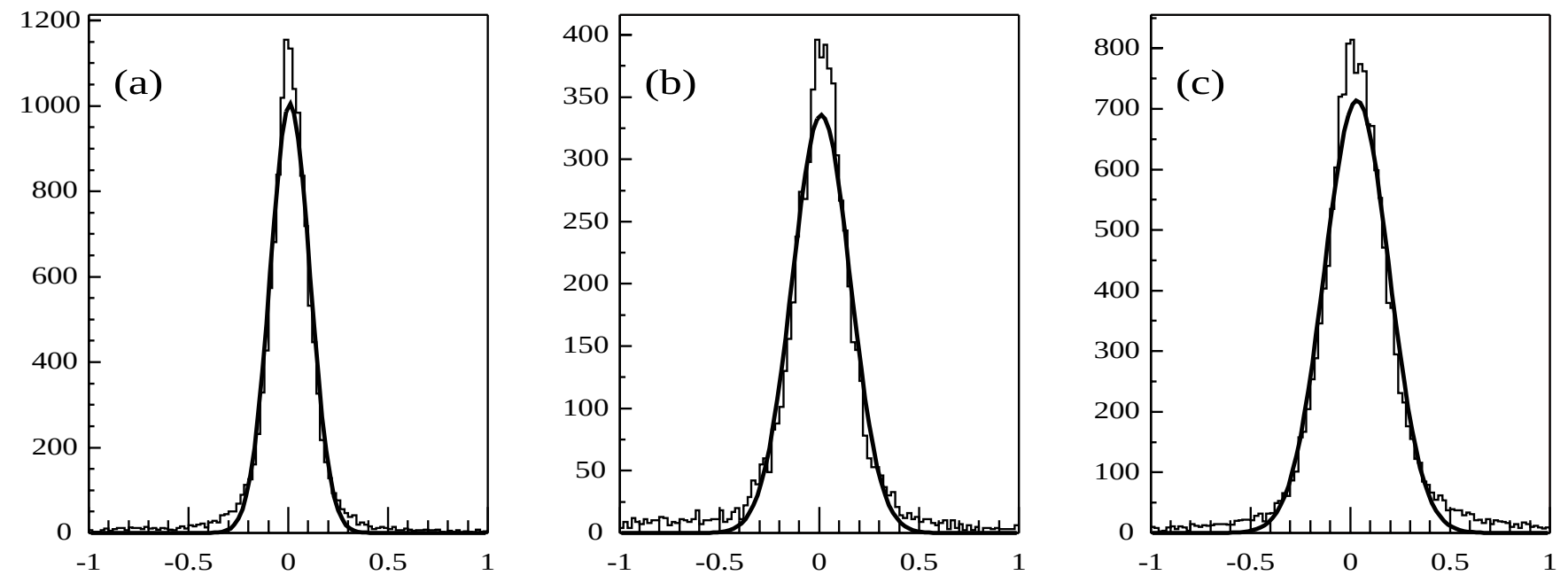

**Figure 12.** Distribution of $(1/p_T{}^{rec}\text{-}1/p_T{}^{gen})/(1/p_T{}^{gen})$, where $p_T{}^{gen}$ and $p_T{}^{rec}$ are the generated and Level-2 reconstructed transverse momenta respectively, shown in three pseudorapidity intervals: a) $|\eta|$ <0.8 b) 0.8< $|\eta|$ <1.2 c) 1.2< $|\eta|$ <2.1

*4.1.2. Inclusion of Tracker Information and Level-3 Selection* The Level-3 muon reconstruction consists of extending the muon trajectories to include hits in the silicon tracker system. Starting from a Level-2 reconstructed muon, the muon trajectory is extrapolated from the innermost muon station to the outer tracker surface, taking into account the muon energy loss in the material and the effect of multiple scattering. As with Level-2, the GEANE package is currently used for the propagation through the iron and calorimeters. Silicon layers compatible with the muon trajectory are then determined, and a region of interest within them is defined to perform regional track reconstruction. The determination of the region of interest is based on the track parameters and uncertainties of the extrapolated Level-2 muon trajectory, obtained with the assumption that the muon originates from the interaction point as described in the previous section. This has a strong impact on the reconstruction efficiency, fake rate, and CPU reconstruction time: well measured muons are reconstructed faster and with higher efficiency than poorly measured ones.

Inside the region of interest, initial candidates for the muon trajectory (regional seeds) are built from pairs of reconstructed hits. The two hits forming a seed must come from two different tracker layers, and all combinations of compatible pixel and double-sided silicon strip layers are used in order to achieve high efficiency. In addition, a beam spot constraint is applied to muon candidates above a given transverse momentum threshold to obtain initial trajectory parameters.

Starting from the regional seeds, a track reconstruction algorithm based on the Kalman filter technique, is used to reconstruct tracks inside the selected region of interest. The track reconstruction algorithm consists of the following steps: trajectory building (seeded pattern recognition), trajectory cleaning (resolution of ambiguities) and trajectory smoothing (final fit). In the first step, the trajectory builder transforms each seed into a set of trajectories. Starting from the innermost layer, the trajectory is propagated to the next tracker layer that is reachable, and updated with compatible measurements found on that layer. In the second step, the trajectory cleaner resolves ambiguities between multiple trajectories that may result from a single seed on the basis of the number of hits and the $\chi^2$ of the track fit. In the final step, all reconstructed tracks are fit once again with reconstructed hits in the muon chambers included from the original Level-2 reconstructed muon, and selected on the basis of a $\chi^2$ cut.

Figure 13 shows the resolution of the transverse momentum determined by the Level-3 constrained fit as expressed by the distribution of the quantity ($1/p_T^{rec}$-$1/p_T^{gen}$) / ($1/p_T^{gen}$), where $p_T^{gen}$ and $p_T^{rec}$ are the generated and reconstructed transverse momenta, respectively. Muons from $W$ decays at high luminosity are used. The distributions are broken up into three pseudorapidity intervals: barrel ($|\eta| <0.8$), overlap ($0.8< |\eta| <1.3$) and endcap ($1.3< |\eta| <2.1$). In these three regions, the fitted $p_T$ resolutions are 1.0%, 1.4%, and 1.7%, respectively. The improvement in resolution over the stand-alone muon measurement from Level-2 is substantial.

The efficiency of the Level-3 tracking algorithm relative to the Level-2 selection is shown in figure 14 as a function of $\eta$ for single muons with $p_T$>10 GeV/c and no pileup.

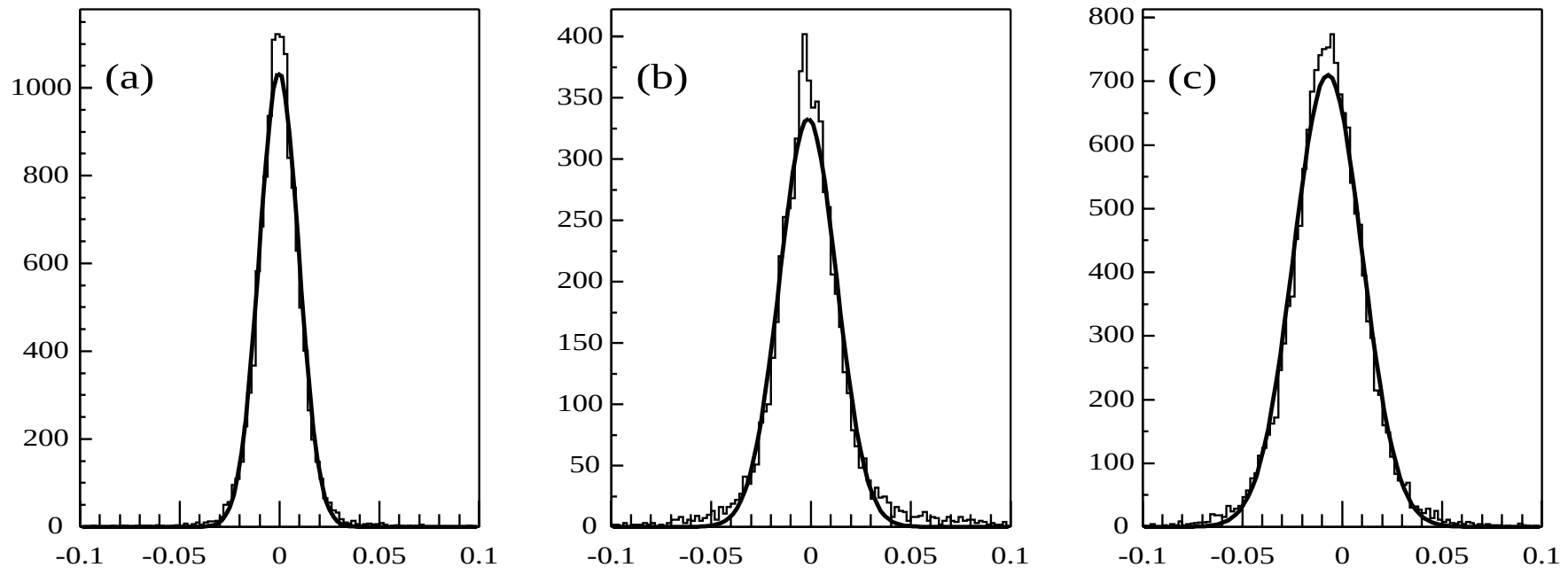

**Figure 13.** Distribution of $(1/p_T^{rec}\text{-}1/p_T^{gen})$ / $(1/p_T^{gen})$ where $p_T^{gen}$and $p_T^{rec}$are the generated and Level-3 reconstructed transverse momenta, respectively, shown in three pseudorapidity intervals: a) $|\eta| < 0.8$, b) $0.8 < |\eta| < 1.2$, and c) $1.2 < |\eta| < 2.1$.

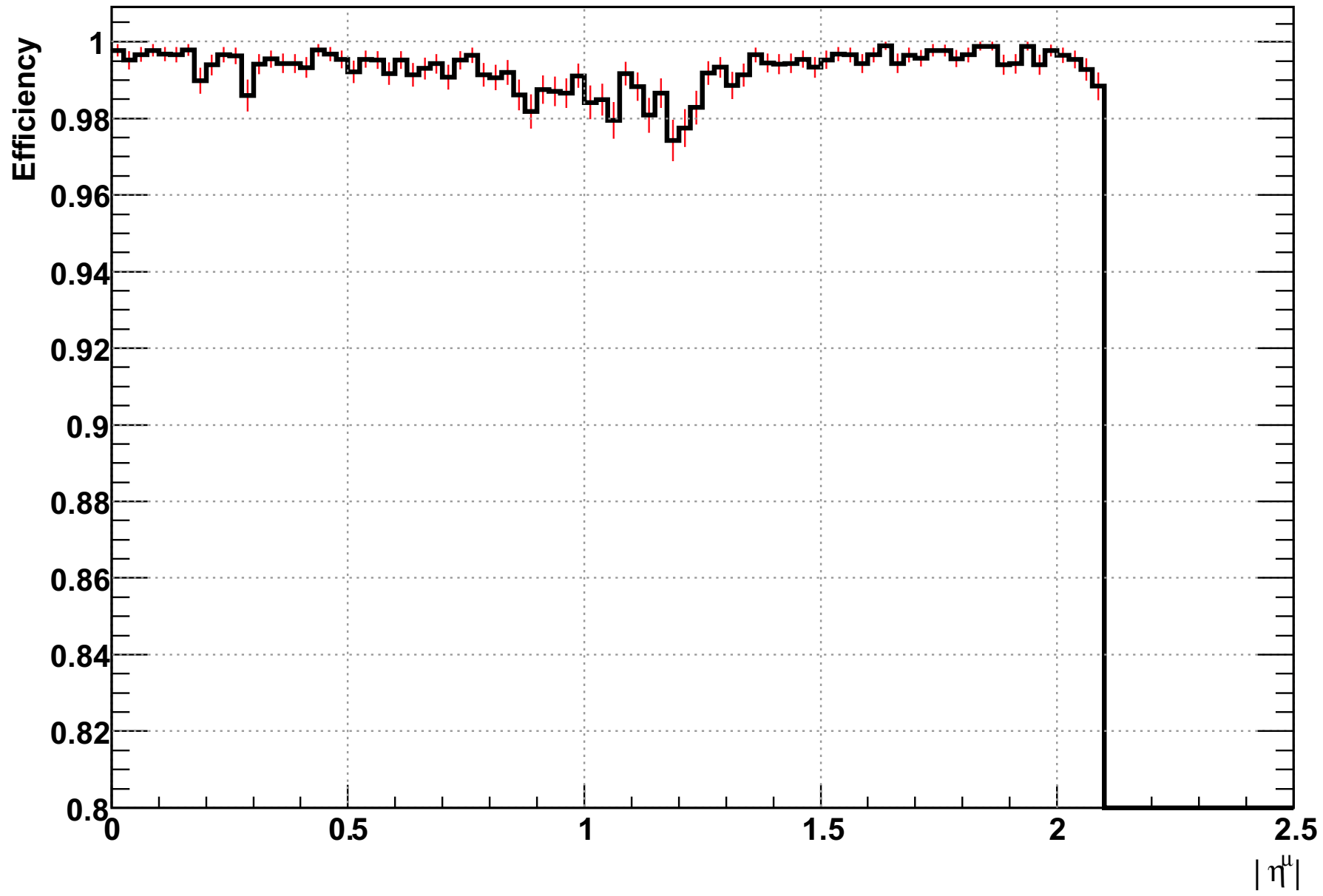

**Figure 14.** Algorithmic efficiency of the Level-3 tracking algorithm as a function of $\eta$ for single muons generated flat over $10 < p_T < 100$ GeV/$c$. No pileup was included.

The muons were generated flat in $p_T$ up to 100 GeV/c. The algorithmic efficiency is typically 99%, except in the pseudorapidity interval 0.8< $|\eta|$ <1.2, where the DT and CSC systems have to be combined to reconstruct muons and the efficiency is about 97%.

*4.1.3. Muon Isolation* The integrated rate of muons at LHC is dominated by muons from $b$, $c$, K, and $\pi$ decays, as shown in figure 15. These muons are generally accompanied by other nearby particles, so they can be suppressed by isolation cuts.
Three isolation techniques have been studied. The first (calorimeter isolation) is based on the standard technique of summing the calorimeter energy in a cone around the muon, and can be used with the stand-alone muon reconstruction at Level-2. However, as it is based on the calorimeter, this technique becomes less effective at high luminosity as more pileup is included in the sum. The second technique (pixel isolation) is based on the partial reconstruction of tracks in the silicon pixel detector; in this case, isolation is determined on the basis of the sum of the transverse momenta of the tracks in a cone around the muon. This method, which can be applied when tracker information is included at Level-3, is less sensitive to pileup, as only tracks originating from the same collision vertex are considered. However, it requires the reconstruction of three pixel hits out of the three layers in the pixel detector for every track; for this reason, it is sensitive to inefficiencies and may not be useful in staging scenarios where only two pixel layers are installed. The third technique, tracker isolation, uses full tracks reconstructed regionally. This method is more robust than pixel isolation, but is more time consuming especially at high luminosity.

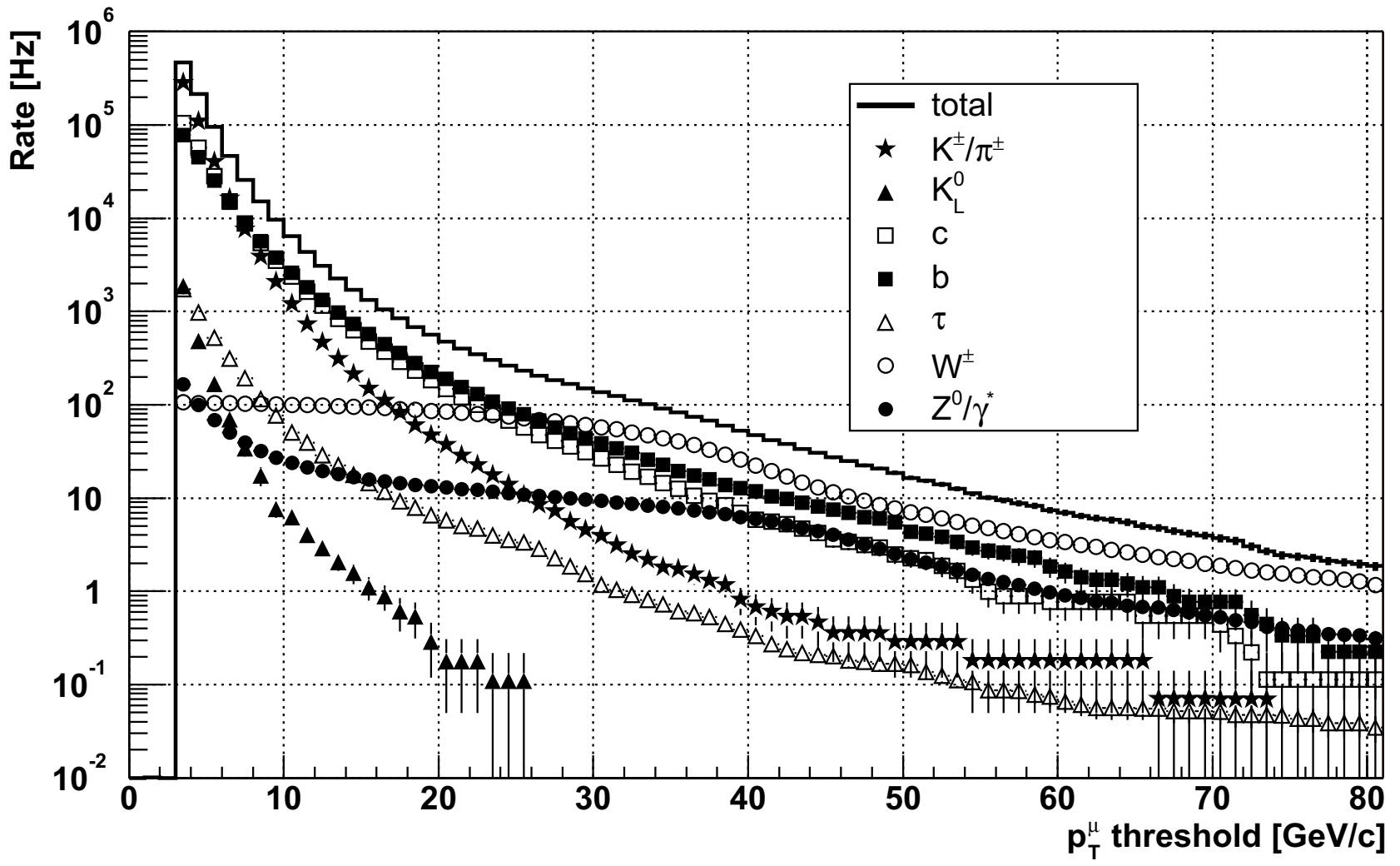

**Figure 15.** Integrated rate of single muons from PYTHIA as a function of the muon $p_T$ threshold and for luminosity $\mathcal{L} = 10^{34}\,\mathrm{cm}^{-2}\mathrm{s}^{-1}$. The breakdown of the muon sources is also shown.

For all three techniques, cones are defined by the condition $\Delta R \leq \Delta R_{max}$, where, $\Delta R = \sqrt{(\Delta\eta)^2 + (\Delta\phi)^2}$ with $\Delta\eta$ and $\Delta\phi$ the distances from the muon direction in pseudorapidity and azimuthal angle, respectively. The $\sum E_T$ deposited in the cone in the case of calorimeter isolation or the $\sum p_T$ of tracks in the cone in the case of pixel and tracker isolation are computed after subtracting the muon contribution (the so called veto value) and compared with a predefined threshold. For each algorithm, both the cone size and the thresholds are chosen by maximizing the rejection for a reference background sample while keeping the efficiency for a reference signal sample above a given nominal value (nominal efficiency). The threshold is determined independently in 52 bins in $\eta$, in order to guarantee a flat signal efficiency as a function of $\eta$.

The rejection power of muon isolation algorithms depends on the $p_T$ of the muon,

since low-$p_T$ muons are usually accompanied by low energy jets that may be below the isolation energy or momentum threshold. This is particularly relevant at Level-2, where the feed-through of low-$p_T$ muons, i.e. muons of low $p_T$ that are wrongly reconstructed as high $p_T$ muons, contaminates the $p_T$ spectrum even for high thresholds. To exclude these muons, the reference background sample is defined as a sample of minimum-bias events containing only muons with $p_T$ above 22 GeV/$c$ (16 GeV/$c$) for high (low) luminosity. The direct $W \rightarrow \mu\nu$ decay is used as reference signal since it contains well isolated muons with adequate $p_T$ spectrum.

The result of the optimization procedure is that for any predefined nominal efficiency value a cone size is chosen, with thresholds defined in bins of pseudorapidity. For all three isolation techniques, typical values of the optimal cone size $\Delta R_{max}$ vary from 0.2 to 0.3.

The main features of the three isolation algorithms are briefly described in the following sections. A detailed description can be found in reference [21].

*4.1.4. Calorimeter Isolation* The calorimeter isolation algorithm uses the muon direction at the impact point for the definition of the cone axis. The extraction of the energy deposits is done independently in the ECAL and the HCAL. The total energy is obtained as a weighted sum, $E_T = \alpha E_T{}^{ECAL} + E_T{}^{HCAL}$, where $\alpha \approx 1.5$ is found to be an optimum value for the isolation technique. Thresholds on $E$ and $E_T$ in individual ECAL crystals and HCAL towers are applied in order to reject noise and pileup deposits. The energy deposit in a small cone around the extrapolated position of the muon on the ECAL/HCAL boundary is used as veto value and subtracted from the measurement in the cone. For ECAL, the transverse energy of crystals within $\Delta R < 0.07$ is subtracted. For HCAL, where the segmentation is much coarser, the transverse energy of a single tower is subtracted, chosen as the tower with highest deposit among those whose center lies at $\Delta R < 0.1$ from the muon extrapolated point. Thresholds on the summed $E_T$ vary from 6.5 to 9 GeV for typical cone sizes of 0.2.

*4.1.5. Pixel Isolation* A pixel reconstruction algorithm looks for pixel hits compatible with tracks with transverse momenta as low as 1 GeV/$c$. The track candidates are used to fit primary vertices; track candidates with no association to reconstructed vertices are rejected. The algorithm returns a list of vertices with the corresponding tracks and their momenta. All pixel tracks contributing to the $\sum p_T$ in the cone are required to come from the same primary vertex as the Level-3 muon, thus reducing the effect of pileup. The veto value is defined as the $p_T$ of the muon candidate, i.e. of the pixel track closest in direction to the muon, within $\Delta R <0.015$. Thresholds on the summed $p_T$ vary from 1.8 to 3.8 GeV/$c$ for typical cone sizes of 0.2. The pixel isolation algorithm can also be applied to Level-2 muon candidates, but because the muon trajectory is less well determined than Level-3 muon candidates, the primary vertex requirement must be dropped and the cone size increased. The performance of the isolation algorithm is therefore different at Level-2 than at Level-3.

*4.1.6. Tracker Isolation* The tracker isolation algorithm is based on the $\sum p_T$ of tracks reconstructed in a cone around the direction of the Level-3 muon, neglecting the contribution from the muon itself. Tracks are reconstructed using regional tracking, i.e. track seeds are created using pairs of pixel hits in a region of interest. The region is defined by a vertex constraint, the minimum transverse momentum for the tracks to be reconstructed as well as a constrain on the track direction at the vertex. Thresholds on the summed $p_T$ varied from 2.0 to 3.0 GeV/c for typical cone sizes of 0.2.

*4.1.7. Performance* A general investigation of the isolation efficiency for signal muons and rejection of background is reported here, whereas the performance with respect to specific Level-1 and Level-2 triggers is included in sections 4.2 and 4.3. The isolation optimization procedure guarantees by construction that the efficiency for the reference signal ($W \to \mu\nu$ events) is flat as a function of the pseudorapidity and equal to (or greater than) the chosen nominal efficiency. The efficiency for the background depends on the $p_T$ of the muon, as shown in figure 16a for the high luminosity case when the efficiency for the reference signal is set to 97%. To show the performance of isolation algorithms, it is necessary to specify the background sample and the minimum generated muon $p_T$ to be taken into account. This is particularly relevant for the Level-2 calorimeter isolation algorithm. In this case, the rate is dominated, for any $p_T$ threshold, by the feed-through of very low-$p_T$ muons, which cannot be rejected by the isolation algorithms but should be rejected by the refined $p_T$ threshold at Level-3. Figure 16b shows the rejection power of the isolation algorithms, expressed as the efficiency for background muons in minimum-bias events with $p_T^{gen}$>22 GeV/$c$ at high luminosity, as a function of the pseudorapidity of the muon.
The efficiency for reference background muons versus the efficiency for reference signal muons is shown in figure 17 at low luminosity with $p_T^{gen}$>16 GeV/c and at high luminosity with $p_T^{gen}$>22 GeV/c. The background rejection can be adjusted by choosing different efficiencies for the reference signal.

## *4.2. Muon HLT Selection*

Here we describe a prototype inclusive muon trigger based on the reconstruction and isolation tools discussed in the previous sections to demonstrate the performance in the muon HLT.

*4.2.1. Single-muon HLT Selection* A single muon inclusive trigger is formed from the following requirements. At Level-1, low quality CSC tracks must be matched with RPC tracks by the Global Muon Trigger in order to ensure a well-measured $p_T$. Low quality here means that there are no strong requirements on the number of muon chambers used, or on which of chambers in the region crossed by the muon are used, to identify the muon. At Level-2, a muon must be reconstructed in the muon system and have a valid extrapolation to the collision vertex. In the barrel region, at least one DT (Drift Tube) track segment reconstructed is required, and the sum of the number of DT segments and RPC hits must exceed three. At Level-3, a muon must have more than 5 silicon hits in total from the pixels and silicon strips.

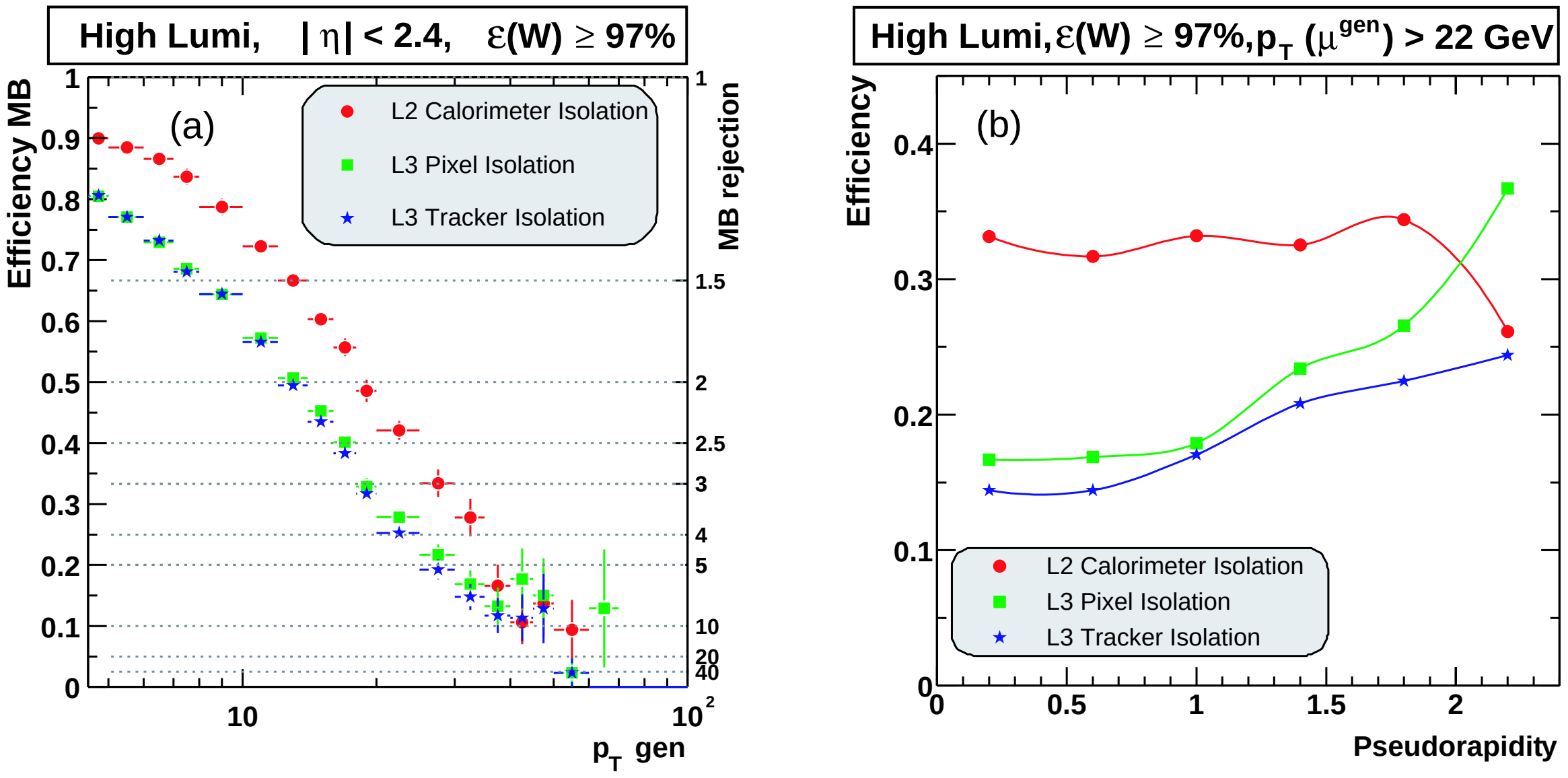

**Figure 16.** Efficiency for background muons to pass the three isolation algorithms as a function of (a) the muon $p_T$ and (b) the muon pseudorapidity, at high luminosity, for a nominal efficiency of 97% to select muons from $W$ decays.

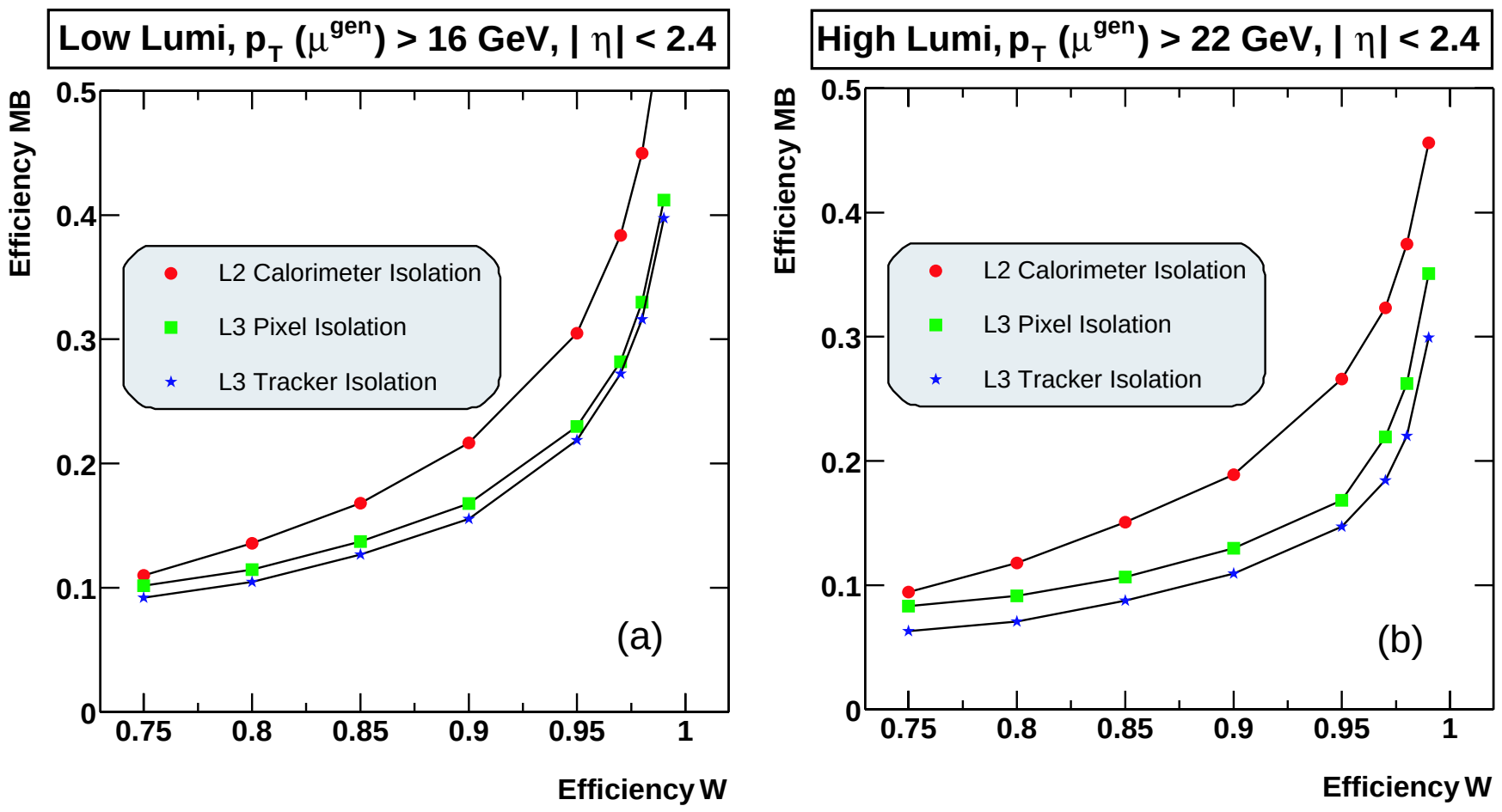

**Figure 17.** Efficiency of the three isolation algorithms on the reference background muons as a function of efficiency for the reference signal muons at (a) low and (b) high luminosity.

The overall efficiency for muons to pass the Level-1 through Level-3 single muon trigger criteria cumulatively as a function of the generated $\eta$ is shown in figure 18. Muons were generated flat in the intervals $5<p_T<100$ GeV/$c$ and $|\eta| < 2.1$ without any pileup. The average combined Level-1 through Level-3 efficiency without any requirements on the reconstructed $p_T$ is 97%, but is lower in some particular regions because of gaps in the geometrical coverage of the chambers.

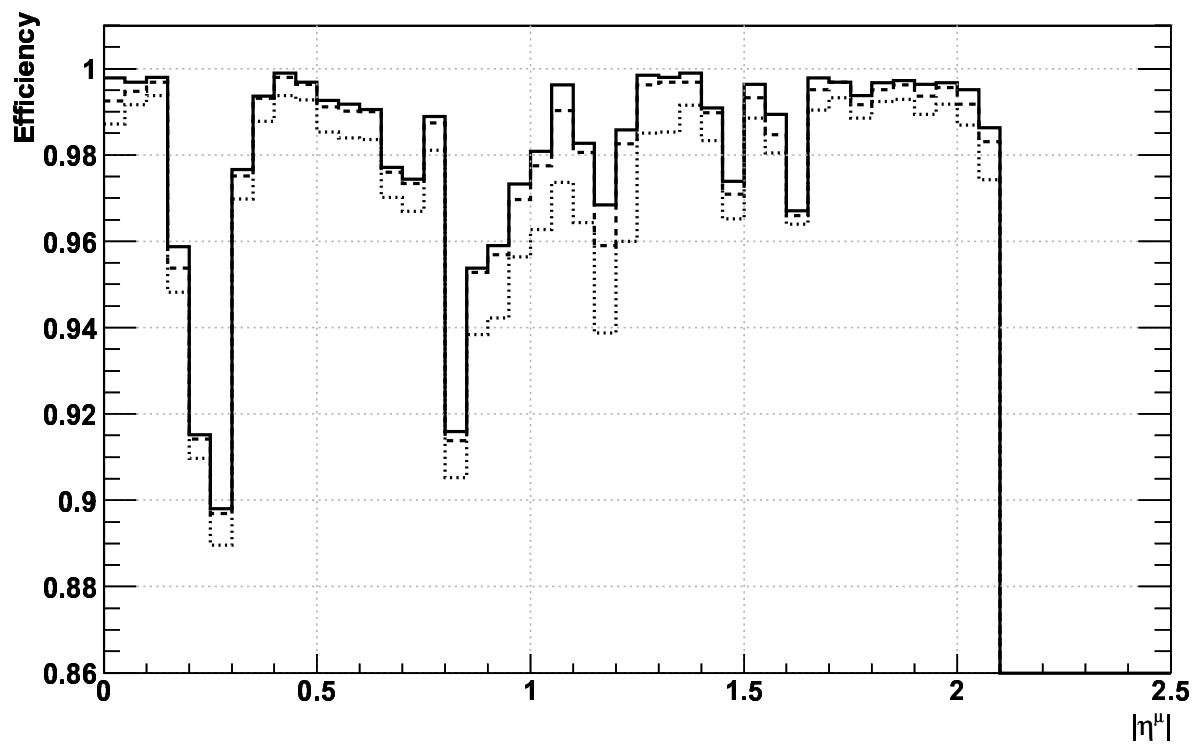

**Figure 18.** Cumulative efficiency for single muons to pass the Level-1 (solid), Level-2 (dashed), and Level-3 (dotted) triggers as a function of the generated muon pseudo-rapidity. No thresholds on $p_T$ are applied. Note the suppressed zero on the $y$-axis. The dips at $|\eta| \sim 0.3$ and 0.8 are due to gaps in the muon chamber coverage.

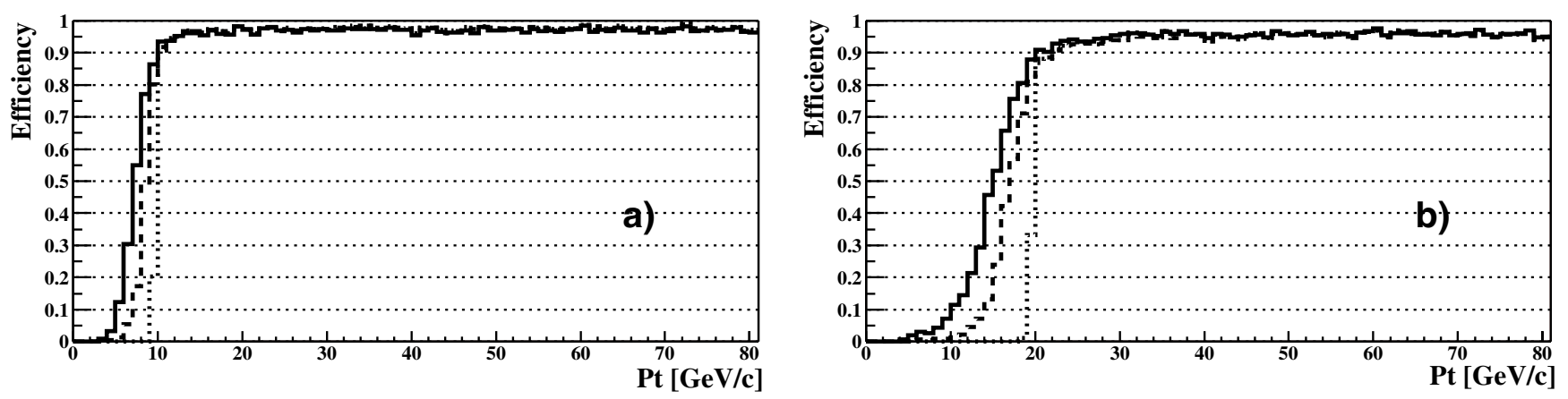

**Figure 19.** Cumulative efficiency for single muons to pass the Level-1 (solid), Level-2 (dashed), and Level-3 (dotted) triggers as a function of the generated $p_T$ for several trigger thresholds: a) $p_T$> 10 GeV/$c$, and b) $p_T$> 20 GeV/$c$.

The efficiency turn-on curves as a function of the generated $p_T$ for two different $p_T$ thresholds are shown in figure 19. The efficiency shown is the cumulative Level-1 through Level-3 efficiency. The threshold at each trigger level is defined at 90% efficiency (relative to the plateau efficiency), and it can be seen that the improved $p_T$ resolution at each successive level sharpens the turn-on curve. The efficiency at Level-3 for high $p_T$ muons is around 95% for the $p_T$values shown.

Additionally, for the HLT trigger, Level-2 muon candidates must satisfy the calorimeter isolation criteria at the 97% efficiency point for the reference signal. At Level-3, candidates must satisfy the tracker and pixel isolation criteria (hereafter collectively referred to as "tracker isolation" in the figures that follow) both at the 97% efficiency point for the reference signal.

The single muon trigger rates as a function of the $p_T$ threshold are shown in figure 20 for both low luminosity ($\mathcal{L} = 2 \times 10^{33}\,\mathrm{cm^{-2}s^{-1}}$) and high luminosity ($\mathcal{L} = 10^{34}\,\mathrm{cm^{-2}s^{-1}}$).

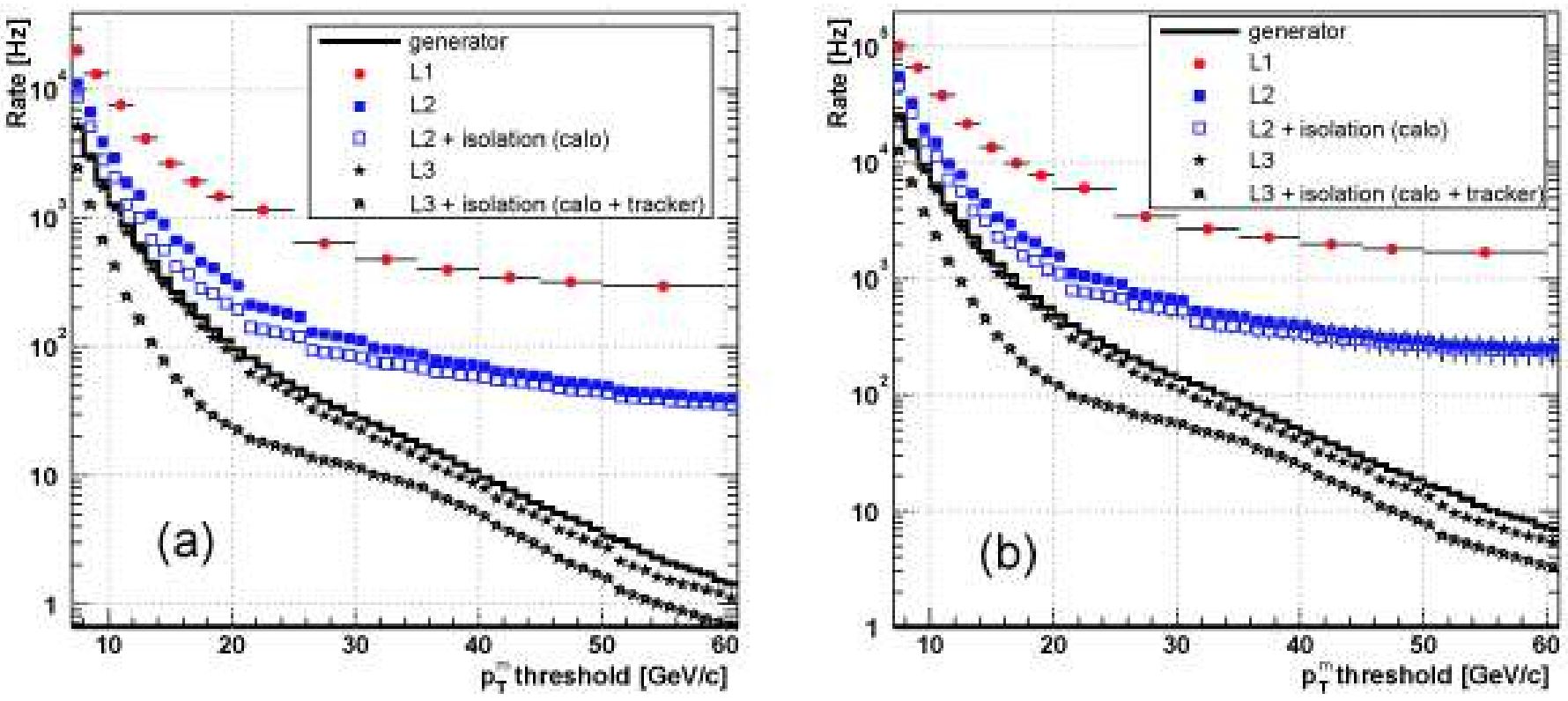

**Figure 20.** The HLT single-muon trigger rates as a function of the $p_T$ threshold for (a) low luminosity and (b) high luminosity. The rates are shown separately for Level-1, Level-2, and Level-3, with and without isolation applied at Levels 2 and 3. The rate generated in the simulation is also shown.

The rates are shown separately for Level-1, Level-2, and Level-3, with and without isolation applied at Levels 2 and 3. Also shown is the single muon rate that was generated in the simulation. The inclusive muon sample consists of QCD background events (e.g. pion and kaons decaying into muons), and genuine prompt muon production processes such as W,Z and heavy flavor production. The details are given in [9]. Due to the $p_T$resolution of the trigger the trigger rates can be higher than the generated rates. At low luminosity, the Level-3 inclusive single-muon trigger rate can be reduced to 30 Hz with a $p_T$ threshold of 18 GeV/$c$ when the isolation criteria are applied. The rate is about 100 Hz without isolation at Level-3 for the same threshold. At high luminosity, a threshold of 38 GeV/$c$ reduces the single-muon Level-3 rate to 30 Hz with isolation (50 Hz without isolation), and a threshold of 31 GeV/$c$ yields a rate of 50 Hz (100 Hz without isolation). The reason why isolation achieves less rejection at higher thresholds can be understood from figure 21, which shows the contributions to the Level-3 trigger rate at high luminosity from all sources of muons before and after all isolation criteria have been applied. The isolation criteria strongly suppress the contributions from $b, c$, K, and $\pi$ decays. This reduces the HLT rate less at high thresholds, however, where the single-muon rate is dominated by $W \rightarrow \mu\nu$ decays. After the isolation criteria, $W$ decays account for 50% (80%) of the inclusive single-muon rate for a threshold of 18 GeV/c (31 GeV/c) for low (high) luminosity.

The efficiency of the HLT single-muon trigger to select $W \rightarrow \mu\nu$ and $t\bar{t} \rightarrow \mu + X$ events, where one of the top quarks is required to decay in $W \rightarrow \mu\nu$, is shown in figure 22 as a function of the $p_T$ threshold at low luminosity. Thresholds are defined as the $p_T$ value for which the efficiency for muons is 90% of the maximum attainable efficiency. Approximately 70% of both the $W$ and top quark decays that have a muon in the fiducial region $|\eta|$ <2.1 are recorded by the isolated single-muon trigger for a $p_T$ threshold of

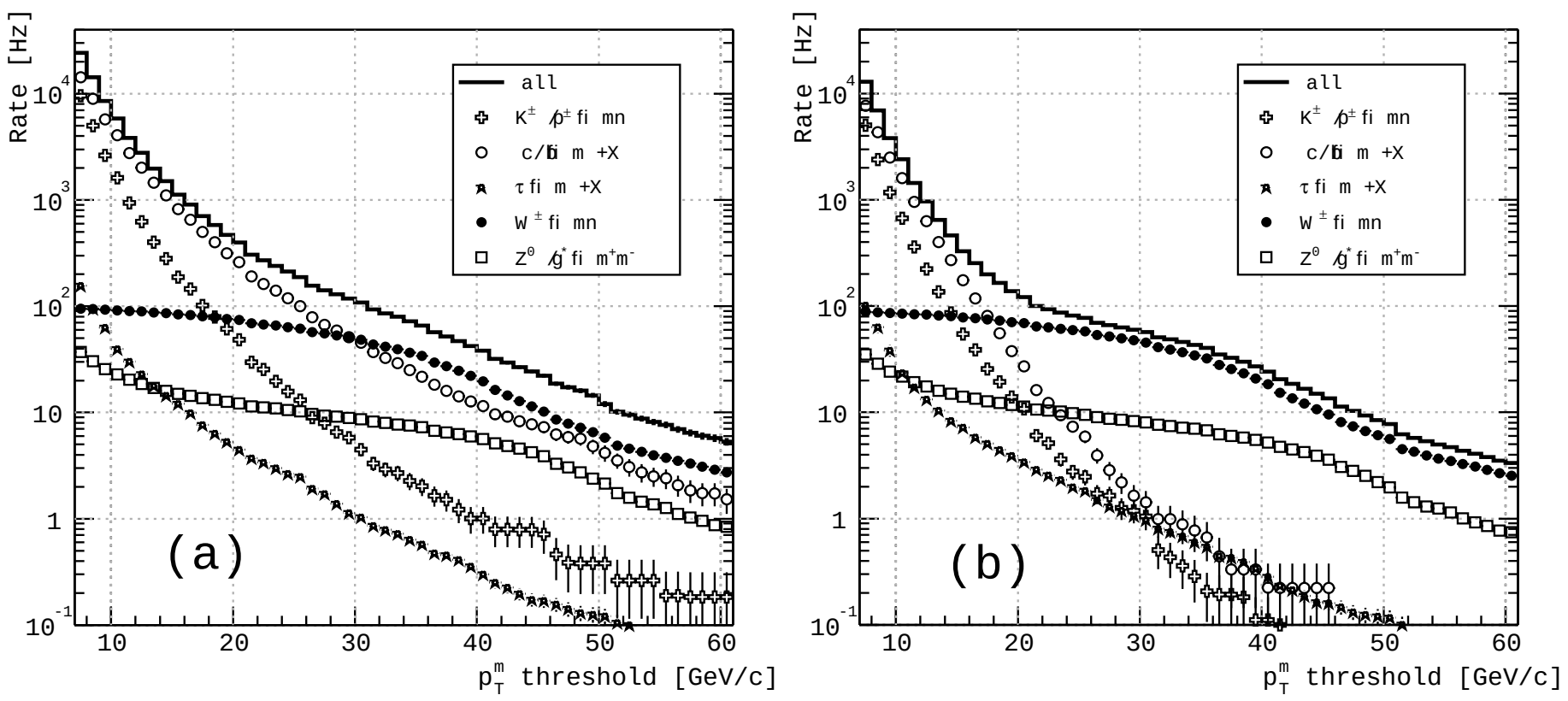

**Figure 21.** Contributions to the Level-3 trigger rate at high luminosity from all sources of muons (a) before and (b) after all isolation criteria have been applied.

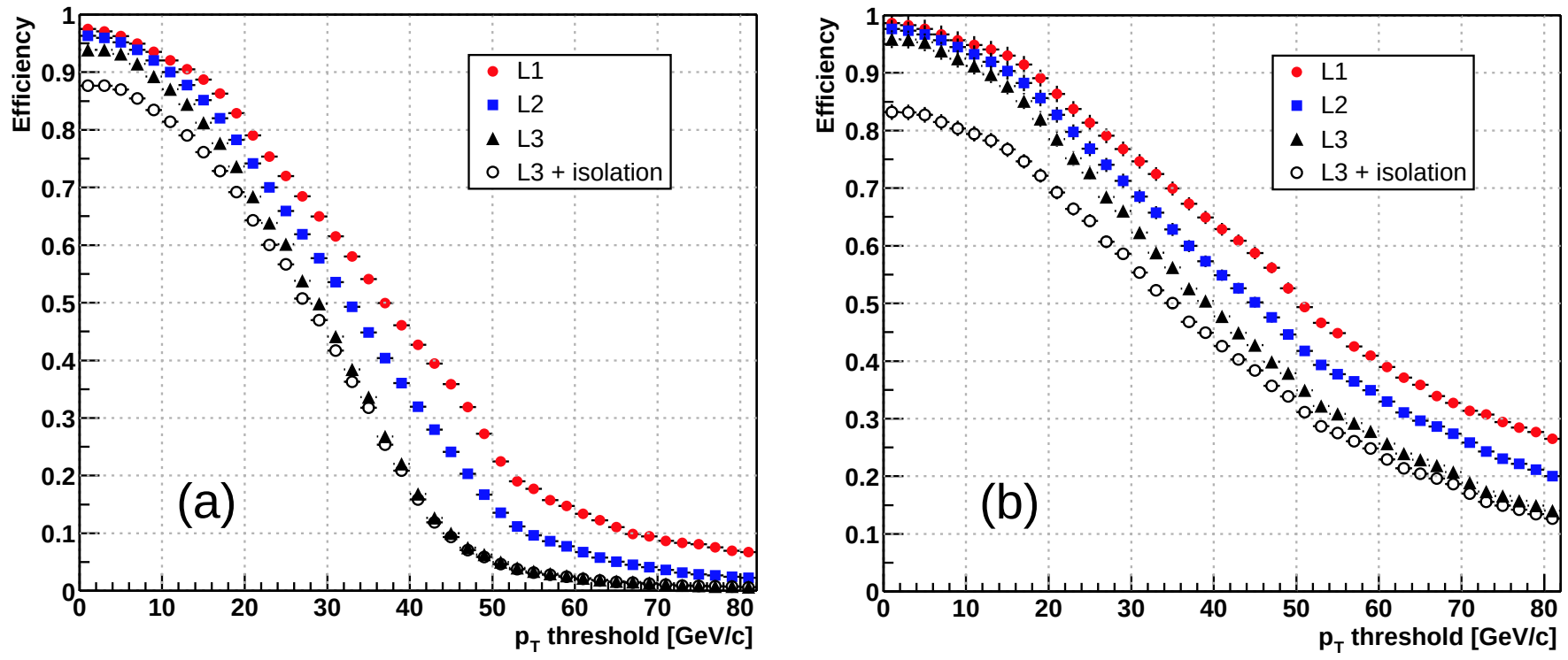

**Figure 22.** Efficiency to select (a) $W \to \mu\nu$ events and (b) $t\bar{t}\mu + X$ events (where one $W \to \mu\nu$ decay is required) as a function of the HLT single-muon $p_T$ threshold. Thresholds are defined at 90% efficiency with respect to the plateau value, and efficiencies shown are for low luminosity.

18 GeV/$c$. The efficiencies are 42% (53%) for $W$ (top) decays for a $p_T$ threshold of 31 GeV/$c$.

*4.2.2. Di-muon HLT Selection* The selection criteria for each muon in an inclusive di-muon trigger are the same as those for the single muon trigger, except that the isolation criteria need only be satisfied by one of the two muons. In addition, at Level-3, both muons are required to have originated from the same vertex in $z$ to within 5 mm (to reduce triggers from muons in separate $pp$ collisions), whereas di-muons that have $\Delta\phi$ <0.05, $|\Delta\eta|$ <0.01, and $\Delta$ $p_T$<0.1 GeV/c are rejected in order to remove fake tracks [10].

The inclusive di-muon trigger rates are studied using an inclusive di-muon sample consisting of QCD background events (e.g. pion and kaons decaying into muons), and genuine prompt muon production processes such as W,Z and heavy flavor production. More details are given in [9]. The combined single and di-muon HLT rates are shown in figure 23 as a function of both the symmetric di-muon $p_T$ threshold and the single muon $p_T$ threshold for both low and high luminosity. Candidate working points at low luminosity for a target rate of 30 Hz is a single muon $p_T$ threshold of 19 GeV/$c$ and a symmetric di-muon threshold of 7 GeV/$c$. Similarly, working points at high luminosity are a single muon $p_T$ threshold of 38 GeV/$c$ and a symmetric di-muon threshold of 12 GeV/$c$ for a combined rate of 33 Hz, or a single muon $p_T$ threshold of 31 GeV/$c$ and a symmetric di-muon threshold of 10 GeV/$c$ for a combined rate of about 55 Hz.

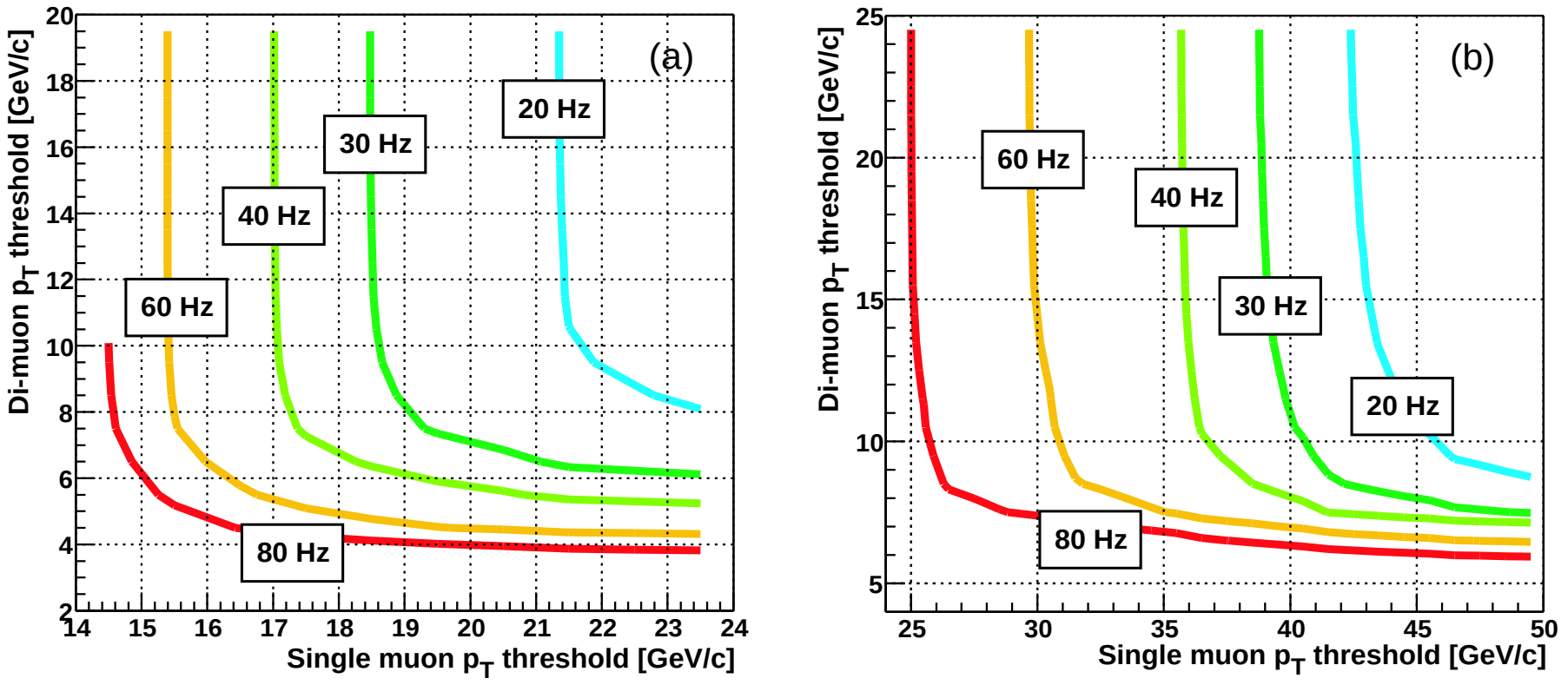

**Figure 23.** Combined single and di-muon trigger rates as a function of both the symmetric di-muon $p_T$ threshold and the single muon $p_T$ threshold for (a) low and (b) high luminosity.

### *4.3. Muon HLT Performance and Timing*

*4.3.1. Efficiencies on Higgs Signals* The single muon and di-muon efficiencies of the HLT selection have been studied as a function of the Higgs mass in the channel $H \to WW^{(*)} \to \mu\mu\nu\nu$. The efficiency versus the single muon and symmetric di-muon thresholds, is shown in figure 24a and 24b, respectively, for Higgs masses of 120, 160 and 200 GeV/$c^2$. At low luminosity, for a single muon threshold of 19 GeV/$c$ and a symmetric di-muon threshold of 7 GeV/$c$, the combined HLT efficiency is 92% for a Higgs mass of 160 GeV/$c^2$. For the channel $H \to ZZ^{(*)} \to \mu\mu\mu\mu$, the combined efficiency of the HLT is 98% (99%) for a Higgs mass of 150 GeV/$c^2$ (200 GeV/$c^2$) for the same thresholds at low luminosity. The efficiency becomes 97% (99%) for the thresholds at high luminosity. These efficiencies are relative to those events with at least one muon inside the geometric acceptance of the trigger, $|\eta| <2.1$, and all final-state muons (two or four, depending on the channel) inside the full acceptance of the muon system: $|\eta| <2.4$.

*4.3.2. Final Rates Written to Permanent Storage* The overall muon HLT rates and efficiencies are summarized in table 10 for low luminosity and table 11 for high luminosity. The Level-1 thresholds used are a single muon $p_T$ threshold of 14 GeV/$c$ and a symmetric di-muon threshold of 3 GeV/$c$ at low luminosity, and a single muon $p_T$ threshold of 20 GeV/$c$ and a symmetric di-muon threshold of 5 GeV/$c$ at high luminosity. The HLT operating point at low luminosity is a single muon $p_T$ threshold of 19 GeV/$c$ and a symmetric di-muon threshold of 7 GeV/$c$, and at high luminosity the operating point is a single muon $p_T$ threshold of 31 GeV/$c$ and a symmetric di-muon threshold of 10 GeV/$c$. The efficiencies for selecting $W$, $Z$, and top-quark decays to muons are also listed. The geometric acceptance is factored out and is 50% for $W$ decays, 71% for $Z$ decays, and 86% for top decays with at least one muon satisfying $|\eta| <2.1$.

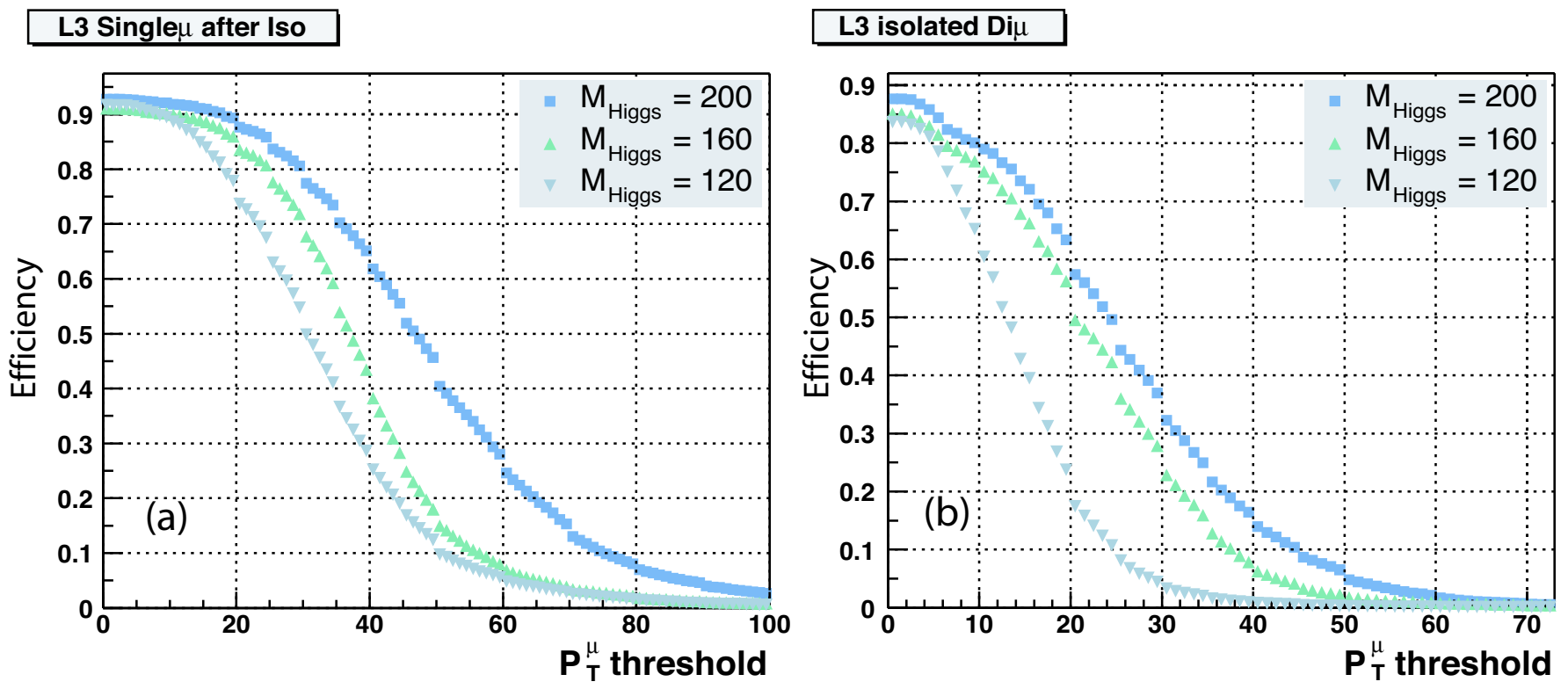

**Figure 24.** Efficiency to select $H \to WW \to 2\mu 2\nu$ decays after Level-3 and isolation cuts are applied as a function of (a) the single muon $p_T$ threshold and (b) the symmetric di-muon $p_T$ threshold. Efficiencies for Higgs masses of 120, 160, and 200 GeV/$c^2$ are shown.

**Table 10.** Muon rates and efficiencies for the low luminosity selection. Both absolute and relative efficiencies are shown, where the relative efficiency is given with respect to the preceding level (except for Level-3, which is respect to Level-2).

| Level | Rate (Hz) Single | Double | Efficiency for $W \to \mu\nu$ Rel. | Abs. | Efficiency for $t\bar{t} \to \mu + X$ Rel. | Abs. | Efficiency for $Z \to \mu\mu$ Rel. | Abs. |
|---|---|---|---|---|---|---|---|---|
| Level-1 | 2700 | 900 | | 0.90 | | 0.94 | | 0.99 |
| Level-2 | 335 | 25 | 0.89 | 0.80 | 0.93 | 0.88 | 0.99 | 0.98 |
| Calo isolation | 220 | 20 | 0.97 | 0.77 | 0.90 | 0.79 | 0.98 | 0.95 |
| Level-3 | 100 | 10 | 0.93 | 0.74 | 0.95 | 0.84 | 0.99 | 0.97 |
| Level-3+calo +tracker isolation | 25 | 4 | 0.94 | 0.69 | 0.86 | 0.72 | 0.95 | 0.92 |
| Total | 29 | | | 0.69 | | 0.72 | | 0.92 |

**Table 11.** Muon rates and efficiencies for the high luminosity selection. Both absolute and relative efficiencies are shown, where the relative efficiency is given with respect to the preceding level (except for Level-3, which is respect to Level-2).

| Level | Rate (Hz) Single | Double | Efficiency for $W \to \mu\nu$ Rel. | Abs. | Efficiency for $t\bar{t} \to \mu + X$ Rel. | Abs. | Efficiency for $Z \to \mu\mu$ Rel. | Abs. |
|---|---|---|---|---|---|---|---|---|
| Level-1 | 6200 | 1700 | | 0.82 | | 0.90 | | 0.97 |
| Level-2 | 700 | 35 | 0.70 | 0.58 | 0.83 | 0.74 | 0.96 | 0.94 |
| Calo isolation | 590 | 25 | 0.97 | 0.56 | 0.91 | 0.68 | 0.97 | 0.91 |
| Level-3 | 100 | 10 | 0.78 | 0.45 | 0.89 | 0.66 | 0.97 | 0.91 |
| Level-3+calo +tracker isolation | 50 | 5 | 0.94 | 0.42 | 0.88 | 0.58 | 0.94 | 0.86 |
| Total | 55 | | | 0.42 | | 0.58 | | 0.86 |

*4.3.3. CPU Usage* The CPU usage of the muon HLT algorithms has been benchmarked on an Intel 1 GHz Pentium-III CPU using a sample of properly weighted minimum-bias events and $W$ decays, which give the dominant contribution to the rate. The results are given in table 12 for low and high luminosity. Each line represents the average time to process an event passing the previous level. For Level-2, this represents the time to process an event passing a Level-1 single muon trigger with a threshold of $p_T$>10 GeV/$c$ ($p_T$>18 GeV/$c$) at low (high) luminosity. This value, which is lower than that proposed in the previous section, leads to a conservative CPU estimate because lower $p_T$ muons take more time to reconstruct. The same $p_T$ thresholds are applied at Level-2 and Level-3.

Also shown in the table, is the time to complete the HLT algorithm excluding the GEANE routine for propagation through iron, which is significantly less that the total time. Clearly there are substantial gains to be made by replacing GEANE with a faster method. For calorimeter isolation GEANE is used for propagation to the ECAL/HCAL boundary. The total time listed in the last row represents the average time spent per Level-1 event by the muon HLT algorithms, factoring in the rejection power at each level that reduces the rate to the next level. This time amounts to approximately 700 ms per Level-1 accept, including the time spent in GEANE.

## 5. Jet Identification

### *5.1. High Level Trigger Jet Selection*

To identify and select a jet object at HLT, a simple and fast iterative seed cone algorithm is used. The algorithm uses all the calorimeter towers and has two parameters: (i) the size of the cone $R=\sqrt{\Delta\eta^2 + \Delta\phi^2}$ in $\eta$-$\phi$ space and (ii) the seed threshold. Partons after hadronization are "clustered" at particle level into generator-jets and matched in $\eta - \phi$ with the reconstructed jets. At the parton level, the jet energy is contained in a reasonably narrow cone in $\eta - \phi$ space. Taking into account soft gluon emission the jet energy resolution is optimized for a larger cone size. However when the jets are reconstructed from calorimeter towers, the larger the cone used, the greater the noise contribution, both due to the electronics and due to the pileup plus underlying event energy depositions.

**Table 12.** CPU usage of the muon HLT algorithms at low and high luminosity on 1 GHz processors. The values given represent the average time to process an event passing the previous trigger level. Also listed is the time without the contribution of the GEANE propagation routine

| | Mean CPU Time (ms/event) $\mathcal{L} = 2 \times 10^{33}\,\mathrm{cm}^{-2}\mathrm{s}^{-1}$ $p_T >$ 10GeV/$c$ | | Mean CPU Time (ms/event) $\mathcal{L} = 10^{34}\,\mathrm{cm}^{-2}\mathrm{s}^{-1}$ $p_T >$ 18GeV/$c$ | |
|---|---|---|---|---|
| HLT Algorithm | Total | Excluding GEANE | Total | Excluding GEANE |
| Level-2 | 640 | 100 | 580 | 100 |
| Calorimeter isolation | 100 | 25 | 90 | 40 |
| Level-3 | 420 | 200 | 590 | 420 |
| Pixel isolation | 65 | 65 | 320 | 320 |
| Tracker isolation | 190 | 190 | 370 | 370 |
| Total/L1 event | 710 | 125 | 660 | 150 |

The jet energy resolution is defined as the R.M.S. of the difference between the generator-level jet $E_T$ and the reconstructed jet $E_T$ divided by the generator-level jet $E_T(s/E_T)$. A comparison at high luminosity conditions of the jet energy resolution is shown in figure 25 for 0.5 and 0.7 cone jets and for two rapidity regions, $|\eta| <1$, and $3.5< |\eta| <4.5$. For low jet $E_T$ and across both rapidity regions a cone size of 0.5 results in better jet energy resolution. The performance of the algorithm at the trigger level for the two different cone sizes is also tested with respect to the di-jet invariant mass resolution. A comparison of the reconstructed di-jet invariant mass, for the jets that most closely match the $W$ decay partons in $t\bar{t}$ events and for two different cone sizes, is shown in figure 26. The di-jet mass resolution, 18 GeV/$c^2$, (s from the Gaussian fit of the core of the distribution) is improved for 0.5 cone jets compared to 22.5 GeV/$c^2$ for 0.7 cone jets. The seed threshold causes an inefficiency for low $E_T$ jets that depends on the details of jet fragmentation. The requirement of a seed at the trigger level helps reducing the number of low $E_T$ jets (typically less than 50 GeV) resulting from noise and pileup, and speeds up the time for jet finding. Figure 27 shows the jet reconstruction efficiency as a function of the jet $E_T$ for different values of the seed threshold. Since none of the current trigger selections use jet thresholds below 40 GeV, a 2 GeV seed is chosen. At high luminosity, low $E_T$ jets from the hard scattering can be promoted to high $E_T$ jets. This can occur when either particles or a low $E_T$ jet from a pileup interaction impact the calorimeter close to one of the low $E_T$ jets of the hard interaction. The resulting higher $E_T$ jet is referred to as a "fake jet". Fake jets can also be found when particles from different interactions impact in the proximity of each other in the calorimeter. In addition, the magnetic field sweeps charged particles with transverse momentum below 0.77 GeV/$c$ out of the barrel and into the endcap region of the calorimeter. This sweeping effect prevents more than half of the charged particles pointed toward the barrel from actually striking the calorimeter, severely altering the transverse energy distribution in the underlying and pileup events. Above 40 GeV, the number of fake jets is found to be small.

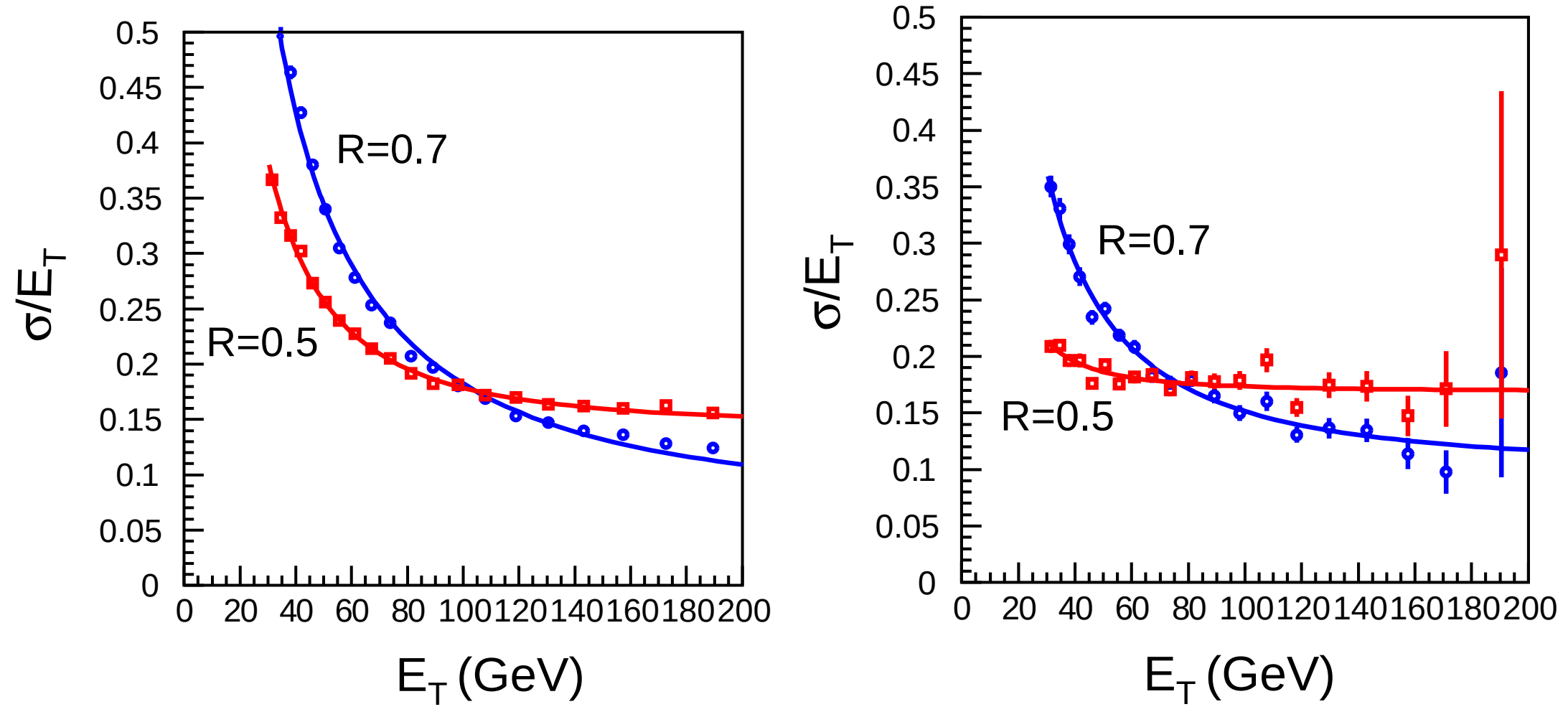

**Figure 25.** Jet $E_T$ resolution as a function of generator jet $E_T$, for two cone sizes (0.5 and 0.7) for jets with $|\eta|$ <1 (left) and 3.5< $|\eta|$ <4.5 (right).

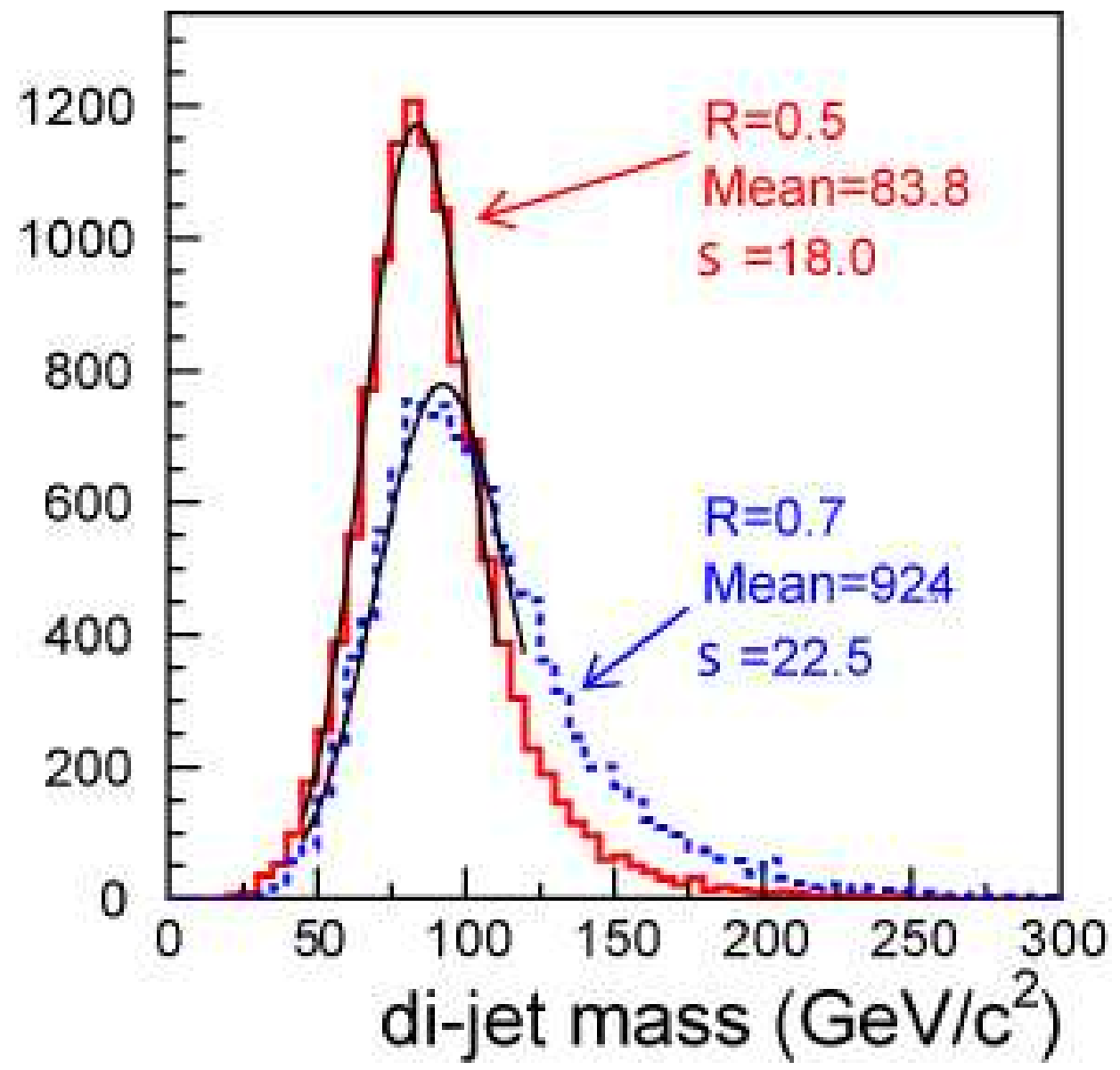

**Figure 26.** The reconstructed di-jet mass for the jets that most closely match the $W$ decay partons in $t\bar{t}$ events for the two different cone sizes as indicated ($\mathcal{L} = 10^{34}\,\mathrm{cm}^{-2}\mathrm{s}^{-1}$).

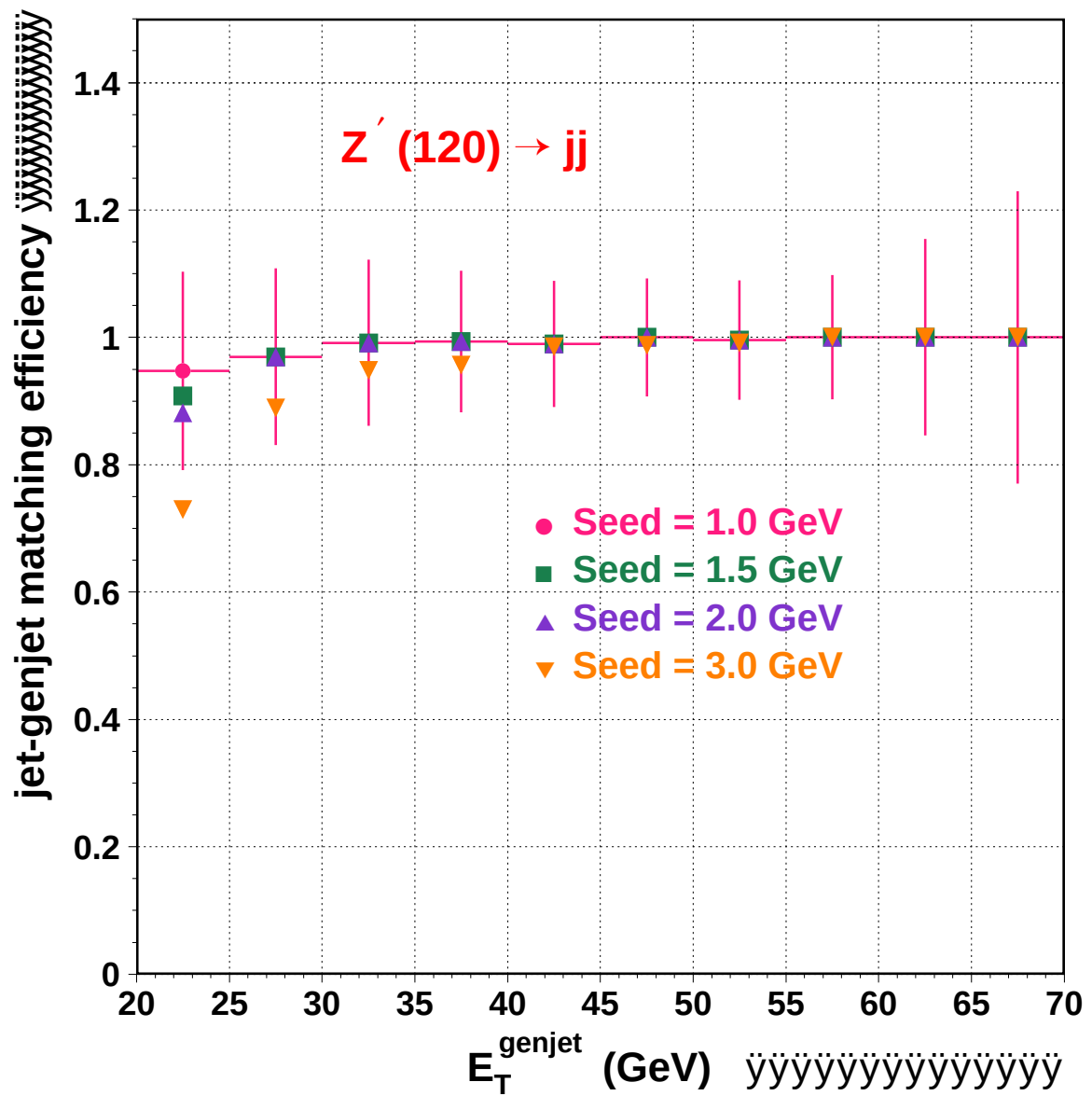

**Figure 27.** Efficiency to find an offline jet that matches a generator-level jet, as a function of the generator-level jet $E_T$ at high luminosity for different seed thresholds.

There is no unambiguous way to classify a jet as "fake" or "real". Most jets found by the jet finder contain some particles from the hard interaction and some particles from the pileup events. The degree of "fakeness" that is acceptable will vary according to the particular trigger and physics channel. In general, fake jets tend to (i) have low $E_T$, (ii) be near the HE/HF boundary and (iii) have a broader transverse profile.

As an example of fake jet rejection, figure 28 shows it is possible to keep real jets with $E_T > 20GeV$ with a reasonable efficiency while suppressing fake jets. For this study the discriminating variable used is the fraction of the jet $E_T$ contained in a 0.25 cone. For an 80% efficiency of selecting real jets (matched in $\eta$-$\phi$ with the generator level jet), about half of the fake jets are rejected.

### *5.2. Jet Energy Scale Corrections*

The CMS calorimeter system is optimized for the precision measurement of electrons and photons. It has a non-linear response to pions. Since the energy of a typical pion in a jet is roughly proportional to $1/\sin\theta$, the response of the calorimeter to jets of a given transverse energy varies with $\eta$ requiring jet energy-scale corrections. The jet energy has to also be corrected for noise in the electronics and for pileup energy.

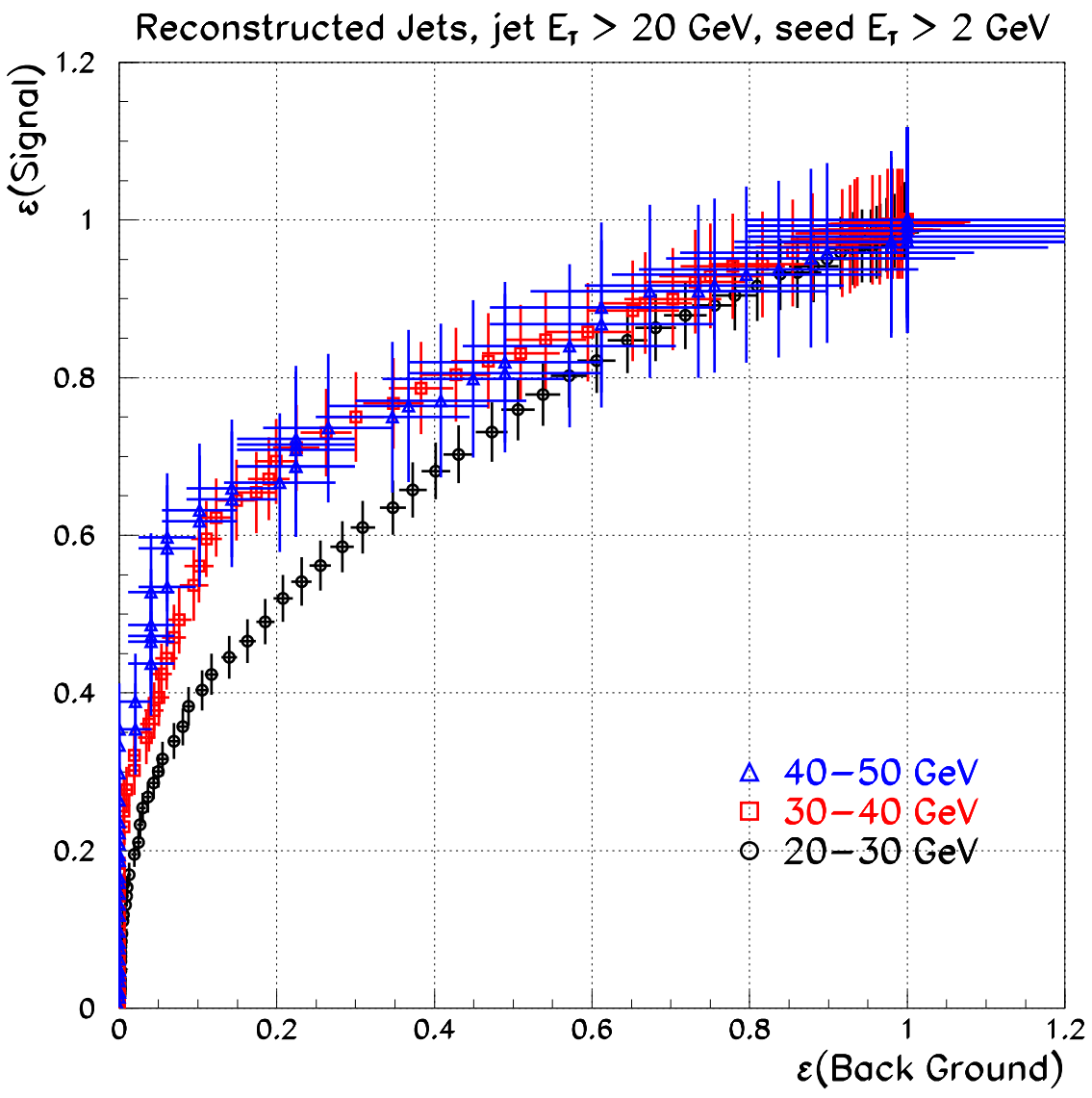

**Figure 28.** Signal efficiency versus background efficiency for removing fake jets. The fraction of the jet $E_T$ in a 0.25 cone is used as a discriminating variable.

The Monte Carlo jets are calibrated on average to the generator jet energy. This technique is expected to yield results that are virtually identical to those that would be obtained from a full calibration using photon-jet balancing in photon plus jet data. Figure 29 shows the size of the typical correction for HLT jets obtained from simulation, for low and high luminosity running conditions. The correction can be as large as 20% for low-$E_T$ (40 GeV) jets. The main effect of the corrections is the removal of the $\eta$ dependence from the trigger efficiency, while the jet energy resolution is improved only slightly. The resolution of the $E_T$ measurement for jets before and after corrections, at both low and high luminosity, is shown in figure 30 for cone 0.5 jets. The resolution for high $E_T$ jets ($E_T$>300 GeV) is similar at low and high luminosity, while the resolution at lower $E_T$ values is significantly worse at high luminosity.

## *5.3. Jet Rates*

The jet rates should be insensitive to the details of the simulation and robust against plausible changes in jet algorithms. Figure 31 shows the single jet trigger rates (at low and high luminosity) for jets reconstructed at the generator level using particles from only the hard scattering (labeled "Signal GenJet") and using generator-level particles from all interactions (labeled "in-time GenJet"). The figure also shows the HLT jet rates both before and after jet energy scale corrections. At fixed $E_T$, the high luminosity rates are about five times larger than the rates at low luminosity, as expected.
Figure 32 shows the HLT rates of single, 3- and 4-jet triggers at low and high luminosity as a function of the calibrated jet $E_T$ threshold.

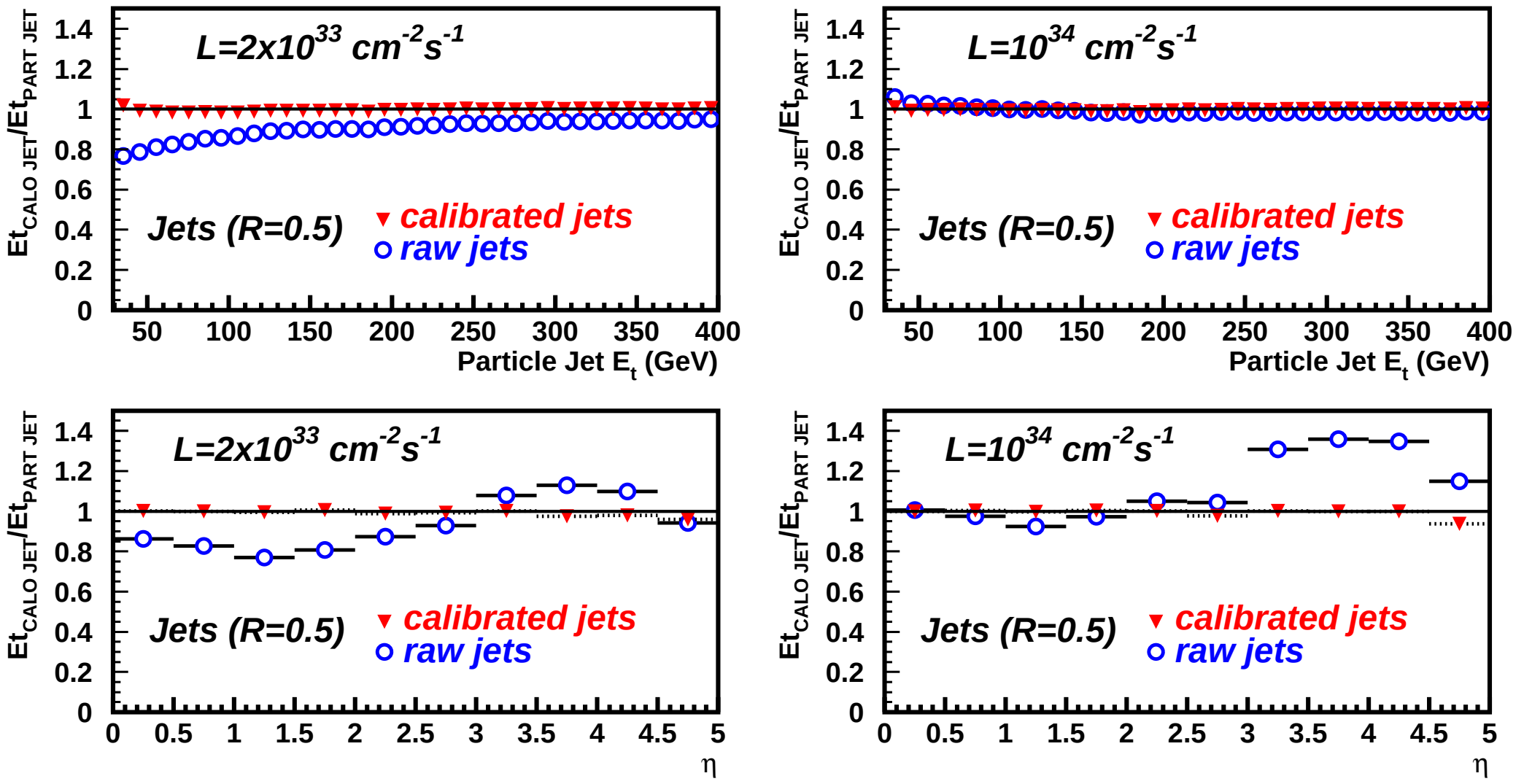

**Figure 29.** Ratio of reconstructed over generated jet $E_T$ for HLT jets versus generated $E_T$ and $\eta$, before and after jet energy scale corrections for (left) low and (right) high luminosity.

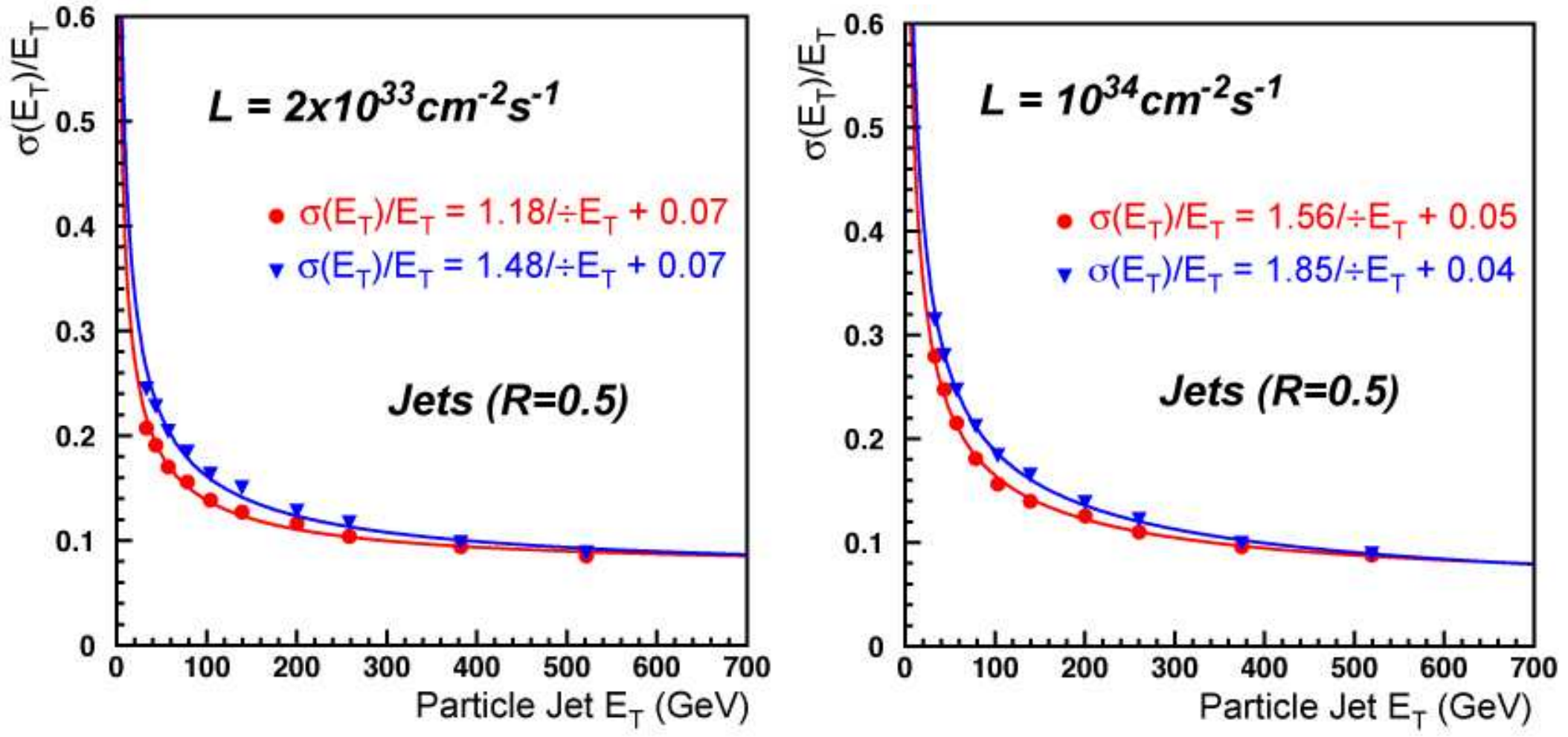

**Figure 30.** Resolution for HLT jets before (triangles) and after (circles) jet energy scale calibration. Low (left) and high (right) luminosity.

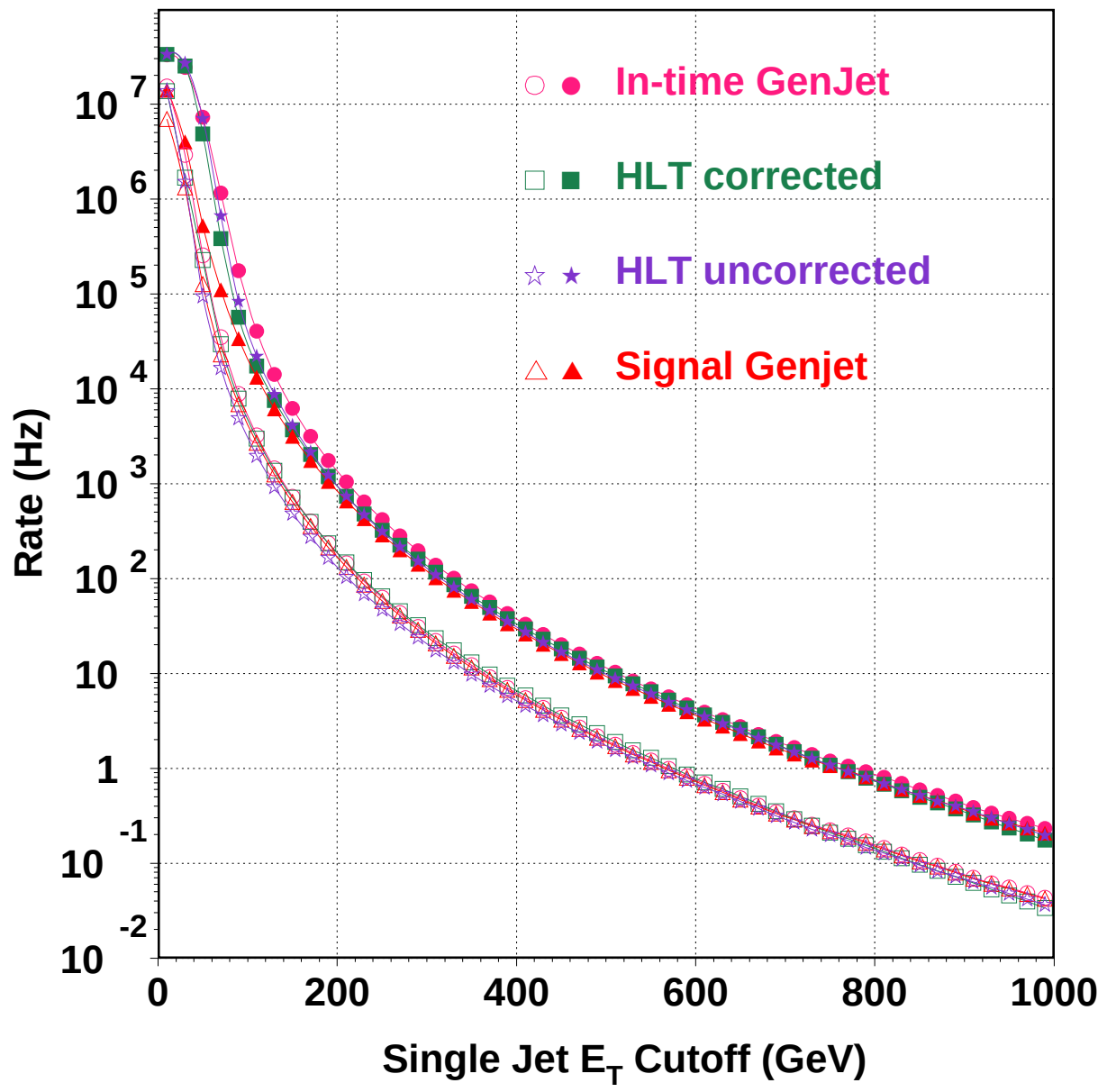

**Figure 31.** Generated single-jet rates as a function of the jet $E_T$ threshold clustering only particles from the hard scattering (GenJet) and all particles– including those from the pileup interactions (in-time GenJet). Also shown the HLT single-jet trigger rates before and after jet energy scale corrections. The open (filled) symbols correspond to the low (high) luminosity.

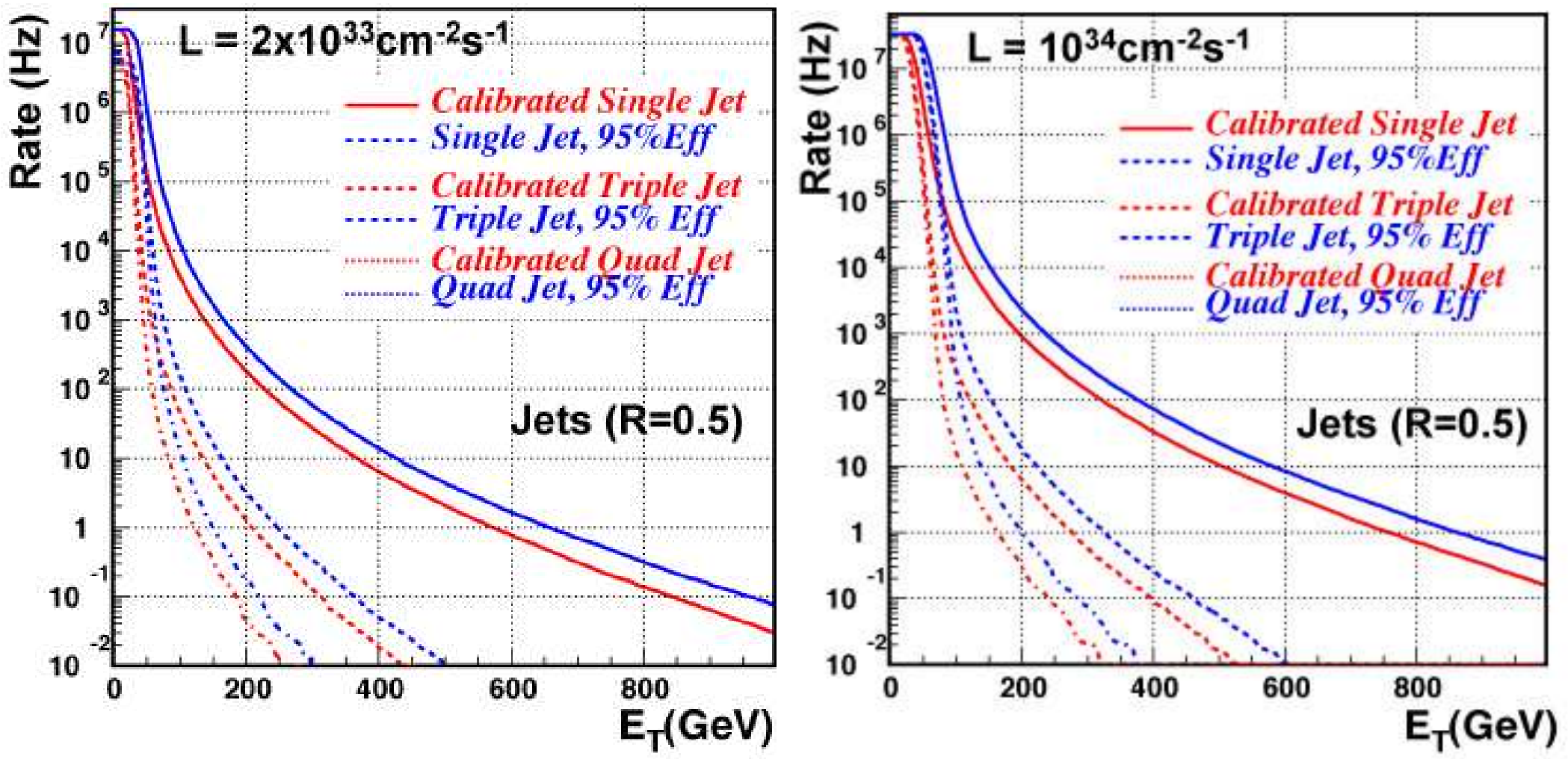

**Figure 32.** HLT rates for single, 3, and 4-jet triggers as a function of the calibrated jet $E_T$ on the $x$ axis (blue/darker curves). The same plot shows also the rate for the threshold that gives 95 % efficiency for the generator-$E_T$ on the $x$ axis (red/lighter curves). Results are for low (left) and high (right) luminosity.

**Table 13.** Jet rate summary table. The table gives the generator-level jet $E_T$ where the threshold (in GeV) on the reconstructed jet $E_T$ gives 95% efficiency. The actual value of the threshold on $E_T$ that corresponds to the 95% efficiency points is given in parenthesis. The threshold is chosen to give a rate of 1 kHz (Level-1) and 1 Hz (HLT).

| | 1-jet trigger GeV | 2-jet trigger GeV | 3-jet trigger GeV | 4-jet trigger GeV |
|---|---|---|---|---|
| low luminosity Level-1 | 177 (135) | 140 (104) | 85 (57) | 70 (45) |
| high luminosity Level-1 | 248 (195) | 199 (153) | 112 (79) | 95 (64) |
| low luminosity HLT | 657 (571) | 564 (489) | 247 (209) | 149 (122) |
| high luminosity HLT | 860 (752) | 748 (652) | 326 (275) | 199 (162) |

The generator-level $E_T$ of jets with 95% trigger efficiency (HLT or L1) that corresponds to a given threshold on calibrated and reconstructed jets is shown in figure 32 and table 13. The relationship between the value of the threshold and the generator-level jet $E_T$ is linear.

Note that the democratic 1 kHz at L1 and 1 Hz at HLT per jet trigger rate is a working example. The $E_T$ thresholds for exclusive jet triggers are very high. Most HLT physics triggers require reasonably low jet $E_T$ thresholds and additional physics objects (e.g. leptons, photons) resulting in acceptable rates.

## 6. Missing Energy Identification

The calorimeter information is used to measure the missing transverse energy ($\not\!\!E_T$, MET or $E_T^{miss}$) and identify neutrinos or other weakly interacting particles that escape detection. A simple algorithm calculates the $\not\!\!E_T$ as the negative vector sum of the $E_T$ of all towers (above a threshold of $E_T(tower) > 500$ MeV). The polar angle of each tower center is calculated with respect to $z$=0. Algorithms that incorporate jet energy scale corrections in the $\not\!\!E_T$ reconstruction have been studied.

In the refered to as "Type-I" $\not\!\!E_T$ corrections uncorrected jets with $E_T$>30 are vectorially subtracted from the $\not\!\!E_T$and corrected jets are vectorially added, while no correction is applied for the unclustered energy. In "Type-II" $\not\!\!E_T$ corrections the jet energy scale corrections for jets with $E_T$>30 GeV are used and in addition the unclustered towers are corrected. Figure 33 shows the mean difference between the generator-level $\not\!\!E_T$ and the reconstructed $\not\!\!E_T$ for three different $\not\!\!E_T$ algorithms as a function of the generator-level $\not\!\!E_T$. The study uses a SUSY sample generated with squark and gluino masses of $\sim$500 GeV/$c^2$ where the final state is multijets and large missing energy from the neutralinos. The same figure also shows the RMS of this difference. The jet energy scale corrections reset the $\not\!\!E_T$ energy scale, but do not improve the $\not\!\!E_T$ resolution.

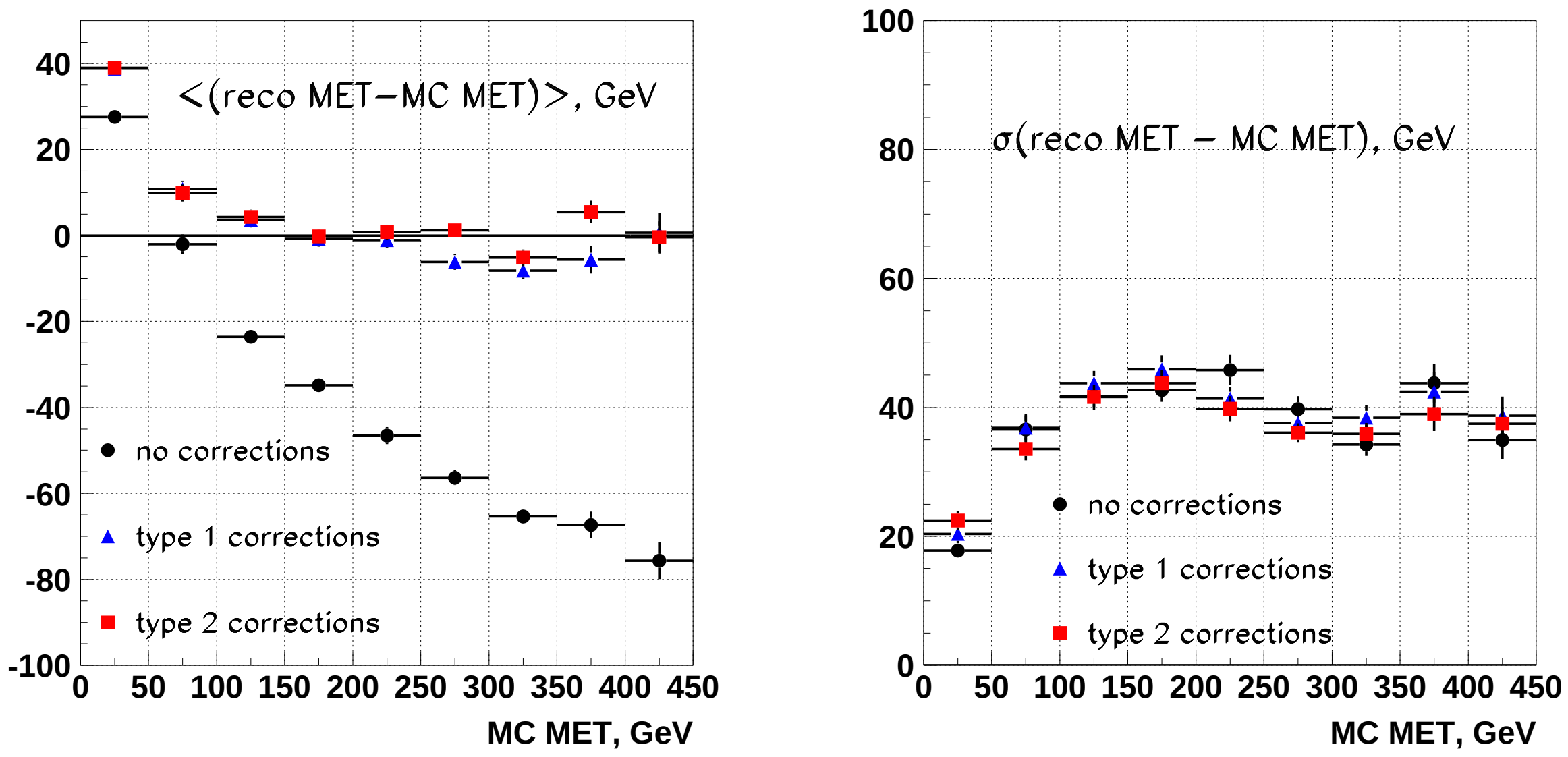

**Figure 33.** (Left) Mean difference between the generator-level $\not\!\!E_T$ and the reconstructed $\not\!\!E_T$ for three different $\not\!\!E_T$ algorithms as a function of generator-level $\not\!\!E_T$. (Right) R.M.S of the difference as a function of the generator-level $\not\!\!E_T$. The analysis is done for $\mathcal{L} = 2 \times 10^{33}\,\mathrm{cm}^{-2}\mathrm{s}^{-1}$.

### *6.1. $\not\!\!E_T$ Rates*

A large source of background for $\not\!\!E_T$ triggers is inclusive di-jet production. Large mismeasurements are expected when one jet impacts in a low response part of the calorimeter in which case the $\not\!\!E_T$ will tend to be along the direction of the under-measured jet. At large $\not\!\!E_T$ efficient rejection of mismeasured QCD events can be achieved using the $\Delta\phi$ correlation between the leading and second leading jet.

Figure 34 shows the rates when a jet with $E_T$ above a given threshold is also required as part of the trigger. To facilitate comparisons between different algorithms and experiments we plot the rates as a function of the requirement that corresponds to 95% efficiency for a given generator-level $\not\!\!E_T$. The reference process used to define the mapping between generator-level $\not\!\!E_T$ and reconstructed $\not\!\!E_T$ is Higgs production via vector-boson fusion and decay in $WW$ with a final state of two leptons and missing energy. The rate plots depend on the signal sample that was used to perform this mapping. Above generator-level $\not\!\!E_T$ of about 50 GeV, the relationship between the cut value and the generator-level $\not\!\!E_T$ is approximately linear. Figure 35 shows these rates at both low and high luminosity for Level-1 and HLT $\not\!\!E_T$. Studies of explicit physics HLT selection with $\not\!\!E_T$ are given in section 9.

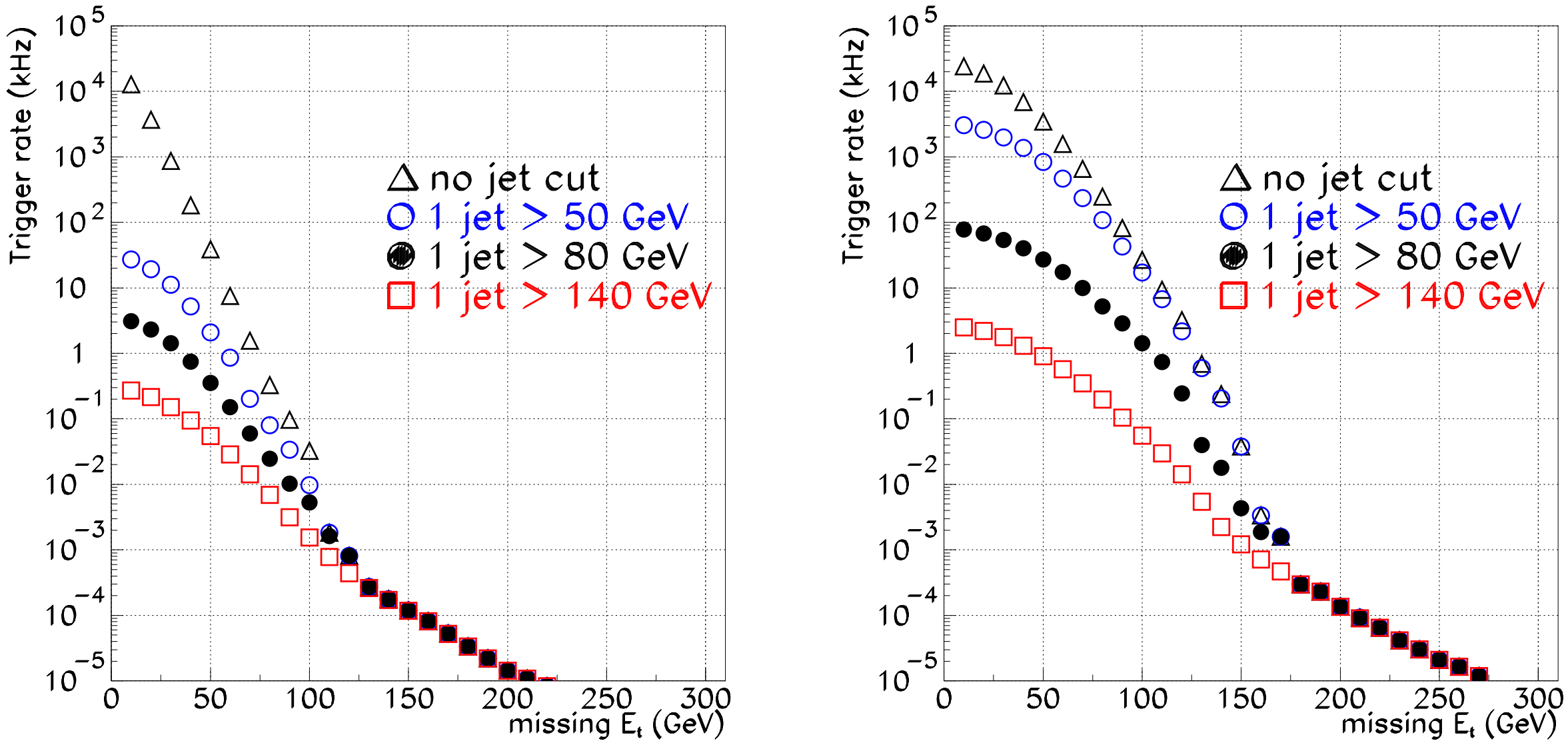

**Figure 34.** Event rates as function of $\not\!\!E_T$ when requiring a jet above a threshold as marked for (left) low luminosity and (right) high luminosity.

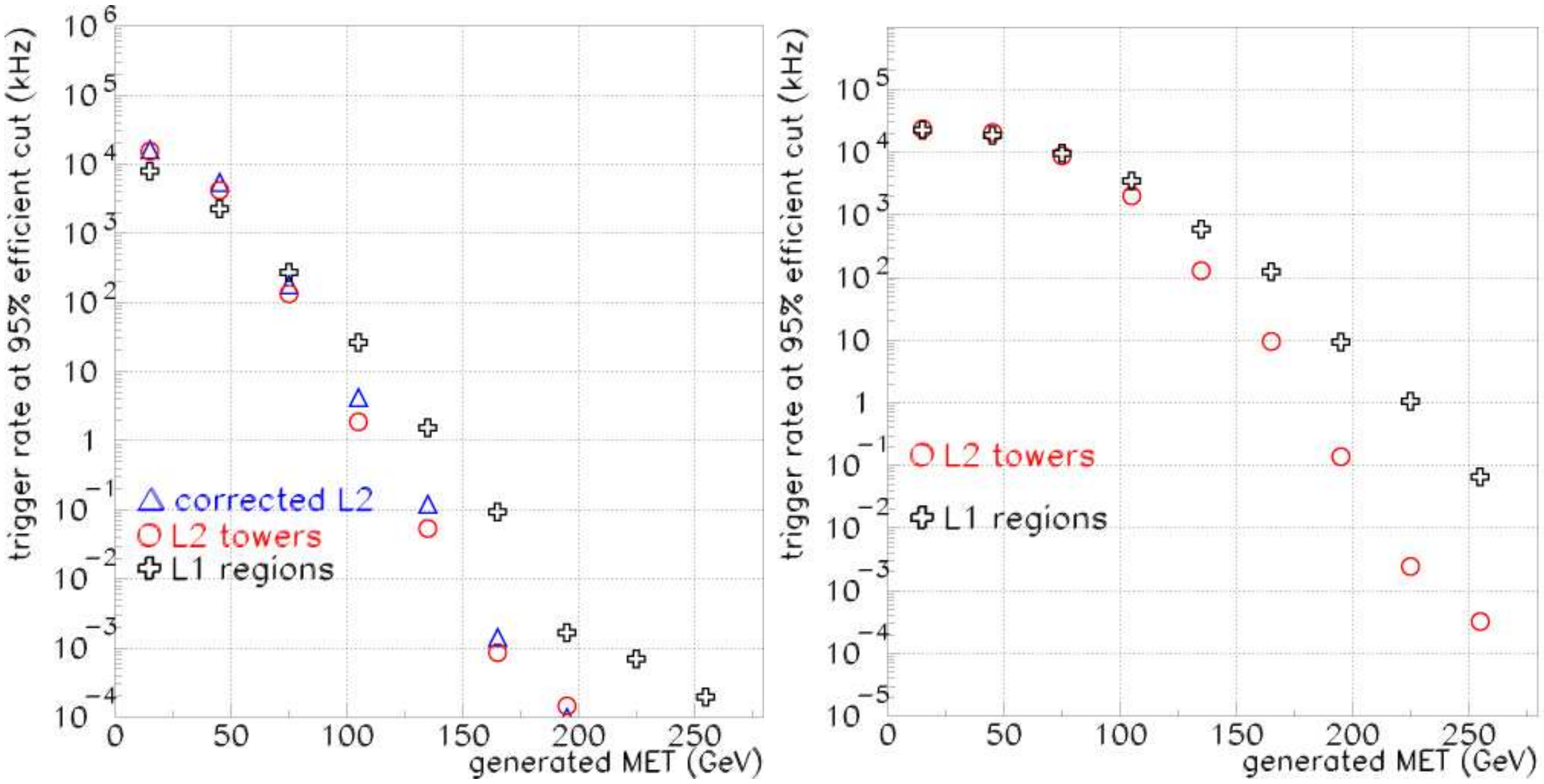

**Figure 35.** Rate versus the $\not\!\!E_T$ requirement that gives 95% efficiency for a given generated $\not\!\!E_T$ at low (left) and high (right) luminosity. The reference sample used for the mapping of the generated and the offline $\not\!\!E_T$ is a Higgs sample with di-leptons plus missing energy in the final state. The crosses correspond to the Level-1 rate, the circles to the HLT rate, and the triangles to the rate for a corrected $\not\!\!E_T$ algorithm applied at the HLT.

**Table 14.** CPU requirements for jet and $\not\!\!E_T$ reconstruction in the HLT. Times are in 1 GHz CPU ms.

| Data Sample | $A/H \rightarrow \tau\tau$ $M_H$=500 GeV/$c^2$ | | inclusive di-jets 50<$p_T{}^{hard}$<80 GeV/$c$ | |
|---|---|---|---|---|
| Luminosity | $2 \times 10^{33}\,\mathrm{cm}^{-2}\mathrm{s}^{-1}$ | $10^{34}\,\mathrm{cm}^{-2}\mathrm{s}^{-1}$ | $2 \times 10^{33}\,\mathrm{cm}^{-2}\mathrm{s}^{-1}$ | $10^{34}\,\mathrm{cm}^{-2}\mathrm{s}^{-1}$ |
| tower-building | 26 | 38 | 24 | 36 |
| jet reconstruction | 9 | 34 | 9 | 45 |
| $\not\!\!E_T$ calculation | 5 | 7 | 5 | 7 |
| total jet/$\not\!\!E_T$ reconstruction | 14 | 42 | 14 | 43 |

### *6.2. CPU Timing Studies for Jet/$\not\!\!E_T$ Reconstruction*

The results of the CPU timing studies (using a 1 GHz CPU) on jet and missing transverse energy reconstruction algorithms are shown on table 14. The CPU time needed to reconstruct jets and $\not\!\!E_T$ is about the same for the signal and background samples. At low luminosity, tower-building takes about 75% of the reconstruction time, and jet and $\not\!\!E_T$ reconstruction about 25%. At high luminosity, tower-building and jet-finding each take about 50% of the time, while the time required for the reconstruction of $\not\!\!E_T$ is negligible.

## 7. Identification of $\tau$-jets

The HLT $\tau$ selection and identification algorithms are designed to select one or two isolated $\tau$-leptons such as those expected from the MSSM Higgs decays $A^0/H^0 \rightarrow \tau\tau$ and $H^+ \rightarrow \tau\nu$. For these benchmark channels the final-states considered are one $\tau$-jet, a lepton ($e/\mu$) plus a $\tau$-jet or two $\tau$-jets. The algorithms are studied at both low and high luminosity. In the former case two detector configurations are considered: the full pixel system and the possible staged pixel system [3] (see also section 1). We assess the performance of the algorithms by measuring the selection efficiencies of the benchmark signal samples $A^0/H^0 \rightarrow \tau\tau$ and $H^+ \rightarrow \tau\nu$ and the corresponding efficiency of the QCD di-jet background. Attention is paid in minimizing the CPU time required for the selection. Both signal and QCD di-jet background samples are required to pass the Level-1 trigger selections discussed in section 1.
The QCD di-jet event samples were generated with PYTHIA 6.2.2 in several different bins of $p_T{}^{hard}$. Most of the contribution to the Level-1 $\tau$-trigger rate (∼89-95%) results from three momentum regions: $50 < p_T{}^{hard} < 80$ GeV/c, $80 < p_T{}^{hard} < 120$ GeV/c and $120 < p_T{}^{hard} < 170$ GeV/c. We use these bins to evaluate the QCD background rejection at the HLT selection. We select signal events at the generator level with loose

The nominal pixel detector configuration with three barrel layers and two forward disks is referred to as the full pixel detector whereas the start-up detector with only two barrel layers and one forward disk is referred to as the staged pixel detector configuration.

offline analysis requirements on leptons ($e$ and $\mu$) and $\tau$-jets. [23, 24, 25]:

- $A^0/H^0 \to \tau\tau \to \ell + \tau$-jet
  - ${p_T}^{\ell} >$ 14 GeV/$c$,
  - ${p_T}^{\tau-\mathrm{jet}} >$ 30 GeV/$c$,
  - $|\eta^{\ell}|$ <2.4,
  - $|\eta^{\tau-\mathrm{jet}}|$ <2.4
- $A^0/H^0 \to \tau\tau \to 2\tau$-jet :
  - ${p_T}^{\tau-\mathrm{jet}} >$ 45 GeV/$c$,
  - $|\eta^{\tau-\mathrm{jet}}|$ <2.4
- $H^+ \to \tau\nu \to \tau$-jet :
  - ${p_T}^{\tau-\mathrm{jet}} >$ 80 GeV/$c$,
  - $|\eta^{\tau-\mathrm{jet}}|$ <2.4

The identification of $\tau$-jets involves information from both the calorimeter and tracking detectors. In the analyses that follow "first (second) Level-1 $\tau$-jet" is the first (second) jet from the $\tau$-jet list provided by the Global Calorimeter Trigger (section Appendix A). Jets in this list are sorted in descending ordered in $E_T$. Correspondingly "first (second) Calo jet" is a jet reconstructed with the calorimeter in a region centered at the first (second) Level-1 $\tau$-jet.

### *7.1. Calorimeter-based $\tau$ selection*

The $\tau$-lepton decays hadronically 65% of the time, producing what we refer to as a $\tau$-jet. This is a jet-like cluster in the calorimeter containing a relatively small number of charged and neutral hadrons. When the $p_T$ of the $\tau$-jet is large compared to the $\tau$ lepton mass, these hadrons have relatively small momentum in the plane transverse to the $\tau$-jet axis. In 77% of hadronic $\tau$ decays, the $\tau$-jet consists of one charged hadron and a number of $\pi^0$'s (one prong decays). Because of these features hadronic $\tau$ decays produce "narrow" jets in the calorimeter. About 90% of the $\tau$-jet energy is contained in a small cone of radius $\Delta R \equiv \sqrt{\Delta\eta^2 + \Delta\phi^2}$ 0.15 to 0.20. In $\tau$-jets with $E_T$> 50 GeV 98% of the transverse energy is contained in a cone of 0.4. Therefore we use as a selection criterion the level of isolation of the $\tau$-candidate jet in the electromagnetic calorimeter. We define $P_{isol}$ as:

$$P_{isol} = \sum_{\Delta R<0.4} E_T - \sum_{\Delta R<0.13} E_T$$

where the sums are over transverse energy deposits in the electromagnetic calorimeter, and $\Delta R$ is the distance in $\eta - \phi$ space from the $\tau$-jet axis. The identification of a $\tau$-jet at the HLT begins with the reconstruction of a jet in a region centered at the Level-1 $\tau$-jet. The iterative cone algorithm ([26]) with a cone size of 0.6 is used. The algorithm uses as input only the calorimeter towers located within a cone of radius 0.8 around the Level-1 $\tau$-jet direction. Restricting the number of towers considered by the jet finder ("regional jet finding") results in a considerable speed-up of the jet finding process. For each jet found, the electromagnetic isolation parameter $P_{isol}$ is calculated. Jets with $P_{isol} < P_{isol}^{cut}$ are considered as $\tau$ candidates.

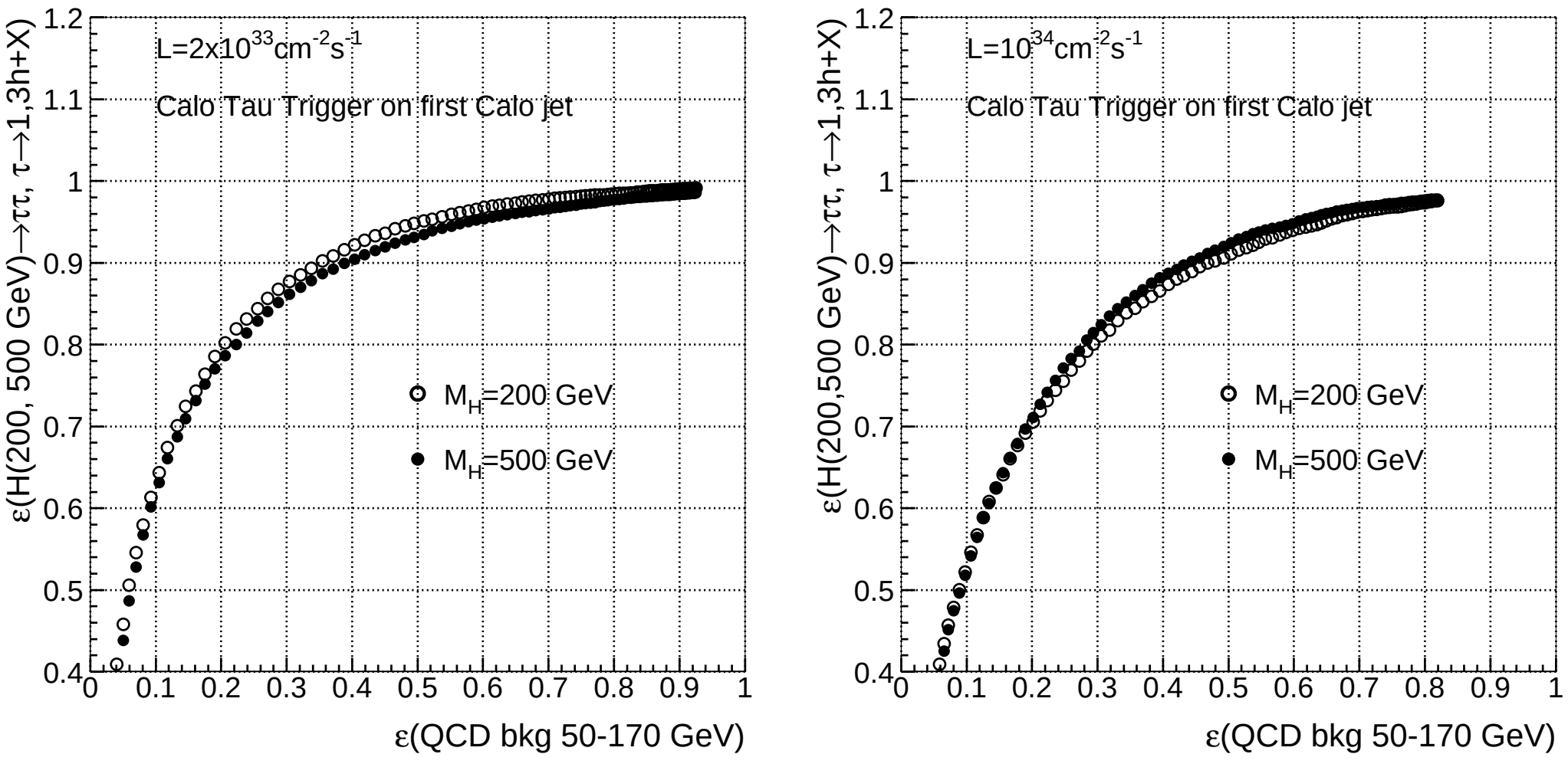

**Figure 36.** Efficiency of the *Calorimeter Tau* trigger when applied to the first calorimeter jet in $A^0/H^0 \rightarrow \tau\tau \rightarrow 2\tau$-jet and QCD di-jet events. $M_H$= 200 and 500 GeV/$c^2$ for low (left) and high (right) luminosity.

**Table 15.** CPU time in ms of a 1 GHz CPU, for $\tau$ -jet identification with the calorimeter in QCD di-jet events.

| | Time (ms) | |
|---|---|---|
| **Reconstruction step** | **$2\times10^{33}$ cm$^{-2}$s$^{-1}$** | **$10^{34}$cm$^{-2}$s$^{-1}$** |
| Build calorimeter towers | 24 | 39 |
| Regional finding of *Calo* jets and $P_{isol}$ calculation | 9 | 15 |
| Total time | 33 | 54 |

We refer to the identification of a $\tau$-jet using the calorimeter as the "Calorimeter Tau" trigger (also "Calo Tau"). The efficiency of the *Calorimeter Tau* trigger for the $A^0/H^0 \rightarrow \tau\tau \rightarrow 2\tau$-jet and QCD di-jet background events for two values of the Higgs mass when the value of the ${P_{isol}}^{cut}$ is varied (between 1 GeV and 20 GeV in 0.2 GeV steps) is shown in figure 36 for low and high luminosity. In the analysis the calorimeter $\tau$ identification is applied to the first calorimeter jet. The *Calorimeter Tau* trigger optimization can be found in [27]. The efficiency of the Higgs selection is largely independent of the Higgs mass when $m_{higgs} > 200$ GeV. The entire selection procedure is very fast, as is shown in table 15 where the times required to build calorimeter towers, for regional jet-finding and for the calculation of the isolation parameter $P_{isol}$ are listed.

### *7.2. Pixel-based $\tau$ selection*

The identification of a $\tau$-jet using charged particle tracks is also based on isolation criteria. Calorimeter triggers provide a region in which isolated groups of tracks that are well matched to the jet axis given by the calorimeter can be searched for. The corresponding isolation criteria require the reconstruction of low-$p_T$ tracks with good efficiency and an acceptably low fake rate. A precise measurement of the track $p_T$ is not necessary. A fast track-finding algorithm using only pixel data meets these requirements. Identification of $\tau$-jets with the pixel detector will be referred to as the "Pixel track Tau" trigger.

The principle of $\tau$-jet identification using the pixel detector is shown in figure 37. The direction of the $\tau$-jet is defined by the axis of the calorimeter jet (denoted "Lvl-2 $\tau$-jet axis"). The track-finding algorithm first reconstructs all track candidates ("pixel lines") and then the interaction vertices from the tracks using a histogramming method. Track candidates in a matching cone $R_m$ around the jet direction and above a threshold ${p_T}^m$ are considered in the search for the signal tracks. The leading signal track ($tr_1$ in figure 37) is defined as the track with the highest $p_T$. Similarly, the vertex from which the leading signal track originates is considered to be the signal vertex ($SV$). Any other track from the $SV$ which is within the narrow signal cone $R_s$ around $tr_1$ is assumed to originate from the $\tau$ decay. Tracks consistent with the $SV$ ( within 2 mm of the $SV$) and with transverse momentum above a threshold ${p_T}^i$ are then searched for inside a larger cone of opening angle $R_i$. If no tracks are found in the $R_i$ cone except for the ones which are already in the $R_s$ cone, the isolation criteria are fulfilled. The optimal values of the cone size $R_m$, signal cone size $R_s$, and the thresholds ${p_T}^m$ and ${p_T}^i$ for 1- and 3-prong $\tau$-jets from $A^0/H^0 \to \tau\tau$ decays and for Higgs mass $M_H \geq 200$ GeV/$c^2$ are: $R_s = 0.07$, $R_m = 0.10$, ${p_T}^m = 3$ GeV/$c$ and ${p_T}^i = 1$ GeV/$c$ [28]. The only remaining free parameter is the size of the isolation cone $R_i$. The efficiency of the *Pixel track Tau* trigger, at both low and high luminosity is shown in figure 38 when the size of the isolation cone is varied in the range of 0.20-0.50. Pixel $\tau$ identification has been applied to the first calorimeter jet in $A^0/H^0 \to \tau\tau \to 2\tau$-jet and QCD di-jet events. The degradation of the performance at high compared to low luminosity is due to high pixel detector occupancy that results in readout inefficiency as well as the contamination of charged-particle tracks from pileup in the isolation cone coming from a vertex other than the hard collision vertex. The selection is independent of the Higgs mass. Comparison between the full and staged pixel systems shows that at constant QCD background rate, the signal efficiency is degraded by approximately 10% in the staged configuration. The timing performance of the *Pixel track Tau* trigger algorithm is shown in table 16. It includes the time to reconstruct clusters from the digitized pixel data and the reconstruction time of the pixel lines and vertices. The time used by the isolation algorithm itself is negligible.

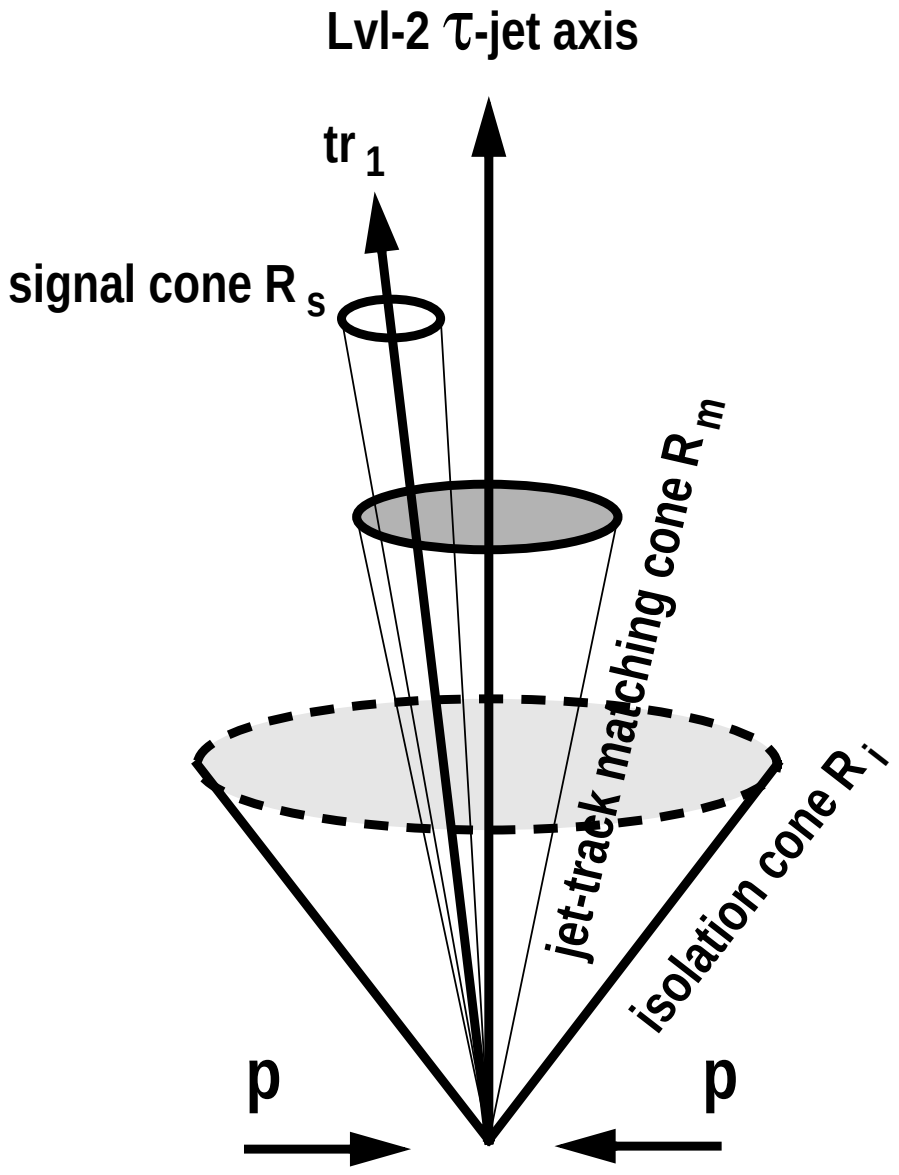

**Figure 37.** Sketch of the basic principle of $\tau$ -jet identification using charged particle tracks.

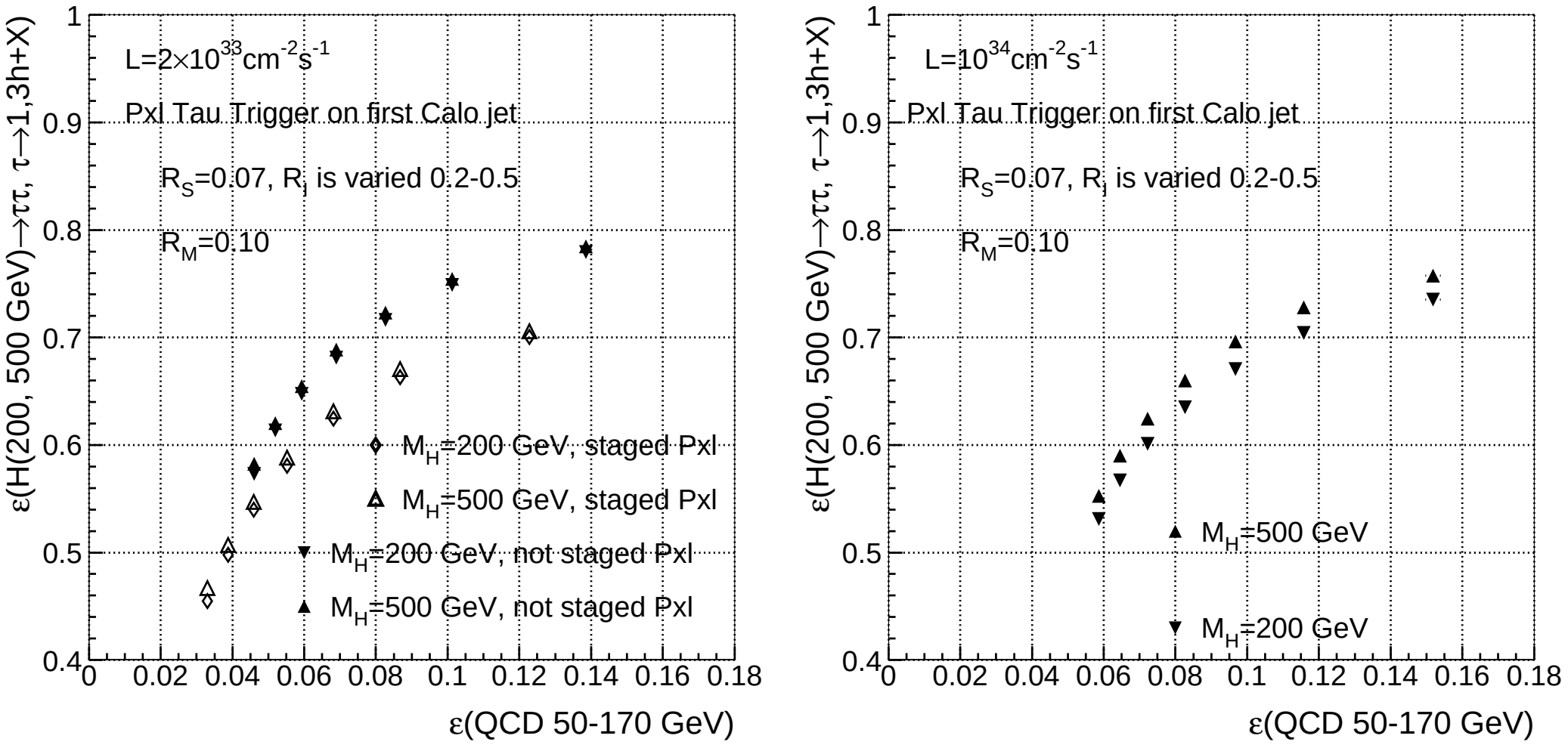

**Figure 38.** Efficiency of *Pixel track Tau* Trigger for the first calorimeter jet in $A^0/H^0 \to \tau\tau \to 2\tau$-jet, for two Higgs masses ($M_H$ = 200 and 500 GeV/$c^2$) versus the efficiency for QCD di-jet background events at (left) $\mathcal{L} = 2 \times 10^{33}\,\mathrm{cm}^{-2}\mathrm{s}^{-1}$ (results for both the full and the staged pixel system configurations are shown) and (right) $\mathcal{L} = 10^{34}\,\mathrm{cm}^{-2}\mathrm{s}^{-1}$.

**Table 16.** CPU time, in ms of a 1 GHz CPU, for $\tau$ -jet identification with the pixel detector for QCD di-jet events.

| | Time (ms) | |
|---|---|---|
| **Reconstruction step** | $\mathbf{2\times\ 10^{33}\ cm^{-2}s^{-1}}$ | $\mathbf{10^{34}cm^{-2}s^{-1}}$ |
| Build clusters from pixel digitized data | 44 | 100 |
| Reconstruct pixel lines and vertices | 34 | 262 |
| Total time | 78 | 362 |

### *7.3. Regional track finding $\tau$ selection*

Identification and selection of $\tau$-jets is also performed using information from tracks reconstructed with the full Silicon Strip Tracker. This selection will be referred to as the "Track Tau" trigger [29]. The *Track Tau* trigger performs the reconstruction of only those tracks in restricted regions of interest (regional tracking), defined by the cones around each jet direction given by the calorimeter jet reconstruction. The signal vertex is first selected using the pixel detector, as the vertex with the maximum $\sum |\vec{P}_T|$ of the associated pixel lines. The regional tracking technique is applied at the seeding level (regional seeding) in order to minimize the seed multiplicity, significantly reducing the CPU time of the algorithm. A special track finder uses pixel lines in a cone of $\Delta R=$ 0.5 around the calorimeter jet direction as track seeds. After track reconstruction, only tracks compatible with the signal vertex are taken into account. A further improvement of the timing performance is obtained by stopping the track reconstruction when six hits have been associated with the track helix. This gives an acceptable resolution on the track parameters and a low fake rate. At the seeding level an option is adopted which uses two pixel hits out of three pixel layers ("2-hit recovery") since it increases the seeding efficiency. The *Track Tau* trigger relies on an isolation requirement similar to the one used in the *Pixel Tau* trigger. To reduce the contamination from soft tracks, only tracks with $p_T>$ 1 GeV/$c$ and $z_0$ compatible with the $z$ position of the signal vertex (within 2 mm) are considered. The number of tracks within a signal cone and within an isolation cone are counted. The cones are defined around the direction of the leading track which is the highest-$p_T$ track found in the matching cone ($R_m$=0.1) around the calorimeter jet direction. The isolation requirement is that there be no tracks in the isolation cone except those contained within the inner, signal, cone. The cone sizes are chosen to optimize the signal efficiency and background rejection and also to minimize the dependence of the efficiency on the jet energy. Higher background reduction can be obtained by requiring the transverse momentum of the leading track, ${p_T}^{LT}$, to exceed a few GeV/$c$. Due to the strong dependence of the track $p_T$ spectrum on the Higgs mass and on the $\tau$ hadronic decay mode, the leading track transverse momentum requirement is carefully optimized.

Figure 39 shows the *Track Tau* trigger performance when applied to the first calorimeter jet in $A^0/H^0 \to \tau\tau \to 2\tau$-jet signal events, and to QCD di-jet background events. The different points correspond to different sizes of the isolation cone $R_i$ which was varied between 0.2 and 0.45. The performance at $\mathcal{L} = 2 \times 10^{33}\,\mathrm{cm}^{-2}\mathrm{s}^{-1}$ is studied for both the full and staged pixel detector configurations. For the same QCD di-jet background efficiency, the signal efficiency is reduced by about 15% in the staged pixel scenario. Since the algorithm has to fulfill stringent CPU time limitations particular attention has been paid to its optimization.

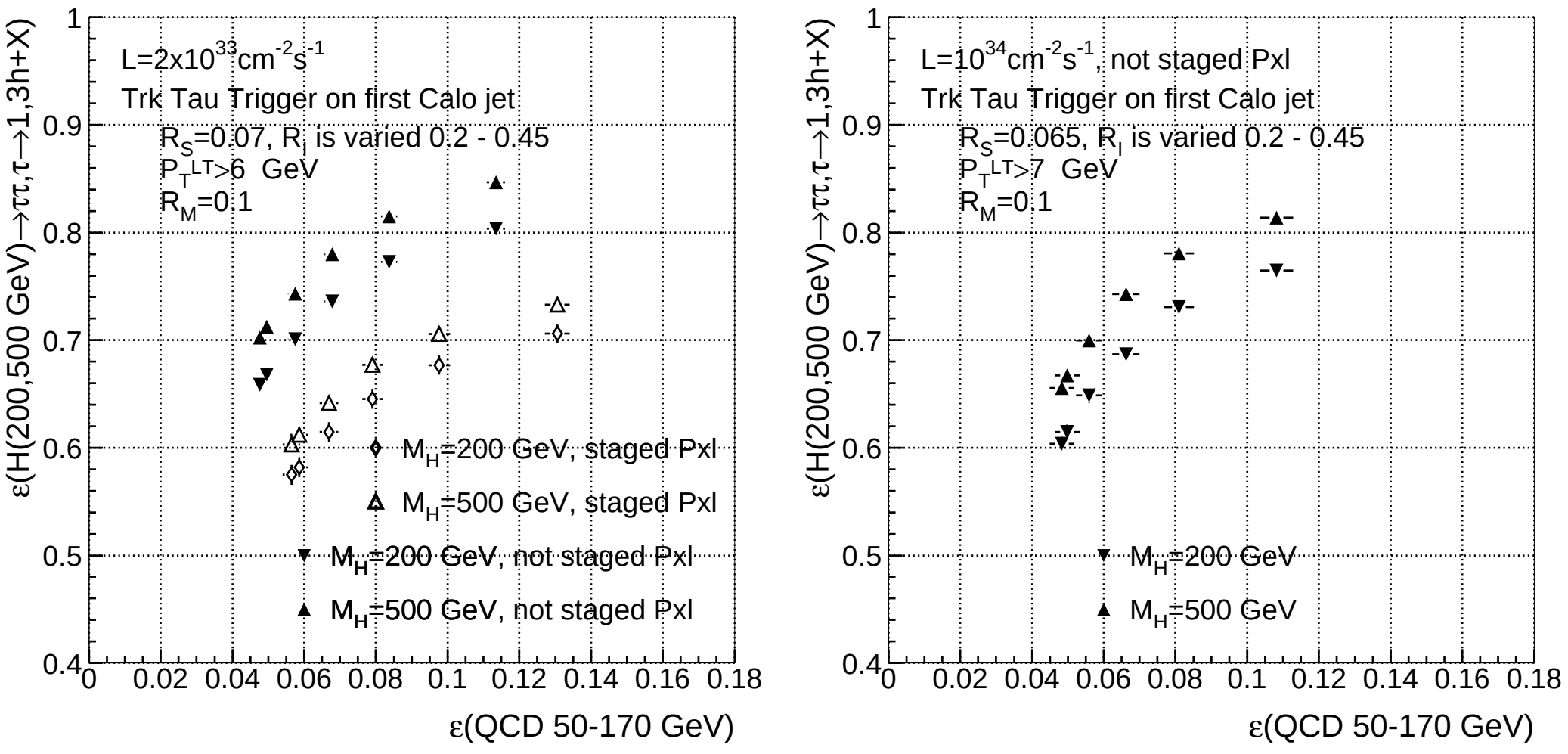

**Figure 39.** *Track Tau* trigger efficiency for the first calorimeter jet in $A^0/H^0 \to \tau\tau \to 2\tau$-jet versus the efficiency for QCD di-jet background events for $M_H = 200$ and 500 GeV/$c^2$. The results are shown both with the full and the staged pixel systems for (left) $\mathcal{L} = 2 \times 10^{33}\,\mathrm{cm}^{-2}\mathrm{s}^{-1}$, and with the full pixel system for (right) $\mathcal{L} = 10^{34}\,\mathrm{cm}^{-2}\mathrm{s}^{-1}$.

### *7.4. MSSM neutral Higgs $2\tau$-jet selection with Calorimeter and Pixel Tau Triggers*

A complete HLT selection for $A^0/H^0 \to \tau\tau \to 2\tau$-jet events can be defined using the *Calorimeter Tau* trigger selection applied to the first calorimeter jet, and the *Pixel track Tau* trigger selection applied to both calorimeter jets. Table 17 gives the purity of the selected jets in $gg \to bbA^0/H^0$, $A^0/H^0 \to \tau\tau \to 2\tau$-jet events with $M_H = 500$ GeV/$c^2$, defined as the fraction of calorimeter jets which correspond to a true $\tau$-jet. To increase the purity of the second jet the following search algorithm is applied: if a second Level-1 $\tau$-jet does not exist in the list or if it exists but the $E_T$ of the second calorimeter jet is less than 50 (70) GeV at low (high) luminosity, a new calorimeter jet in the region centered on the first Level-1 central jet is reconstructed. The purity of the second jet when chosen in such a way is increased to about 90% (the numbers in parentheses in table 17 correspond to the purity of the selection when this re-definition algorithm is not applied). Usage of the *Calorimeter Tau* trigger as a pre-selector before applying the *Pixel track Tau* trigger allows considerable reduction of the total CPU time per event. The requirements are optimized by examining the background rejection of the *Calorimeter Tau* trigger step ($S_{calo}$), the efficiency for the signal, and the CPU time usage, while keeping a suppression factor of the full HLT selection of $\sim 10^3$. Results of such a study for $M_H = 200$ GeV/$c^2$ are presented in table 18 for low luminosity running conditions with the full pixel system. It has been found that for a total suppression factor of $10^3$, at both low and high luminosity, a calorimeter suppression factor of three yields the highest signal efficiency. At this operating point ($P_{isol} = 5.6$ GeV) the total time $T_{tot}$ of the full path is 59 ms for low luminosity and 174 ms for high luminosity. Figure 40 shows the efficiency of the *Calorimeter+Pixel track Tau* trigger selection for the signal and for QCD di-jet background events at low and high luminosity. The size of the isolation cone $R_i$ is varied in the range between 0.20 and 0.50, and the optimal suppression factor of 3 for the *Calorimeter Tau* trigger is used. For a total suppression factor of about $10^3$, there is little difference in the efficiency between the staged and full pixel configurations at low luminosity.

**Table 17.** Purity of calorimeter jets in $A^0/H^0 \to \tau\tau \to 2\tau$-jet events at low and high luminosity. Numbers for the second jet without (with) parentheses are the purity after (before) re-definition of the second jet (see text).

| $\mathcal{L} = 2 \times 10^{33}\,\mathrm{cm^{-2}s^{-1}}$ | | $\mathcal{L} = 10^{34}\,\mathrm{cm^{-2}s^{-1}}$ | |
|---|---|---|---|
| 1st jet | 2nd jet | 1st jet | 2nd jet |
| 0.98 | 0.90(0.63) | 0.98 | 0.88(0.65) |

**Table 18.** Efficiency for $A^0/H^0 \to \tau\tau \to 2\tau$-jet events, and total CPU time ($\mathrm{T}_{tot}$), as a function of the calorimeter isolation requirement ($P_{isol}$) and its background rejection factor ($\mathrm{S}_{calo}$) at $\mathcal{L} = 2 \times 10^{33}\,\mathrm{cm^{-2}s^{-1}}$. An overall suppression factor $10^3$ for background events is maintained. The bolded column corresponds to the operating point.

| $P_{isol}$ GeV | - | 10.4 | 7.6 | **5.6** | 4.6 | 4.0 | 3.4 | 3.2 | 2.6 |
|---|---|---|---|---|---|---|---|---|---|
| $\mathrm{S}_{calo}$ | 1.0 | 1.5 | 2.0 | **3.0** | 4.0 | 5.0 | 6.2 | 7.5 | 10.0 |
| $\mathrm{T}_{tot}$ (ms) | 110 | 85 | 72 | **59** | 52 | 50 | 45 | 43 | 41 |
| $\epsilon_{calo+pixel}$ % | 35 | 37 | 40 | **41** | 40 | 39 | 37 | 36 | 35 |

### *7.5. MSSM neutral Higgs $2\tau-$jet selection with the Track Tau trigger.*

This section presents the performance of a complete HLT selection for $A^0/H^0 \to \tau\tau \to 2\tau$-jet, when the *Calorimeter Tau* trigger selection, applied on the first calorimeter jet, is followed by the *Track Tau* trigger on both calorimeter jets (*Calo+Track Tau* trigger path). Figure 41 shows the signal versus background efficiency when the *Track Tau* trigger selection is applied to both calorimeter jets, at low luminosity (left), for both the complete and staged pixel detector, and at high luminosity (right). The different points correspond to different sizes of the isolation cone, $R_i$, which is varied between 0.2 and 0.45. A 6 (7) GeV/$c$ requirement on the $p_T$ of the leading track is applied at low (high) luminosity to reach a background rejection factor of $\sim 10^3$. At a QCD di-jet background efficiency of $\sim 10^{-3}$ the signal efficiency with the staged pixel detector is reduced by $\sim$20%. Figure 42 shows the signal versus background efficiency when the *Calorimeter Tau* trigger selection applied on the first calorimeter jet, is followed by the Track Tau trigger on both calorimeter jets (*Calo+Track Tau* trigger path). Table 19 summarizes the efficiency of the *Track Tau* trigger at the operating points where the background rejection factor is $\sim 10^3$. These points have been chosen taking into account the overall performance and minimizing the $\tau$-jet energy dependency of the signal efficiency. The third and fifth rows of table 19 list the efficiency at low and high luminosity for the full *Calo+Track Tau* trigger selection. The *Track Tau* trigger CPU time distribution is shown in figure 43 for Higgs and QCD di-jet background events at low luminosity. The time needed for a double tag is only slightly larger than the time needed for a single tag, since the second calorimeter jet is analyzed only in the $\sim$ 6% of background events which pass the single tag on the first calorimeter jet. Only 10% of the QCD events require more than 400 ms to be analyzed. An important advantage of the *Calo+Track Tau* trigger selection over the simple *Track Tau* trigger is its better time performance, which is due to the fact that only those events which pass the calorimeter $\tau$ selection on the first jet need to be analyzed by the *Track Tau* trigger. Following the same method described in section 7.4 and using the numbers from table 18 ($S_{calo} = 3$, $T_{calo} \sim 33$ ms) an average time of about 130 ms/event is expected at low luminosity.

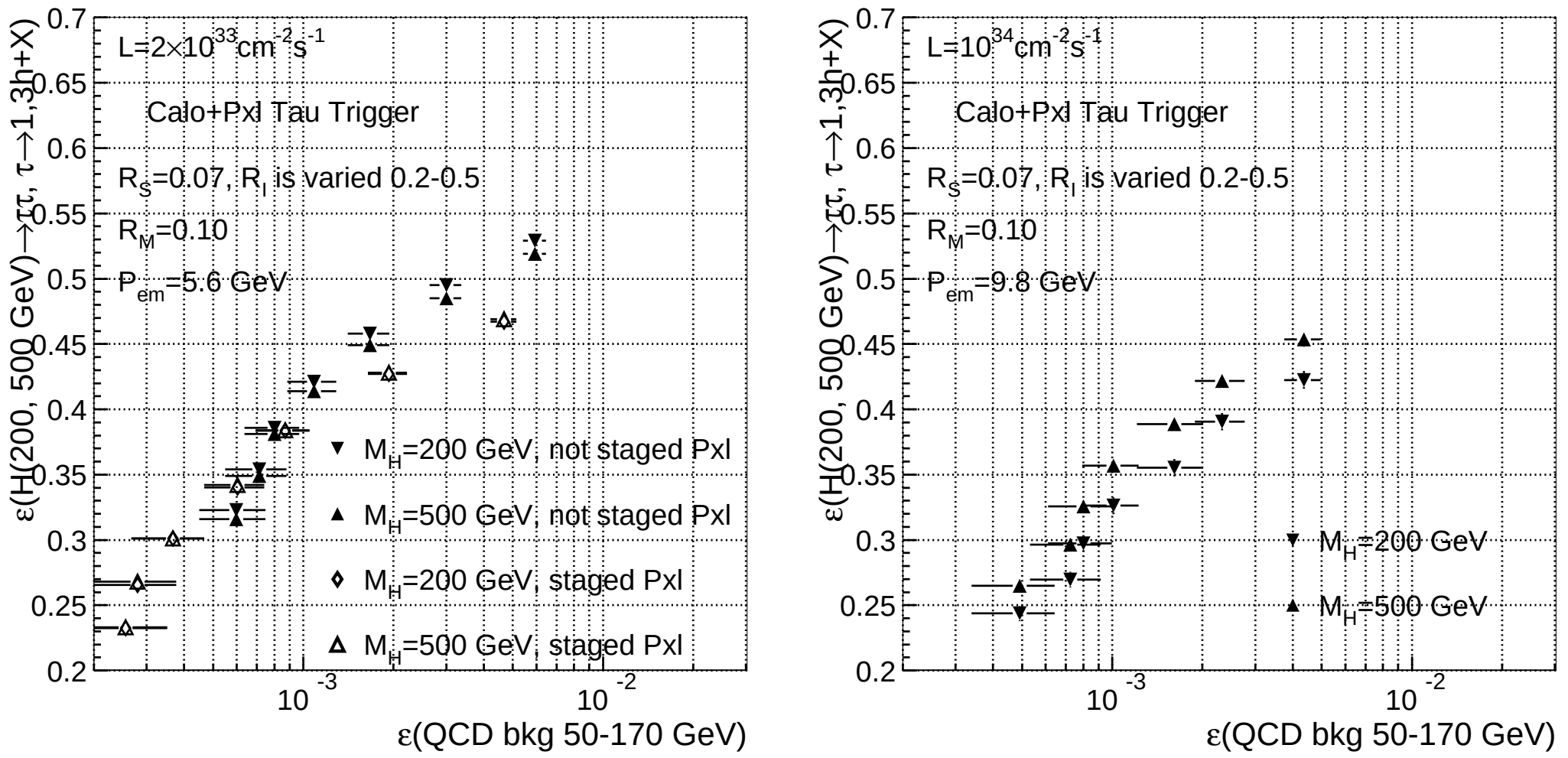

**Figure 40.** Efficiency of the *Calo+Pixel Tau* trigger path for $A^0/H^0 \rightarrow \tau\tau \rightarrow 2\tau$-jet and QCD 2-jet background events when the size of the isolation cone $R_i$ is varied in the range 0.20-0.50. The optimal suppression factor of 3 for the *Calo Tau* trigger is taken. Results are shown for two Higgs masses, $M_H$ =200 and 500 GeV/$c^2$ , and for (left) $\mathcal{L} = 2 \times 10^{33}\,\mathrm{cm}^{-2}\mathrm{s}^{-1}$ with the full and staged pixel systems, and (right) $\mathcal{L} = 10^{34}\,\mathrm{cm}^{-2}\mathrm{s}^{-1}$.

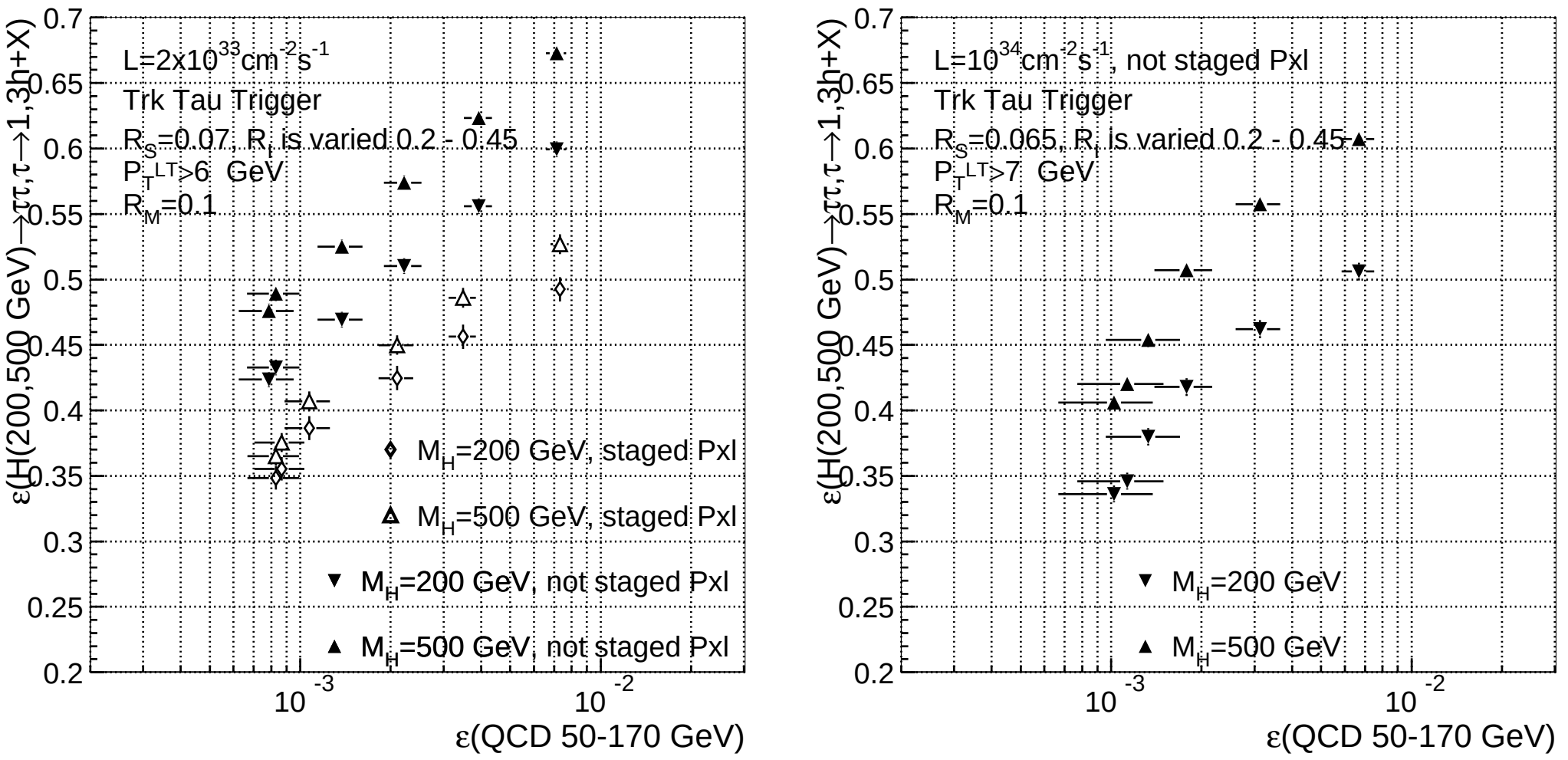

**Figure 41.** Efficiency of the *Track Tau* trigger applied to both *Calo* jets in $A^0/H^0 \to \tau\tau \to 2\tau$-jet versus that in QCD di-jet events. Results are shown for two Higgs masses, $M_H$ =200 and 500 GeV/$c^2$ , and for (left) $\mathcal{L} = 2 \times 10^{33}\,\mathrm{cm}^{-2}\mathrm{s}^{-1}$ with the full and staged pixel systems, and (right)$\mathcal{L} = 10^{34}\,\mathrm{cm}^{-2}\mathrm{s}^{-1}$.

**Table 19.** Summary of *Track Tau* trigger efficiency when two *Calo* jets are tagged in $A^0/H^0 \to \tau\tau \to 2\tau$-jet events. The third and fifth rows show the results when the *Calorimeter Tau* trigger selection applied on the first *Calo* jet is followed by the *Track Tau* trigger on both *Calo* jets. A QCD di-jet background rejection (last column) of ~$10^3$ is required. Due to the limited Monte Carlo statistics some statistical errors for the QCD background are large.

| **Luminosity** $cm^{-2}s^{-1}$ | **Configuration/ Trigger** | $M_H$ 200 GeV/$c^2$ | $M_H$ 500 GeV/$c^2$ | **QCD** |
|---|---|---|---|---|
| $2\times10^{33}$ | Staged pixels | | | |
| | *Track Tau* | 0.355±0.006 | 0.375±0.005 | $(8.6\pm1.6)\times10^{-4}$ |
| $2\times10^{33}$ | Full pixels | | | |
| | *Track Tau* | 0.433±0.006 | 0.489±0.005 | $(8.3\pm1.6)\times10^{-4}$ |
| $2\times10^{33}$ | Full pixels | | | |
| | *Calo + Track Tau* | 0.446±0.006 | 0.486±0.005 | $(1.0\pm0.2)\times10^{-3}$ |
| $10^{34}$ | *Track Tau* | 0.346±0.006 | 0.420±0.005 | $(1.1\pm0.4)\times10^{-3}$ |
| $10^{34}$ | *Calo + Track Tau* | 0.361±0.006 | 0.427±0.005 | $(9.4\pm3.0)\times10^{-4}$ |

The time required by the *Track Tau* trigger at high luminosity is much larger and is currently estimated at ~1 s/event. However, making use of the *Calo+Track Tau* Trigger selection, this time will be reduced to less than 400 ms/event. A further improvement is expected with the use of a regional pixel reconstruction.

To summarize, the *Calo+Pixel track Tau* trigger and the *Calo+Track Tau* trigger are the two optimal HLT paths for the selection of MSSM $A^0$ and $H^0$ bosons produced in the process $gg \to bbA^0/H^0,\ A^0/H^0 \to \tau\tau \to 2\tau$-jet events ($M_H = 200$ and 500 GeV/$c^2$).

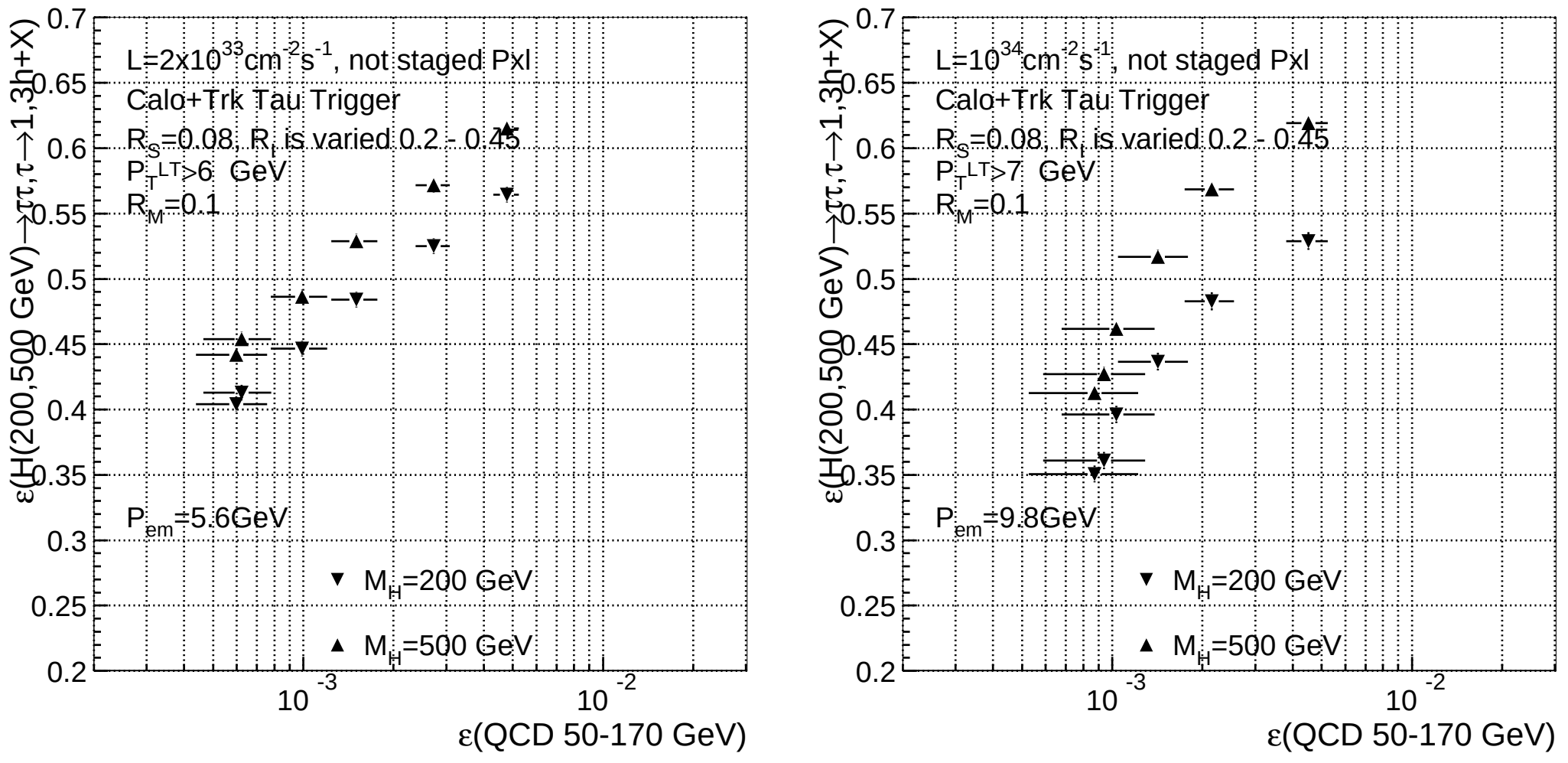

**Figure 42.** *Calo+Track Tau* trigger efficiency for $A^0/H^0 \to \tau\tau \to 2\tau$-jet versus that in QCD di-jet events. The *Calo Tau* trigger selection applied on the first *Calo* jet is followed by the *Track Tau* trigger on both calorimeter jets. Results are shown for two Higgs masses $M_H =$ 200 and 500 GeV/$c^2$ for low (high) luminosity on the left (right).

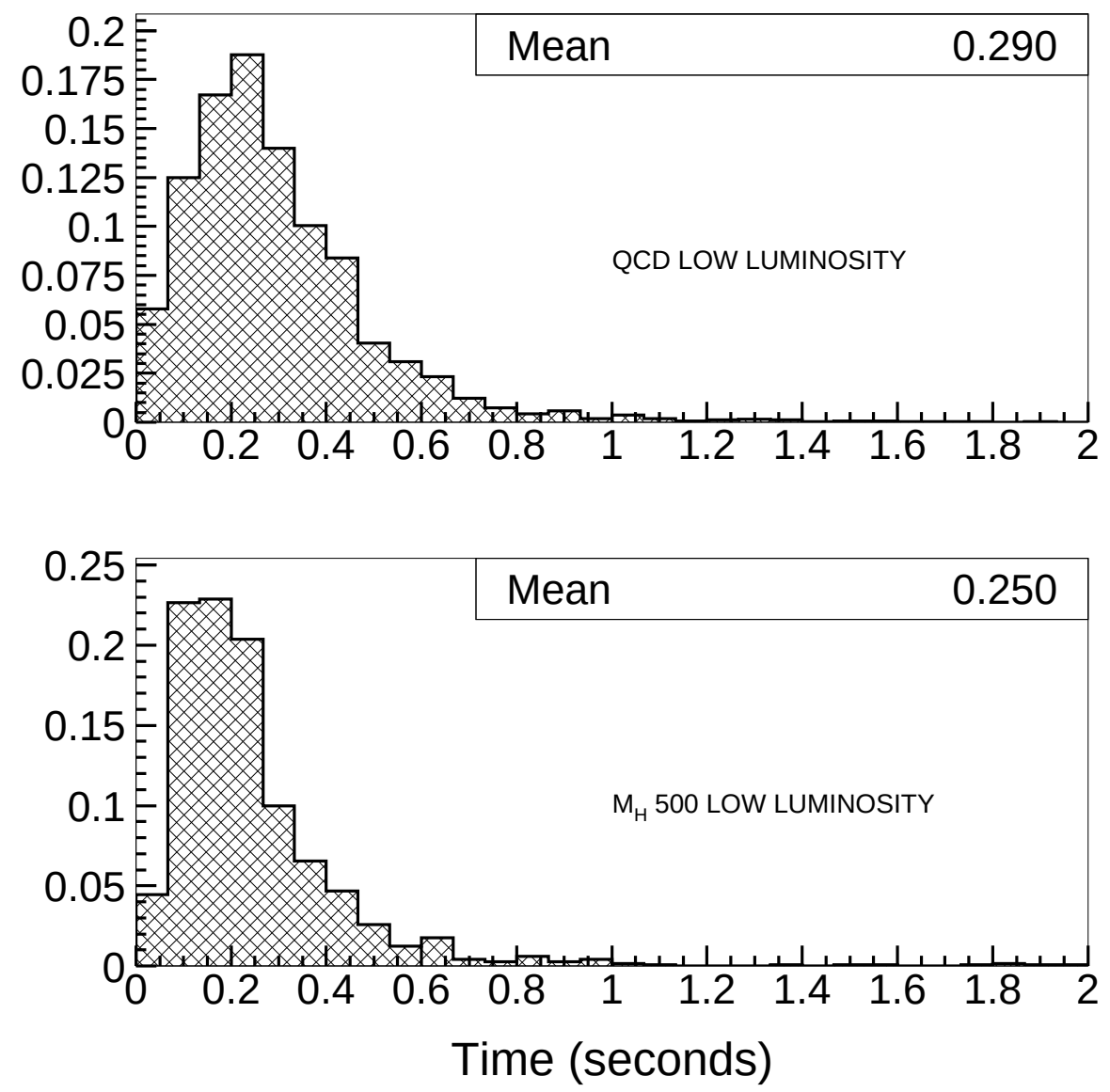

**Figure 43.** *Track Tau* trigger reconstruction time (in seconds) for double tagging at low luminosity. (Upper) QCD di-jet events. (Lower) $A^0/H^0 \to \tau\tau \to 2\tau$-jet events.

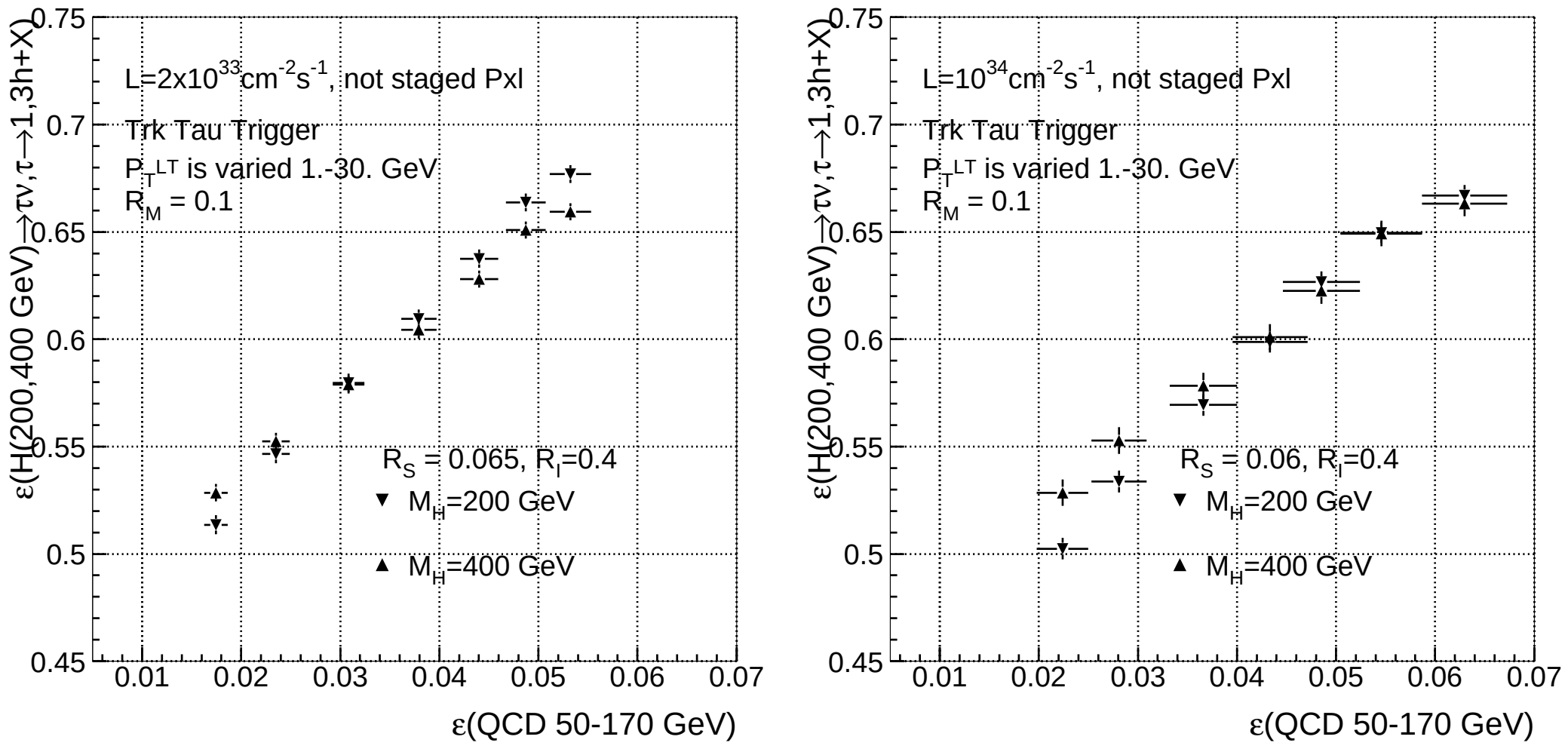

**Figure 44.** *Track Tau* trigger efficiency for $H^+ \to \tau\nu \to \tau$-jet events versus the QCD di-jet background efficiency. The results are given for $M_H = 200$ GeV/$c^2$ and $M_H = 400$ GeV/$c^2$. and for (left) $\mathcal{L} = 2 \times 10^{33}$ cm$^{-2}$s$^{-1}$, with the full pixel system, and (right) $\mathcal{L} = 10^{34}$ cm$^{-2}$s$^{-1}$.

*The Calo+Pixel track Tau* trigger is approximately twice as fast as the *Calo+Track Tau* trigger (59 versus 130 ms at low luminosity and 170 versus ∼400 ms at high luminosity), but is ∼15% less efficient. The loss of *Calo+Pixel track Tau* trigger efficiency could be compensated for by an increase of the Level-1 bandwidth allocated to the single and double $\tau$ trigger.

## *7.6. MSSM charged Higgs τ-jet selection with the Track Tau trigger.*

The *Track Tau* trigger will work as the final part of the selection MSSM charged Higgs bosons in the process $gb(g) \to H^+t(b)$, $H^+ \to \tau\nu \to \tau$-jet, $t \to bjj$. The Level-1 single-$\tau$ trigger followed by an HLT selection of events with $\not\!\!E_T >$ 65 (95) GeV at low (high) luminosity gives an output rate of about 30 (70) Hz. The selection must provide a suppression factor of ∼30 to match the output bandwidth requirements. The *Track Tau* Trigger selection is applied to the calorimeter jet reconstructed in the region of the first Level-1 $\tau$-jet. The purity of the calorimeter jet in the signal events is 85%. The isolation criteria used in the *Pixel* or *Track Tau* triggers cannot provide the required suppression factor. Additional background rejection is obtained by applying a cut on the $p_T$ of the leading track in the *Track Tau* trigger. Figure 44 shows the *Track Tau* trigger efficiencies for the signal and QCD background events passing the Level-1 single-$\tau$ trigger. Since the cut on $\not\!\!E_T$ and the *Track Tau* selection efficiency are uncorrelated [29], a cut on $\not\!\!E_T$ has not been applied due to limited Monte Carlo statistics. The different points in figure 44 correspond to varying the leading track, $p_T$ requirement, ${p_T}^{LT}$, from 1 to 30 GeV/$c$. The isolation parameters used are $R_m = 0.1$, $R_s = 0.065$ (0.060) for low (high) luminosity, and $R_i = 0.4$. A rejection factor of 30 can be reached with a 20 (25) GeV/$c$ requirement on ${p_T}^{LT}$ at low (high) luminosity, which is well below the foreseen offline analysis requirement [25]. Table 20 summarizes the efficiency of the *Track Tau* trigger at this working point with a background rejection factor of 30.

**Table 20.** *Track Tau* Trigger efficiency for the process $gb(g) \to H^+t(b)$, $H^+ \to \tau\nu \to \tau$-jet, $t \to bjj$ and for the QCD background. The efficiencies for the signal are presented for a background suppression factor of ∼30.

| Signal and background samples | | Efficiency of *Track Tau* trigger | |
|---|---|---|---|
| | | Low Luminosity<br>${p_T}^{LT} > 20$ GeV/$c$<br>$R_s$ =0.065 | High Luminosity<br>${p_T}^{LT} > 25$ GeV/$c$<br>$R_s$ =0.06 |
| $H^+ \to \tau\nu$ | $M_H$ = 200 GeV/$c^2$ | 0.580 ± 0.004 | 0.534 ± 0.005 |
| | $M_H$ = 400 GeV/$c^2$ | 0.579 ± 0.004 | 0.553 ± 0.006 |
| QCD | 50 < $p_T$< 170 GeV/$c$ | 0.031 ± 0.002 | 0.028 ± 0.003 |

### *7.7. MSSM neutral Higgs e+τ-jet HLT selection*

The triggering scheme for $gg \to bbA^0/H^0$, $A^0/H^0 \to \tau\tau \to e + \tau$-jet has been studied for $M_H$ = 200 GeV/$c^2$. The trigger accepts events which pass either a single-$e$ or a combined $e + \tau-$jet ("eTau") trigger. In what follows the combined trigger will be referred to as the "e+eTau" trigger. The *eTau* trigger requires the presence of both an electron and a $\tau$-jet, with thresholds lower than those used in the single-$e$ and single-$\tau$-jet triggers. The $\tau$-jet candidate at Level-1 is defined as the most energetic $\tau$-jet that is not collinear with the electron candidate ($\Delta R > 0.3$). This condition avoids misidentification in signal events (the $\tau$ purity in signal events increases from 61% to 99%) with negligible effect on the overall efficiency. Figure 45 shows the Level-1 *e+eTau* trigger rate at $\mathcal{L} = 10^{34}\,\text{cm}^{-2}\text{s}^{-1}$as a function of the electron and $\tau$-jet thresholds used in the *eTau* trigger. The dots with numbers on the *e+eTau* trigger iso-rate curves indicate the Level-1 selection efficiency for the signal. For a given Level-1 trigger rate, the efficiency increases when the electron threshold is reduced and the $\tau$-jet threshold is raised. This results from the steeply falling $p_T$ spectrum of the electron in the signal channel. Figure 46 shows the increase in Level-1 efficiency obtained by using the *e+eTau* trigger, as opposed to just the single electron trigger, as a function of the extra bandwidth which one allocates to the *eTau* trigger. The curves are each obtained by fixing the threshold for the electron in *eTau* trigger and varying the threshold on $\tau$-jet. The HLT selection is applied independently on the electron stream and the *eTau* stream at Level-2.0 and Level-2.5. At Level-2.0, a threshold is applied only on the electron candidate (section 3.4 ). At Level-2.5, pixel/super-cluster matching is used for the electron candidate (section 3.5 ) and the $\tau$-jet identification is applied as described in section 7.2. Table 21 shows the details of the full selection for four scenarios. In all cases, the *eTau* trigger uses a 20 GeV electron threshold (corresponding to 25.5 GeV on the Level-1 95% efficiency scale) while the $\tau$ threshold is varied in such a way that the rate added by the *eTau* trigger to the single-electron trigger rate is 0.14 kHz, 0.39 kHz, 0.85 kHz and 1.28 kHz respectively. Accepting an additional rate of 0.85 kHz at Level-1 leads to a relative improvement in efficiency at Level-2.5 of about 10%, at a price of a 7 Hz rate increase. This is illustrated in figure 47 which shows the $p_T$ spectrum of the electron that is recovered at Level-2.5 when the *eTau* trigger is added to the single electron trigger at Level-1.
Since the single electron thresholds are lower in the low luminosity scenario, less is gained by adding the combined *eTau* trigger. The HLT selection scheme is the same as at high luminosity but the Level-1 thresholds are different. For a scenario where 0.82 kHz for the *eTau* trigger is added at Level-1, the relative gain in efficiency at Level-2.5 is ∼4%.

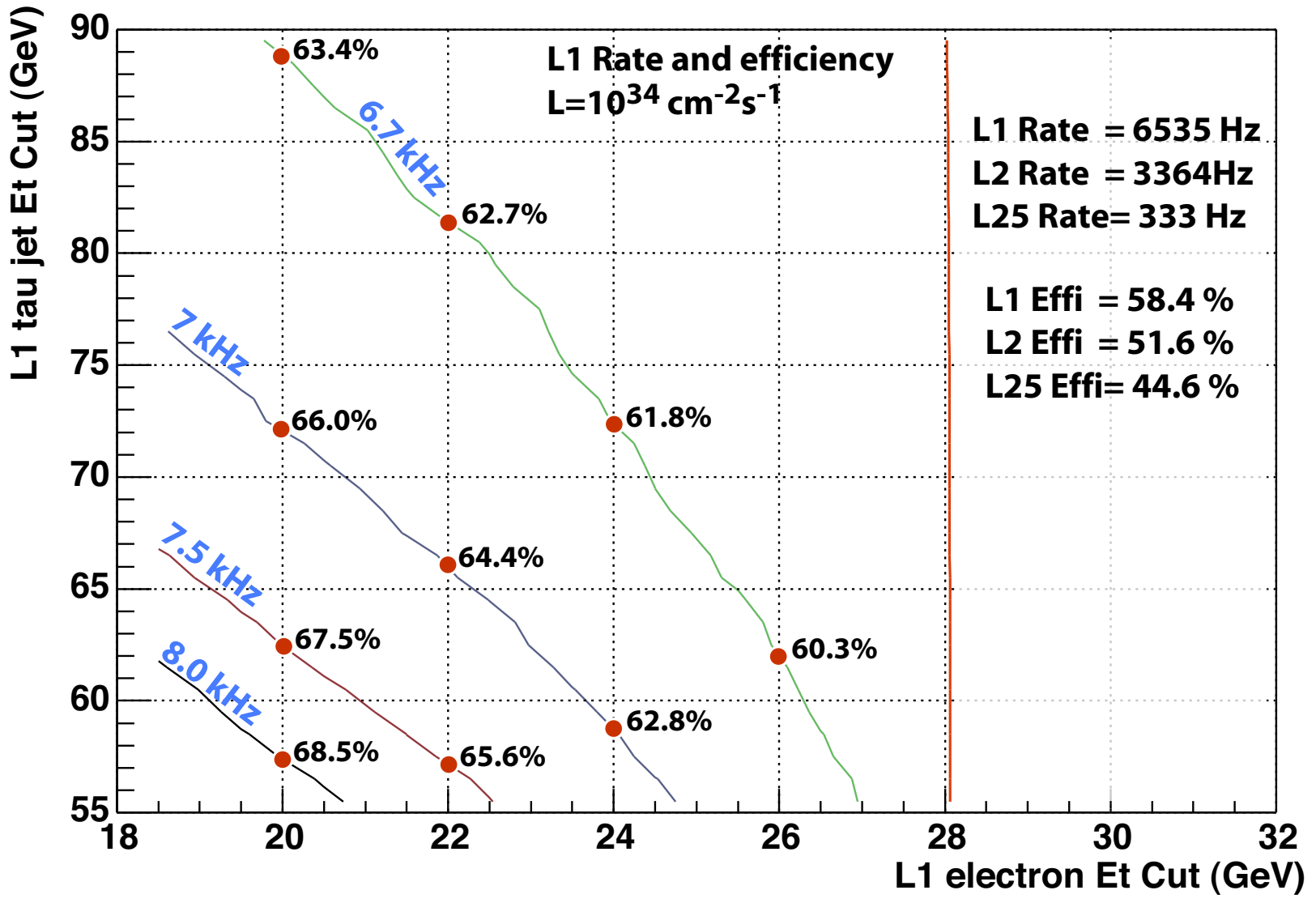

**Figure 45.** Level-1 $e$+$eTau$ trigger rate at $\mathcal{L} = 10^{34}\,\mathrm{cm}^{-2}\mathrm{s}^{-1}$ as a function of the $e$ and $\tau$-jet thresholds, for a fixed single-$e$ trigger threshold. A constant 6.54 kHz bandwidth is allocated to the single-$e$ trigger. The four curves correspond to the additional bandwidths of 0.14 kHz, 0.39 kHz, 0.85 kHz and 1.28 kHz for the four scenarios, allocated to the $eTau$ trigger. The vertical line at 28 GeV corresponds to the single-$e$ trigger. The values listed in the upper right corner are the rates and efficiencies for the single-$e$ trigger at Level-1, Level-2.0 and Level-2.5.

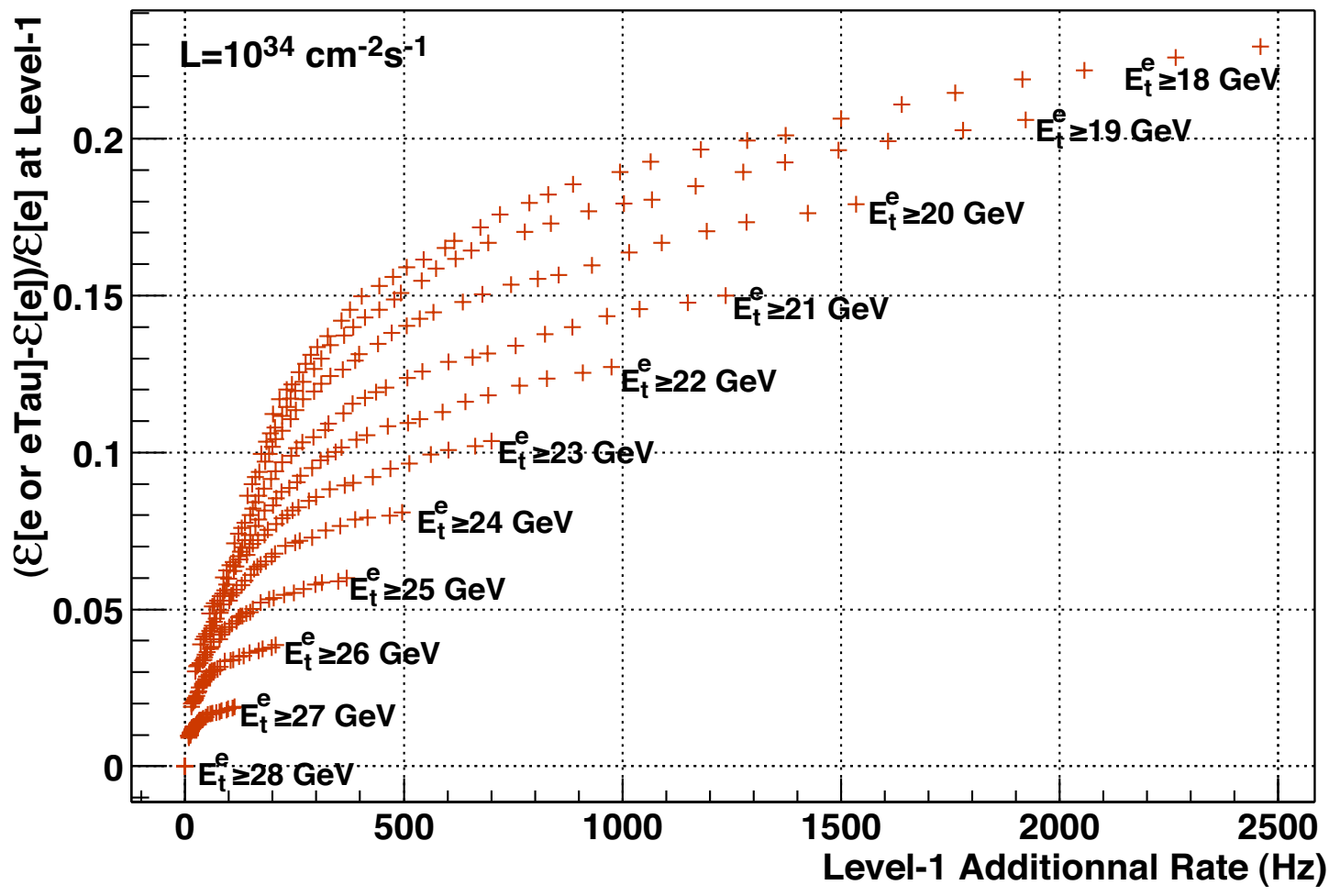

**Figure 46.** The increase in Level-1 efficiency, at $\mathcal{L} = 10^{34}\,\mathrm{cm}^{-2}\mathrm{s}^{-1}$ , obtained using the $e$+$eTau$ trigger, as opposed to just the single electron trigger, as a function of the extra bandwidth allocated to the $eTau$ trigger. Each curve is obtained by fixing the electron threshold ($E_t^e$ in the plot) and varying the threshold for the $\tau$ -jet.

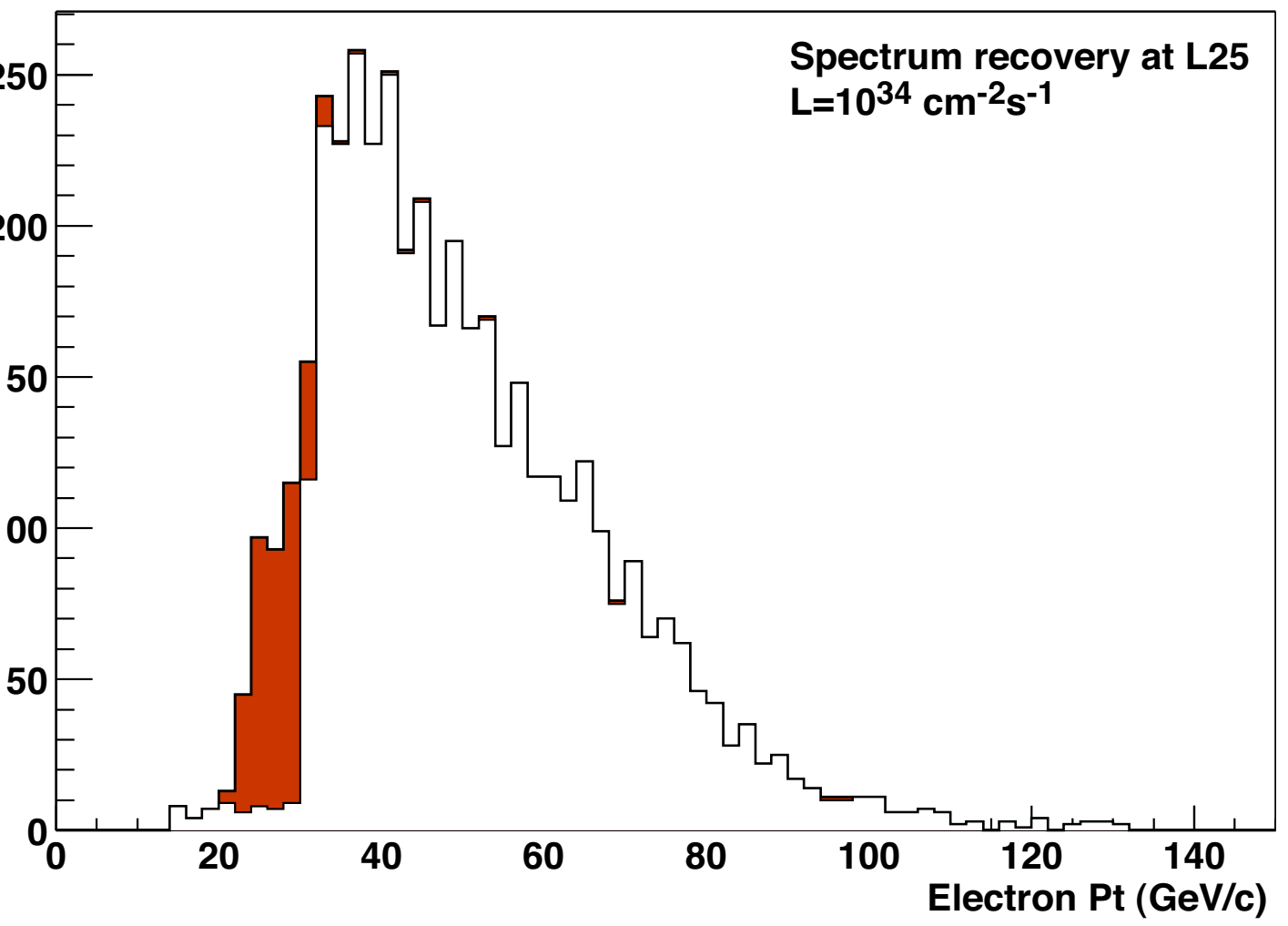

**Figure 47.** Electron $p_T$ in $A^0/H^0 \to \tau\tau \to e + \tau$-jet events at $\mathcal{L} = 10^{34}\,\mathrm{cm}^{-2}\mathrm{s}^{-1}$. The filled portion of the histogram corresponds to the gain observed at Level-2.5 by adding the Level-1 $e + \tau$ trigger, the remaining unfilled part corresponds to electrons passing the single electron trigger.

### *7.8. MSSM neutral Higgs $\mu$+$\tau$-jet HLT selection*

The reconstruction efficiencies and online background rejection performance for $A^0/H^0 \to \tau\tau \to \mu + \tau$-jet have been studied for the case when the Higgs particle is produced in association with $b$-quarks and the mass of Higgs particle is 200 GeV/$c^2$. The studies focus on high-luminosity running conditions where the event selection and background reduction are more difficult. The quoted efficiencies are given with respect to the baseline Monte Carlo sample generated with ${p_T}^{\mu} > 14$ GeV/$c$, ${p_T}^{\tau-\mathrm{jet}} > 30$ GeV/$c$, $|\eta^{\mu}| < 2.4$), $|\eta^{\tau}| < 2.4$)). This sample contains 45% of all generated $H \to \mu + \tau$-jet decays. Due to the relatively low threshold of the Level-1 single-$\mu$ trigger (${p_T}^{cut} = 20$ GeV/$c$ at high luminosity), about 66% of the baseline events are accepted by the Level-1

**Table 21.** Evolution of the rate and the efficiency of the different trigger levels at high luminosity. Results for four different $\tau$ thresholds in the $eTau$ trigger are shown, as are results obtained with no $eTau$ trigger (just the single electron trigger).

| **L1 $eTau$ thresh (GeV)** | **(20, 57)** | $e$+$eTau$ **(20, 62)** | **trigger (20, 72)** | **(20, 89)** | $e$ **trigger (28)** |
|---|---|---|---|---|---|
| L1 Rate (Hz) | 7819 | 7389 | 6933 | 6677 | 6535 |
| L1 Additional Rate (Hz) | 1284 | 854 | 398 | 142 | |
| L1 Efficiency | 0.685 | 0.675 | 0.661 | 0.634 | 0.584 |
| L1 Additional Efficiency | 0.101 | 0.091 | 0.077 | 0.050 | |
| L2.0 Rate (Hz) | 4219 | 3945 | 3631 | 3452 | 3364 |
| L2.0 Additional Rate (Hz) | 855 | 581 | 267 | 88 | |
| L2.0 Efficiency | 0.614 | 0.605 | 0.590 | 0.562 | 0.5165 |
| L2.0 Additional Efficiency | 0.098 | 0.089 | 0.074 | 0.046 | |
| L2.5 Rate (Hz) | 343 | 339 | 338 | 333 | 332 |
| L2.5 Additional Rate (Hz) | 11 | 7 | 6 | 1 | |
| L2.5 Efficiency | 0.492 | 0.489 | 0.483 | 0.469 | 0.446 |
| L2.5 Additional Efficiency | 0.046 | 0.043 | 0.037 | 0.023 | |

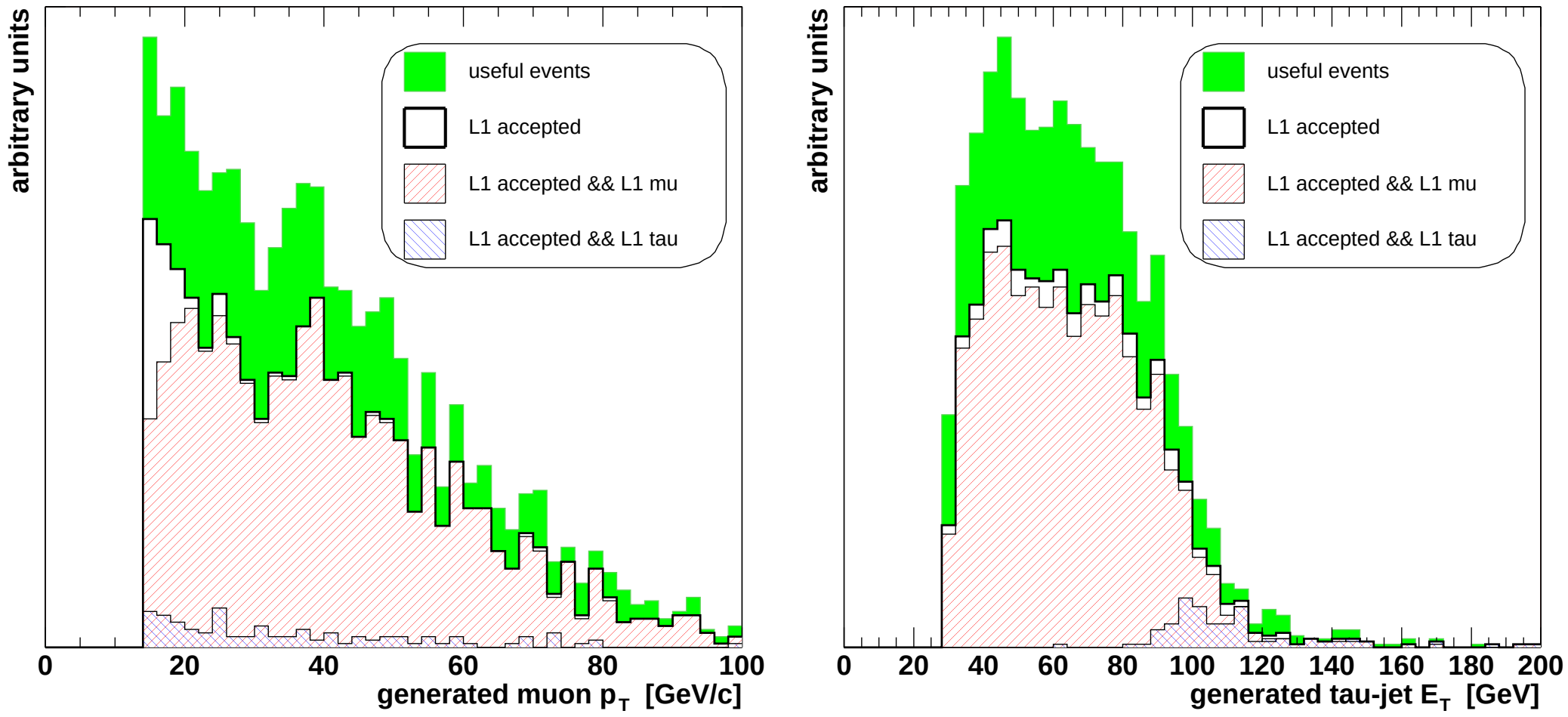

**Figure 48.** The $p_T$ spectrum of generated muons (left) and the $E_T$ spectrum of generated $\tau$-jets (right). The fraction of Level-1 events accepted by single-$\mu$ or single-$\tau$ trigger is indicated. The histogram marked as "L1 accepted" contains the additional 4% accepted by the $\mu - \tau$ Level-1 Trigger (with combined threshold $p_T \geq$ 14 GeV/$c$, $E_T \geq$ 45 GeV). The baseline sample is marked as "useful" events.($\mathcal{L} = 10^{34}\,\mathrm{cm}^{-2}\mathrm{s}^{-1}$.)

single-$\mu$ or Level-1 single-$\tau$ triggers as shown in figure 48, while there is a additional 4% acceptance when the combined $\mu$+$\tau$-jet selection is used. In figure 49, the efficiency of the Level-1 $\mu$ trigger is shown as a function of the muon $p_T$ and the $\tau$ $E_T$. Only events where both the $\tau$-jet and the $\mu$ are found by the trigger (irrespective of their $p_T$ and $E_T$) contribute to the plot.

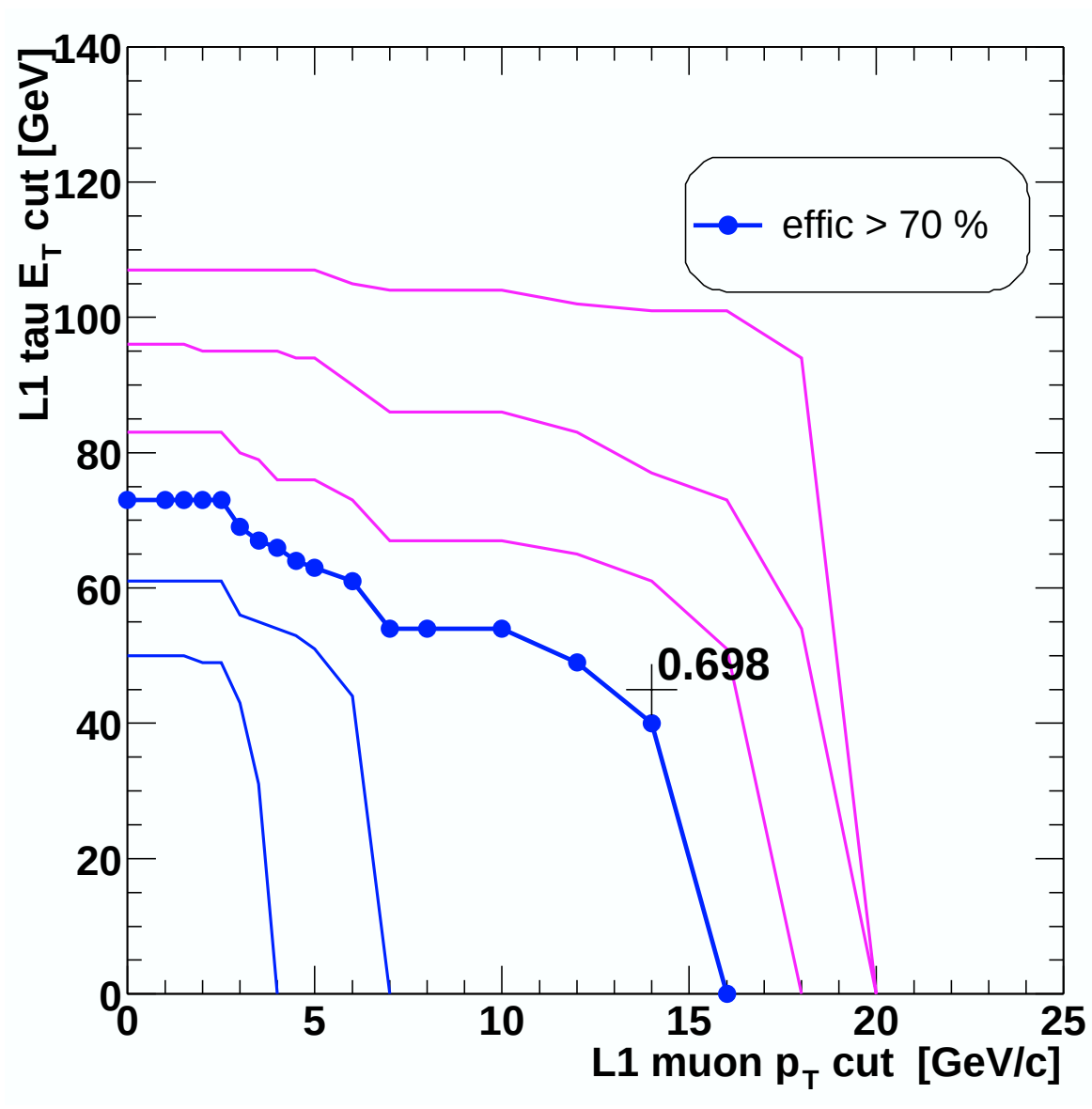

**Figure 49.** Level-1 trigger efficiency as a function of the muon $p_T$ and $\tau$-jet $E_T$ requirements for high luminosity. $\mu - \tau$ events accepted by the single-$\mu$ or single-$\tau$ trigger are included. The difference between contour lines represents a change in efficiency of 1%. The Level-1 threshold corresponding to the offline requirement selects ~70% of the baseline events.

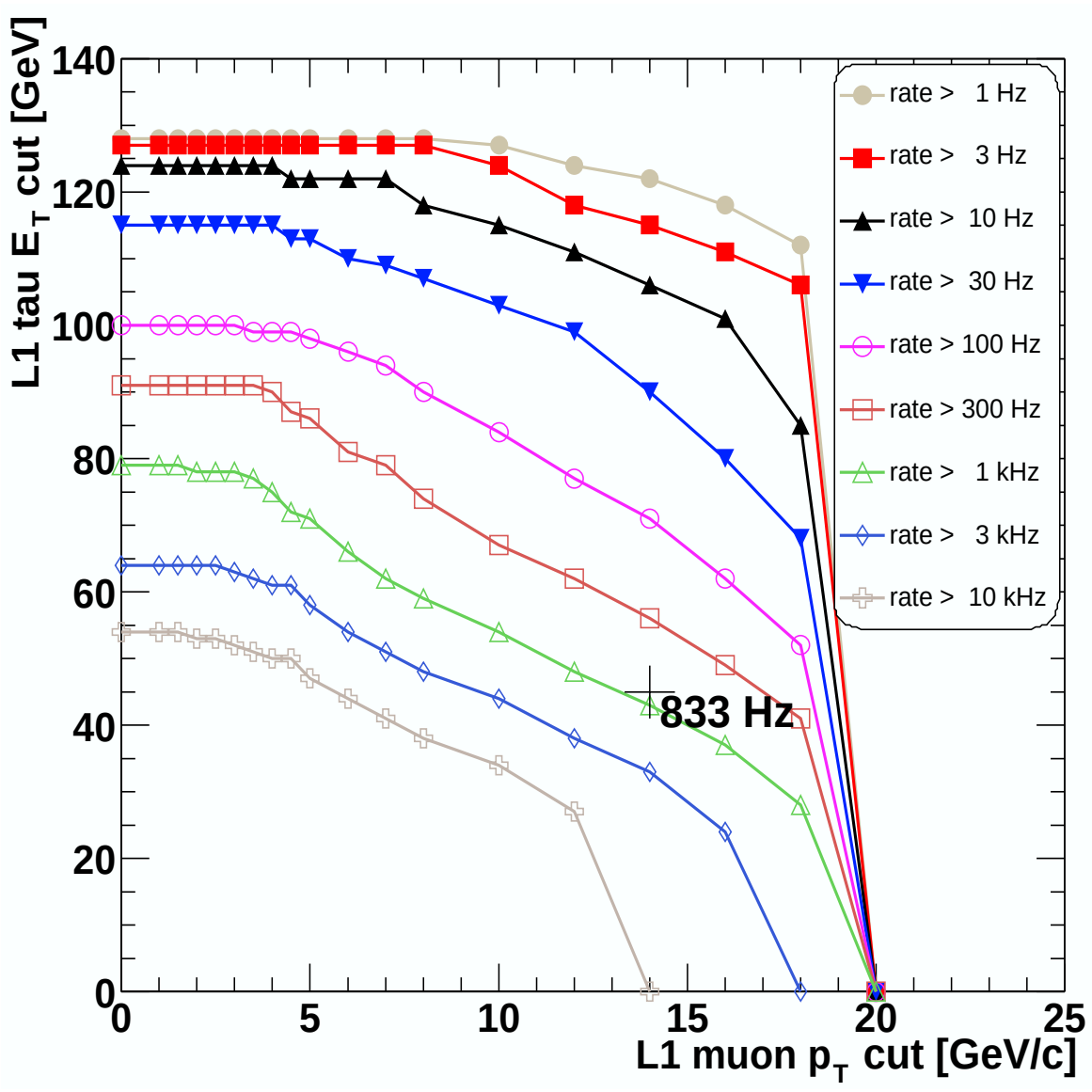

**Figure 50.** Level-1 Trigger rate at high luminosity from $\mu - \tau$ events. This rate is in addition to the Level-1 single-$\mu$ and Level-1 single-$\tau$ trigger rates. The additional rate, for the Level-1 cut that corresponds to the offline cut, is 0.83 kHz.

The additional rate taken by the Level-1 combined $\mu + \tau$-jet trigger over that of the single-$\mu$ trigger (∼6 kHz) and single $\tau$-jet trigger (∼2 kHz) is 0.83 kHz as shown in figure 50. The full rate for events selected by the combined trigger (both Level-1 $\mu$ and Level-1 $\tau$ found) passing the single-$\mu$, single $\tau$-jet or combined thresholds is 5.8 kHz. The HLT analysis path for this channel proceeds as follows:

- identification of a Level-2 jet corresponding to the Level-1 $\tau$-jet with an $E_T$ requirement
- calorimeter isolation requirement for the $\tau$
- Level-2 $\mu$ identification and $p_T$ requirement
- calorimeter isolation requirement for the $\mu$
- Level-3 $\mu$ identification and $p_T$ requirement
- $\tau$-jet identification and isolation requirement in the pixel detector
- $\mu$ isolation requirement with full tracker (or with pixel detector)

The resulting HLT efficiencies and rates are shown in figure 51 as the function of the HLT requirements. Only events passing the proposed Level-1 thresholds are included. Setting the HLT thresholds at the offline values preserves 32% of the baseline sample events. The corresponding rate from the muon minimum-bias samples is about 1 Hz. The detailed list of rejection factors and efficiencies for each HLT step is given in table 22 for the high luminosity case. The low luminosity case is simpler. The Level-1 single-$\mu$ threshold of 12 GeV/$c$ is below the offline requirement and there is no need to allocate bandwidth to the $\mu+$ $\tau$-jet channel. The Level-1 single-$\mu$ and single-$\tau$ triggers ($E_T$>93 GeV) select about 72% of the events in the sample. The $\mu$ and $\tau$ identification and selection criteria in the HLT reduce this efficiency to 39% with a background rate of 0.2 Hz.

### *7.9. Summary of Level-1 and HLT selection for Higgs Channels with $\tau$ -leptons.*

The Level-1 and HLT paths used to trigger on the MSSM $A^0/H^0$ and $H^+$ Higgs bosons with mass greater than 200 GeV/$c^2$ are:

- $A^0/H^0 \to \tau\tau \to 2\tau$-jets.
  - Level-1: single and double $\tau$-jet.
  - HLT: calorimeter + tracker isolation.
- $H^+ \to \tau\nu \to \tau - jet + \not{E}_T$.
  - Level-1: single $\tau$-jet.
  - HLT: calorimeter $\not{E}_T$ and tracker isolation on the $\tau$-jet.
- $A^0/H^0 \to \tau\tau \to e + \tau$-jet.
  - Level-1: single-$e$ and combined $e$+$\tau$-jet triggers.
  - HLT: electron selection (section 3) and $\tau$-jet isolation with the tracker.
- $A^0/H^0 \to \tau\tau \to \mu + \tau$-jet.

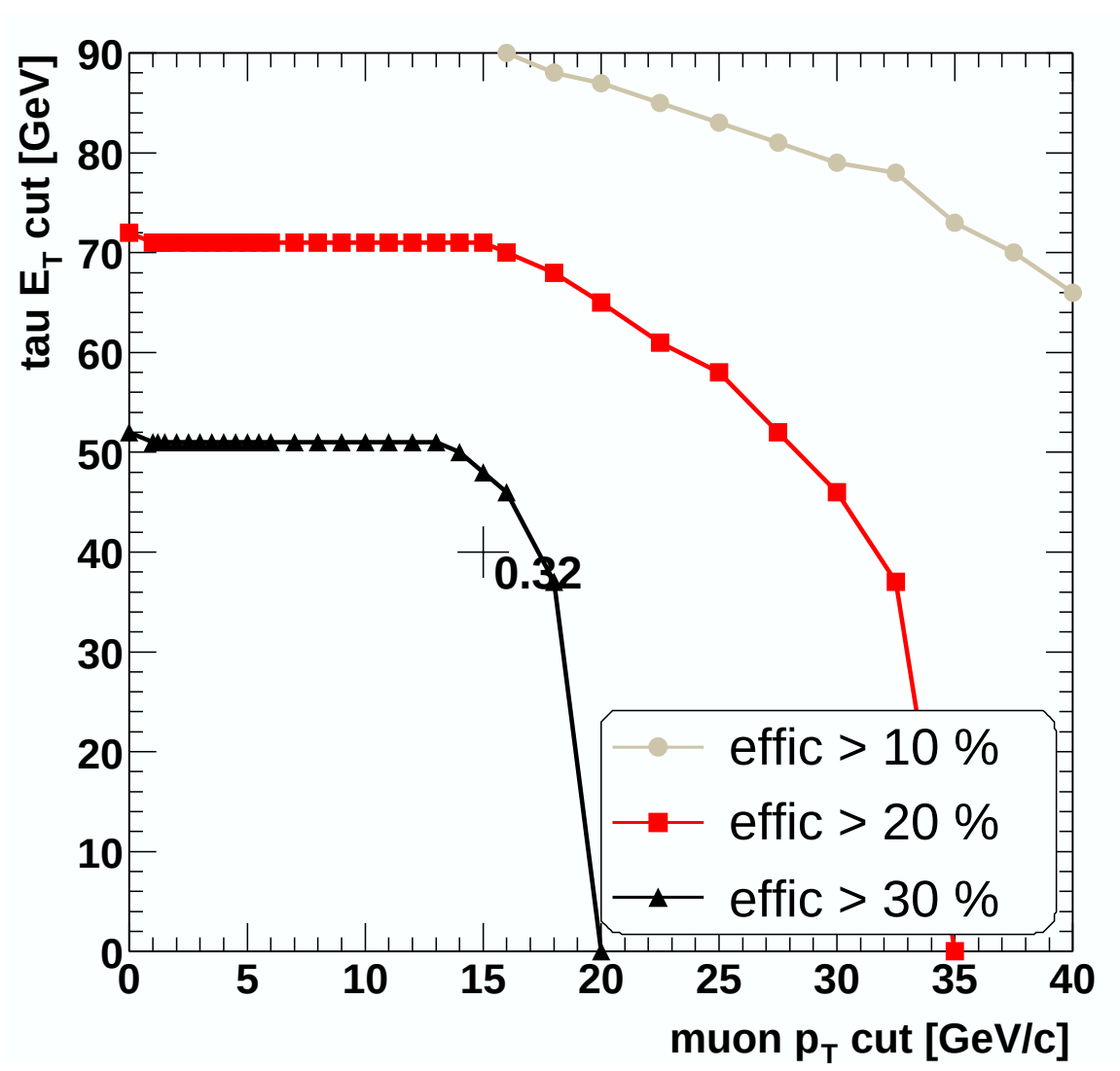

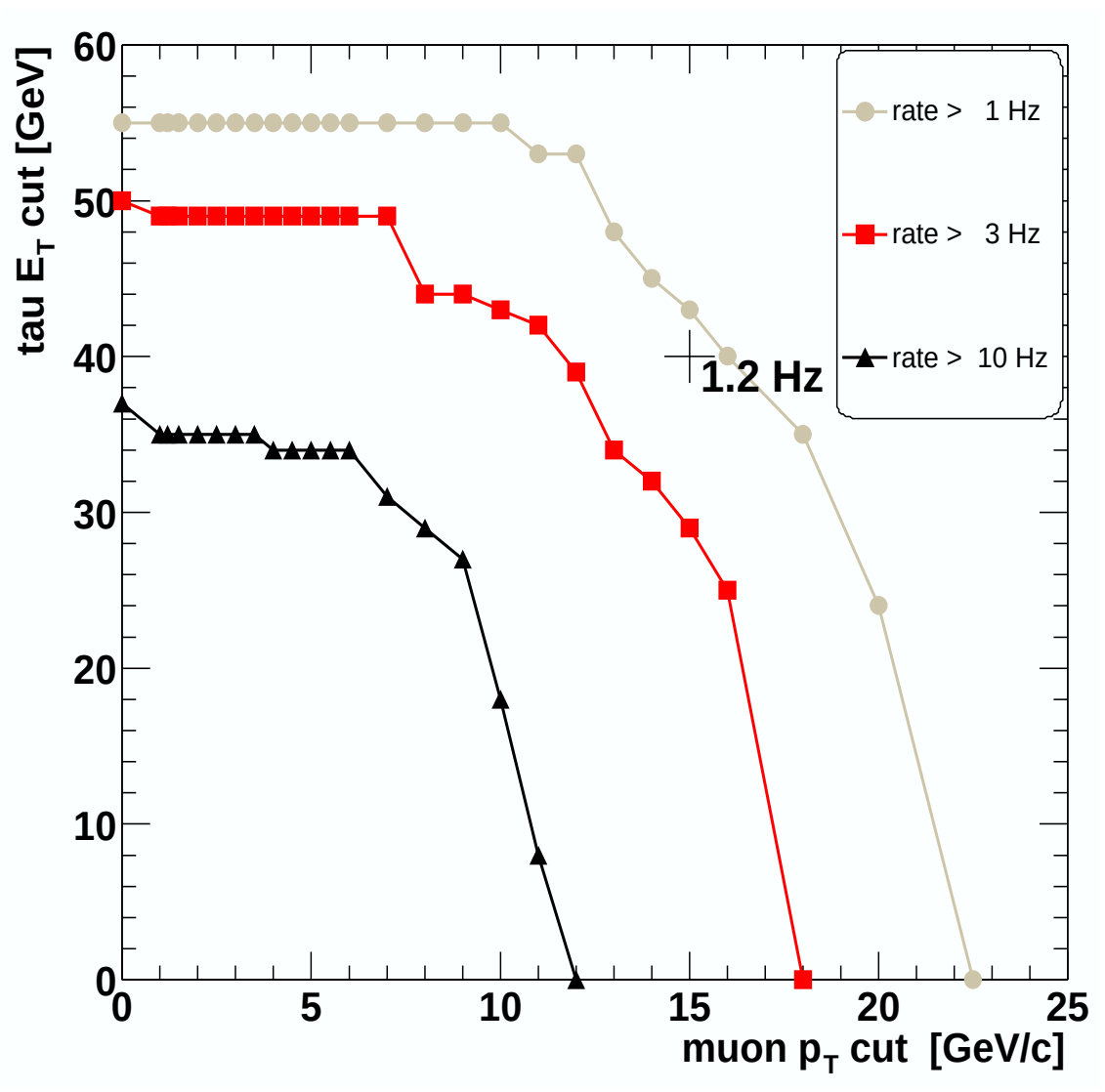

**Figure 51.** Full selection (Level-1+HLT) in $A^0/H^0 \to \tau\tau \to \mu + \tau$-jet channel. The selection efficiency (left) and corresponding background rate (right) are shown. The Level-1 combined $\mu - \tau$ requirement is at Level-1 $p_T{}^{\mu} = 14$ GeV/$c$ and Level-1 $\tau$-jet $E_T$= 45 GeV. Events must pass Level-1 and all HLT $\mu$ and $\tau$ selection requirements. The $\tau$ 95% efficiency scale is used for the $\tau$ $E_T$. The $\mu$ 90% efficiency scale is used for $p_T{}^{\mu}$ .The efficiency and rate for the offline threshold ($p_T$= 15 GeV/$c$, $E_T$= 40 GeV) are marked. $\mathcal{L} = 10^{34}\,\mathrm{cm}^{-2}\mathrm{s}^{-1}$.

**Table 22.** Summary of Level-1 and HLT selection efficiencies and background rates in the $A^0/H^0 \to \tau\tau \to \mu + \tau$-jet channel for the thresholds corresponding to offline requirement of $\mu$ $p_T$= 15 GeV/$c$ and $\tau$-jet $E_T$= 40 GeV. The efficiency is defined with respect to the baseline sample.

| | Efficiency | Rate [Hz] |
|---|---|---|
| Events passing Level-1 single $\mu$ single $\tau$, or combined trigger | 0.70 | $5.8\times10^{33}$ |
| Events passing Level-1 combined trigger* | 0.04 | 830 |
| L2 identification with $E_T$ and $p_T$ requirements | 0.63 | 990 |
| L2 and calo tau isolation | 0.53 | 380 |
| L2 and muon calo isolation | 0.61 | 420 |
| **L2 combined** | **0.51** | **150** |
| L3 identification with $\mu$ $p_T$ cut | 0.49 | 59 |
| L3 and tau isolation | 0.33 | 3.4 |
| L3 and muon isolation | 0.48 | 25 |
| **L3 combined (HLT)** | **0.32** | **1.2** |

* not selected by single $\mu$ or single $\tau$ trigger

**Table 23.** The efficiency of the Level-1 and HLT selections, the HLT output rates and CPU time for low (high) luminosity for the MSSM Higgs decays with $\tau$-leptons in the final state. The CPU time is given only for the low luminosity study.

| Channel | Level-1 $\epsilon$ (%) | HLT output (Hz) | HLT $\epsilon$ (%) | HLT CPU (ms) |
|---|---|---|---|---|
| $2\tau$-jet | 78 (62) | 3 (8) | 45 (36) | 130 |
| $\tau$-jet+$\not\!\!E_T$ | 81 (76) | 1 (2) | 58 (53) | 38 |
| $\mu + \tau$-jet | 72 (70) | 0.2 (1.2) | 54 (46) | 660 |
| $e + \tau$-jet | 80 (69) | 0.4 (1.8) | 70 (71) | 165 |

- Level-1: single- $\mu$ and combined $\mu$ +$\tau$-jet triggers.
- HLT: muon selection (section 4) and $\tau$-jet isolation with the calorimeter and the tracker.

The HLT output rates, the signal efficiency (for $M_H$=200 GeV/$c^2$) of the Level-1 and HLT selections as well as the CPU time at both low and high luminosity are listed in table 23. The efficiency of the combined triggers used for $A/H \to \tau\tau \to \ell + \tau$-jet channels is about 2-5 % higher than those of the single-lepton trigger and a function of the Higgs mass. For Higgs masses around 120 GeV/$c^2$, the combined triggers are expected to contribute significantly to the efficiency of the fusion channel $qq \to qqH$, $H \to \tau\tau$.

## 8. Identification of $b$-jets

Inclusive $b$-tagging of jet triggers can be used for the HLT selection of physics channels with $b$-jets in the final state. The algorithms used for $b$-tagging rely on the $b$-hadron proper life time ($c\tau$ ~450 $\mu$m), which gives rise to tracks with large impact parameter with respect to the production vertex.

### *8.1. b-tagging Algorithm*

A wide range of algorithms have been developed within CMS to tag $b$-jets [30]. The tagging method chosen for the studies presented here relies on the track impact parameter.

The track impact parameter can be calculated either in the transverse view (*2D* impact parameter) or in three dimensions (*3D* impact parameter). In the former case it is not affected by the uncertainty on the $z$-component of the primary vertex position while in the latter case a larger set of information can be used. In both cases the calculation is performed starting from the trajectory parameters at the innermost measurement point. In the *2D* impact parameter case the estimate can be done analytically since the trajectory is circular in the transverse view. In the three-dimensional case the extrapolation is performed by iteration. Figure 52 shows the main ingredients of the three-dimensional impact parameter calculation: first the point of closest approach of the track to the jet direction, $S$, is found. This point approximates the decay point of the $B$ hadron. The tracks are then linearized and their three-dimensional impact parameter is computed as the minimum distance from the primary vertex $V$. The $VQ$ segment in figure 52 is called the decay length and approximates the flight path of the B hadron. The impact parameter is signed as positive if $Q$ is upstream of $V$ in the jet direction (as in the example shown in figure 52), and negative otherwise. The tracks from a $B$ decay should have a positive impact parameter, while those coming from the primary vertex have an impact parameter comparable to the experimental resolution. The tag makes use of the track impact parameter significance, which is defined as the ratio of the value of the impact parameter with its uncertainty. A jet is tagged as a $b$-jet if there exist a minimum number of tracks with impact parameter significance above a given threshold.

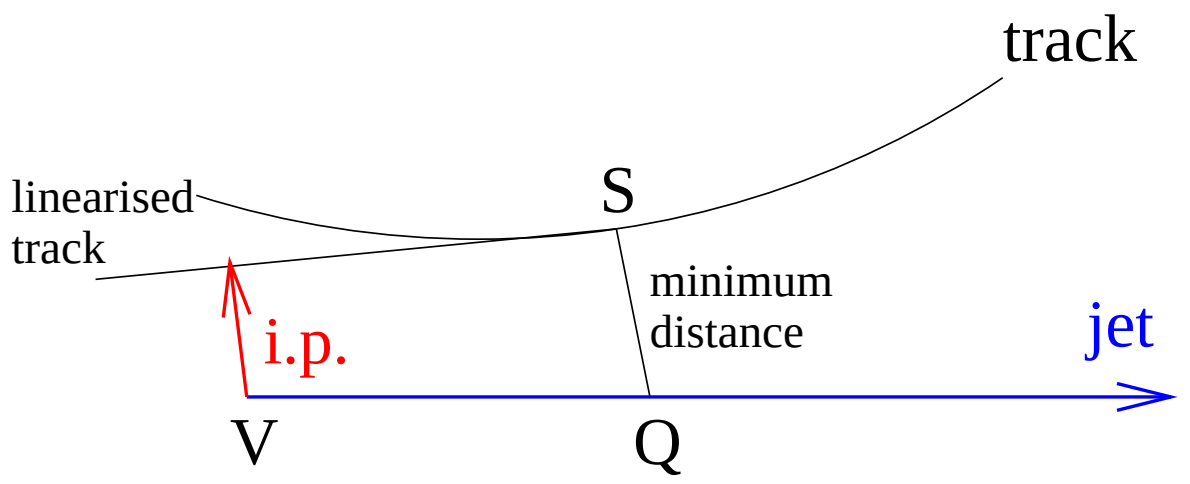

**Figure 52.** Representation (not to scale) of the track three-dimensional impact parameter.

In order to speed up the reconstruction, only tracks within a jet cone are used. The performance of the $b$-tagging algorithm (tagger) depends crucially on the quality of the tracks and the jet direction.

Tracks resulting from secondary interactions with the material, $K^0_S$ and $\Lambda^0$ decays are reduced by requiring the *2D* impact parameter be less than 2 mm and imposing a maximum on the decay length $VQ$ which depends on the jet energy and rapidity and varies between 1.5 to 10 cm. Optimization of these requirements was performed to maximize the $b$-tag signal efficiency at a fixed mis-tagging rate of 1%.

### *8.2. Tagging region*

Tracks are reconstructed in a cone around the Level-1 calorimeter jet. The cone apex is taken as the pixel reconstructed primary vertex with the algorithm presented in [32]. The optimal cone width depends on the reconstructed jet $E_T$. The number of tracks from $b$-decays inside the jet cone is largely a function of the cone size. The fraction of tracks coming from $b$-decays reaches a plateau value [31] at a $\Delta R \sim 0.25$. Beyond this point only tracks from the hadronization process are added. In the case of light flavour jets the number of tracks increases almost linearly. At high luminosity the ratio of non-$b$ tracks to $b$-tracks increases, requiring a harder $p_T$ cut.

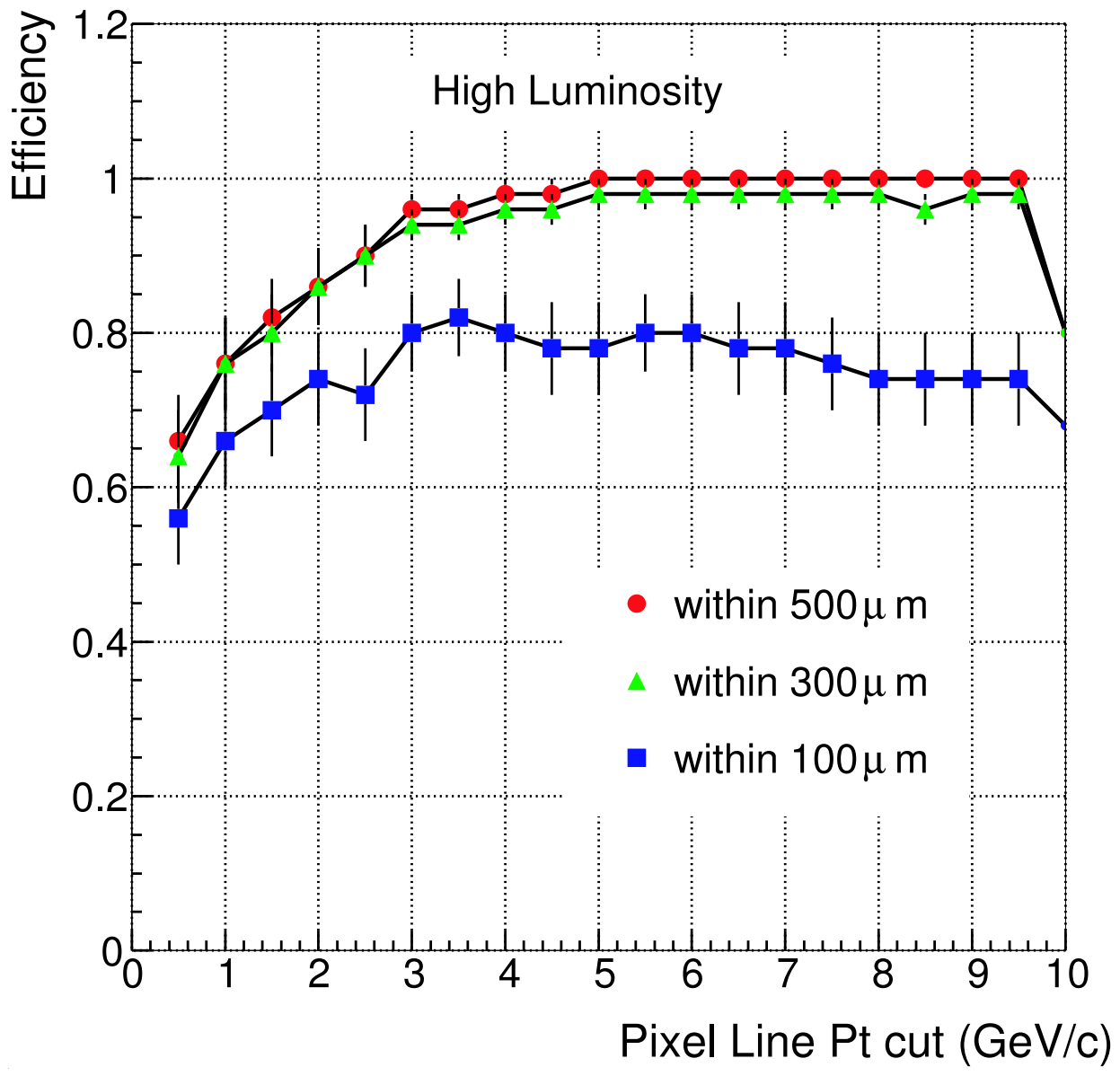

**Figure 53.** Efficiency of the pixel algorithm to correctly determine the primary vertex of the event within 100, 300 and 500 $\mu$m, at high luminosity, as a function of the minimum $p_T$ cut on the pixel lines.

The primary vertex is taken as the vertex having associated to it pixel lines with the largest summed $p_T$. At low luminosity the algorithm has high efficiency. At high luminosity the algorithm is modified so that only pixel lines above a high $p_T$ threshold are used. The high efficiency is maintained with a small loss of precision. Figure 53 shows the efficiency of the pixel algorithm to correctly assign the primary vertex within 100, 300 and 500 $\mu$m, at high luminosity, as a function of the minimum $p_T$ cut on the pixel lines. Primary vertex reconstruction requires about 50 msec on a 1 GHz Pentium-III CPU for both the low and high luminosity cases.

## *8.3. Track Reconstruction*

Track reconstruction is based on the partial reconstruction of tracks using the regional approach (section Appendix A): starting from pixel seeds, additional hits compatible with the track $p_T$, are sought in a region around the jet axis. The reconstruction is stopped after an adequate number of hits is found along the trajectory.
Two different regional seeding algorithms have been studied. The first (referred to as "pixel selective seeds") uses the pixel lines found by the pixel reconstruction which are contained inside a cone of $\Delta R < 0.4$ around the jet direction, and whose extrapolated $z$-impact point along the beam line is within 1 mm from the primary vertex. The second algorithm (referred to as the "combinatorial seed generator") uses all combinations of pixel hits which form an angle with respect to the jet direction of $\Delta\phi < 0.2$ and $\Delta\eta < 0.2$, centered in the primary vertex, with a tolerance of $\pm 1$ cm. The first method is faster. The efficiency of the second is comparable with that obtained at offline reconstruction.

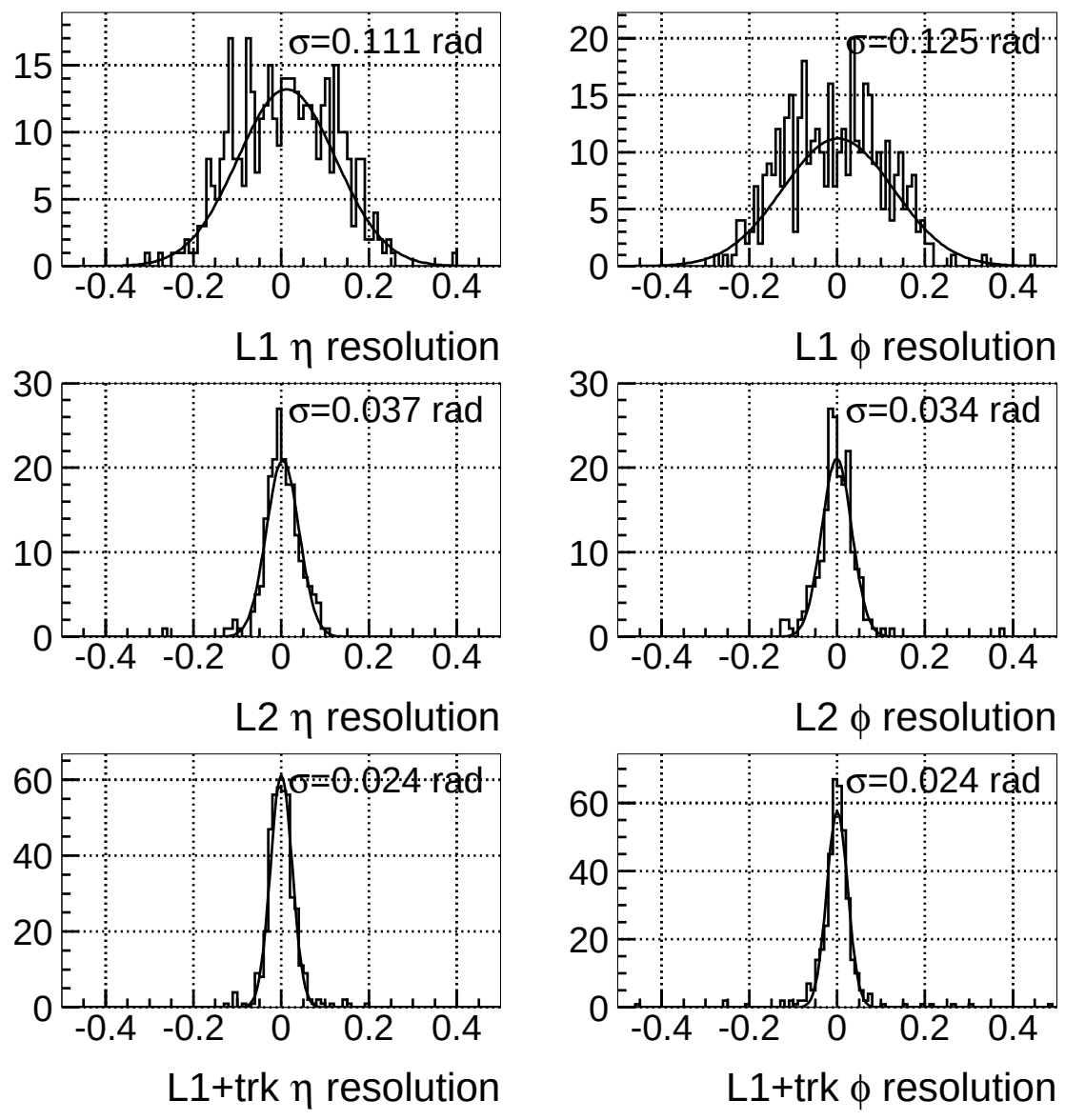

**Figure 54.** Jet angular resolution with respect to the generator information, using Level-1 jet reconstruction, HLT jet reconstruction, and by adding tracking information.

Tracks which have been reconstructed within the jet cone can be used to refine the jet direction measurement. Due to the coarse granularity of the calorimeter trigger cells, the direction resolution of the Level-1 jets is rather poor. Reduced angular resolution on the jet direction can cause a sign flip of the track impact parameter, deteriorating the performance of the tagger. The re-computed direction is determined as the $p_T$-weighted sum of the track directions. Figure 54 shows the difference in direction of the jet found at the generator level and of the jet reconstructed at Level-1, HLT and after including the tracks.

### *8.4. Performance and Timing*

We study HLT selections for two samples of events: back-to-back di-jets of different transverse energies and an inclusive QCD sample. The di-jet sample was produced in two different $p_T$ bins: $|\eta| < 1.4$ and $1.4 < |\eta| < 2.4$, corresponding to the central and forward regions of the tracker. Three bins with $E_T$= 50, 100 and 200 GeV were used. For $E_T$= 50 GeV the track spectrum is softer, with multiple scattering limiting the performance of the tag, while for $E_T$= 200 GeV the performance is limited by the high particle density. In the generation of these events, all the $pp \rightarrow qq$ processes were included, but only events with jets within the $|\eta|$ and $E_T$ range in question were selected. For the QCD sample, events generated with PYTHIA 6.152 [5] were retained if 50 $<{p_T}^{hard} <$ 170 GeV/$c$. A total of about 50,000 events in each ${p_T}^{hard}$ bin were analyzed. The performance of the selection is given in terms of the efficiency to tag $b$-jets versus the efficiency for $u$-jets, as well as execution time.

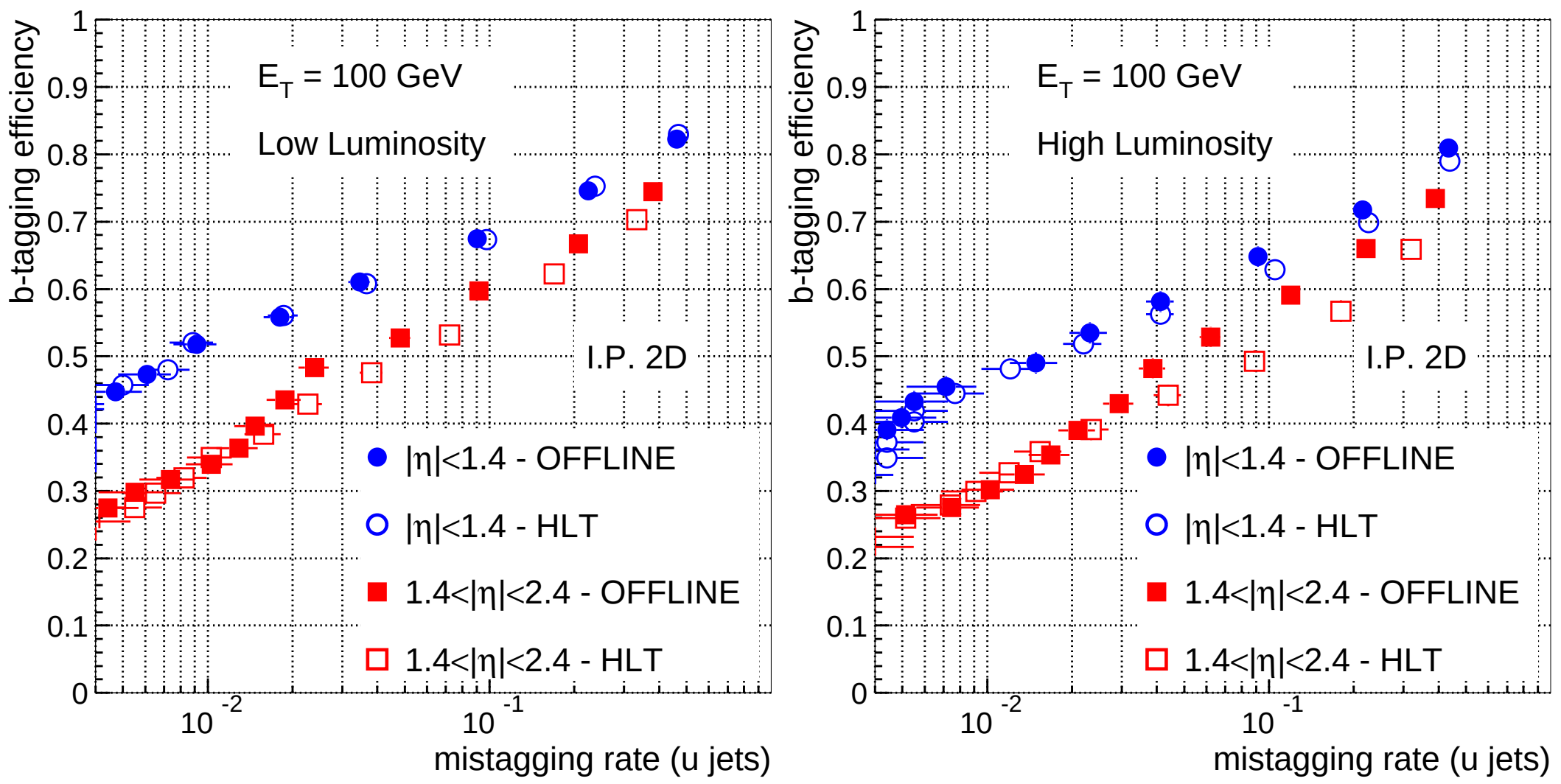

**Figure 55.** $b$-tagging efficiency versus mis-tagging rate for $E_T$=100 GeV jets and for (left) low and (right) high luminosity running conditions.

Tracks are reconstructed using a regional approach as explained in section 8.3, using both seeding algorithms, with a cut on $p_T$ of 1 GeV/$c$ (2 GeV/$c$) at low (high) luminosity. Track reconstruction is stopped when up to seven hits along the track are found. The primary vertex is reconstructed using the algorithm described in [32], with the only exception being that the default cut on pixel lines is 1 GeV/$c$ (5 GeV/$c$) for low (high) luminosity. A jet is tagged as a $b$-jet if it has two tracks exceeding a threshold on impact parameter significance. Tracks are required to have at least three (two) pixel hits for the full (staged) pixel detector configuration. Figure 55 shows the performance of the algorithm for 100 GeV $E_T$ jets and for two different $\eta$ regions, at low and high luminosity. The performance varies as a function of luminosity due to the different track $p_T$ requirement and the increased pixel readout inefficiency at high luminosity [33]. The performance is also compared with the one which could be obtained by using fully reconstructed tracks. No degradation is visible, so the online selection does not result in a reduced performance. A comparison of the performance with the staged and full pixel detector is shown in figure 56. Figure 57 compares the tagging performance in the different bins of jet $E_T$. The higher performance of the tagging algorithm for 100 GeV $E_T$ jets (compared to the 50 GeV jets) is due to the reduction of the multiple scattering component of the impact parameter uncertainty. The 200 GeV jets have higher mis-tagging rate due to the increased track multiplicity. A better performance can be achieved using the three-dimensional impact parameter, as demonstrated in the performance comparison for the 100 GeV $E_T$ jets shown in figure 57. An inclusive trigger capable of efficiently selecting $b$-jets originating for example from Higgs decays, can be implemented by demanding the presence of one or two $b$-tagged jets within the tracker acceptance. We have investigated the performance of a selection based on the leading or the next-to-leading jet which has been tagged as having at least 2 tracks with a transverse impact parameter significance larger than 2. Figure 58 shows the rate after this selection as a function of the corrected jet $E_T$ for the first two leading jets in the event. The execution time of the HLT algorithm is measured on a 1 GHz CPU and is studied as a function of the number of hits along the track for 100 GeV $E_T$ jets at low luminosity for the pixel selective seeds. The timing is measured for three different stages:

- Pixel and primary vertex reconstruction
- Track seeding, building, cleaning and smoothing
- $b$-tagging

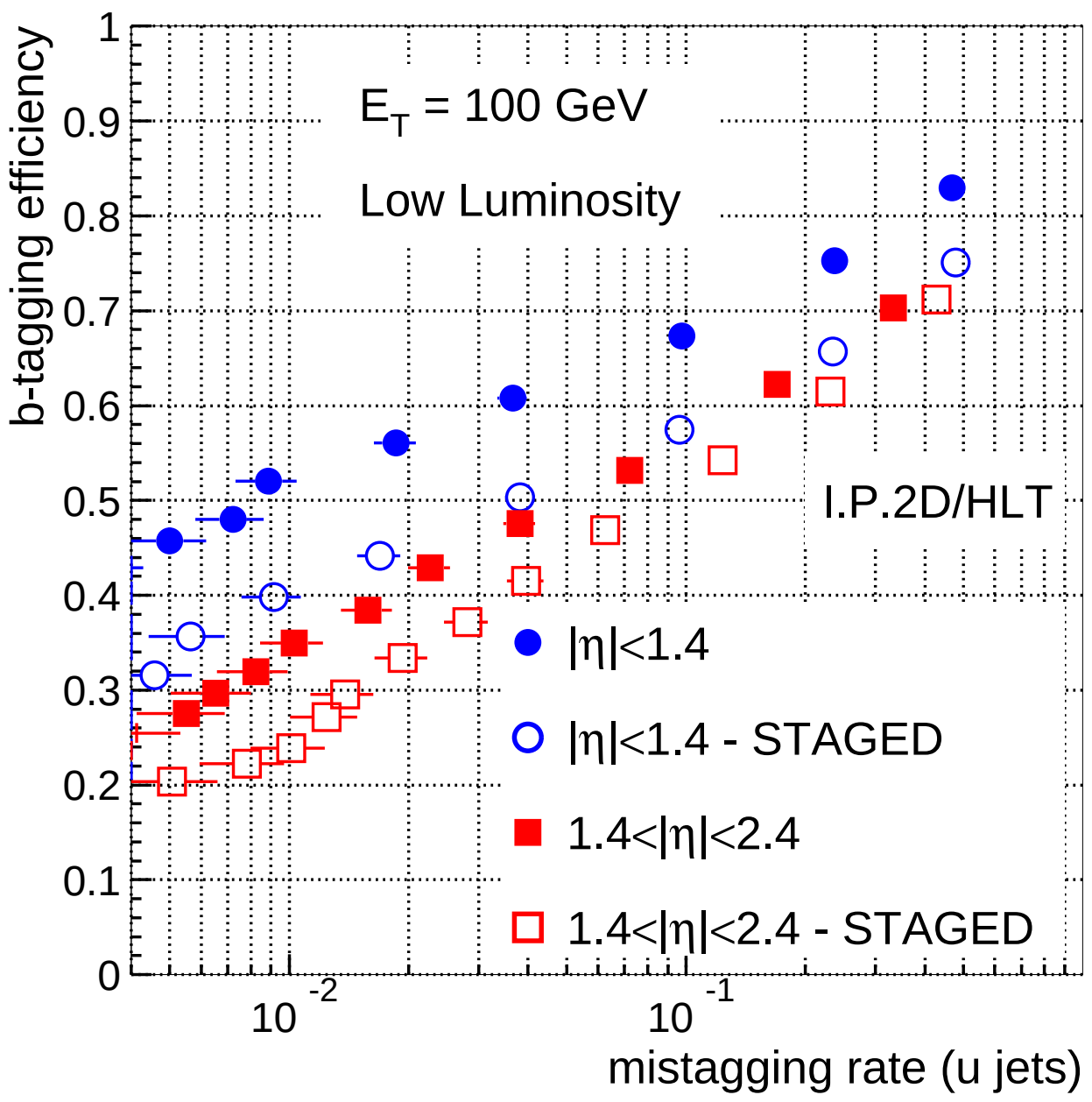

**Figure 56.** *b*-tag performance: comparison of the staged and full pixel detector configurations.

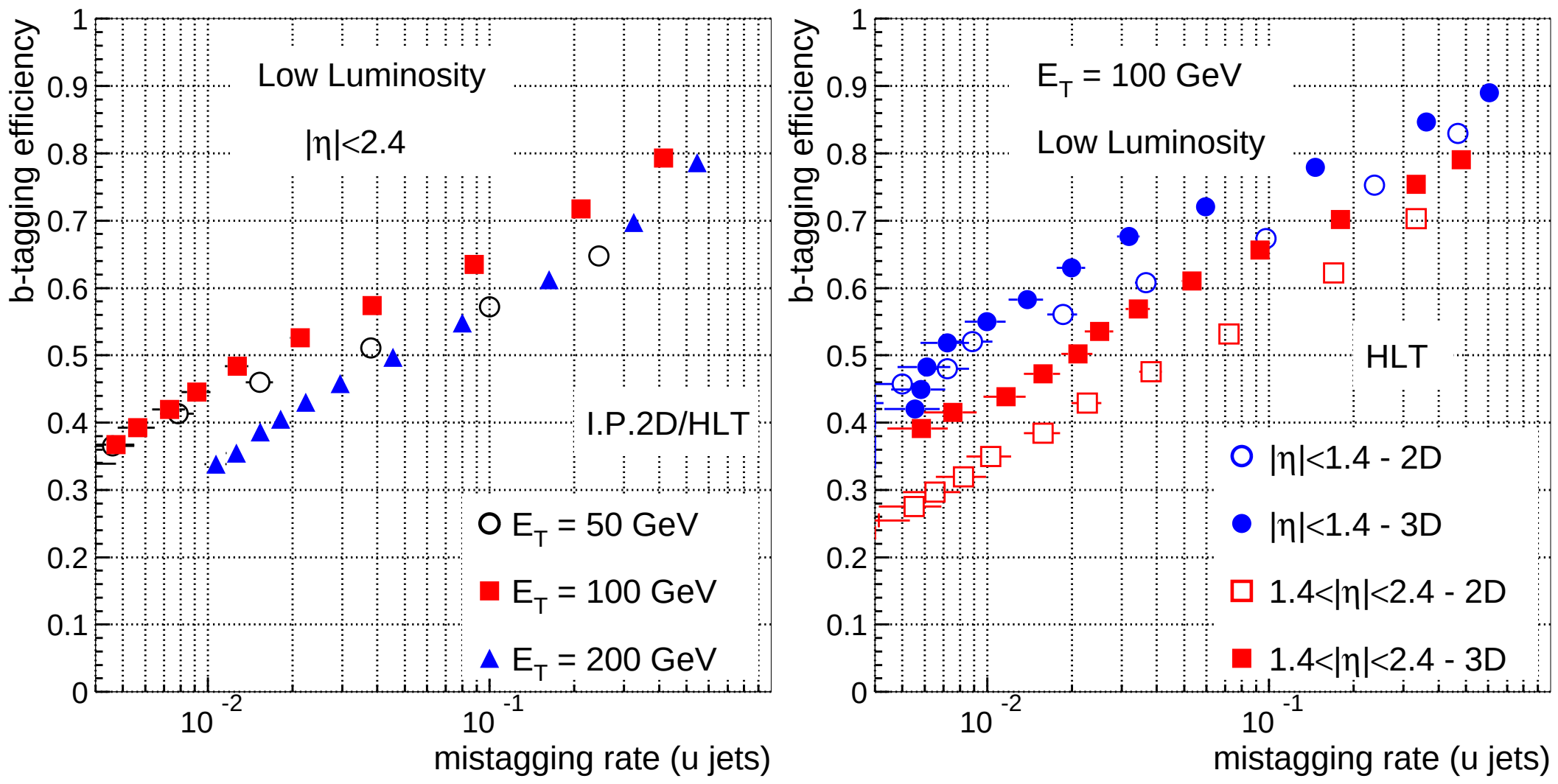

**Figure 57.** Efficiency for *b*-tagging versus the mis-tagging rate for different jet energies (left) and for *3D* versus *2D* impact parameter (right).

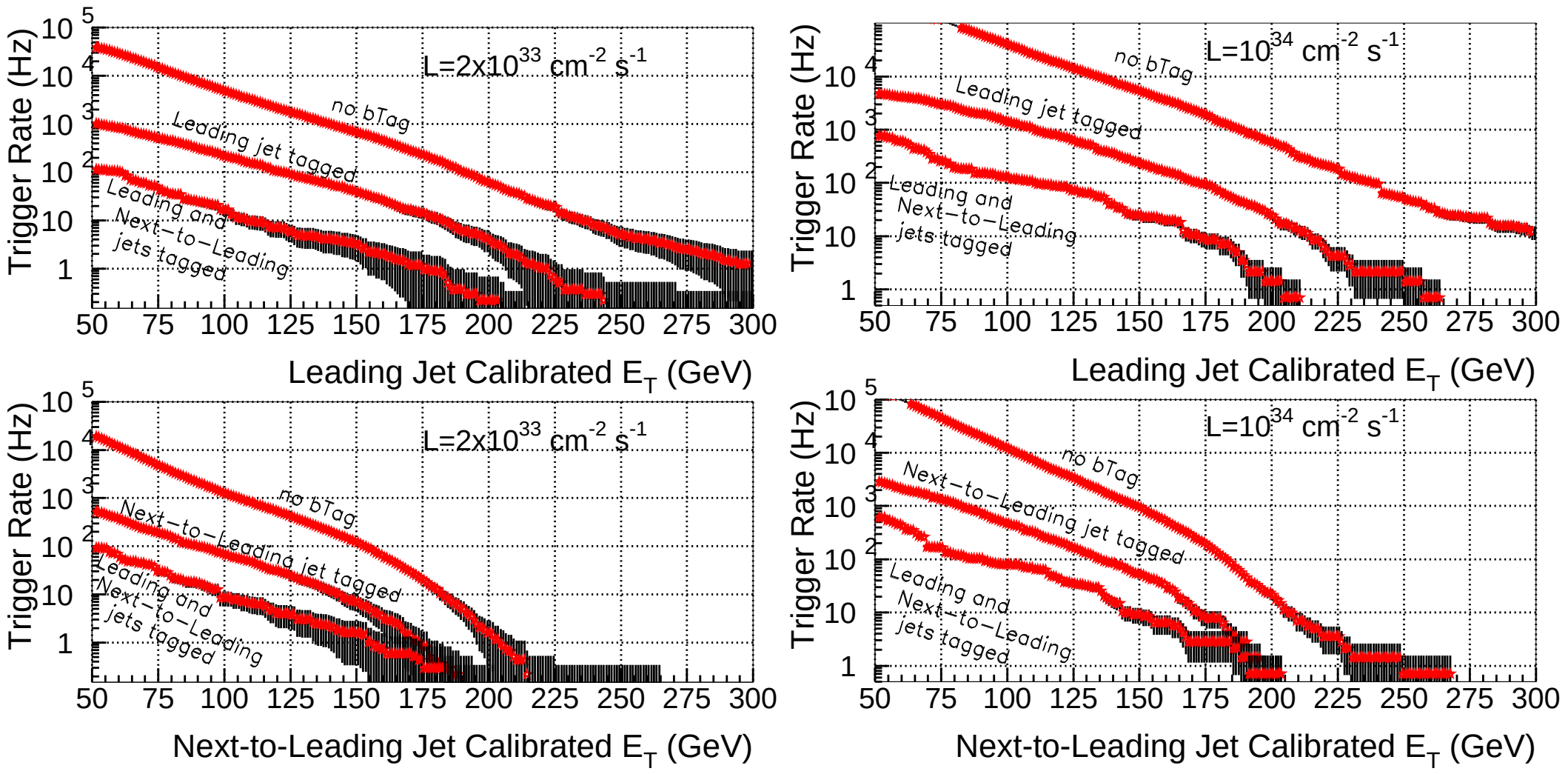

**Figure 58.** Rate after the $b$-tag selections for low (left) and high (right) luminosity for the leading (top) and next-to-leading (below) jets ordered in calibrated $E_T$, within the tracker acceptance. The upper curves indicate the trigger rate at Level-1. The middle curves refer to the case where the leading jet is tagged (top) or the next-to-leading jet (below) is tagged, while the lower curves refer to the case when both the two leading jets are tagged.

Track reconstruction is the most time consuming stage, and scales almost linearly with the number of hits. The $b$-tagging part is negligible, suggesting that it might be possible to use more sophisticated algorithms with increased performance. At high luminosity, the larger number of seeds increases significantly the execution time. As a comparison, for the same jet transverse energy and pseudorapidity bin, the number of seeds which need to be considered increases from 7 to ∼44 in going from low to high luminosity, even after taking into account the larger $p_T$ cut applied. The execution time also depends strongly on the jet energy when using the combinatorial seeding method. The execution time is considerably smaller if the pixel selective seeds algorithm is used.

### *8.5. Summary*

The HLT can benefit from efficient $b$-jet tagging with reasonable execution times at both low and high luminosity. As an example, the requirements of a possible dedicated $b$-tagged HLT selection with at least one jet inside the tracker acceptance and a $b$-tag based on the *2D* impact parameter significance $s$ are:

- For jets with $E_T < 80$ GeV: two tracks with $s > 1.5$.
- For jets at the $E_T$ range $80 < E_T < 150$ GeV: two tracks with $s > 2.0$.
- For jets with $E_T > 150$ GeV: two tracks with $s > 2.5$.

At low luminosity running conditions such a selection is 55% efficient for $b$-jets with a background rejection factor of about 10, almost independent of the jet $E_T$. It corresponds to a rate of 5 Hz for a cut on the leading jet $E_T$ of about 200 GeV or 160 GeV on the next-to-leading jet. This selection can be used in the case of SUSY searches, where the number of $b$-jets in the final state is large (section 9).

## 9. HLT Selection and Performance Overview

This section summarizes the physics object selection, the total estimate for the CPU requirements, and the physics performance of a prototype HLT table for the start–up luminosity of $2 \times 10^{33}$ cm$^{-2}s^{-1}$. The current set of thresholds and rates to storage for each physics object, described in the preceding sections, is listed in table 24. The values of the thresholds shown in table 24 are indicative of the type of event mixture that would yield an output event rate of $\mathcal{O}(100)$ Hz. The CMS start–up HLT allocated total rate is 150 Hz to allow for (i) trigger contingency (additional physics and back–up triggers) and consequently (ii) meeting the goals of a rich physics program. A 1.5 MB/event imposes a requirement on the data recording rate of ∼225 MB/s.

### *9.1. CPU Requirement*

A key issue for the HLT is the CPU power required for the execution of the physics objects selection algorithms. The algorithms were timed on a Pentium-III 1 GHz processor, and the requirements varied from ∼ 50 ms for jet reconstruction, to ∼700 ms for muon reconstruction.

The first step is to weight the CPU needs of the algorithms by the frequency of their application, which is the Level-1 trigger rate of the corresponding channels. This is shown in table 25 and yields a total of 4092 CPU seconds as the total need to cover the 15.1 kHz of events output from the Level-1 trigger for low luminosity conditions. In the full Level-1 trigger rate budget of table 2 there is an additional 0.9 kHz of minimum bias events that will be used for calibration and monitoring. These events are assumed to require the same amount of CPU as the mean of the ∼15 kHz of events for which CPU time estimates are available. The average processing time per event at Level-1 is 271 ms.

**Table 24.** High-Level Trigger requirements at low luminosity. The thresholds correspond to the values in $E_T$ or $p_T$ with 95% efficiency (90% efficiency for muons). There is no actual threshold in the HLT selection for $\tau$-jets, so the threshold shown is that of the corresponding Level-1 Trigger requirement. (*)Calibration percentage allocated corresponds to well understood detector performance. At start-up, this can be as high as 30% of the total rate.

| Trigger | Threshold (GeV or GeV/c) | Rate (Hz) | Cumulative Rate (Hz) |
|---|---|---|---|
| inclusive electron | 29 | 33 | 33 |
| di-electron | 17 | 1 | 34 |
| inclusive photon | 80 | 4 | 38 |
| di-photon | 40, 25 | 5 | 43 |
| inclusive muon | 19 | 25 | 68 |
| di-muon | 7 | 4 | 72 |
| $\tau$-jet * $\not{E}_T$ | 86 * 65 | 1 | 73 |
| di-$\tau$-jets | 59 | 3 | 76 |
| 1-jet * $\not{E}_T$ | 180 * 123 | 5 | 81 |
| 1-jet OR 3-jets OR 4-jets | 657, 247, 113 | 9 | 89 |
| electron * $\tau$-jet | 19 * 45 | 0.4 | 89.4 |
| muon * $\tau$-jet | 15 * 40 | 0.2 | 89.6 |
| inclusive b-jet | 237 | 5 | 94.6 |
| calibration and other events (10%)* | | 10 | 105 |
| **TOTAL** | | | **105** |

**Table 25.** Summary of CPU time required for the selection of each physics objects in the HLT. The CPU figures refer to a 1 GHz Intel Pentium-III CPU.

| Physics Object | CPU time per Level-1 event (ms) | Level-1 Trigger rate (kHz) | Total CPU time (s) |
|---|---|---|---|
| electron/photon | 160 | 4.3 | 688 |
| muon | 710 | 3.6 | 2556 |
| tau | 130 | 3.0 | 390 |
| jets and $\not{E}_T$ | 50 | 3.4 | 170 |
| electron + jet | 165 | 0.8 | 132 |
| $b$-jets | 300 | 0.5 | 150 |

We estimate the CPU power to carry out the physics program at the start-up of the LHC, when the luminosity will not have reached its full value. The current scenario for CMS is to provide a DAQ system capable of reading a maximum of 50 kHz of events accepted by the Level-1 Trigger. The CPU requirement for this system is 15,000 CPUs as in those available in a standard commercial Personal Computer (PC). Since these timing measurements in 2002 the power per CPU has been increased by a factor 3-3.5 so in 2005 this corresponds to about 4500-5000 CPUs. To extrapolate these figures from early 2005 to the year 2007 the basic thesis of Moore's Law, i.e. that CPU power increases by a factor two every 1.5 years, is used. This implies that a total of $\sim$ 2,000 CPUs will be needed for the system at the LHC start-up.

### *9.2. Efficiency of the HLT selection for Major Physics Channels*

The discovery of the long-sought Higgs boson is the focus of the physics program of the LHC. This section summarizes the studies of some of the expected Higgs signals, both within the standard model and in the context of supersymmetry. The mass range to discover the standard model Higgs at the LHC ranges from the upper limit of direct searches at LEP, namely 114.4 GeV/$c^2$ to approximately 1 TeV/$c^2$. In the heavy mass range (up to 800 GeV/$c^2$) the channels with the best sensitivity for the Higgs is the $H \to ZZ$. In particular the $H \to ZZ \to$ four leptons, e.g. $\mu\mu\mu\mu$ channel is referred to as the "gold-plated" one, for masses above 180 GeV/$c^2$. For lower values of the Higgs mass the decay channels $H \to WW^* \to ll\nu\nu$, $H \to \gamma\gamma$, $WH \to W\gamma\gamma$, $qqH \to qqWW^*$, and $qqH \to qq\tau\tau$ become important for the Higgs study. The channels $t\bar{t}H \to t\bar{t}\gamma\gamma$, and $t\bar{t}H \to t\bar{t}b\bar{b}$ can also be explored to discover and study the Higgs.

For the region around 120 GeV/$c^2$ the most promising channels to date are the fusion processes or the decays into two photons. The decay $H \to \gamma\gamma$ will be triggered at the HLT requiring two photons detected in the ECAL and validated by the tracker, using the asymmetric cuts of $E_T$>40 and 25 GeV for the photons. This requirement persists in the final offline selection for this channel. It yields a final background rate to tape of 5 Hz. For low-luminosity running the combined Level-1/HLT trigger efficiency for a Higgs with mass of 115 GeV/$c^2$ is 77% for all the decays where both photons are within the ECAL fiducial volume ($|\eta| < 2.5$). With respect to all events where the decay photons have the offline minimum $E_T$, the efficiency is 83.7%. The geometrical acceptance for the two photons to be within the fiducial volume is 65%. The trigger efficiency for high luminosity running is typically 5% lower.

The channel $H \to WW^* \to \mu\mu\nu\nu$ has been studied for Higgs mass values of 120, 160 and 200 GeV/$c^2$. The combined low-luminosity HLT trigger consists of a single muon $p_T$ threshold of 19 GeV/$c$ and a symmetric double muon threshold of 7 GeV/$c$ and has a total output event rate of 29 Hz. The total efficiency within the fiducial acceptance is 92% at $M_H = 160$ GeV/$c^2$. The efficiency increases with increasing Higgs mass. The efficiency for this trigger decreases by 10-15% at high luminosity, because the thresholds have to be increased to 31 and 10 GeV/$c$ for the single and double muon trigger respectively. The golden channel mentioned above $H \to ZZ \to 4\mu$ has near-full efficiency since the efficiency for $Z \to \mu\mu$ is about 92%. These efficiencies are relative to having at least one muon within the geometric acceptance of the trigger, $|\eta| < 2.1$, and all final state muons (two or four, depending on the channel) within the full acceptance of the muon system ($|\eta| < 2.4$).

These examples show that working HLT trigger schemes with good efficiency have been identified for the Higgs search. Note that a heavy standard model Higgs will be easier to trigger on than a light one (~120 GeV/$c^2$). For the minimum sypersymmetric standard model (MSSM) neutral high-mass Higgs particles, $A/H$, and the charged Higgs, the "golden" decays to heavy gauge bosons are closed or strongly suppressed. Therefore alternative decay channels have been studied, in particular the decay modes $A/H \to \tau\tau$ and $H^+ \to \tau^+\nu$. The main background to these channels are QCD jets. The $\tau$-jets are identified as described in section 7. The identification can be performed by calorimeter $\tau$-jet candidates supplemented either by a fast track finding algorithm in the pixel detectors or by using regional reconstructed track finding algorithm. For the low luminosity runs the latter algorithm is found to be more efficient, giving an efficiency of about 45% (49%) for $M_H = 200$ GeV/$c^2$ ($M_H = 500$ GeV/$c^2$), while achieving the required background reduction of a factor ~1000. Both $\tau$-jets are required to pass the trigger criteria. The corresponding efficiencies for high luminosity running are about 15% lower.

A similar trigger can be defined for the charged heavy Higgs decay $H^+ \to \tau\nu$, requiring one $\tau$-jet and missing $E_T$ at the HLT selection. The efficiencies are 58% ($M_H =$ 200-400 GeV/$c^2$) for low luminosity and about 10% lower for high luminosity conditions.

For the neutral MSSM Higgs the combined electron/$\tau$-jet trigger channels have been studied for the decay $A/H \to \tau\tau \to e + \tau$-jet. The combined trigger has a reduced threshold on the electron, compared to the single-electron trigger, and obtains an HLT efficiency of 70% at both low and high luminosity. The overall efficiency of the combined trigger is only about 2-5% higher than the one obtained by the single electron trigger, but this is mostly due to the choice of a relatively high mass for the Higgs. For Higgs masses around 120 GeV/$c^2$, the fusion channel $qqH \to qq\tau\tau$ benefits from such a trigger. Similar studies were performed for the channel $A/H \to \mu + \tau$-jet and result in an efficiency of 54% (46%) at low (high) luminosity.

As a specific example of a more exotic Higgs study, we describe in detail the trigger strategy for so called "invisible Higgs" detection. Some extensions of the Standard Model (SM) contain Higgs bosons which can decay into stable neutral weakly interacting particles, therefore giving rise to final states with large missing energy. In recent work [39, 40], it was shown that the potential for discovering a Higgs boson decaying invisibly at the LHC can be extended considerably by studying Higgs production via weak boson fusion (WBF). The presence of the two forward-backward tagging jets that accompany the Higgs production via weak boson fusion is a powerful tool for separating signal from the very large backgrounds. The large rapidity gap between the two jets is used in the trigger definition. The following offline cuts for an invisible Higgs analysis were proposed in reference [39] and studied using a detailed simulation [40]:

**Table 26.** Acceptance of the CMS HF calorimeter to tagging jets ($E_T$> 30 GeV) in the $qq \rightarrow qqH$ process

| | no jets in HF | one jet in HF | 2 jets in HF |
|---|---|---|---|
| no cut on $\vert\eta_{j1} - \eta_{j2}\vert$ | 49% | 45% | 6% |
| $\vert\eta_{j1} - \eta_{j2}\vert$ >4.4 | 22% | 65% | 13% |

(i) $E_T^j > 40$ GeV, $|\eta_j| < 5.0$, $|\eta_{j1} - \eta_{j2}| > 4.4$, $\eta_{j1}\eta_{j2} < 0$,

(ii) $\not\!\!E_T$> 100 GeV

(iii) $M_{jj} > 1200$ GeV/$c^2$

(iv) $\Delta\phi_{jj} < 1$

The efficiencies of the Level-1 trigger and the HLT selections discussed below are the efficiencies for events that pass the above offline selections. The HF calorimeters play a crucial role in the selection of the invisible Higgs because of the two forward-backward tagging jets. The relative acceptance in the HF calorimeter (3.0< $|\eta|$ <5.0) for these jets after requiring $E_T$>30 GeV for each jet is shown in table 26 before and after the cut on the rapidity gap between the jets. After the rapidity gap constraint, almost 80% of the Higgs events will have at least one tagging jet in the HF. The Level-1 jet trigger covers the entire calorimeter acceptance, including the HF calorimeter. At Level-1 a jet+$\not\!\!E_T$ trigger can therefore be used for the invisible Higgs selection. Figure 59 shows the transverse energy of the highest $E_T$ jet (left) and calorimeter $\not\!\!E_T$ (right) reconstructed at the HLT and at Level-1 at $\mathcal{L} = 10^{34}\,\mathrm{cm}^{-2}\mathrm{s}^{-1}$ for Higgs events passing the WBF requirements and for a Higgs mass of 120 GeV/$c^2$. The Level-1 trigger was optimized by examining the trigger rate for a single jet plus $\not\!\!E_T$ trigger versus the signal efficiency by changing the $\not\!\!E_T$ thresholds for the fixed set of single jet thresholds. Figure 60 shows the Level-1 jet+$\not\!\!E_T$ trigger rate and Higgs efficiency for single jet thresholds of 70(60), 90(70), 110(80) at high(low) luminosity when the $\not\!\!E_T$ threshold is varied.

Table 27 lists the Level-1 jet+$\not\!\!E_T$ trigger thresholds and the Higgs efficiency for Level-1 jet+$\not\!\!E_T$ trigger rates of 0.2, 0.5 and 1.0 kHz at low and high luminosity. A rate of 0.5 kHz gives a high Higgs efficiency, namely 98% (80%) for low(high) luminosity.

For the HLT selection, the offline requirements $E_T^j > 40$ GeV, $|\eta_j| < 5.0$, and $|\eta_{j1} - \eta_{j2}| > 4.4$ are used along with the requirement of $M_{jj} > 1$ TeV/$c^2$, and a $\not\!\!E_T$ requirement. Figure 61 (left plot) displays the rate for QCD multi-jet events as a function of the $\not\!\!E_T$ threshold with only the $E_T$ and $\eta$ requirements applied and with the additional $M_{jj}$ requirement also applied. A rate of 0.1 Hz can be reached at low luminosity for $\not\!\!E_T$>110 GeV without the $M_{jj}$ cut. The signal efficiency in this case is close to 100% since the offline cut on $\not\!\!E_T$ is 100 GeV (figure 59). At high luminosity a rate of 0.2 Hz can be reached with the $M_{jj}$ cut and an ~150 GeV threshold on $\not\!\!E_T$. The total Level-1 jet+$\not\!\!E_T$ and HLT efficiency for Higgs at high luminosity is shown in figure 61 (right plot) as a function of the $\not\!\!E_T$ cutoff. For 0.2 Hz rate ($\not\!\!E_T$> 150 GeV and $M_{jj} > 1$ TeV/$c^2$) the

**Figure 59.** Higgs events in the process $qq \to qqH$, with the Higgs decaying invisibly ($M_H$ = 120 GeV/$c^2$), that satisfy the WBF requirements. (Left) Transverse energy of the highest-$E_T$ jet reconstructed in offline (solid histogram) and at Level-1 (dashed histogram). (Right) $\not\!\!E_T$ reconstructed offline (solid histogram) and at Level-1 (dashed histogram). (for )

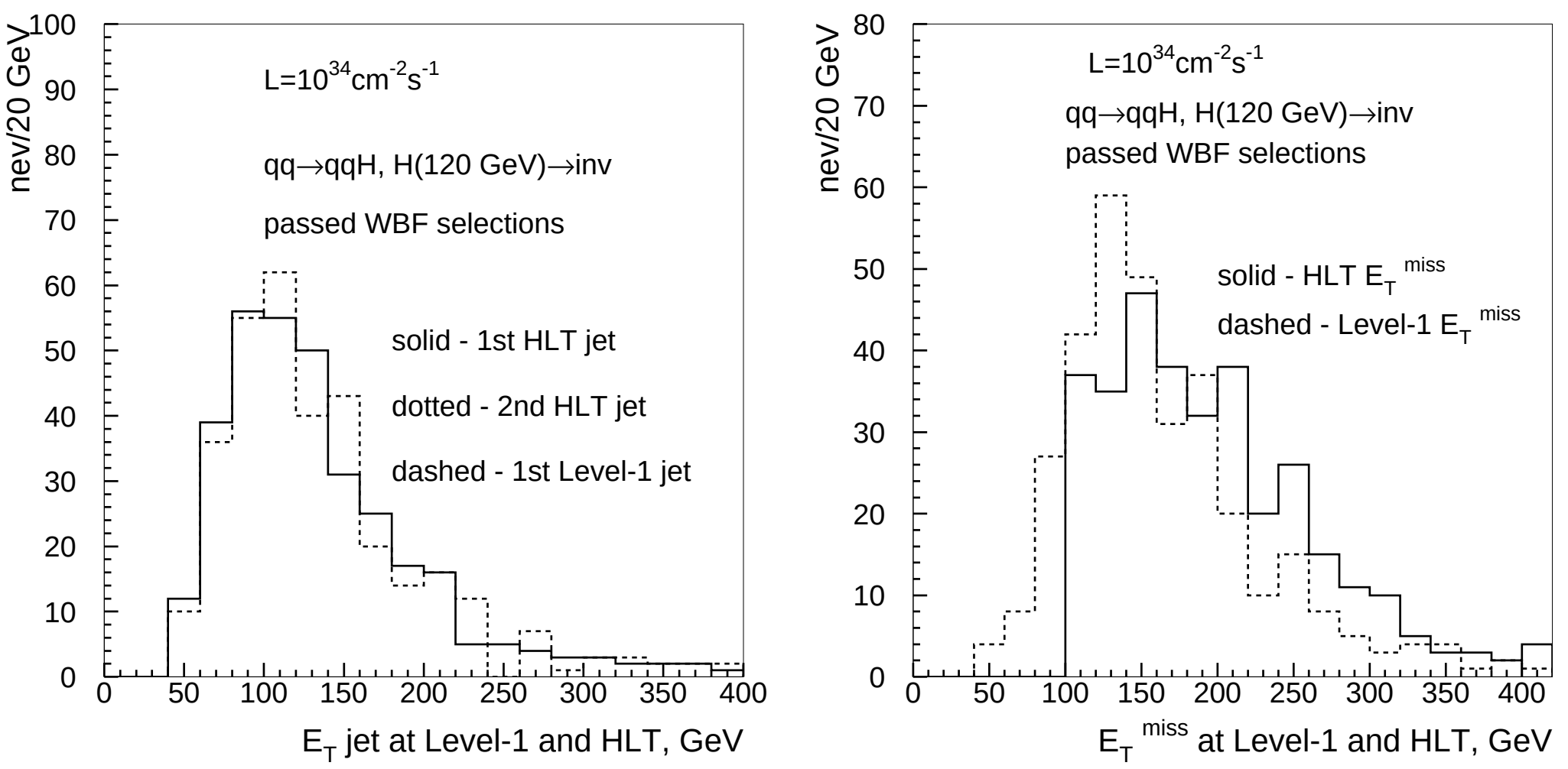

.

**Figure 60.** QCD 2-jet rate (in kHz) for a jet+$E_T^{miss}$ trigger versus efficiency for $qq \to qqH$, with the Higgs decaying invisibly, for a Higgs mass of $M_H$ = 120 GeV/$c^2$, and for events which satisfy the WBF selection when the $\not\!\!E_T$ threshold is varied and with single jet thresholds as labeled, low (left) and high (right) luminosity.

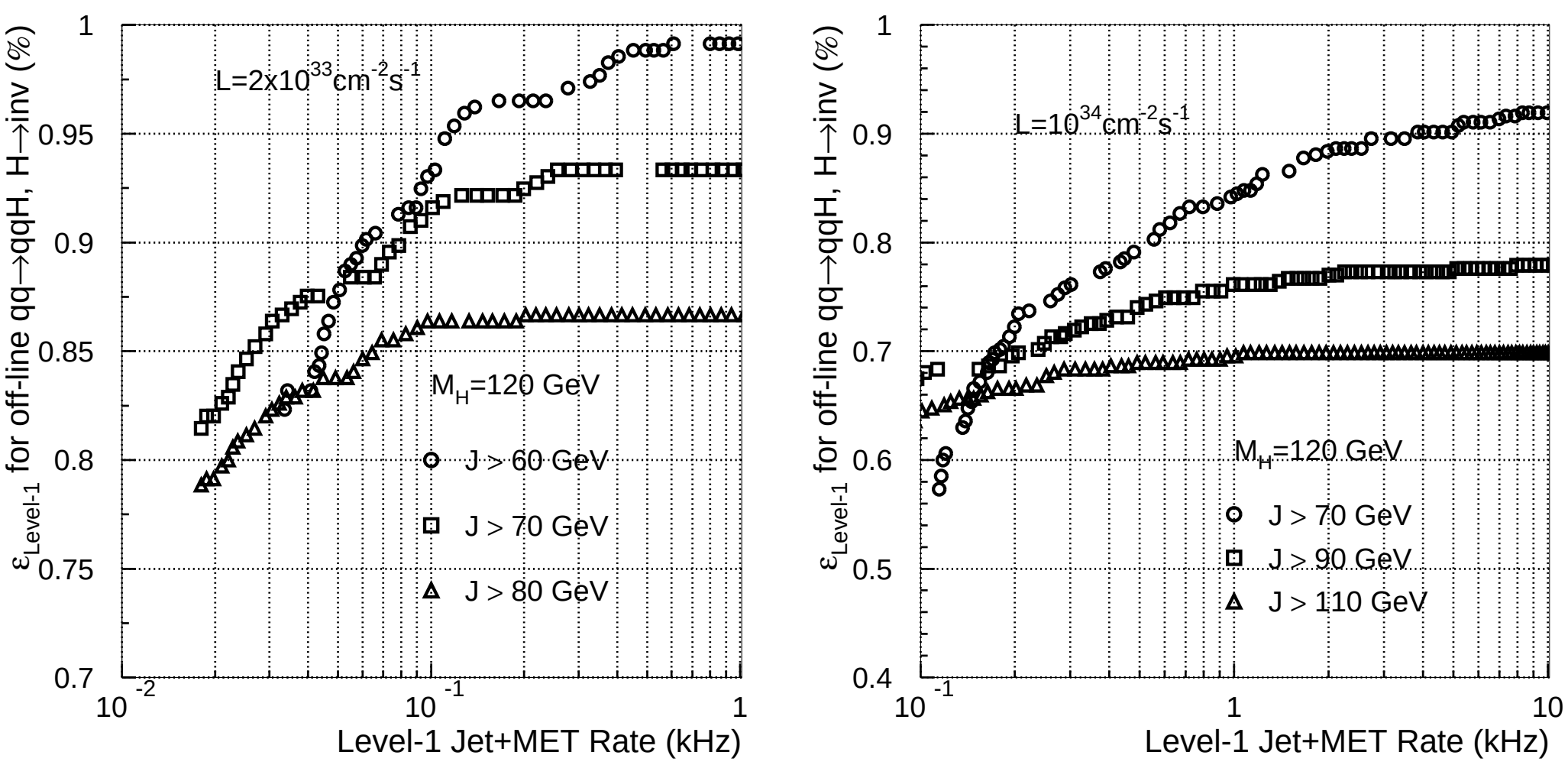

**Table 27.** Summary of the single-jet plus $\not{E}_T$ Level-1 trigger thresholds for a total trigger rate of 0.2, 0.5, and 1.0 kHz at low and high luminosity. Shown are the jet and $\not{E}_T$ thresholds, the efficiency for $qq \to qqH$ with the Higgs ($M_H = 120$ GeV/$c^2$) decaying invisibly for events which pass the WBF cuts.

| Rate for Level-1 jet+$\not{E}_T$ trigger | | 0.2 kHz | 0.5 kHz | 1.0 kHz |
|---|---|---|---|---|
| low luminosity | single-jet threshold | 60 | 60 | 60 |
| | $\not{E}_T$ threshold (GeV) | 73 | 64 | 56 |
| | efficiency | 0.96 | 0.98 | 0.99 |
| high luminosity | single-jet threshold, | 70 | 70 | 70 |
| | $\not{E}_T$ threshold (GeV) | 122 | 112 | 73 |
| | efficiency | 0.72 | 0.79 | 0.84 |

**Figure 61.** Left: QCD 2-jet background rate after topological WBF requirements as a function of the $\not{E}_T$ cutoff for $\mathcal{L} = 10^{34}\,\mathrm{cm}^{-2}\mathrm{s}^{-1}$(solid histogram) and $\mathcal{L} = 2 \times 10^{33}\,\mathrm{cm}^{-2}\mathrm{s}^{-1}$(dashed histogram). Right: total Level-1 single jet plus $\not{E}_T$and HLT efficiency for $qq \to qqH$, with the Higgs decaying invisibly, $M_H = 120$ GeV/$c^2$, and for events which passed WBF cuts, as a function of $\not{E}_T$ cutoff for $\mathcal{L} = 10^{34}\,\mathrm{cm}^{-2}\mathrm{s}^{-1}$.

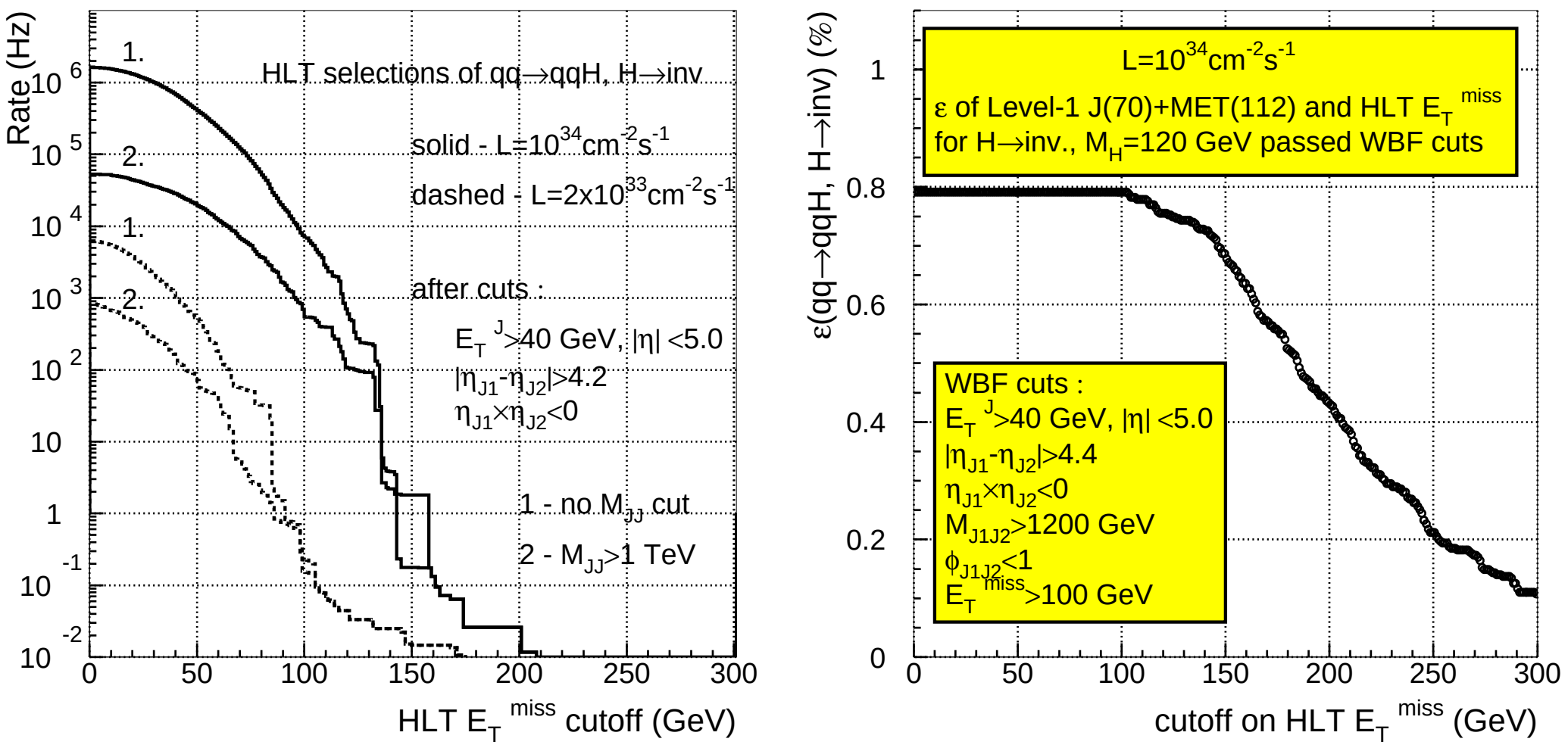

Higgs efficiency is 0.7.

## *9.3. Supersymmetry Searches*

One of the main goals of the LHC is to search for evidence for supersymmetry (SUSY), the most powerful extension of the standard model. If SUSY exists, large amounts of supersymmetric particles (sparticle) are expected to be produced shortly after the LHC turn-on. However, unless SUSY is discovered at the Tevatron before the LHC start-up, the signature of SUSY will not be known in advance. The most popular SUSY models invoke the conservation of R-parity (RP), which makes the lightest Supersymmetric particle stable and in some cases an excellent candidate for cold dark matter. In these models, squark and gluino events, which are produced strongly and therefore have very large production cross section, would appear in the detector as events with multiple jets and large $\not\!\!E_T$. Due to cascade decays of charginos and neutralinos the final state usually also contains a number of leptons. In some points of the parameter space the direct chargino neutralino production provides striking tri-lepton signatures.

**Table 28.** Parameters used for the generation of the SUSY mSUGRA samples used in this paper. For all points $A_0$= 0, $\tan\beta$ = 10, and $\mu > 0$. Masses are in GeV/$c^2$ and cross section in pb.

| Point | $M_0$ | $M_{1/2}$ | $\sigma$ | $m\tilde{g}$ | $m\tilde{u}_L$ | $m\tilde{\chi}_1^0$ | $M_h$ |
|---|---|---|---|---|---|---|---|
| 4 | 20 | 190 | 181 | 466 | 410 | 70 | 110 |
| 5 | 150 | 180 | 213 | 447 | 415 | 66 | 110 |
| 6 | 300 | 150 | 500 | 349 | 406 | 45 | 106 |
| 7 | 250 | 1050 | 0.017 | 2235 | 1986 | 445 | 122 |
| 8 | 900 | 930 | 0.022 | 2032 | 1962 | 391 | 121 |
| 9 | 1500 | 700 | 0.059 | 1625 | 1975 | 293 | 120 |

Supersymmetry models have a large number of free parameters. There have been several studies to identify points in the SUSY parameter space that will in some way span the range of signatures and predictions that apply to the start of the LHC. Reference [41] was used to select the points studied here. These points all use the mSUGRA parametrization of the SUSY parameter space. Other parametrizations have not been considered, since the purpose of this study is not to provide an exhaustive study of SUSY but to give examples of the prototype Level-1 anf HLT selection efficiency for supersymmetric signatures.

At low luminosity, the greatest challenge comes from the points with the lowest sparticle masses just above the reach of the Tevatron, because the transverse energies of the jets and $\not\!\!E_T$ are relatively low. At high luminosity, the challenge is to maximize the acceptance for the highest mass points, since they have the smallest cross section.

Table 28 lists the parameters and the masses of some sparticles as well as the production cross sections for the points used to exercise and test the appropriate HLT selection paths. The SUSY mass spectra and branching ratios were calculated using `ISAJET 7.51`[42]. This information was imported into `HERWIG 6.301`[43], which was used to generate the samples. The points were chosen to give a variety of potential SUSY signatures. Point 4 has enhanced slepton (especially stau) production. Point 5 is a "typical" SUSY point with squarks lighter than gluinos resulting in large $\not\!\!E_T$. At point 6 the gluinos are lighter than the squarks resulting in large jet multiplicity final states with a smaller $\not\!\!E_T$ than a typical point. At point 7 stau and sneutrino production is enhanced. Point 8 is characterized by an enhanced $b$-jet yield compared to point 7, due to neutralino decays to higgs while the $\not\!\!E_T$ is similar to that of point 7. For point 9, the gluinos are lighter than all squarks, except for the lighter stop, thus allowing the decay of gluino to the lightest stop which dominates and therefore events have many jets and smaller $\not\!\!E_T$. While these exact points (points 4, 5, and 6 in particular) may be excluded by LEP Higgs searches, Higgs production and the exact mass for the Higgs in the decays does not play an important role in the observability or the characteristics of these events. The same points are used to simulate SUSY with R-parity violation with $\chi_1^0 \rightarrow jjj$.

**Figure 62.** HLT rate versus efficiency for SUSY signals, for events that pass the Level-1 jet+$\not\!\!E_T$ trigger. (Left) HLT rate-efficiency contours for each point and for a range of jet thresholds using the 4-jet trigger path. (Right) HLT rate-efficiency contours for each point and for a range of $\not\!\!E_T$ thresholds using the jet+$\not\!\!E_T$trigger path.

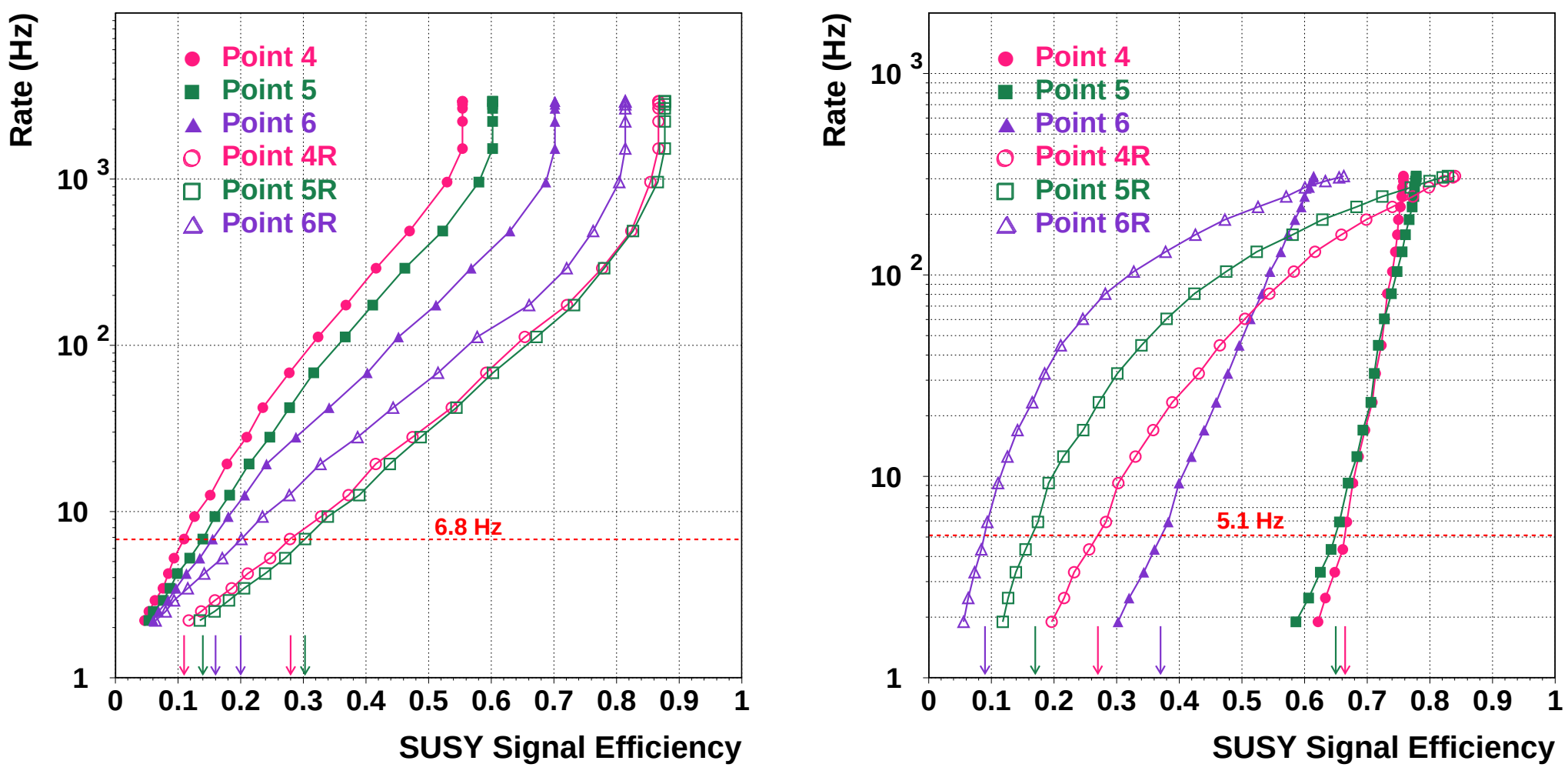

For the low-mass points, simple triggers with jets and $\not\!\!E_T$ were considered. At low luminosity a 3-jet trigger and a jet+$\not\!\!E_T$ trigger are considered at Level-1. For the HLT, the jet+$\not\!\!E_T$ and the 4-jet channel are considered. Figure 62(left) shows the 4-jet HLT trigger rate versus the signal efficiency for events that pass the Level-1 jet+$\not\!\!E_T$ trigger for the six SUSY points as the threshold on the leading jet is varied. Figure 62(right) shows the rate versus efficiency for the jet+$\not\!\!E_T$trigger as the threshold on the $\not\!\!E_T$is varied. The arrows on the plots indicate the thresholds chosen for the low luminosity trigger table as a compromise between efficiency and bandwidth. Table 29 summarizes the Level-1 and HLT thresholds values, the trigger rates and signal efficiencies for a set of points and for low luminosity running. Points 4R, 5R and 6R are the corresponding $R$-parity violating ones. The HLT efficiencies shown are with respect to events that pass the Level-1 trigger. After the first few runs, and once the actual trigger conditions are known, more triggers will be added to increase the efficiency for SUSY signals.

For the high luminosity case, the high mass SUSY points 7, 8, and 9 are considered (as well as the corresponding $R$-parity violating ones). It is straightforward to design highly efficient triggers for these samples. Table 30 summarizes the optimized Level-1 and HLT requirements and output rates. The Level-1 trigger used are the single jet+$\not\!\!E_T$ and a 3-jet trigger, while the HLT uses a $\not\!\!E_T$ and a 4-jet selection.

**Table 29.** The Level-1 and HLT thresholds, rates and efficiencies for six supersymmetry points at low luminosity. The HLT efficiencies are with respect to events that pass the Level-1 trigger. All thresholds refer to the values with 95% efficiency, with the exception of the Level-1 $\not\!\!E_T$ which is the actual threshold value. For a definition of the SUSY points, see text.

| SUSY point | Level-1 Trigger | | High-Level Trigger | |
|---|---|---|---|---|
| | 1 jet >88 GeV+ $\not\!\!E_T$>46 GeV efficiency (%) | 3 jets, $E_T$>86 GeV efficiency (%) (cumulative) | 1 jet >180GeV+ $\not\!\!E_T$>123 GeV efficiency (%) | 4 jets, $E_T$>113 GeV efficiency (%) (cumulative) |
| 4 | 88 | 60 (92) | 67 | 11 (69) |
| 5 | 87 | 64 (92) | 65 | 14 (68) |
| 6 | 71 | 68 (85) | 37 | 16 (44) |
| 4R | 67 | 89 (94) | 27 | 28 (46) |
| 5R | 58 | 90 (93) | 17 | 30 (41) |
| 6R | 47 | 84 (87) | 9 | 20 (26) |
| Background | rate (kHz) | rate (kHz) (cumulative rate) | rate (Hz) | rate (Hz) (cumulative rate) |
| | 2.3 | 0.98 (3.1) | 5.1 Hz | 6.8 (11.8) |

**Table 30.** The Level-1 and High-Level Trigger threshold values, rates and efficiencies for six supersymmetry points at high luminosity. The HLT efficiencies are with respect to events that pass the Level-1 Trigger. All thresholds refer to the values with 95% efficiency, with the exception of the Level-1 $\not\!\!E_T$ which is the actual cut value. For a definition of the SUSY points, see text.

| SUSY point | Level-1 Trigger | | High-Level Trigger | |
|---|---|---|---|---|
| | 1 Jet >113 GeV+ $\not\!\!E_T$>70 GeV efficiency (%) | 3 jets, $E_T$>111 GeV efficiency (%) (cumulative) | $\not\!\!E_T$>239 GeV efficiency (%) | 4 Jets, $E_T$>185 GeV efficiency (%) (cumulative) |
| 7 | 90 | 62 (90) | 85 | 18 (85) |
| 8 | 97 | 76 (98) | 90 | 28 (92) |
| 9 | 91 | 67 (94) | 72 | 28 (76) |
| 7R | 91 | 99 (100) | 70 | 75 (90) |
| 8R | 86 | 100 (100) | 58 | 78 (88) |
| 9R | 75 | 99 (100) | 41 | 52 (64) |
| Background | rate (kHz) | rate (kHz) (cumulative rate) | rate (Hz) | rate (Hz) (cumulative rate) |
| | 4.5 | 1.1 (5.4) | 1.6 | 1.5(3.0) |

**Table 31.** Expected mass limits on new particles that decay to di-jets at LHC turn-on [45]

| Particle | Limit for 2 $fb^{-1}$ ($GeV/c^2$) | Limit for 100 $fb^{-1}$ ($GeV/c^2$) |
|---|---|---|
| W′ | 720 | 920 |
| Z′ | 720 | 940 |
| E6 di-quarks | 570 | 780 |
| Axigluons | 1160 | 1300 |
| excited quarks | 910 | 1180 |

## *9.4. Other New Particle Searches*

Many scenarios of new physics, such as technicolour, "Little Higgs" models [44], and grand unified theories, predict new particles that decay to two jets. Table 31 lists some of the expected limits on various particles at LHC turn-on [45].

Because the contribution to the measured width of a di-jet resonance from the calorimeter resolution is large compared to the intrinsic width of most of these new particles, the results for different particles scale according to their production cross sections. The search for the $Z'$ is used as an example here. The results for other particles can be estimated by scaling these results.

The search for low-mass di-jet resonances at the LHC will be challenging due to the large backgrounds from standard model processes. Significant amounts of data below the resonance will be needed to be fitted to obtain the free parameters in the ansatz. A very approximate estimate for the luminosity needed for a $5\sigma$ discovery can be obtained by estimating the number of events due to the signal in a window with a width $\pm 2\sigma$ of the Gaussian part of the resonance and demanding a $5\sigma$ excess. This yields a lower limit on the required luminosity because it does not take into account systematic uncertainties and the problem of fitting for the free parameters.

The results are listed in the second column of table 32. The analysis assumes that data will be needed down to an $E_T$ cutoff of at least at $M/4$. As an example, the discovery of a $Z'$ with mass 600 GeV/$c^2$, will require data down to an $E_T$ cutoff of 150 GeV. Table 32 lists the rate at $E_T = M/4$, the prescaling factor and resulting rate that would be needed to discover this particle in 1 year of low luminosity running (20 $fb^{-1}$), and in 5 years of low luminosity running (100 $fb^{-1}$).

One can also consider the time it takes to discover the $Z'$ as a function of the $Z'$ mass for a constant rate to storage. The instantaneous luminosity in LHC is expected to

**Table 32.** Minimum requirements to discover a $Z'$ decaying to two jets. Listed are: the integrated luminosity needed to have a $5\sigma$ $Z'$ signal,the threshold on the jet trigger $E_T$ used to determine the luminosity, the rate for the single jet trigger at that threshold, the rate needed to acquire the events within 1 year (20 $\text{fb}^{-1}$) (and also, in parenthesis, the prescale factor required and the number of events), and the rate and prescale to acquire that number of events in 5 years.

| $Z'$ mass ($\text{TeV}/c^2$) | Luminosity $\text{fb}^{-1}$ | $M/4$ GeV | rate Hz | rate (prescale) (events) 1 year Hz | rate (prescale) 5 years Hz |
|---|---|---|---|---|---|
| 0.6 | 1.4 | 150 | 800 | 56 (14.2) ($40 \cdot 10^6$) | 11(71) |
| 0.8 | 2.3 | 200 | 200 | 23 (8.7) ($26 \cdot 10^6$) | 4.5 (443) |
| 1.0 | 4.3 | 250 | 55 | 12 (4.7) ($26 \cdot 10^6$) | 2.3 (24) |
| 1.2 | 7.3 | 300 | 25 | 9 (2.7) ($33 \cdot 10^6$) | 1.9 (13) |
| 1.4 | 14 | 350 | 11 | 7.8 (1.4) ($55 \cdot 10^6$) | 1.6 (7) |
| 1.6 | 20 | 400 | 6 | 6 (1) ($60 \cdot 10^6$) | 1.2 (5) |
| 1.8 | 31 | 450 | 3.5 | - | 1.1 (3.2) |
| 2.0 | 52 | 500 | 2 | - | 1.1 (1.9) |

have an exponential decay, and therefore the rates at a given threshold also decrease exponentially. Applying this model, and using dynamic prescaling to keep the rate to storage constant, the prescaling factor also decreases exponentially with time.
Furthermore, one can assume several dynamic prescale scenarios

- an exponential luminosity decay with time constant of 10 hours, during a beam fill of 10 hours, and a fixed rate to storage for the single jet trigger of 20 Hz;
- a fixed prescale, with a rate to tape for the single jet trigger of 20 Hz at $\mathcal{L} = 2 \times 10^{33}\,\text{cm}^{-2}\text{s}^{-1}$;
- an exponential luminosity decay with a 10-hour lifetime, during a beam fill of 10 hours, a fixed rate to storage for all triggers of 100 Hz, taking an initial rate for the single-jet triggers of 5 Hz.

All three scenarios assume 20 $\text{fb}^{-1}$ per year for low luminosity running. Figure 63 shows the time to discovery for these three options.
Further topics of new physics include for example extra dimensions and little Higgs models. In general massive objects are produced in these scenarios, for which the CMS trigger can perform efficiently. A few examples are given for illustration.
Extra dimensions may be an explanation, via geometry, for the large disparity between the electroweak and Planck scale as well as the flavour structure we observe in the standard model. The energy scale where the extra dimensions operate is not theoretically known. The most likely scale is the GUT/Planck scale around $10^{15}$ -$10^{19}$ GeV. There is no strong experimental constraints to date, that excludes the existence of large extra dimensions that have an effect in the physics at the TeV scale. Examples of signatures of extra dimensions within the most popular models are:

- $q\bar{q}, gg \rightarrow gG, ZG, \gamma G$ (ADD models), where $G$ is the graviton that escapes detection. These signatures imply the presence of missing $E_T$ in the final state as in the case of supersymmetry events in association with a jet or a boson. For the most promising channel the signal may become visible over the standard model backgrounds for $\not\!\!E_T$ values of 500 GeV [47]. These events can be efficiently triggered at Level-1 using the single jet plus $\not\!\!E_T$, or the single lepton triggers.
- $q\bar{q}, gg \rightarrow G \rightarrow WW, ZZ, \gamma\gamma$ (ADD and RS models). These signatures use isolated photons, electrons and muons and sometimes jets from weak boson decays. For ADD models there is a continuum of graviton mass states over the whole energy range, while in case of the RS models a series of resonances is expected with the first one in the TeV mass range. The massive resonances will be easily triggered by the lepton and jet triggers.
- $q\bar{q}, gg \rightarrow \gamma^1/Z^1, G, ee, \mu\mu$, jet-jet ; $q\bar{q} \rightarrow W \rightarrow l\nu$ (ADD,RS and TeV$^{-1}$ models). Except for the ADD case again one expects resonances with masses larger than about 1 TeV. The final states with leptons can be triggered with high efficiency by the lepton triggers. For the ADD two fermion final states, the di-jet one is least efficient and has the same acceptance as discussed for the di-jet $Z'$ trigger above.
- In Universal Extra Dimensions models [48], the Kaluza-Klein (KK) particle spectra resemble the supersymmetric spectra, with the special characteristic that in general the mass differences between the KK particle states are small, leading to jets and leptons produced in decays with relatively small $E_T$ , typically of the order of a few tens of GeV. In model variations that invoke KK-parity violation [49], spectacular signatures with di-leptons, di-photons and di-jets plus missing energy become important. These can be retained using the single or double lepton/photon triggers as well as the jet plus $\not{E}_T$ triggers.

**Figure 63.** Time of obtaining 5 $\sigma$ significance over the background for the 3 prescale schemes described in the text as a function of the $Z'$ mass.

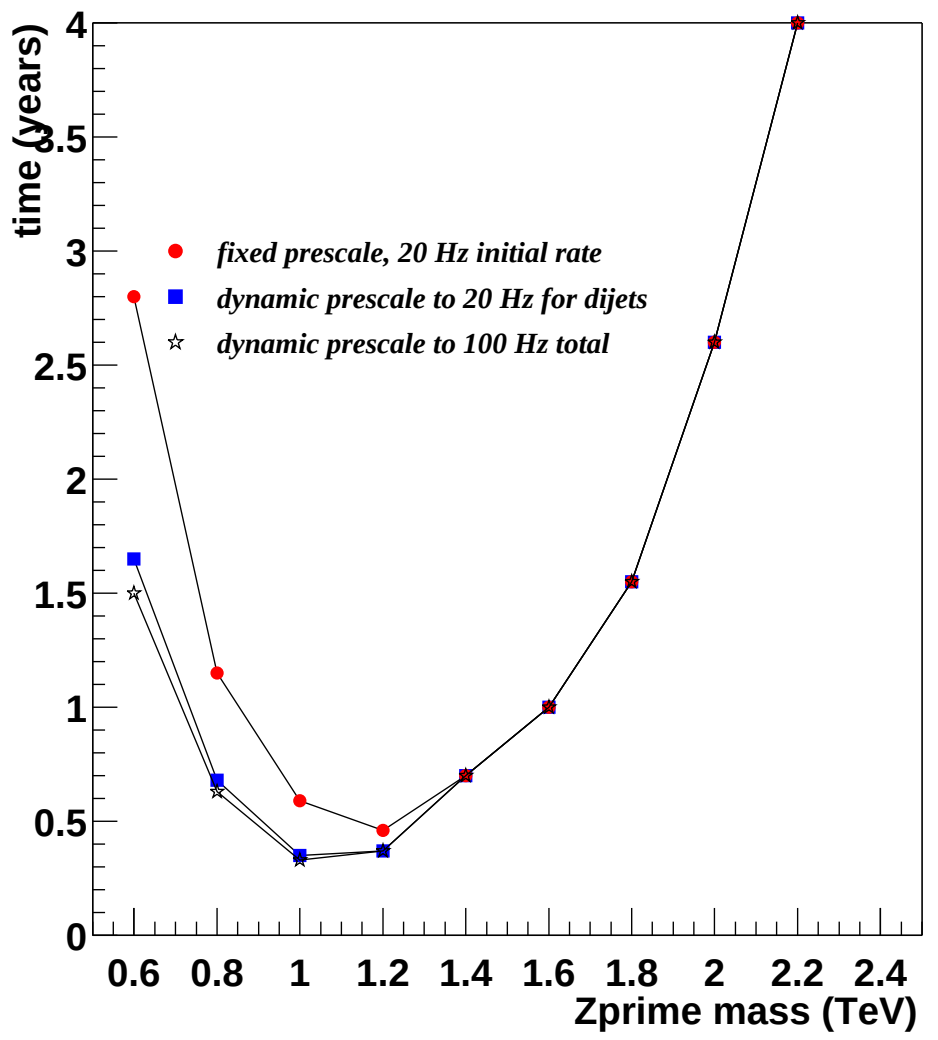

Within Little Higgs models, new particles can be expected in the TeV range such as heavy top quarks $T$ and new gauge bosons. The $T$ quarks will decay via channels such as $T \to Zt \to ZWb; T \to Wb; T \to ht \to hWb$. The new gauge bosons $A_H, Z_H$ and $W_H$ can decay in leptons or the SM gauge bosons. Most analyses [50] use lepton channels either from the $W$'s in the decays of the $T$, or standard model gauge bosons, with $E_T$ cuts of typically $\mathcal{O}(100)$ for leptons produced with $|\eta| < 2.5$. The CMS lepton triggers can efficiently trigger on the production of these new particles using the lepton trigger paths.

*9.4.1. Standard Model Physics* The measurement of $W$ and $Z$ boson production properties, and especially their couplings, will be one of the topics which will be studied at the LHC. Deviations may hint at new physics. First manifestations of supersymmetry may have to be discriminated against a background of $W$+jets and $Z$+jets events. The same holds for heavy flavour physics and, in particular, studies of the top quark, where couplings, rare decay modes, spin measurements and correlations have to be studied. $W$, $Z$ and top-quark production also provide key tests of QCD.
The main channels for analysis of $W$ and $Z$ bosons at LHC will be their leptonic decays. The efficiencies for $W$ and $Z$ bosons have been determined in the previous sections of this report for electrons and muons. The production rates for $W \to e\nu$ and $Z \to ee$ are approximately 20 nb and 2 nb respectively and therefore lead to a rate of events of 40 and 4 Hz respectively at low luminosity.
About 60% of the produced $W$-bosons will have an electron in the fiducial volume of the ECAL $|\eta| < 2.5$. Using the single electron trigger the overall efficiency is 67% (59%) at low (high) luminosity.
Similar numbers are obtained for the muon decay channel. Here the geometrical fiducial acceptance of $|\eta| < 2.1$ is about 50%, and the trigger acceptance using the single muon trigger is 69% at low luminosity and 42% at high luminosity. For the channel $Z \to \mu\mu$ with muons in the fiducial volume, the trigger acceptance is larger and amounts to 92 % (86 %) at low (high) luminosity.
The efficiency for top quarks via the decay $t\bar{t} \to \mu + X$ amounts to 72% at low luminosity. Detailed studies of top production and decay properties will be among the main physics topics of the first years of running. Top-quark production is often the main background in various searches, foremost in SUSY searches, and for this reason it will have to be understood thoroughly early on in the LHC physics program.

*9.5. Summary*

A prototype trigger table for the Level-1 and the HLT selection at a start-up luminosity of $2 \times 10^{33} \mathrm{cm}^{-2}\mathrm{s}^{-1}$ has been presented and studied in detail. The assumption of this table is a total DAQ bandwidth of 50 kHz. It has been shown that high efficiencies for most physics objects are attainable with a selection that remains inclusive and avoids topological or other biases on the event. The overall CPU requirement of this selection is approximately 300 ms on an Intel 1 GHz Intel Pentium-III CPU.
Much more sophisticated trigger requirements can, and will be used. As an example, at a minimum, as the instantaneous luminosity drops throughout a fill of the LHC bandwidth will be freed from the triggers discussed here. This additional bandwidth can be reallocated to the same triggers by decreasing the thresholds, as in the example of the di-jet resonance dynamic prescale factors discussed in this chapter.
The additional bandwidth may also be used in introducing new triggers for example non-top heavy flavour specific. Introduction of such triggers is then purely an issue of whether there are adequate CPU resources for the selection of the relevant events. The systematic optimization of the track reconstruction code and the extensive use of regional and conditional track reconstruction allow for the very fast search and the full reconstruction of $B$-meson decays. Furthermore, the optimization of the tracking code indicates that it can be applied to the full Level-1 event rate at both low and high luminosity. This would extend and complement the current Level-2 selections described in this section. There is ongoing work in the area of optimization of the tracking code and of its application in various parts of the selection.
The selection presented in this paper indicates that the CMS trigger system – which has been designed and is presently being built– has sufficient level of sophistication and flexibility to provide the HLT selection of 1:1000 in a single processor farm. Furthermore, the full event record is available, and the software that implements all algorithms can be changed and extended. The CMS HLT architecture allows the implementation of further improved selection algorithms to be applied on the various physics channels, as well as for adjusting to unforeseen circumstances resulting from the beam conditions, high background levels or new physics channels not previously studied. With the robust and reliable HLT architecture, CMS is looking forward to collecting the data from the LHC collisions in 2007.

## Appendix A. Level-1 Trigger at CMS

The CMS Level-1 trigger system is organized into three major subsystems: *(i)* the Level-1 calorimeter trigger, *(ii)* the Level-1 muon trigger, and *(iii)* the Level-1 global trigger. The muon trigger is further organized into subsystems representing the three different muon detector systems, the Drift Tube trigger (DT) in the barrel, the Cathode Strip Chamber (CSC) trigger in the endcap and the Resistive Plate Chamber (RPC) trigger covering both barrel and endcap. The Level-1 muon trigger includes a global muon trigger that combines the trigger information from the DT, CSC and RPC trigger systems, and sends this to the Level-1 global trigger.
The data used as input to the Level-1 trigger system as well as the input data to the global muon trigger, global calorimeter trigger and the global trigger are transmitted to the data acquisition system (DAQ) for storage along with the event readout data. In addition, all trigger objects found, whether they were responsible for the Level-1 trigger accept decision or not, are also sent. The decision whether to trigger on a specific crossing or to reject that crossing is transmitted to all of the detector subsystem front-end and readout systems.

*Appendix A.1. Level-1 Calorimeter Trigger*

The CMS calorimeter trigger has 4176 trigger towers. Of these 2448 are in the barrel, 1584 in the end cap and 144 in the forward calorimeters (figure A1). Each ECAL half-barrel is segmented in $17\eta\times74\phi$-towers resulting in individual trigger towers of dimension in $\eta$-$\phi$ of 0.087×0.087. A trigger tower in the barrel is formed by 5 × 5 crystals. The ECAL trigger towers are divided in strips of $1\eta\times5\phi$ crystals (figure A2). The strip information allows for a finer analysis of the lateral energy spread of the electromagnetic showers. The strips are arranged along the bending plane in order to collect in one or two adjacent strips almost all the energy of electrons with bremsstrahlung and converted photons. In the ECAL endcap where the crystals are arranged in a $x-y$ geometry, the trigger towers do not follow exact $(\eta, \phi)$ boundaries. The trigger tower average $(\eta, \phi)$ boundaries are 0.087×0.087 up to $\eta\sim$1.74. The trigger tower size in $\eta$ is growing with $\eta$ as shown in figure (figure A1). The number of crystals per trigger tower varies between 25 at $\eta\sim$1.5 and 10 at $\eta\sim$2.8. Both in the barrel and in the endcap the boundaries of ECAL and HCAL trigger towers follow each other. Each trigger tower in the barrel corresponds to the $\eta, \phi$ size of an HCAL physical tower and the HCAL tower trigger energy is the sum of the first two inner longitudinal segments.
In the end cap ($\eta>$1.479) two ECAL trigger towers correspond to one HCAL physical tower in $\phi$. In this region the HCAL energy of one tower is equally divided between the two ECAL trigger towers that correspond to it. In the barrel-endcap transition region, barrel and endcap segments are summed together.
The trigger segmentation of the forward hadron calorimeter (HF) does not have fine $\phi$ binning because this detector does not participate in the electron or photon triggers. However the coverage needs to be seamless for the jet and missing energy triggers. The segmentation in the forward region matches the boundaries of the 4 × 4 trigger regions in the rest of the calorimetry. The resulting HF trigger tower segmentation of $4\eta\times18\phi$ is used in the jet and missing energy triggers. The $\phi$ bins are exactly 20° (4× 0.087) and the $\eta$ divisions are approximately the size of the out end cap divisions. The jet trigger extends seamlessly to $|\eta|=5$. The missing transverse energy is computed using 20° divisions for the entire $\eta, \phi$ plane.
The trigger towers are organized in calorimeter regions, each one formed by $4\times4$ trigger towers. Each of the HF towers is treated as $4 \times 4$ region since it is segmented in 20° $\phi$ bins. The calorimeter $4 \times 4$ regions are the basis of the jet and energy triggers. The $\eta - \phi$ indexes of the calorimeter regions are used to identify the location of L1 calorimeter trigger objects (electron/photons and jets) in the upper stages of the trigger chain.

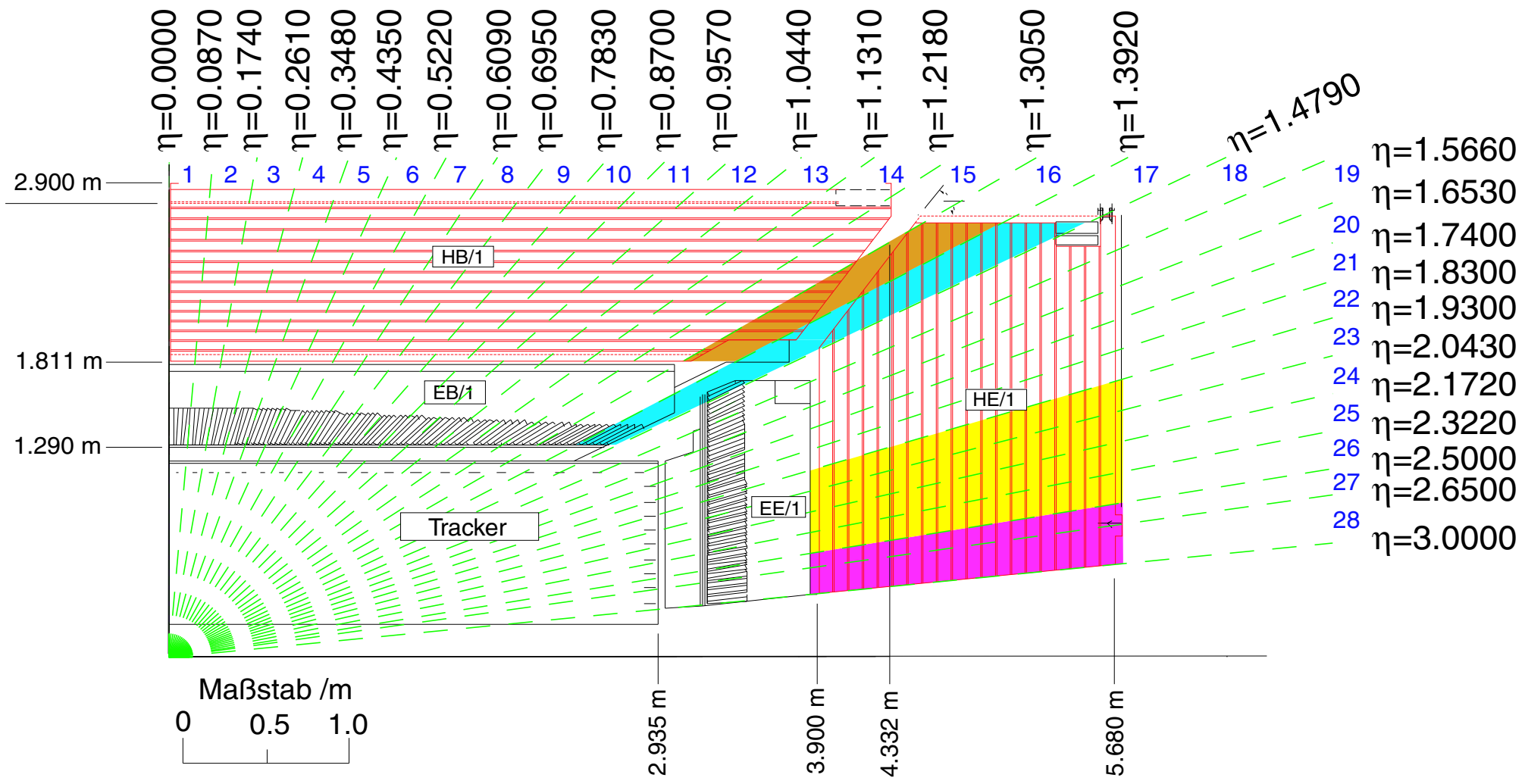

**Figure A1.** Layout of the calorimeter trigger towers in the $r - z$ projection

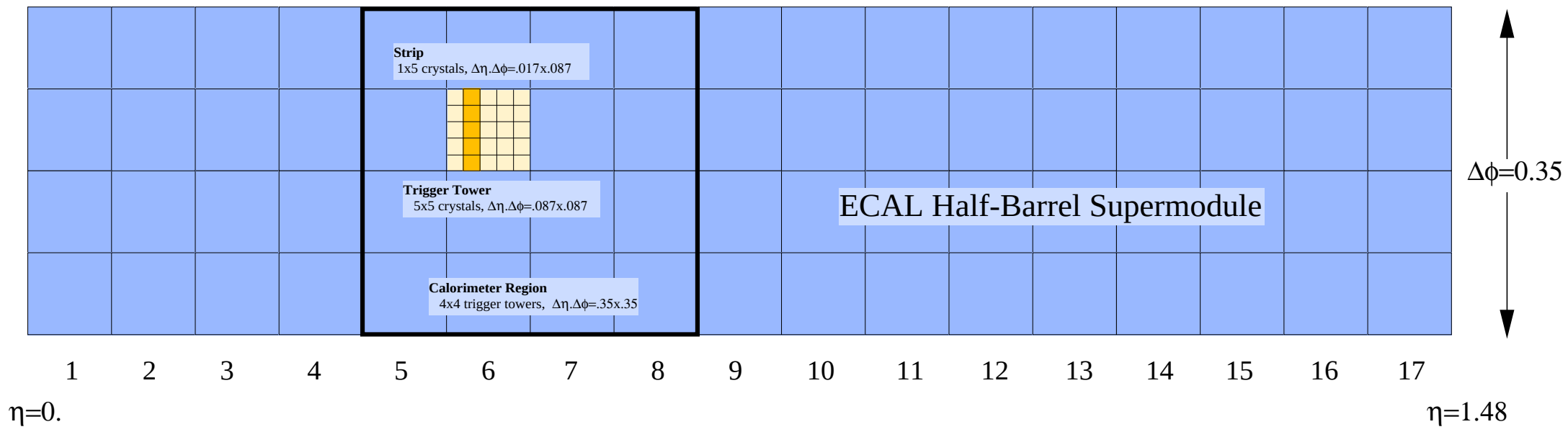

**Figure A2.** Calorimeter trigger tower layout in one ECAL half-barrel supermodule. The trigger towers are organized in calorimeter regions of $4 \times 4$ towers.

The transverse energy sum is computed for each calorimeter trigger tower. The ECAL trigger cell $E_T$ is the sum of the $E_T$ of $5 \times 5$ crystals in the barrel and a variable number of crystals in the endcap. The HCAL trigger cell $E_T$ is the sum of the $E_T$ of the longitudinal compartments of the inner hadron calorimeter.

For every ECAL trigger tower the information that reflects the lateral extension of the electromagnetic shower (referred to as "Fine Grain" or FG veto bit) is used to improve the rejection of background in the electron trigger. The FG veto bit is active when the highest energy adjacent strip pair has less than a programmable fraction $R$ (typically 90%) of the tower energy. Electrons and photons (converted or non-converted), in the presence of noise and high luminosity pileup, have $R$ <0.9 in 2% of the cases. The energy for the trigger tower is sent in $E_T$ scale rather than $E$ scale so the trigger scale is not sensitive to minimum ionizing particle energy deposit in the HCAL. The fine grain bit is used to identify minimum ionizing particles requiring the HCAL tower energy to be inside a programmable energy range. The data is transmitted to the Regional Calorimeter Trigger (RCT), which finds candidate electrons, photons, taus, and jets. The RCT separately finds both isolated and non-isolated electron/photon candidates and transmits them along with sums of transverse energy to the Global Calorimeter Trigger (GCT). The GCT sorts by $E_T$ the candidate electrons, photons, taus, and jets and forwards the top four of each type to the global trigger. The GCT also calculates the total transverse energy and total missing energy vector. It transmits this information to the global trigger. The RCT transmits an $(\eta, \phi)$ grid of "quiet" regions to the global muon trigger for muon isolation cuts.

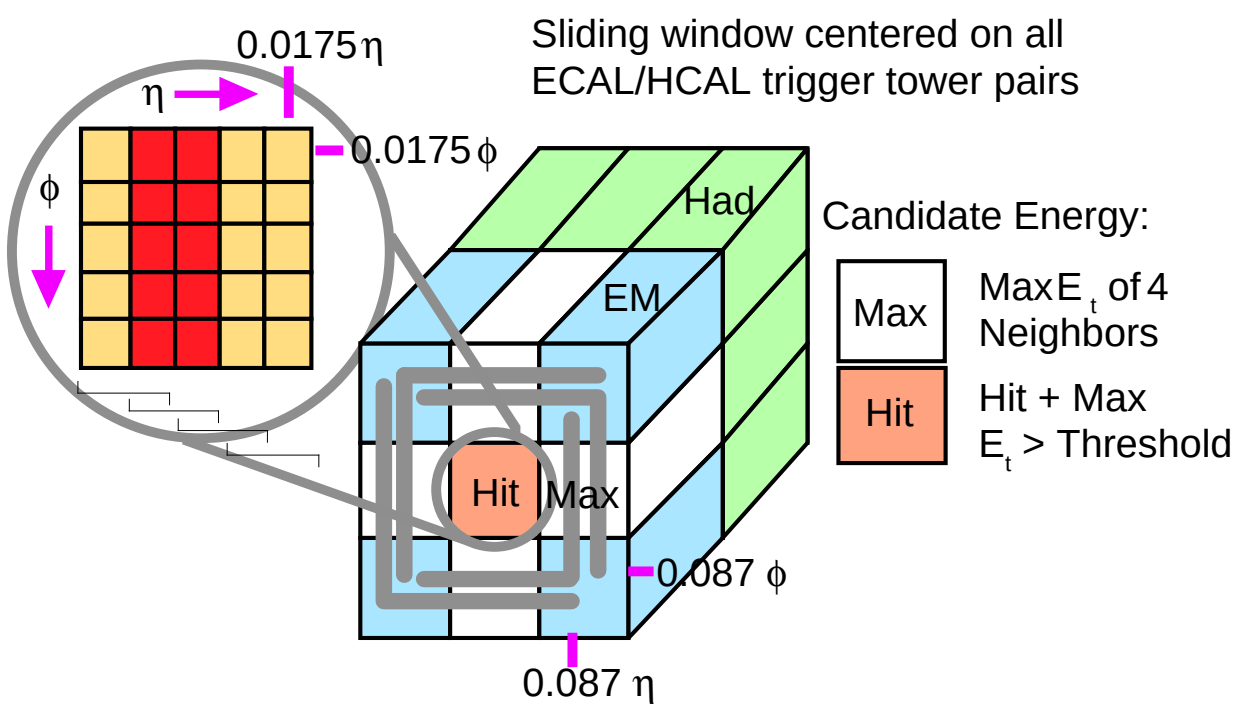

**Figure A3.** Electron/photon trigger algorithm.

### *Appendix A.2. Level-1 Electron and Photon Triggers*

The electron/photon trigger uses a $3 \times 3$ trigger tower sliding window technique which spans the complete $\eta, \phi$ coverage of the CMS electromagnetic calorimeter. Two independent streams are considered, non-isolated and isolated electrons/photons. The isolated stream requires electromagnetic and hadronic energy isolation criteria. The implementation of longitudinal and lateral shower profile selection cuts, as well as electromagnetic and hadronic isolation programmable criteria provides safety and flexibility for the calorimeter electron/photon trigger.

An overview of the electron/photon isolation algorithm is shown in figure A3. This algorithm involves the eight nearest trigger tower neighbors around the central hit trigger tower and is applied over the entire $(\eta, \phi)$ plane. The electron/photon candidate $E_T$ is determined as follows: The $E_T$ of the "hit trigger tower" (electromagnetic plus hadronic, indicated as HitMax in figure A3) is summed with the highest of the four broad side neighbor towers (indicated as MaxE$_t$ in figure A3). The summed transverse energy of the two towers provides a sharper efficiency turn-on with the true $E_T$ of the particles. The non-isolated candidate requires passing of two shower profile vetoes, the first of which is based on the fine-grain ECAL crystal energy profile (FG veto). The second is based on the HCAL to ECAL energy comparison, e.g. $Had/Em$ less than 5% (HAC veto). The isolated candidate requires passing two additional vetoes the first of which is based on passing the FG and HAC Vetoes on all eight nearest neighbors, and the second on the existence of at least one quiet corner, *i.e.*, one of the five-tower corners has all crystals below a programmable threshold, *e.g.*, 1.5 GeV. Each candidate is characterized by the $\eta, \phi$ indices of the calorimeter region where the hit tower is located.

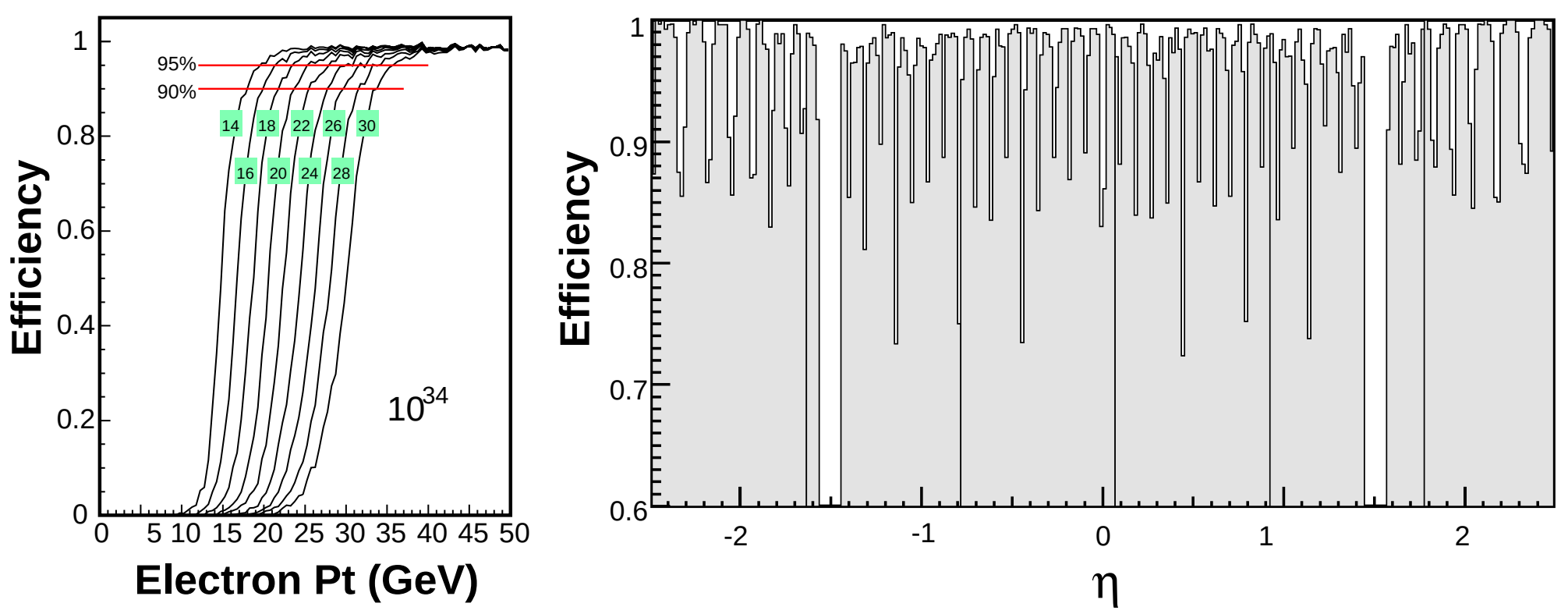

**Figure A4.** The efficiency of the Level-1 trigger for single electrons as a function of the electron $p_T$. On the right, the efficiency, as function of $\eta$, for electrons with $p_T$=35 GeV/$c$.

In each calorimeter region (4 $\times$ 4 trigger towers) the highest $E_T$ non-isolated and isolated electron/photon candidates are separately found. The 16 candidates of both streams found in a wider trigger region corresponding to 16 calorimeter regions (covering $\eta \times \phi$=3.0$\times$0.7) are further sorted by transverse energy. The four highest-$E_T$ candidates of both categories are transferred to the Global Calorimeter Trigger (GCT) and retained for processing by the CMS global trigger. The nominal electron/photon algorithm allows both non-isolated and isolated streams. The non-isolated stream uses only the hit tower information including any leakage energy from the maximum neighbor tower. This stream will be used at low luminosity to provide the electron trigger from $b$ semileptonic decays. The isolation and shower shape trigger cuts are programmable and can be adjusted to the running conditions. For example, at high luminosity the isolation cuts could be relaxed to take into account higher pileup energies. The specification of the electron/photon triggers also includes the definition of the $\eta - \phi$ region where it is applicable. In particular, it is possible to define different trigger conditions (e.g. energy thresholds and isolation cuts) in different rapidity regions. The efficiency of the electron/photon algorithm, as a function of the electron transverse momentum, for different thresholds applied at Level-1, is shown in Figure A4. Also shown is the efficiency, as function of pseudorapidity for electrons with $p_T$=35 GeV/$c$. To connect the Level-1 threshold to an effective requirement on the electron transverse momentum, the electron $p_T$ at which the Level-1 trigger is 95% efficient, is determined as function of the Level-1 threshold. This is shown in Figure A5. From this result, the rate for electron/photon triggers as a function of the effective cut on the $E_T$, *i.e.* of the point at which the trigger is 95% efficient, can be computed. Figure A6 shows the rates for single electrons as a function of the $E_T$ of the electron (95% point). Double–, triple– and quad–electron/photon triggers can be defined. The requirements on the objects of a multi-electron/photon trigger, namely the energy threshold, the cluster shape and isolation cuts and the $(\eta, \phi)$ region, are set individually. Requirements on the $(\eta, \phi)$ separation between objects can also be defined.

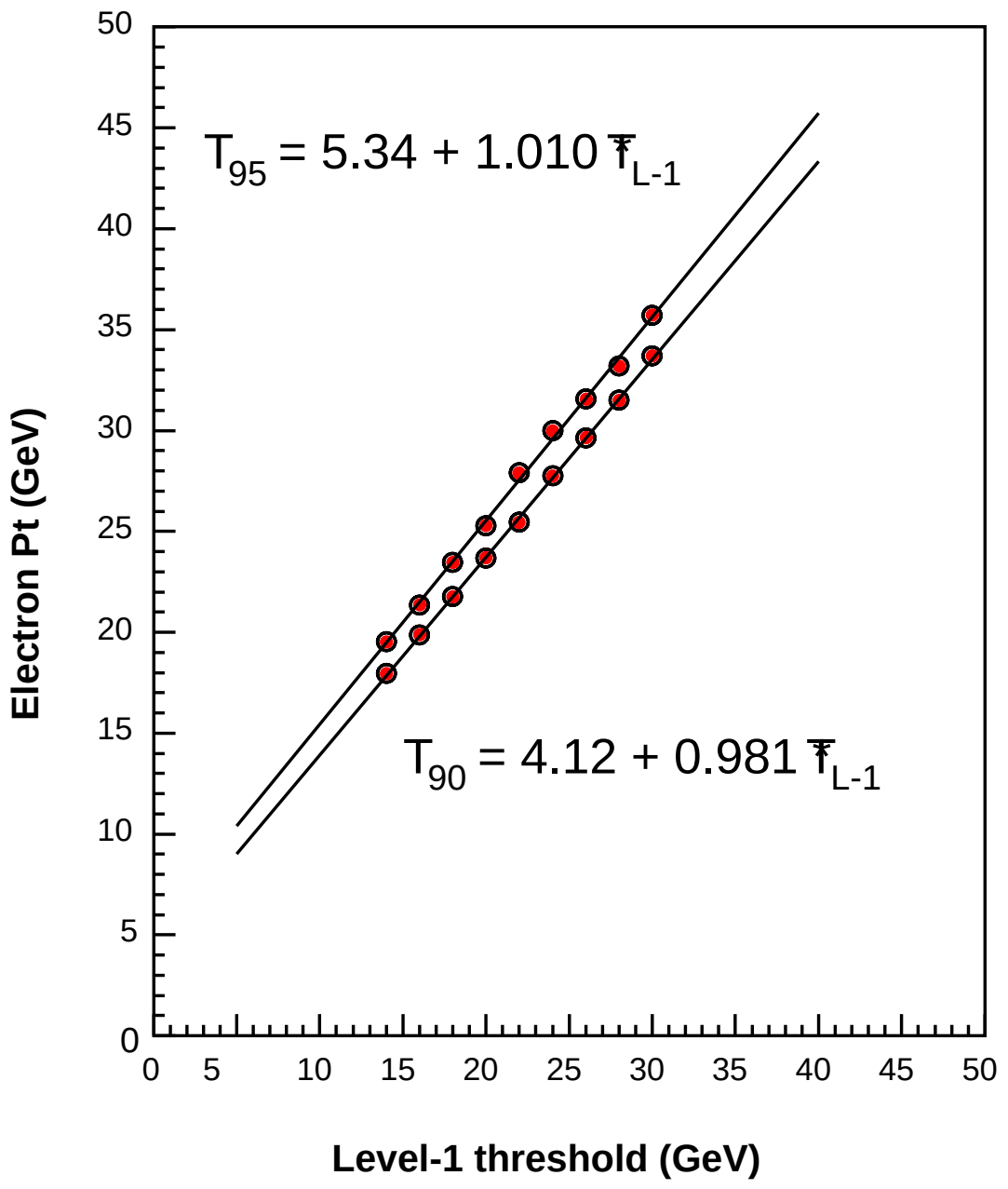

**Figure A5.** The electron $p_T$ at which the Level-1 Trigger is 95% efficient as a function of the Level-1 threshold.

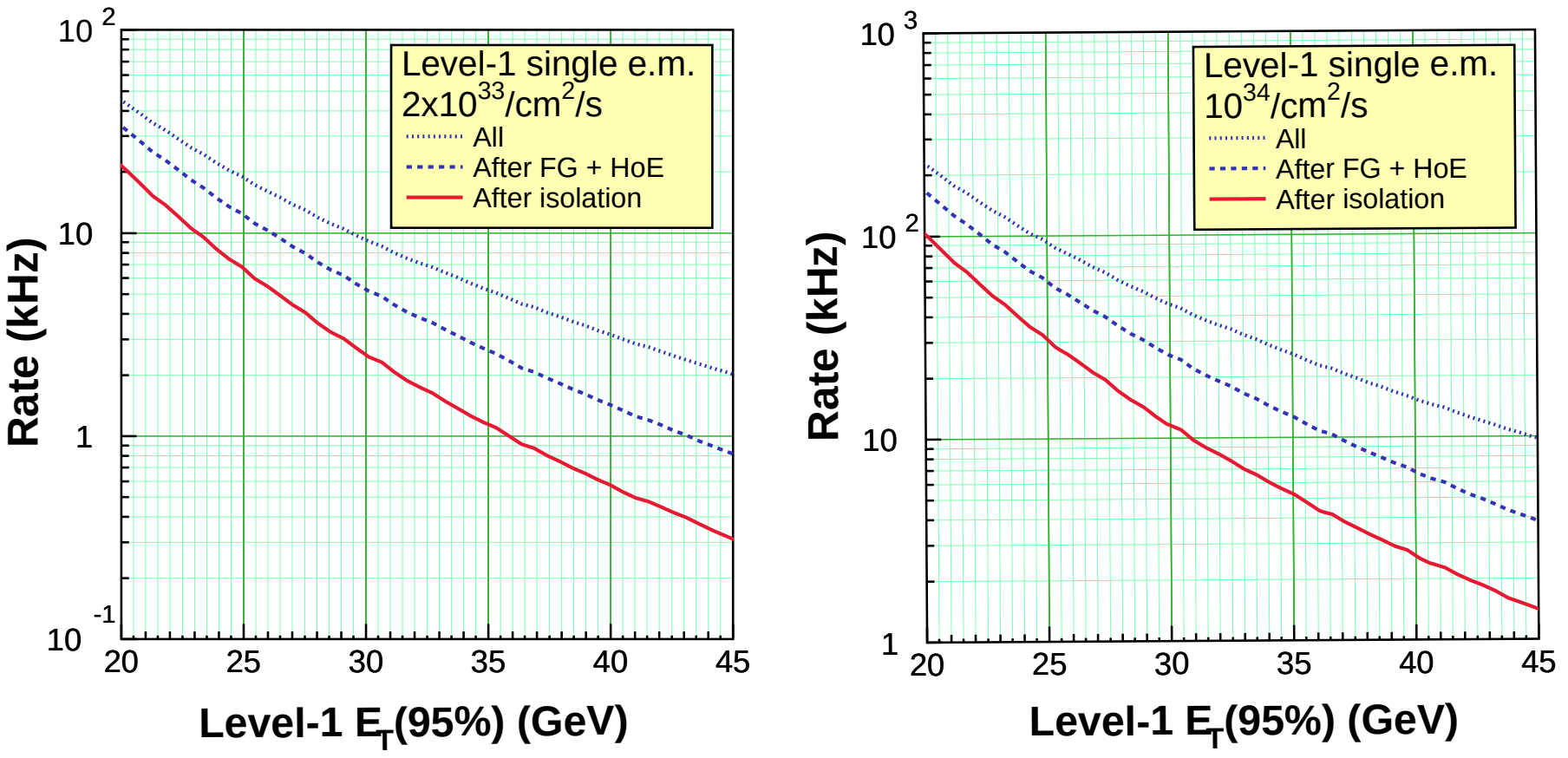

**Figure A6.** The rate of the single electron/photon Level-1 trigger at low (left) and high (right) luminosity. FG and HoE refer to the shower profile and hadronic over electromagnetic energy isolation criteria.

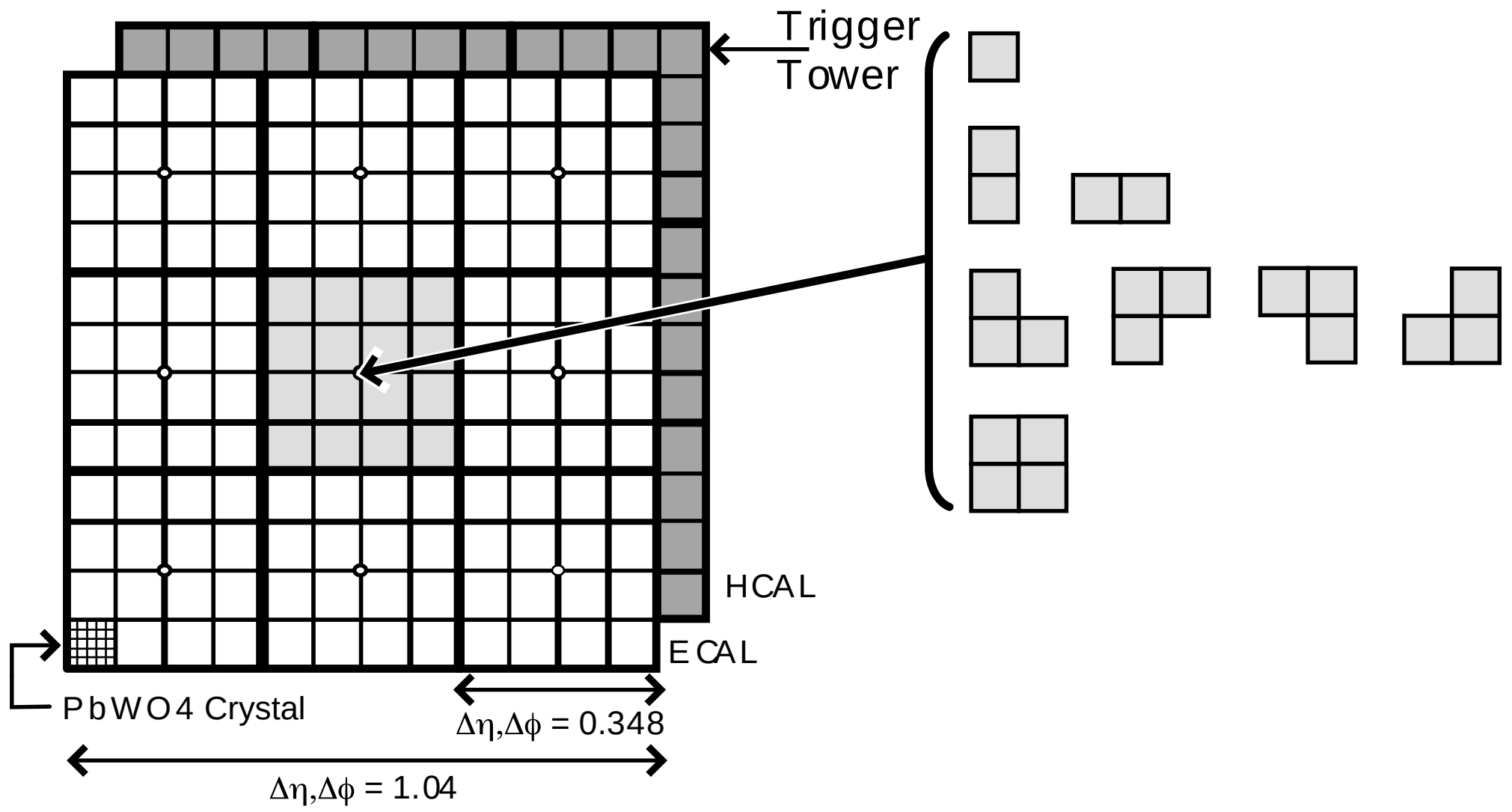

**Figure A7.** Jet and $\tau$ trigger algorithms.

*Appendix A.3. Jet and $\tau$-jet triggers*

The jet trigger uses the transverse energy sums (electromagnetic plus hadronic) computed in calorimeter regions (4×4 trigger towers as shown if figure A7). In the forward hadron calorimeter the region is the trigger tower itself. The jet trigger uses a 3×3 calorimeter region sliding window technique which spans the complete $(\eta, \phi)$ coverage of the CMS calorimeters seamlessly. The central region $E_T$ is required to be higher than the eight neighbor region $E_T$ values. In addition, the central region $E_T$ is required to be greater than a fixed value, to suppress noise.

The jets and $\tau$–jets are characterized by the transverse energy $E_T$ in 3×3 calorimeter regions. Therefore the summation spans 12×12 trigger towers in the barrel and the endcap or 3×3 towers in the forward hadron calorimeter. The $\phi$ size of the jet window is the same everywhere (60°) while the $\eta$ binning is increasing as a function of $\eta$ according to the calorimeter and trigger tower segmentation. The jets are labeled by their $(\eta, \phi)$ indices.

Single and three–prong decays of $\tau$–leptons form narrow clusters of energy deposits in the calorimeter. Since the decays involve charged pions which deposit energies in the hadron calorimeter, the electron/photon trigger does not capture them. Therefore, the transverse profiles of active tower patterns are analyzed to tag narrow jets as potential $\tau$–lepton decays. An active tower is defined as a trigger tower with ECAL or HCAL $E_T$ above a separately programmable threshold. The energy deposit in each trigger tower, ECAL and HCAL separately, is compared to a programmable threshold to obtain two 4×4 single-bit activity patterns. The energy deposit pattern in the 4×4 region is examined and if the pattern does not match any of the 1-, 2-, 3- and 4-tower patterns shown in figure A7, this region cannot include a $\tau$-cadidate therefore, its "tau-veto" bit is set. At the next stage of processing, overlapping 3×3 regions, *i.e.*, 1212 trigger towers, is considered. These 1.044×1.044 $\eta - \phi$ sums define jets if the central region has more energy than its 8 neighbors. The logical OR of the tau-bits of these 9 regions constitute the ultimate tau-veto for the jet. If this jet does not have tau-veto set, it is redefined as a tau-jet and is sorted in $p_T$ separately. The $\tau$-veto bits can be used both for $\tau$-like energy deposit identification and stringent isolation. Counters of the number of jets above programmable thresholds in various $\eta$ regions are provided to give the possibility of triggering on events with a large number of low energy jets. Jets in the forward and backward hadron calorimeters are sorted and counted separately (due to background $\eta$-dependence) but the global trigger uses them seamlessly. The four highest energy central and forward jets, and central taus in the calorimeter are selected. The selection of the four highest energy central and forward jets and the four highest energy taus provides enough flexibility for the definition of combined triggers. Single, double, triple and quad jet (including $\tau$-jet ) triggers are possible. The single jet ($\tau$-jet ) trigger is defined by the transverse energy threshold, the $(\eta, \phi)$ region and by a prescaling factor. Prescaling will be used for low energy jet ($\tau$-jet) triggers, necessary for efficiency measurements. The multi-jet ($\tau$-jet) triggers are defined by the jet multiplicity and the jet transverse energy thresholds, by a minimum separation in $(\eta - \phi)$, and by a prescaling factor. The global trigger accepts the definition, in parallel, of different multi-jet ($\tau$-jet) trigger conditions.

### *Appendix A.4. Transverse Energy Triggers*

The $E_T$ triggers use the transverse energy sums (Em+Had) computed in calorimeter regions (4×4 trigger towers in barrel and endcap). $E_x$ and $E_y$ are computed from $E_T$ using the coordinates of the calorimeter region center. The computation of missing transverse energy from the energy in calorimeter regions does not affect significantly the resolution for trigger purposes. The missing $E_T$ is computed from the sums of the calorimeter regions $E_x$ and $E_y$. The sum extends up to the end of forward hadronic calorimeter, *i.e.*, $|\eta|$=5. The missing $E_T$ ($\not{E}_T$) triggers are defined by a threshold value and by a prescaling factor. The global trigger accepts the definition, in parallel, of different missing $E_T$ triggers conditions. The total $E_T$ is given by the sum of the calorimeter regions $E_T$. The sum extends up to the end of forward calorimeter. The total $E_T$ triggers are defined by a threshold value and by a prescaling factor. The global trigger accepts the definition, in parallel, of different total $E_T$ triggers conditions. The total energy trigger is implemented with a number of thresholds which are used both for trigger studies and for input to the luminosity monitor. Some of these thresholds are used in combination with other triggers. Other thresholds are used with a prescale and one threshold is used for a stand-alone trigger. The lower threshold $E_T$ trigger provides a good calorimeter and trigger performance diagnostic.

The trigger is defined as the scalar sum of the $E_T$ of jets above a programmable threshold with a typical value of jet $E_T >$ 10 GeV. This trigger is not as susceptible as the total $E_T$ given by the sum of the calorimeter regions $E_T$ deposits to both noise and pileup effects. Although the total $E_T$ is a necessary technical trigger, it has limited use from the physics point of view. The trigger can capture high jet multiplicity events such as those from fully hadronic top decay, hadronic decays of squarks and gluinos. Although these events have several hundred GeV/$c^2$ energy, they may actually fail the jet triggers because the $E_T$ of individual jets could be softer than the thresholds. In addition, the trigger can use individually calibrated jet energies unlike the total $E_T$ trigger which cannot be easily calibrated.
For each calorimeter region of 4×4 trigger towers a "Quiet" and "MIP" bit is computed. The Quiet bit is "on" if the transverse energy in the calorimeter region is below a programmable threshold. The MIP bit in a calorimeter region requires on top of the Quiet bit condition, that at least one of the 16 trigger towers has the HCAL Fine Grain bit "on". The quiet and MIP bits are used in the Global Muon Trigger.

### *Appendix A.5. The Level-1 muon trigger*

The muon measurement at CMS is performed by Drift Tubes (DT) located outside the magnet coil in the barrel region and cathode Strip Chambers (CSC) in the endcap region. The CMS muon system is also equipped with Resistive Plate Chambers (RPC) both in the barrel and endcap regions used in triggering and reconstruction. The Drift Tube system is comprised of four muon stations interleaved with the iron of the yoke to make full use of the magnetic return flux. Each station in comprised of two or three superlayers (SL). Each DT superlayer is split in four layers of staggered drift tubes, while each CSC station is comprised of six layers of cathode strip chambers. The Drift Tube and Cathode Strip Chamber triggers systems process the information from each chamber locally and are refereed to as *local triggers*. They provide one vector (position and angle) per muon per station. Track Finders (TF) collect these vectors from the different stations and combine them to form muon tracks. The Track Finders play the role of a regional trigger. Up to four best (highest $p_T$ and quality) muon candidates from each system are selected and sent to the Global Muon Trigger. In the case of RPC there is no local processing apart from synchronization and cluster reduction. Hits from all stations are collected by the Pattern Comparator Trigger (PACT) which detects the muon candidates based on the occurrence of predicted hit patterns. Muon Sorters select the top four muons from the barrel and the top four from the endcaps and send them to the Global Muon Trigger (GMT). The GMT compares the information from the TF (DT/CSC) and the PACT (RPC) and attempts to correlate the CSC and DT tracks with RPC tracks. If two candidates are matched their parameters are combined to give optimum precision. The GMT correlates the muon candidate tracks with the corresponding calorimeter towers, based on the position in $\eta - -\phi$, to determine if these muons are isolated. Quiet and MIP bits delivered by the Calorimeter Trigger are used to form an isolated muon trigger and to confirm the muon trigger using the calorimeter information. The CSC and Drift Tube Track Finders exchange track segment information in the region where the chambers overlap. Coarse RPC data can be sent to the CSC trigger to help resolve spatial and temporal ambiguities in multimuon events. The final ensemble of muons are sorted based on their initial quality, correlation and $p_T$ and the four top candidates are sent to the Global Trigger. Transverse momentum thresholds are applied by the Global Trigger for all trigger conditions.

*Appendix A.5.1. Drift Tube (DT) Trigger* The drift chambers deliver data for track reconstruction and for triggering on different data paths at the local trigger level. The trigger front-end (Bunch and Track Identifier or BTI), is used in $\phi$ and $\eta$ to perform a rough muon track fit. It uses the four layers of DTs in one superlayer to measure the position and direction of trigger candidate tracks with at least three hits out of four. The algorithm fits a straight line within programmable angular acceptance. The BTI performs the bunch crossing assignment of every found muon track candidate.
Since this method must foresee alignment tolerances and needs to accept alignments of only three hits, the algorithm can generate false triggers. Hence in the bending plane a system composed of a Track Correlator (TRACO) and a chamber Trigger Server (TS) is used to filter the information of the two $\phi$ superlayers of a chamber. The TRACO/TS block selects, at every cycle among the trigger candidates, at most two track segments with the smallest angular distances (i.e. higher $p_T$) with respect to the radial direction to the vertex.
Track segments found in each station are then transmitted to a regional trigger system called Drift Tube Track Finder (DTTF). The task of the Track Finder is to connect track segments delivered by the stations into a full track and assign a transverse momentum value to the finally resolved muon track. The system is comprised of sectors (72 in total), each of them covering 30° in the $\phi$ angle, and five wheels in the $z$-direction. Each Sector Processor is logically divided in three functional units - the Extrapolator Unit (EU), the Track Assembler (TA) and the Assignment Units (AU).
The Extrapolator Unit attempts to match track segments pairs of distinct stations. Using the spatial coordinate $\phi$ and the bending angle of the source segment, an extrapolated hit coordinate is calculated. The two best extrapolations per each source are forwarded to the Track Assembler. The Track Assembler attempts to find at most two tracks in a detector sector with the highest rank, :*i.e.* exhibiting the highest number of matching track segments and the highest extrapolation quality. Once the track segment data are available to the Assignment Unit, memory-based look–up tables are used to determine the transverse momentum and the $\phi$. The $\eta$ coordinates, are assigned separately using hits in the $\eta$-superlayers of the three innermost station and applying a pattern method.

*Appendix A.5.2. CSC Trigger* The CSC Local Trigger finds muon segments, also referred to as Local Charged Tracks (LCTs), in the 6-layer endcap muon CSC chambers. Muon segments are first found separately by anode and cathode electronics and then time correlated, providing precision measurement of the bend coordinate position and angle, approximate measurement of the non-bend angle coordinate, and identification of the correct muon bunch crossing with high probability.
The primary purpose of the CSC cathode trigger electronics is to measure the $\phi$

coordinate precisely to allow a good muon momentum measurement up to high momentum. The charge collected on an anode wire produces an opposite-sign signal on several strips, and precision track measurement is obtained by charge digitization and precise interpolation of the cathode strip charges. The six layers are then brought into coincidence in LCT pattern circuitry to establish position of the muon to an RMS accuracy of 0.15 strip widths. Strip widths range from 6-16 mm.

The primary purpose of the CSC anode trigger electronics is to determine the exact muon bunch crossing with high efficiency. Since the drift time can be longer than 50 ns, a multi-layer coincidence technique in the anode "Local Charged Track" (LCT) pattern circuitry is used to identify a muon pattern and find the bunch crossing.

The task of the Cathode Strip Chamber Track-Finder is to reconstruct tracks in the CSC endcap muon system and to measure the transverse momentum ($p_T$), pseudo-rapidity ($\eta$), and azimuthal angle ($\phi$) of each muon. The measurement of $p_T$ by the CSC trigger uses spatial information from up to three stations to achieve a precision similar to that of the DT Track-Finder despite the reduced magnetic bending in the endcap.

Cathode and anode segments are brought into coincidence and sent to the CSC Sector Processor electronics which links the segments from the endcap muon stations. Each Sector Processor unit finds muon tracks within 60°. A single extrapolation unit forms the core of the Sector Processor trigger logic. It takes the three dimensional spatial information from two track segments in different stations, and tests if those two segments are compatible with a muon originating from the nominal collision vertex with a curvature consistent with the magnetic bending in that region. Each CSC Sector Processor can find up to three muon candidates within 60°. A CSC muon sorter module selects the four best CSC muon candidates and sends them to the Global Muon Trigger.

*Appendix A.5.3. RPC Trigger* The RPC Pattern Trigger Logic (PACT) is based on the spatial and time coincidence of hits in four RPC muon stations. Because of energy loss fluctuations and multiple scattering there are many possible hit patterns in the RPC muon stations for a muon track of defined transverse momentum emitted in a certain direction. Therefore, the PACT should recognize many spatial patterns of hits for a given transverse momentum muon. In order to trigger on a particular hit pattern left by a muon in the RPCs, the PACT performs two functions: it requires time coincidence of hits in patterns ranging from 3 out of 4 muon stations to 4 out of 6 muon stations along a certain road and assigns a $p_T$ value. The coincidence gives the bunch crossing assignment for a candidate track. The candidate track is formed by a pattern of hits that matches with one of many possible pre-defined patterns for muons with defined transverse momenta. The pre-defined patterns of hits have to be mutually exclusive *i.e.* a pattern should have a unique transverse momentum assignment. The patterns are divided into classes with a transverse momentum value assigned to each of them. PACT is a threshold trigger; it gives a momentum code if an actual hit pattern is straighter than any of pre-defined patterns with a lower momentum code. The patterns will depend on the direction of a muon *i.e.* on its $\phi$ and $\eta$.

*Appendix A.5.4. Global Muon Trigger (GMT)* The GMT receives the best four barrel DT and the best four endcap CSC muons and combines them with 4+4 muons sent by the RPC PACT. It performs a matching based on the proximity of the candidates in ($\eta-\phi$) space. If two muons are matched, their parameters are combined to give optimum precision. The GMT also contains logic to cancel "ghost" tracks that arise when a single muon is found by more than one muon system and is not otherwise matched, such as at the boundary between the DT and CSC muon systems. The selected muon candidates are ranked based on their transverse momentum, quality and to some extent pseudorapidity and the best four muon candidates in the entire CMS detector are sent to the Global Trigger.

The Global Muon Trigger also receives information from the calorimeters. The Regional Calorimeter Trigger sends two bits based on energy measurements representing isolation and compatibility with a minimum ionizing particle in $\Delta\eta \times \Delta\phi$=0.35×0.35 trigger regions. The GMT extrapolates the muon tracks back to the calorimeter trigger towers and the vertex and appends the corresponding isolation and minimum ionizing bits (ISO and MIP) to the track data indicating isolation or confirmation of the muon by the calorimeter. The muon track data sent to the GT are the , the sign of the charge, the $\eta$ and $\phi$ as well as the ISO and MIP bits.

### *Appendix A.6. The Level-1 Global Trigger*

The Global Trigger accepts muon and calorimeter trigger information, synchronizes matching sub-system data arriving at different times and computes up to 128 trigger algorithms in parallel. The trigger decision is communicated to the Trigger and Control System (TCS) for distribution to the sub-systems to initiate the readout. The global trigger decision is made using logical combinations of the trigger data from the Global Calorimeter and Global Muon Triggers. The Level-1 Trigger system sorts ranked trigger objects, rather than histogramming objects over a fixed threshold. This allows all trigger criteria to be applied and varied at the Global Trigger level rather than earlier in the trigger processing. All trigger objects are accompanied by their coordinates in $\eta - \phi$ space. For muon candidates the charge is also delivered. This allows the Global Trigger to vary thresholds based on the location of the trigger objects. It also allows the Global Trigger to require trigger objects to be close or opposite from each other. In addition, the presence of the trigger object coordinate data in the trigger data (which is read out first by the DAQ after a Level-1 accept decision) permits a quick determination of the regions of interest where the more detailed HLT analysis should focus. Besides handling physics triggers, the Global Trigger provides for test and calibration runs, not necessarily in phase with the machine, and for prescaled triggers, as this is an essential requirement for computing trigger efficiencies.

The Global Level-1 Trigger is responsible for deciding whether to accept or reject an event and for generating the corresponding L1 Accept signal (L1A). The final L1A decision is the logical OR of all algorithms used at L1. This decision is transmitted through the Trigger Control System (TCS) to the Timing Trigger and Control system (TTC). The TCS automatically prescales or shuts off the L1A case the detector readout buffers are at risk of overflow.

# References

[1] The LHC Study Group: *The Large Hadron Collider Conceptual Design*, CERN-AC-95-05 (1995).

[2] See E.g. B. Heinemann, Plenary talk at ICHEP04, Beijing, to appear in proceedings.

[3] J.G. Branson et al., *High Transverse Momentum Physics at the Large Hadron Collider*, Eur. Phys. J Direct **C4** (2002) N1, hep-ph/0110021.

[4] F. Gianotti and M. Pepe-Altarelli, *Precision Physics at the LHC*, Nucl. Phys. Proc. Suppl. **89** (2000) 177, hep-ex/0006016.

[5] T. Sjöstrand, L. Lonnblad and S Mrenna, *PYTHIA6.2: Physics and Manual*, hep-ph/0108264; T. Sjöstrand, P. Edén, C. Friberg, L. Lönnblad, G. Miu, S. Mrenna and E. Norrbin, Computer Physics Commun. **135** (2001) 238.

[6] G. A. Schuler T. and Sjöstrand, Phys. Rev. **D49** (1994) 2257.

[7] CMS Coll., *The Trigger and Data Acquisition project, Volume I, The Level-1 Trigger, Technical Design Report*, CERN/LHCC 2000-038, CMS TDR 6.1, 15 December 2000.

[8] H. Baer et al., *ISAJET, a Monte Carlo for pp $\overline{p}p$ and $e^+e^-$ reactions*, hep-ph/0001086.

[9] N. Amapane et al., *Monte Carlo Simulation of Inclusive Single-Muon and Di-Muon Samples*, CMS Note 2002/041.

[10] H. Sakulin, *Improved di-muon trigger rate simulation methods for the CMS experiment at the LHC*, J. Phys. **G30** (2004) N45.

[11] CMS Simulation Package CMSIM, *The Users Guide and Reference Manual*, http://cmsdoc.cern.ch/cmsim/cmsim.html.

[12] *GEANT3, version 3.21/13: Detector Description and Simulation Tool*, CERN program library long writeup W5013.

[13] E. Meschi et al., *Electron Reconstruction in the CMS Electromagnetic Calorimeter*, CMS Note 2001/034.

[14] P. Aspell et al., *Results from the 1999 Beam Test of a Preshower Prototype*, CMS Note 2001/001.

[15] I. Puljak, These de Doctorat de l'Universite Paris VI, 21 Sept 2000.

[16] CMS Coll., *The Electromagnetic Calorimeter Project, Technical Design Report*, CERN/LHCC 97-33, CMS TDR 4, 15 December 1997.

[17] G. Daskalakis and K. Lassila-Perini, *Jet rejection using the pixel matching for the low and the high luminosity*, CMS Note 2002/039.

[18] P. Billoir, R. Früwirth and M. Regler, Nucl Instr. Meth. **A241** (1985) 115.

[19] V. Innocente, M. Maire, E. Nagy, *GEANE*, CERN program library long writeup W5013-E.

[20] G. Bruno et al., *Local reconstruction in the muon detectors*, CMS Note 2002/043.

[21] N. Amapane, M. Fierro, M. Konecki, *High-Level Trigger Algorithms for Muon Isolation*, CMS Note 2002/040.

[22] D. Green et al., *Energy Flow Objects and Usage of Tracks for Energy Measurement in CMS*, CMS Note 2002/036.

[23] R. Kinnunen and A. Nikitenko, *Study of H(SUSY) $\rightarrow \tau\tau \rightarrow l + \tau_{Jet} + \not{E}_T$ in CMS*, CMS Note 1997/106.

[24] D. Denegri and R. Kinnunen, *Study of H(SUSY) $\rightarrow \tau\tau \rightarrow h^+ + h^- + X$ in CMS*, CMS Note 1999/037.

[25] R. Kinnunen, *Study for Heavy Charged Higgs in pp $\rightarrow tH^+$ with $H^+\tau + \nu$ in CMS*, CMS Note 2000/045.

[26] J. E. Huth et al., Proceeding of Research Directions for the decade, Snowmass 1990.

[27] S. Eno et al., *A Study of a First and Second Level Tau Trigger*, CMS Note 2000/055.

[28] D. Kotlinski, A. Nikitenko and R. Kinnunen, *Study of a Level-3 Tau Trigger with the Pixel Detector*, CMS Note 2001/017.

[29] G. Bagliesi, S. Gennai and G. Sguazzoni, *A L2 Trigger for Tau Hadronic Decays with Tracker Isolation in the Low Luminosity Scenario*, CMS Note 2002/018.

[30] G. Segneri and F. Palla, *Lifetime-based b-tagging with CMS*, CMS Note 2002/046.

[31] The LEP Heavy Flavour Group, "Input Parameters for the LEP/SLD Heavy Flavour Analyses",

LEPHF/2001-01, http://www.cern.ch/LEPEWWG/heavy/lephf0101.ps.gz.

[32] D. Kotlinski and A. Starodumov, *High Level Tracker Triggers for CMS*, Presented at: Vertex 2001, Brunnen Switzerland, September 2000,CMS Conference Report 2002/003.

[33] D. Kotlinski, *The CMS Pixel Detector*, Nucl. Phys. Proc. Suppl. **120** (2003) 249.

[34] M. Winkler et al, *Estimation of Alignment Parameters, Using the Kalman Filter with Annealing*, CMS Note 2002/008.

[35] A. Ostaptchouk et al, *The Alignment System of the CMS Tracker*, CMS Note 2001/053.

[36] D. Futyan and C. Seez, *Intercalibration of ECAL Crystals in Phi Using Symmetry of Energy Deposition*, CMS Note 2002/031.

[37] CMS Coll., *The Hadronic Calorimeter Project, Technical Design Report*, CERN/LHCC 97-31, CMS TDR 2, 20 June 1997.

[38] CMS Coll., *The Muon Project, Technical Design Report*, CERN/LHCC 97-32, CMS TDR 3, 15 December 1997.

[39] O.J.P. Eboli and D. Zeppenfeld, Phys. Lett. **B495** (2000) 147.

[40] B. Di Girolamo et al., *Experimental Observation of an Invisible Higgs Boson at the LHC*, Workshop on Physics at TeV Colliders, Les Houches, 2001, the Higgs Working Group: summary report, hep-ph/0203056.

[41] H. Baer et al., Phys. Rev. **D58** (1998) 075008.

[42] H. Baer, F.E. Paige, S.D. Protopopescu and X. Tata, hep-ph/0001086.

[43] G.Corcella, I.G.Knowles, G.Marchesini, S.Moretti, K.Odagiri, P.Richardson, M.H.Seymour and B.R.Webber, JHEP **01** (2001) 010.

[44] N. Arkani-Hamed, A.G. Cohen, T. Gregoire, E. Katz, A.E. Nelson, J.G. Wacker, *The Minimal Moose for a Little Higgs*, JHEP **0208** (2002) 021; M. Schmaltz, *Introducing the Little Higgs*, hep-ph/0210415.

[45] D. Amidei and R. Brock, *Future Electroweak Physics at the Fermilab Tevatron: Report of the TeV_2000 Study Group*, FERMILAB-PUB-96/08.

[46] CDF coll., F. Abe et al., Phys. Rev. Lett. **74** (1995) 3538.

[47] L. Vacavant, I Hinchliffe, *Signals of Models with Large Extra Dimensions in ATLAS*, J. Phys. **G27** (2001) 1839.

[48] H. Cheng, K. Matchev and M. Schmalz, *Bosonic Supersymmetry? Getting fooled at the LHC*, Phys. Rev. **D66** (2002) 056006, hep-ph/0205314.

[49] C. Macesanu, C.D. McMullen, S. Nandi, *New Signal for Universal Extra Dimensions*, Phys.Lett. **B546** (2002) 253.

[50] G. Azuelos et al., *Exploring Little Higgs models with ATLAS at the LHC*, hep-ph/0402037.